\documentclass{aa}  

\usepackage{graphicx}
\usepackage{graphics}
\usepackage{natbib}
\usepackage{float}
\usepackage{caption}
\bibpunct{(}{)}{;}{a}{}{,} % to follow the A&A style

\usepackage{txfonts}
\begin{document}

\newcommand{\sigmahtwo}{$\Sigma_{\rm H_2}$}

\title{Radial distribution of dust, stars, gas, and star-formation rate in DustPedia\thanks{
DustPedia is a project funded by the EU under the heading 
`Exploitation of space science and exploration data'. 
It has the primary goal of exploiting existing data in the 
\textit{Herschel} Space Observatory and Planck Telescope databases. 
} 
face-on galaxies}
\author{V. Casasola$^{1}$, 
L.~P. Cassar\`a$^{2}$, 
S. Bianchi$^{1}$, 
S. Verstocken$^{3}$, 
E. Xilouris$^{2}$, 
L. Magrini$^{1}$, 
M.~W.~L.~Smith$^{4}$, 
I.~De~Looze$^{5}$, 
M.~Galametz$^{6}$, 
S.~C.~Madden$^{6}$, 
M.~Baes$^{3}$, 
C.~Clark$^{4}$,
J.~Davies$^{4}$, 
P.~De~Vis$^{8}$,
R.~Evans$^{4}$,
J.~Fritz$^{7}$,
F.~Galliano$^{6}$, 
A.~P.~Jones$^{8}$, 	
A.~V.~Mosenkov$^{3}$, 
S.~Viaene$^{3}$,
N.~Ysard$^{8}$\\ 
}
\institute{
$^{1}$ INAF-Osservatorio Astrofisico di Arcetri, Largo E. Fermi 5, I-50125, Florence, Italy\\
\email{casasola@arcetri.astro.it}  \\
$^{2}$ National Observatory of Athens, Institute for Astronomy, Astrophysics, Space Applications and Remote Sensing, 
Ioannou Metaxa and Vasileos Pavlou GR-15236, Athens, Greece\\
$^{3}$ Sterrenkundig Observatorium, Department of Physics and Astronomy, Universiteit Gent Krijgslaan 281 S9, B-9000 Gent, Belgium \\
$^{4}$ School of Physics and Astronomy, Cardiff University, The Parade, Cardiff CF24 3AA, UK \\
$^{5}$ Department of Physics and Astronomy, University College London, Gower Street, London WC1E 6BT, UK\\
$^{6}$ Laboratoire AIM, CEA/DSM - CNRS - Universit\'e Paris Diderot, IRFU/Service d'Astrophysique, CEA Saclay, 91191, Gif-sur- Yvette, France \\
$^{7}$ Instituto de Radioastronom\'\i a y Astrof\'\i sica, UNAM, Campus Morelia, A.P. 3-72, C.P. 58089, Mexico\\
$^{8}$ Institut d'Astrophysique Spatiale, CNRS, Univ. Paris-Sud, Universit\'e Paris-Saclay, B\^{a}t. 121, 91405, Orsay Cedex, France \\
}
\date{Received ; accepted}

\titlerunning{Scale-lengths in DustPedia}
\authorrunning{Viviana Casasola et al.}

\abstract
{}
{The purpose of this work is the characterization of the radial distribution of dust, stars, gas, and star-formation rate (SFR) in a sub-sample of 18 face-on spiral galaxies extracted from the DustPedia sample.
}
{This study is performed by exploiting the multi-wavelength, from ultraviolet (UV) to sub-millimeter bands, DustPedia database, in addition
to molecular ($^{12}$CO) and atomic (H{\sc i}) gas maps and metallicity abundance information available in the literature.
We fitted the surface brightness profiles of the tracers of dust and stars, the mass surface density profiles of dust, stars, molecular gas, and total gas, and the SFR 
surface density profiles with an exponential curve and derived their scale-lengths.
We also developed a method to solve for the CO-to-H$_2$ conversion factor ($\alpha_{\rm CO}$) per galaxy by using dust and gas mass profiles.
}
{Although each galaxy has its own peculiar behaviour, we identified a common trend of the exponential scale-lengths vs. wavelength. 
On average, the scale-lengths normalized to the $B$-band 25~mag/arcsec$^2$ radius decrease from UV to 70~$\mu$m, from 0.4 to 0.2, 
and then increase back up to $\sim$0.3 at 500 microns.
The main result is that, on average, the dust mass surface density scale-length is about 1.8 times the stellar one derived from IRAC data and 
the 3.6~$\mu$m surface brightness, and close to that in the UV.
We found a mild dependence of the scale-lengths on the Hubble stage T: 
the scale-lengths of the $\textit{Herschel}$ bands and the 3.6~$\mu$m scale-length 
tend to increase from earlier to later types,  
the scale-length at 70~$\mu$m  tends to be smaller than that at longer sub-mm wavelength 
with ratios between longer sub-mm wavelengths and 70~$\mu$m 
that decrease with increasing T. 
The scale-length ratio of SFR and stars
shows a weak increasing trend towards later types.
Our $\alpha_{\rm CO}$ determinations are in the range $(0.3 - 9)$~M$_\odot$~pc$^{-2}$~(K km s$^{-1}$)$^{-1}$, almost invariant by using a fixed dust-to-gas ratio mass (DGR) 
or a DGR depending on metallicity gradient. 
}
{} 
% 5 {} token are mandatory
 
%\abstract{
%aaa
%}


 \maketitle

%
%________________________________________________________________

\section{Introduction}
\label{sec:intro}
The mass of the interstellar medium (ISM) is composed of gas for $\approx$99$\%$ ($\approx$75$\%$ as hydrogen and $\approx$25$\%$ as helium), 
and primarily of dust for the remaining $\approx$1$\%$. 
The ISM is of vital importance for the formation and evolution of galaxies since it is in the environment from which stars are formed.
Stars indeed form inside dense molecular clouds, from the conversion of gas into stars \citep[e.g.,][]{krumholz11}, 
while this gas is also ejected through stellar winds and supernova explosions \citep[e.g.,][]{veilleux05}. 
The formation of hydrogen molecules (H$_2$), the raw material for 
star formation (SF), takes place on the surfaces of dust grains \citep[][]{gould63} that are formed from supernova/stellar ejecta and/or in the ISM 
\citep[e.g.,][]{draine03}.

Although the dust constitutes a small percentage of the ISM, understanding its spatial distribution is
of particular importance for two reasons.
First, the dust affects our view of galaxies at different wavelengths, by absorbing ultraviolet (UV) and
optical light and reemitting it in the infrared (IR).
Correcting for dust extinction is usually the main source of uncertainty when deriving properties such
as the star-formation rate (SFR), age, or metallicity \citep[e.g.,][]{calzetti94,buat96,calzetti01,perez-gonzalez03}.
Secondly, the dust constitutes an important element in the chemical evolution of the ISM.
Metals resulting from stellar nucleosynthesis are returned to the ISM, either as gas and solid grains 
condensed during the later stages of stellar evolution; they can later be destroyed and incorporated into new generations of stars.
Elements are injected into the ISM at different rates \citep[see e.g.,][]{dwek09}.
For instance, atomic bounds of carbon are most likely responsible for the emission features,
whose carriers are ofter identified with Polycyclic Aromatic Hydrocarbons (PAHs); 
they might have been originally
produced in AGB stars, whose typical lifetimes are a few Gyr \citep[e.g.,][]{marigo08,cassara13,villaume15,simonian16}. 
Other elements may be synthesized in more massive stars, which die as supernovae in shorter timescales.
Thus, the abundance of the MIR emission feature carriers 
and the dust-to-gas mass ratio (DGR) are expected to vary with the age of the stellar populations and correlate 
with the metal (e.g., oxygen and nitrogen) abundance of the gas \citep[see e.g.,][]{galliano08}.
However, dust formation results from a long chain of poorly-understood processes, from the formation and injection into the ISM of 
dust seeds formed in the atmospheres of AGB stars or in SNe ejecta, to grain growth and destruction in the ISM \citep[e.g.,][]{valiante09,asano13,mattsson14,remy-ruyer14,bocchio16}. 
Based on \citet{draine09}, only $\sim$10$\%$ of interstellar dust is directly formed in stellar sources, with the remaining
$\sim$90$\%$ being later condensed in the ISM. 
However, \citet{jones11} found that the destruction efficiencies might have been severely overestimated. 
They concluded that the current estimates of global dust lifetimes
could be uncertain by factors large enough to call into
question their usefulness \citep[see also][]{ferrara16}. 

The connection between dust, gas, and stars (and SF processes) has been
investigated mainly through spatially unresolved studies of the ISM components and stars/SF \citep[e.g.,][]{buat89,buat92,kennicutt89,kennicutt98a,deharveng94,cortese12+,clark15,remy-ruyer15,devis17}.
Although these works are useful for characterizing global disk properties, understanding intimately 
the structure of galaxies requires resolved measurements.
Thanks to IR and sub-millimeter (sub-mm) data from space telescopes (e.g., $\textit{Spitzer}$ and $\textit{Herschel}$) 
together with ground-based facilities (e.g., IRAM-30m, IRAM-PdBI now NOEMA, ALMA), we also have resolved measurements of
dust and gas for large samples of nearby galaxies even beyond the Local Group
\citep[e.g.,][]{wong02,bigiel08,schruba11,munoz09a,casasola15}.     

Although the ISM is distributed in an irregular manner between the stars,
following a variety of structures (e.g, clouds, shells, holes, and filaments), the dust and molecular component of the ISM, 
and stars are all generally distributed in a disk, often with an exponential decline of surface density with radius 
\citep[e.g.,][]{alton98,bianchi07,bigiel08,munoz09a,hunt15}.
The relative distribution of the dust components emitting at different infrared (IR) and sub-mm wavelengths gives information
on the dominant heating mechanism of dust.
In general, dust emission within galaxies is powered by radiation coming from both sites of recent SF
and from more evolved stellar populations, as well as from the presence of Active Galactic Nuclei \citep[AGN, e.g.][]{draineli07}.  
There is a long-standing debate about the exact fraction of dust heating contributed
by each stellar population \citep[e.g.,][]{law11,boquien11,bendo12,bendo15,hunt15}, 
which depends on several factors such as the  intrinsic  spectral  energy  distribution (SED)  
of  the stellar populations, the dust mass and optical properties, as well as the relative dust-star geometry.
   
Radial profiles  of dust and stars are strong diagnostics to distinguish between the relative distributions
and to understand how dust is heated.
Based on this, \citeauthor{alton98} (\citeyear{alton98}; see also \citealt{davies99}), by comparing the 
ISO 200~$\mu$m observations (Full Width Half Maximum, FWHM $\simeq$ 2$^{\prime}$) 
with improved resolution IRAS 60 and 100~$\mu$m data (FWHM$\simeq$1-2$^{\prime}$) for 8 nearby spiral galaxies, 
found that cold dust ($\sim$18--21~K) detected by ISO becomes more dominant at larger radii with respect to warm 
dust ($\sim$30~K) revealed by IRAS. The comparison of the 200~$\mu$m images with $B-$band data also showed that 
the cold dust is radially more extensive than stars (the latter traced, in some measure, by the blue light).  
With higher resolution {\em Spitzer} observations of 57 galaxies in the SINGS sample, \citet{munoz09a} instead 
found that the distribution of the dust mass (derived from SED fits up to the longest available wavelength of 
160 $\mu$m; FWHM = 38\arcsec) has a similar scale-length to that measured for stellar emission at 3.6~$\mu$m.
However, the result is not incompatible with that of \citet{alton98}, since near-IR (NIR) surface brightness profiles are steeper 
than in the optical.

The advent of {\em Herschel} allowed to reach a competitive resolution beyond the peak of thermal dust emission.
In a preliminary, qualitative analysis of
the surface brightness profiles of the two grand-design spirals NGC~4254 (M~99) and NGC~4321 (M~100),
\citet{pohlen10} showed that cold dust detected with \textit{Herschel}-SPIRE data (250~$\mu$m--FWHM~=~18$^{\prime\prime}$, 350~$\mu$m--FWHM~=~24$^{\prime\prime}$, 
500~$\mu$m--FWHM~=~36$^{\prime\prime}$) extends to at least the optical radius of the galaxy and shows 
features at similar locations as the stellar distribution.
Scalelengths at 250~$\mu$m were derived for the 61 galaxies in the sample of the project {\em Key Insights 
into Nearby Galaxies: Far-Infrared Survey with Herschel} \citep[KINGFISH,][] {kennicutt11} by \citet{hunt15}.
They confirmed the results of \citet{munoz09a}, i.e. of the similarity between NIR and FIR scale-lengths. They did 
not provide, however, scale-lengths at the larger wavelength accessible to SPIRE in the sub-mm, nor for the 
dust mass distribution that could be derived by resolved SED fitting. The scale-length of an averaged dust-mass 
profile was recently determined by \citet{smith16}, from a stacking analysis of 45 low-inclination galaxies part of 
the Herschel Reference Survey \citep[HRS, ][]{boselli10}. Since the main focus of the work was the detection of
dust in the outskirt of spiral disks, scale-lengths for individual objects were not analysed and shown.

In this paper, we study the spatial variations of dust, stars, gas, and SFR of a sample of 18 nearby, spiral, face-on galaxies 
extracted from the DustPedia sample\footnote{http://www.dustpedia.com/} by analyzing their surface brightness and 
mass surface density profiles \citep[see][for a detailed description of the DustPedia project and sample]{davies17}.
To perform this we exploit the multi-wavelength (from UV to sub-mm wavelengths) DustPedia database\footnote{http://dustpedia.astro.noa.gr/}. 
This paper covers -with continuity- a wider wavelength range, that is homogeneously treated, with respect to previous works dedicated 
to scale-lengths of stars, dust and other ISM components in galaxies.
     
The paper is organized as follows.
In Sect.~\ref{sec:sample} we outline the sample selection and in Sect.~\ref{sec:data} we present the different data used in this work.
In Sect.~\ref{sec:treatment} we describe the adopted procedure to homogenize the dataset, and in Sect.~\ref{sec:masses} 
the methods adopted to derive the mass of dust, stars, and gas and the SFR.
Sec.~\ref{sec:prof} presents the surface brightness profiles and mass and SFR surface density profiles, their exponential fits and
corresponding scale-lengths, while Sec.~\ref{sec:xco} describes the method we developed to solve for single CO-to-H$_2$ conversion factors by using     
dust and gas mass profiles.      
Finally, in Sect.~\ref{sec:conclusions} we summarize our main results.

\begin{sidewaystable*}
\caption{\label{sample} Main properties of the galaxy sample.}
\centering
\begin{tabular}{lccccccccccccccccccc}
\hline
\hline
Galaxy & $\alpha_{\rm J2000}$ & $\delta_{\rm J2000}$ & Ref. & T & RC3 type & D$_{\rm 25}$ &  D$_\mathrm{sub-mm}$ 
& (d/D)$_\mathrm{sub-mm}$ & Dist. & $i$ & PA  & $b/a$ & Ref. & Nuclear  \\
&	       &                                    &                                    $\alpha_{\rm J2000}$, $\delta_{\rm J2000}$  & & & & & & 
& & & &  $i$, PA, $b/a$ & Activity$^{a}$ \\
& [$^{\rm h}$ $^{\rm m}$ $^{\rm s}$]  & [$^{\circ}$ $^{\prime}$ $^{\prime\prime}$]  & & & [$^{\prime}$] & [$^{\prime}$] & [$^{\prime}$] 
& & [Mpc] & [deg]  & [deg]\\
\hline
NGC~5457 (M~101) 	& 14 03 12.6  & +54 20 57  & 1 & 6 & SAB(rs)cd   	& 24.0 	& 18.2 	& 1.00 & 7.0  	& 18.0     	&  39.0    	& 0.95 	&  1 	& H  \\
NGC~3031 (M~81)  & 09 55 33.1  & +69 03 55   & 1 & 2 & SA(s)ab       	& 21.4 	& 21.0 	& 0.54 & 3.7 	&  59.0 	& 330.2  	& 0.52	& 5   & L/S1.8 \\
NGC~2403               & 07 36 51.1  & +65 36 03   & 1 & 6 & SAB(s)cd    	& 20.0 	& 16.6 	& 0.67 & 3.5  	& 62.9  	& 123.7  	& 0.46	& 5	& L  \\
IC~342                    & 03 46 48.5  &  +68 05 47   & 3 & 6 & SAB(rs)cd   	& 20.0 	& 21.8 	& 0.80 & 3.1   	& 31.0   	&  37.0    	& 0.86	&  6 	& H   \\
NGC~300                & 00 54 53.4  & -37 41 03    & 2 & 7 & SA(s)d         	& 19.5 	& 19.4 	& 0.74 & 2.0  	& 43.0    	& 290.0    	& 0.73	& 2   	& -- \\
NGC~5194 (M~51) & 13 29 52.7  & +47 11 43   & 1 & 4 & SA(s)bc pec 	& 13.8 	& 12.6 	& 0.74 & 7.9  	& 42.0   	& 172.0    & 0.74	& 1	& S2   \\
NGC~5236 (M~83) & 13 37 00.9  & -29 51 57   & 1 & 5 & SAB(s)c    	 	& 13.5 	& 15.2 	& 0.89 & 6.5    	&   24.0   	& 225.0    & 0.91	&  1	& H   \\
NGC~1365            & 03 33 36.4  & -36 08 25    & 4  & 3 & SB(s)b 		& 12.0 	& 11.6 	& 0.57 & 17.7  	& 40.0   	& 220.0    & 0.77	& 4	& S1.8    \\
NGC~5055 (M~63) & 13 15 49.2  & +42 01 45   & 1 & 4 & SA(rs)bc      	& 11.8 	& 13.1 	& 0.49 & 8.2   	&  59.0  	& 101.8  	& 0.52	& 5	& H/L   \\
NGC~6946            & 20 34 52.2  & +60 09 14    & 1 & 6 & SAB(rs)cd 	& 11.5 	& 11.0 	& 0.84 & 5.6  	& 32.6 	& 242.7  	& 0.84	& 5	& S2/H    \\
NGC~925               & 02 27 16.5  & +33 34 44   & 1 & 7 & SAB(s)d 	 	& 10.7 	& 10.1 	& 0.50 & 8.6    	& 66.0 	& 286.6   	& 0.41	& 5 	& H  \\
NGC~1097            & 02 46 19.0  & -30 16 30    & 3  & 3 & SB(s)b 		& 10.5 	& 10.0 	& 0.60 & 19.6  	& 46.0   	& 135.0    & 0.69	& 8	& L     \\
NGC~7793           & 23 57 49.7  & -32 35 28     & 1 & 8 & SA(s)d 		& 10.5 	& 10.4 	& 0.58 & 3.8   	& 49.6 	& 290.1  	& 0.65	& 5	& H     \\
NGC~628 (M~74)  & 01 36 41.8  & +15 47 00   & 1 & 5 & SA(s)c 		& 10.0 	& 9.6 	& 1.00 & 9.0  	& 7.0   	& 20.0     & 0.99 	& 1	& H   \\
NGC~3621             & 11 18 16.5  & -32 48 51     & 1 & 7 & SA(s)d         	& 9.8 	& 11.2 	& 0.46 & 6.9   	&  64.7 	& 345.4  	& 0.43	&  5 	& H   \\
NGC~4725            & 12 50 26.6  & +25 30 03   & 3  & 2 & SAB(r)ab pec 	& 9.8 	& 10.2 	& 0.55 & 13.6 	& 54.0   	& 36.0      & 0.59	& 7	& S2    \\
NGC~3521             & 11 05 48.6  & -00 02 09    & 1 & 4 & SAB(rs)bc    	& 8.3 	& 9.6 	& 0.52 & 12.0  	& 72.7 	& 339.8  	& 0.30	& 5	& H/L   \\
NGC~4736 (M~94) & 12 50 53.0  & +41 07 13   & 1 & 2 & (R)SA(r)ab   	& 7.8 	& 12.8 	& 0.69 & 5.2 	& 41.4 	& 296.1  	& 0.75	& 5	& S2/L   \\
\hline
\hline
\end{tabular}
\tablefoot{
References:  
Hubble stage T is from \citet{tully88};  
galaxy morphology (RC3 type) and redshift-independent distances (Dist.) are from NED database; 
D$_{\rm 25}$ is from HyperLeda database;
sub-mm diameter (D$_\mathrm{sub-mm}$) and sub-mm minor-to-major axis ratio ((d/D)$_\mathrm{sub-mm}$)
have been measured in the present work.
Coordinates ($\alpha_{\rm J2000}$,  $\delta_{\rm J2000}$),
inclination angle ($i$), position angle (PA), and minor-to-major axis ratio ($b/a$) are from:
1) \citet{walter08} (THINGS), 
2) \citet{westmeier11},
3) NED,
4) \citet{jorsater95},
5) \citet{deblok08},
6) \citet{crosthwaite00},
7) \citet{leroy12},
8)  \citet{ondrechen89}. 
$^{a}$ Classification of nuclear activity is from NED:  H = H~{\sc ii} nucleus, S = Seyfert nucleus, and L = LINER.
The number attached to the class letter designates the Seyfert type, classification based on the appearance of the optical spectrum 
\citep[][]{osterbrock81}. 
Two classifications are given for some ambiguous cases.
}
\end{sidewaystable*}

\section{The sample selection}
\label{sec:sample}
For the purpose of this work, we selected large spiral galaxies from the DustPedia sample with a small (or moderate) disk inclination, imaged
over their whole extent with both PACS and SPIRE
in $\textit{Herschel}$. 
We first searched for objects with D$_\mathrm{25}> 6'$~\footnote{D$_\mathrm{25}$ is defined as the length 
the projected major axis of a galaxy at the isophotal level 25 mag/arcsec$^{2}$ in the $B-$band (this is the diameter of the galaxy if it is a disk).}
in the $B-$band and 
with Hubble stage T ranging from 1 to 8  (including galaxies with classifications from Sa to Sdm, 
as listed in the Nearby Galaxy Atlas; \citealt{tully88}). 
We then refined the search after measuring the extent of the galaxy in the 250 $\mu$m images at 
the isophote of 12~mJy/beam  \citep[i.e.\ at twice the confusion noise, a sensitivity achieved in 
all SPIRE images;][]{nguyen10}. 
A total of 18 galaxies were found with PACS and SPIRE data and sub-mm minor-to-major axis ratio 
(d/D)$_\mathrm{sub-mm} \ge 0.4$ and D$_\mathrm{sub-mm}\ge 9\arcmin$, corresponding to
$\approx$15 resolution elements in the SPIRE 500~$\mu$m maps.

The final sample includes galaxies with all sorts of peculiarities, such the presence of bars and/or signatures of interaction with companions.
All galaxies are indeed present in the sample of interacting galaxies of \citet{casasola04} that includes 
more than 1000 galaxies  appearing to be clearly interacting with nearby objects and presenting
tidal tails or bridges, merging systems and galaxies with disturbed structures.  
We also checked the presence of AGN in our sample.
Ten of the eighteen sample galaxies are classified as low-luminosity AGN (L$_{\rm X} < 10^{42}$~erg~s$^{-1}$), including Seyferts and
low-ionization nuclear emission-line region (LINER) galaxies, 
while the reaming sources have H~{\sc ii} nuclei (except for NGC~300 for which there is no nuclear classification, to our knowledge). 
Table~\ref{sample} collects the main properties of the sample galaxies, that are listed in order of decreasing D$_{25}$.

\section{The data}
\label{sec:data}
In this section we present the data collected for the studied galaxy sample, whose main properties are listed in Tables~\ref{table:data}
and \ref{table:gas}.
Most of the used data  have been retrieved from the DustPedia database except for molecular ($^{12}$CO) and atomic (H{\sc i}) gas maps, 
a fraction of the optical images, and information on metallicity.

\subsection{$\textit{Herschel}$ data and their reduction}
\label{herscheldata}
We used both PACS (70, 100, and 160~$\mu$m) and SPIRE (250, 350, and 500~$\mu$m) $\textit{Herschel}$ maps.
PACS images have FWHM of 6, 8, and 12\arcsec\ at 70, 100, and 160~$\mu$m, respectively, and
SPIRE images of 18, 24, and 36\arcsec\ at 250, 350, and 500~$\mu$m, respectively.
The reduction of these data for the entire DustPedia sample, including galaxies studied in this paper,
has been performed by the DustPedia team.  
\citet{davies17} describe in detail the data reduction adopted for both PACS and SPIRE observations.  
Final reduced $\textit{Herschel}$ images as well as the other ones present in the DustPedia database are in units of Jy~pix$^{-1}$
\citep[for details on the database, see][]{clark17}.
The $\textit{Herschel}$ images have been used to derive the FIR/sub-mm light distribution and the dust mass surface density profile
(Sect.~\ref{sec:dustmass}).

\begin{table*}
\caption{\label{table:data} 
Main properties of the \textit{Herschel} images and the ancillary ones used for this work.
These images are from the DustPedia database, except for the SINGS ones.}
\centering
\begin{tabular}{lcccc}
\hline\hline
Instrument/Survey & Wavelength  & FWHM & Ref. FWHM &Pixel size\\
&  & [\arcsec] & & [\arcsec]	\\
\hline
\textit{GALEX}-FUV\tablefootmark{a} & 1516~\AA  & 4.0--4.5 	& 1 	& 3.2\\
\textit{GALEX}-NUV\tablefootmark{a} & 2267~\AA  & 5.0--5.5 	& 1	& 3.2 \\
SDSS-$g$\tablefootmark{b} & 4686~\AA  & 1.4 			& 2 	& 0.45\\
SDSS-$i$\tablefootmark{b} & 7480~\AA   & 1.2 				& 2	& 0.45\\
SINGS-$B$\tablefootmark{c} & 4360~\AA  & $\sim$1--2 		& 3	& 0.433\\
SINGS-$R$\tablefootmark{c} & 6440~\AA   & $\sim$1--2 		& 3 	& 0.433\\
\textit{Spitzer}-IRAC-1\tablefootmark{a} & 3.6~$\mu$m  & 1.7 	& 4	& 0.75\\
\textit{Spitzer}-IRAC-2\tablefootmark{a} & 4.5~$\mu$m  & 1.7  	& 4	& 0.75\\
WISE-W3\tablefootmark{a} & 12~$\mu$m  & 6.8 			& 5	& 1.375\\
WISE-W4\tablefootmark{a} & 22~$\mu$m  & 11.8 			& 5	& 1.375\\
\textit{Herschel}-PACS\tablefootmark{d} & 70~$\mu$m  & 6 	& 6	& 2\\
\textit{Herschel}-PACS\tablefootmark{e} & 100~$\mu$m  & 8 	& 6	& 3\\
\textit{Herschel}-PACS\tablefootmark{a} & 160~$\mu$m  & 12	& 6	& 4\\
\textit{Herschel}-SPIRE\tablefootmark{a} & 250~$\mu$m  & 18 	& 7	& 6\\
\textit{Herschel}-SPIRE\tablefootmark{a} & 350~$\mu$m  &  24 & 7	& 8\\
\textit{Herschel}-SPIRE\tablefootmark{a} & 500~$\mu$m  &  36 & 7	& 12\\
\hline
\hline
\end{tabular}
\tablefoot{
\tablefoottext{a}{Available data for the whole sample.}
\tablefoottext{b}{Available data for 9 sample galaxies: NGC~5457~(M~101), NGC~2403, NGC~3031~(M~81), NGC~5194~(M~51), 
NGC~5055~(M~63), NGC~4736~(M~94), NGC~3521, NGC~4725, and NGC~628~(M~74).}
\tablefoottext{c}{Data for NGC~3621 and NGC~1097. For these two galaxies SDSS-$g$ and SDSS-$i$ data are not available.}
\tablefoottext{d}{Available data for the whole sample except  for NGC~300.}
\tablefoottext{e}{Available data for the whole sample except  for NGC~2403, NGC~3031~(M~81), NGC~5194~(M~51), and NGC~5236~(M~83).
The numbers in column (4) refer to the following references:
1) \citet{morrissey07}; 2)  \citet{bramich12}; 3) http://irsa.ipac.caltech.edu/data/SPITZER/SINGS/; 4) see ``IRAC Instrument Handbook'', http://irsa.ipac.caltech.edu/data/SPITZER/docs/irac/iracinstrumenthandbook/; 5) see ``Explanatory Supplement to the WISE
All-Sky Data Release Products '', http://wise2.ipac.caltech.edu/docs/release/allsky/expsup/index.html; 6) see ``PACS Observer's Manual'', 
http://herschel.esac.esa.int/Docs/PACS/html/pacs$_{-}$om.html; 7) see ``The Spectral and Photometric Imaging Receiver (SPIRE) Handbook'',  
http://herschel.esac.esa.int/Docs/SPIRE/html/spire$_{-}$om.html.}
}
\end{table*}

\subsection{GALEX data}
\label{sec:GALEX}
We used ultraviolet (UV) images of our sample galaxies observed by the \textit{Galaxy Evolution Explorer} (\textit{GALEX}) satellite in 
its far-ultraviolet (FUV, $\lambda_{\rm eff} = 1516$~\AA) and near-ultraviolet (NUV, $\lambda_{\rm eff} = 2267$~\AA) bands.
The \textit{GALEX}  images have FWHM in the range 4\farcs0--4\farcs5 and 5\farcs0--5\farcs5 for the FUV and NUV bands, 
respectively.

Massive, young stars emit most of their energy in this part of the spectrum and at least in star-forming galaxies they outshine the emission from any other stage
of the evolution of a composite stellar population \citep[e.g.,][]{bruzual03}.
Therefore, the flux emitted in the UV in spiral (and irregular) galaxies is an excellent measure of the current SFR \citep[e.g., ][]{kennicutt98b}.
Additionally, the light emitted in the UV can be very efficiently absorbed by dust and then re-emitted at far-IR (FIR) wavelengths.
Therefore, an analysis of the energy budget comparing the IR and UV emission is a powerful tool to
determine the dust attenuation of light at all wavelengths \citep[e.g.,][]{buat05,cortese06,mao12}. 
However, an analysis on the energy budget is beyond the scope of this paper. 
The \textit{GALEX} images have been used to derive the UV stellar light distribution and the SFR surface density profiles
together with IR emission (Sect.~\ref{sec:ssfr}).

\subsection{SDSS and SINGS data}
\label{sec:sdss}
We used two optical datasets. 
The first one consists of images originally from the Sloan Digital Sky Survey (SDSS), Data Release 12 \citep[DR12\footnote{DR12 
is the final data release of the SDSS-III, containing all SDSS observations through July 2014
and encompassing more than one-third of the entire celestial sphere, https://www.sdss3.org/},][]{york00}.
We used images in $g$ (4686~\AA) and $i$ (7480~\AA) wavebands for 9 sample galaxies.
The FWHM is of $\sim$1\farcs4 and $\sim$1\farcs2 for $g$ and $i$ waveband, respectively.
The second optical dataset we used is not contained in the DustPedia database. 
These images come from ``The \textit{Spitzer} Infrared Nearby Galaxies Survey - Physics of the Star-Forming ISM and Galaxy Evolution'' 
(SINGS\footnote{http://irsa.ipac.caltech.edu/data/SPITZER/SINGS/, see also \citet{dale17} for the new recalibration of the whole dataset.}) collaboration, that obtained optical BVRI data for the galaxies in the SINGS sample.
We used $B$ (4360~\AA) and $R$ (7480~\AA) wavebands for other two sample galaxies.
The FWHM is of $\sim$1--2\arcsec\ in both wavebands.

The optical light revealed in $g$ and $i$ SDSS and $B$ and $R$ SINGS images traces the young blue stellar content of a galaxy
\citep[e.g.,][]{bruzual93,maraston05,piovan06,cassara13}.
These images have been used both to derive the optical stellar light and stellar mass surface density profiles of the sample galaxies (see Sect.~\ref{sec:starmass}).

\subsection{$\textit{Spitzer}$ data}
\label{sec:irac}
We used NIR  images, originally obtained  with the IRAC camera  on $\textit{Spitzer}$ at  3.6 and 4.5~$\mu$m. 
These images have a median FWHM of $\sim$1\farcs7  for both wavebands. 

The IRAC 3.6 and 4.5~$\mu$m bands are sensitive mainly to the dust penetrated old stellar component.
Since these NIR channels are nearly minimally sensitive to dust absorption and emission
\citep[e.g.,][]{fazio04,willner04}, the corresponding images are used as tracers of the stellar mass distribution 
\citep[e.g.,][]{kennicutt03,helou04}.
Younger (hotter) stars  are  not  expected  to  contribute significantly to the observed stellar emission. 
However, due to their strong UV fluxes, these younger stars can heat their surrounding dust,  which,  in  turn,  re-radiates  at  longer
wavelengths and can also account for a significant fraction of the light at 3.6~$\mu$m \citep[e.g.,][]{meidt12}.
The IRAC 3.6 and 4.5~$\mu$m images have been used both to derive the NIR stellar light and stellar mass 
surface density profiles of the sample galaxies (see Sect.~\ref{sec:starmass}).

\subsection{WISE data}
\label{sec:wise}
We used mid-IR (MIR) images at 12~$\mu$m and 22~$\mu$m bands originally taken with the Wide-field  Infrared  Survey  Explorer \citep[WISE,][]{wright10}.
The median FWHM is 6\farcs8 and 11\farcs8 for the 12~$\mu$m and 22~$\mu$m images, respectively.

The W3 band covers the complex of emission bands at 8--13~$\mu$m imputed to carbonaceous material as well as
the broad silicate absorption feature at 10~$\mu$m. The W4 band, on the other hand, is free of these emission features: it therefore 
serves as a good measure of the pure dust continuum, sampling warm dust in the high temperature tail of the typical distribution of grains and/or 
non-thermal emission from stochastically heated grains. 
The W3 and W4 WISE images have been used to derive the corresponding MIR emission profiles and the 22~$\mu$m--W4 images have been combined with 
the GALEX--FUV ones to correct the FUV luminosity for dust attenuation, and subsequently, derive maps tracing the current SFR surface density profiles
(see Sect.~\ref{sec:ssfr}).

\begin{table*}
\tiny
\caption{\label{table:gas} 
Collected H{\sc i} 21~cm and $^{12}$CO (1--0 or 2--1) emission line data 
used to derive the mass surface density of the atomic and molecular gas, respectively.}
\centering
\begin{tabular}{lcccccc}
\hline\hline
Galaxy          & Instrument H{\sc i} (FWHM) & Ref. H{\sc i} & Pixel size H{\sc i} & Instrument $^{12}$CO  (FWHM) & Ref. $^{12}$CO ($^{12}$CO line) & Pixel size $^{12}$CO   \\
\hline
NGC~5457 (M~101) 		& 	VLA (6\arcsec)		& 	1 	& 11\arcsec	& IRAM 30~m (11\arcsec)			&	4 (2--1) 	& 2\arcsec   	\\
NGC~3031 (M~81) 		& 	VLA (6\arcsec) 		&	1 	& 11\arcsec	& BIMA (6\arcsec)				&	5 (1--0)	& 1\arcsec \\
NGC~2403        		& 	VLA (6\arcsec) 		&	1	& 11\arcsec	& IRAM 30~m (11\arcsec)			&	4 (2--1)	& 2\arcsec	\\
IC~342          			&  	VLA (38\arcsec)	& 	2	& 6\arcsec	& NRO 45m (15\arcsec)			& 	6 (1--0)	& 1\arcsec \\
NGC~300         			\\
NGC~5194 (M~51) 		&	VLA (6\arcsec) 		& 	1	& 11\arcsec	& IRAM 30~m (11\arcsec)			& 	4 (2--1)	& 2\arcsec	\\
NGC~5236 (M~83) 		&	VLA (6\arcsec) 		&	1	& 11\arcsec	&   \\
NGC~1365        		\\
NGC~5055 (M~63) 		&	VLA (6\arcsec) 		&	1	& 11\arcsec	& IRAM 30~m (11\arcsec)	 		&	4 (2--1)	& 2\arcsec \\
NGC~6946        		&	VLA (6\arcsec) 		&	1	& 11\arcsec	& IRAM 30~m (11\arcsec)	 		&	4 (2--1)	& 2\arcsec \\
NGC~925         			&	VLA (6\arcsec) 		&	1	& 11\arcsec	& IRAM 30~m (11\arcsec)	 		&	4 (2--1)	& 2\arcsec \\
NGC~1097        		&					&		&			& ATFN Mopra 22~m (30\arcsec) 	&	7 (1--0)	& 1\arcsec \\
NGC~7793        		&	VLA (6\arcsec) 		& 	1  	& 11\arcsec \\
NGC~628 (M~74)  		&	VLA (6\arcsec) 		&	1	& 11\arcsec	& IRAM 30~m (11\arcsec)			&	4 (2--1)	 & 2\arcsec \\
NGC~3621        		&	VLA (6\arcsec)		& 	1  	& 11\arcsec	& 		\\
NGC~4725        		&  WSRT (13\farcs22)	&	3	& 13\farcs22	& IRAM 30~m (11\arcsec)	 		&	4 (2--1)	& 2\arcsec \\
NGC~3521        		&	VLA (6\arcsec) 		&	1	& 11\arcsec	& IRAM 30~m (11\arcsec)	 		&	4 (2--1)	& 2\arcsec \\
NGC~4736 (M~94) 		&	VLA (6\arcsec) 		&	1	& 11\arcsec	& IRAM 30~m (11\arcsec)	 		&	4 (2--1)	& 2\arcsec \\
\hline
\hline
\end{tabular}
\tablefoot{
References for the H{\sc i} and $^{12}$CO FWHM values are:
1) THINGS survey \citep{walter08};
2) NED \citep{crosthwaite01}; 
3) WHISP survey from NED \citep{swaters02}; 
4) HERACLES survey \citep{leroy12};
5) BIMA SONG survey \citep{helfer03};
6) Nobeyama CO Atlas of Nearby Spiral Galaxies \citep{kuno07};
7) Private communication (PI's proposal: M. Smith).
}
\end{table*}

\subsection{H{\sc i} and $^{12}$CO images}
\label{sec:gasdata}
We derived gas masses profiles for the galaxies in our sample by using images of the H{\sc i} 21~cm emission line and 
of different $^{12}$CO emission lines for tracing the atomic and molecular gas component, 
respectively.
The data were available through various surveys, catalogues, and individual projects.
For 15 galaxies we retrieved H{\sc i} data, for 13 galaxies $^{12}$CO data, and
for 12 of these galaxies we have data for both H{\sc i} and $^{12}$CO emission lines.
In Table~\ref{table:gas} we list the collected H{\sc i} and $^{12}$CO images we used.
These images are already reduced and details on the data reduction procedures 
are contained in the case-by-case references quoted in Table~\ref{table:gas}.
In the following, we give main information on these images.   

Concerning the atomic gas component, for 13 galaxies H{\sc i} maps are from
``The H{\sc i} Nearby Galaxy Survey'' 
\citep[THINGS,][]{walter08}\footnote{http://www.mpia.de/THINGS/Overview.html}.    
THINGS is a program undertaken at the NRAO Very Large Array (VLA) to perform 
21~cm H{\sc i} observations of 34 nearby galaxies at the angular resolution of 6\arcsec.
From THINGS data repository we extracted reduced integrated H{\sc i} intensity maps (``moment 0'') in unit
of Jy~beam$^{-1}$~m~s$^{-1}$.
For IC~342 we used the 21~cm H{\sc i} map available from NED catalogue in unit
of Jy~beam$^{-1}$~m~s$^{-1}$ 
and obtained with the NRAO VLA at the angular resolution of 
38\arcsec.
For NGC~4725 the 21~cm H{\sc i} map is from ``Westerbork observations of neutral Hydrogen in Irregular and SPiral galaxies'' 
\citep[WHISP,][]{swaters02}\footnote{http://www.astro.rug.nl/~whisp/},      
a survey undertaken at the Westerbork Synthesis Radio Telescope (WSRT)
an resolution of 13\farcs22.  
We downloaded the reduced integrated H{\sc i} intensity map from NED in Westerbork Unit 
(1~W.U. = $2.2 \times 10^{20}$~H~atoms~cm$^{-2}$).

Concerning the molecular gas component, for 10 sample galaxies we used 
$^{12}$CO(2--1) line intensity (230~GHz/1.3~mm, rest frame) 
maps from ``The HERA CO-Line Extragalactic Survey'' 
\citep[HERACLES,][]{leroy09}\footnote{http://www.mpia.de/HERACLES/Data.html}.
HERACLES is a Large Program that used the IRAM~30~m  telescope
to map the $^{12}$CO(2--1) emission from 48 nearby galaxies at the
angular resolution of 11$^{\prime\prime}$.
From HERACLES data repository we extracted the reduced integrated intensity maps 
and associated uncertainty, in units of K~km~s$^{-1}$.
For NGC~3031 (M~81) we used the $^{12}$CO(1--0) line intensity (115~GHz/2.6~mm, rest frame) 
map from  BIMA SONG 
and available from NED in units of Jy~beam$^{-1}$~m~s$^{-1}$.
BIMA SONG is a systematic imaging study of the $^{12}$CO(1--0) molecular line emission within the centers and disks
of 44 nearby spiral galaxies performed with the the 10-element Berkeley-Illinois-Maryland Association (BIMA) 
millimeter interferometer at the angular resolution
of 6\arcsec.
For IC~342 we downloaded the $^{12}$CO(1--0) line intensity map from  
``Nobeyama CO Atlas of Nearby Spiral Galaxies''\citep{kuno07}\footnote{http://www.nro.nao.ac.jp/~nro45mrt/html/COatlas/}
in units of K~km~s$^{-1}$,
a $^{12}$CO(1--0) survey of 40 nearby spiral galaxies performed with the NRO 45m telescope  
at the angular resolution of 15\arcsec.
For NGC~1097, we used the $^{12}$CO(1--0) line intensity map in units of K~km~s$^{-1}$
observed by us with the  ATFN Mopra 22~m telescope (PI: M.~W.~L.~Smith) at the angular resolution of 30\arcsec.

\subsection{Metallicity abundances}
\label{sec:metallicity}
We also searched for metallicity abundance information for the face-on DustPedia sample.
\citet{pilyugin14} compiled [O~{\sc ii}]~$\lambda$3727+$\lambda$3729, [O~{\sc iii}]~$\lambda$5007,  [N~{\sc ii}]~$\lambda$6584, and [S~{\sc ii}]~$\lambda$6717+$\lambda$6731
emission lines (all normalised to H$\beta$) for around 3740 published spectra of H~{\sc ii} regions across the optical disks of 130 nearby late-type galaxies. 
Using these emission lines they determined homogeneous oxygen (O/H) and nitrogen (N/H) abundance distributions and fitted radial profiles within the isophotal radius in every galaxy following this equation:
\begin{eqnarray}
\rm {12 + log(O/H)} &=& {\rm{12 + log(O/H)}}_{r_0} + C_{\rm O/H} \times (r/r_{25})
\label{eq:oh}
\end{eqnarray}
\noindent
where 12~+~log(O/H)$_{r_0}$ is the oxygen abundance at $r_0 = 0$, i.e., the extrapolated central oxygen abundance, $C_{\rm O/H}$ 
is the slope of the oxygen abundance gradient expressed in terms of dex~$r_{\rm 25}$$^{-1}$, and $r/r_{25}$ is the fractional radius 
(the galactocentric distance normalized to the disk's isophotal radius $r_{25}$).   
All galaxies of our sample are contained in \citet{pilyugin14} and we used their metallicity values (see their Table~2) 
to study the variation of the CO-to-H$_2$ conversion factor as a function of O/H abundance
(see Sects.~\ref{sec:gasmass} and \ref{sec:xco}).

We stress that different strong-line diagnostics and calibration methodologies (e.g., photoionization models versus empirical electronic temperature, ${T}_{{\rm{e}}}$, derivations) 
yield substantial systematic offsets in the inferred gas metallicities \citep[e.g.,][]{kennicutt03,moustakas10,lopez-sanchez12,bresolin16}, 
reaching values up to 0.7~dex \citep[][]{kewley08}.
Methods calibrated from ${T}_{{\rm{e}}}$ measurements tend to occupy the bottom of the abundance scale.
Since to our knowledge abundances derived with the ${T}_{{\rm{e}}}$-method are not available for the whole sample, we take advantage of using metallicity values 
derived from the same calibration for our entire sample is to have an homogeneous treatment of data.

\section{Image treatment}
\label{sec:treatment}
As described in \citet{clark17}, UV, optical, and NIR images in the DustPedia database have Galactic foreground stars 
removed.  The same procedure was used to remove foreground stars in the additional SINGS images used in this work. 
Sky subtraction was instead performed by defining a few (at least five) rectangular regions of blank sky far away from everything, 
containing no detectable stars and other features. The average brightness level of these regions has therefore been subtracted 
from the image. While the choice of the correct sky value might affect the derivation of surface brightness profiles and their
scale-lengths, we checked that it has little influence for the inner disk profiles we analyse here. Even for the outer parts of the
disk in the 70 and 100~$\mu$m defining the extent of our analysis (see Sect.~\ref{sec:scalel}), changes in the adopted 
sky values, within the uncertainty of their estimate, produce negligible variations in the results.

All images from GALEX-FUV at 1516~\AA\ to \textit{Herschel}-SPIRE at 350~$\mu$m (Table~\ref{table:data}) and H{\sc i} and $^{12}$CO maps
(Table~\ref{table:gas}) were convolved to the resolution of the 500~$\mu$m (\textit{Herschel}-SPIRE) maps (36$^{\prime\prime}$). 
We chose the resolution of 500~$\mu$m, which is used in our SED fitting procedures and defines the resolution of the derived dust maps (Sect.~\ref{sec:dustmass}).
For most cases, convolution kernels to the 500~$\mu$m PSF are available in \citet{aniano11}. The higher resolution gas maps, instead, were 
convolved to the lower resolution using simple gaussian kernels. The H{\sc i} map of IC~342 was left unaltered, since its angular resolution 
is close to that of 500~$\mu$m.

Then, all maps have been resampled to the 500~$\mu$m map pixel size (12$^{\prime\prime}$, see Table~\ref{table:data}).
These procedures were performed by using IDL\footnote{http://idlastro.gsfc.nasa.gov/, Landsman (1993).},
IRAF\footnote{IRAF is the Image Reduction and Analysis Facility. 
IRAF is written and supported by the National Optical Astronomy Observatories (NOAO) in Tucson, 
Arizona. NOAO is operated by the Association of Universities for Research in Astronomy (AURA), Inc. 
under cooperative agreement with the National Science Foundation.}, 
and IRAM/GILDAS\footnote{http://www.iram.fr/IRAMFR/GILDAS/} \citep{guilloteau} software packages.\\

\section{Surface density of galaxy properties}
\label{sec:masses}
In this section we describe the adopted techniques to derive the mass surface density of dust, stars, and gas and the SFR surface density.

\subsection{Dust mass}
\label{sec:dustmass}
The dust mass surface density has been derived by comparing a modelled SED, convolved with the {\em Herschel} filter 
response functions, with the observed data at each position within a galaxy. The parameters of the model were retrieved 
using a standard $\chi^2$ minimization technique.

We adopted the optical properties and grain size distributions of the THEMIS\footnote{The 
Heterogeneous Evolution Model for Interstellar Solids, http://www.ias.u-psud.fr/themis/index.html} 
dust model, as described in \citet{jones13} and successive updates \citep{kohler14,ysard15,jones17}. 
We assumed that the  emission is due to grains exposed to a radiation field with the same spectrum as that of the local interstellar 
radiation field \citep[LISRF,][]{mathis83}, and intensity scaled up and down via the parameter $U$ to account for 
different heating environments within the galaxy ($U$=1 for the same conditions as in the Solar neighborhood). 
The dust emission for the various $U$ values, including both thermal and stochastic processes, was computed
with the DustEM software \citep{compiegne11}. 

Our approach is a simplified version of the more complex \citet{draineli07} procedure, which include both a radiation 
field of intensity $U=U_\mathrm{min}$, responsible for most of the thermal peak at $\lambda \ge 100$~$\mu$m, and 
a power-law distribution of higher intensity radiation fields ($U>U_\mathrm{min}$). This second component, though 
necessary to describe the emission at shorter wavelengths, contributes little to the dust mass determination. In fact, 
applications of the method to both global and resolved SEDs \citep[see, e.g.\,][respectively]{dale12,aniano12} show 
that only a minor fraction of the dust mass (smaller than a few percent) is exposed to the more intense fields.
Thus, fits to $\lambda \ge 100$~$\mu$m \textit{Herschel} SEDs assuming that dust is heated by a single radiation field
with $U=U_\mathrm{min}$ can retrieve the bulk of the dust mass \citep[see also][]{magrini11,bianchi13}. 

\begin{figure}
\includegraphics[width=9.5cm]{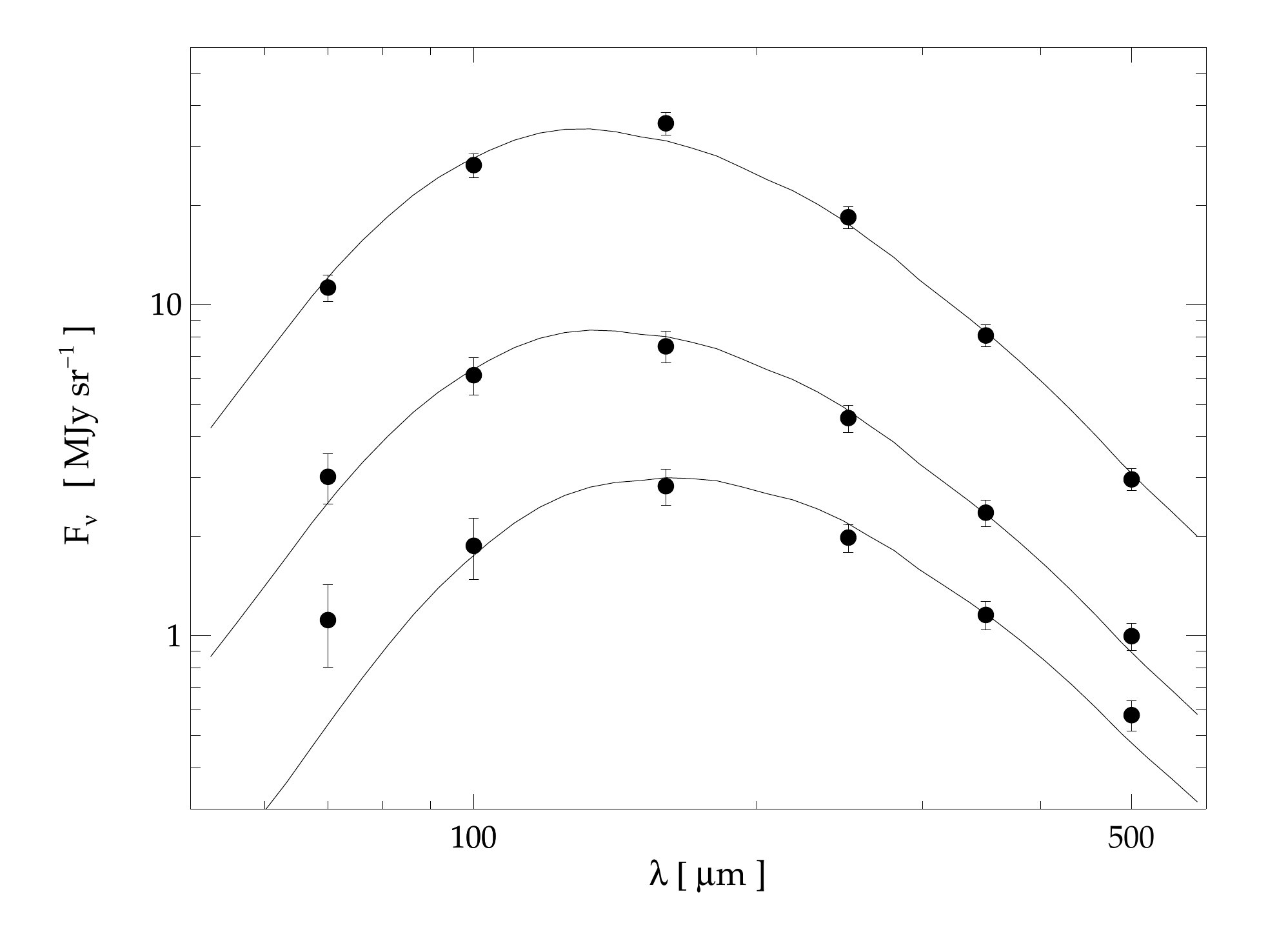}
\caption{
Example SED fits: the datapoints are from three different positions on NGC~5457 (M~101) at $R\approx$ 0.3, 0.5, and 0.7 $\times$~$r_\mathrm{25}$, 
corresponding to the local average heating conditions (fitted $U\approx$ 3, 2 and 1, respectively,  from top to bottom).
}
\label{fig:sedfits}
\end{figure} 

Initially, we derived the dust mass surface density $\Sigma_\mathrm{dust}$ and radiation field intensity $U$ for all the 
pixels where the {\em Herschel} surface brightness was larger than 2$\sigma$ in all bands with $\lambda \ge 160 \mu$m.
We thus produced  $\Sigma_\mathrm{dust}$ and $U$ maps, in a way analogous to \citet{magrini11}. However, this procedure 
could have biased the results to flatter gradients: in about half of the
galaxies fits in the outer parts of the objects were possible only for a fraction of the pixels, and only those
were counted in the profile determination.
Conversely, by assuming that the non-fitted pixels had null values for $\Sigma_\mathrm{dust}$ and $U$, the results 
would have been biased to steeper gradients. Thus, we resorted to use the method of \citet{hunt15}: the dust mass 
surface density profiles is directly derived by fitting, for each radial bin, the SED resulting from the azimuthally averaged 
surface brightness profiles. The method was not our first choice, because of other potential biases induced by the
possible temperature mixing along the isophotal ellipses. However, for the regions where there is a full pixel coverage
of the azimuthal bins, there is little differences with our initial, pixel-by-pixel, determinations; instead, for the external
part of a galaxy the  \citet{hunt15} procedure allows to extend the profiles to larger distances from the center. Errors
on the fit results were estimated using a Monte Carlo bootstrapping technique, assuming as errors for the surface brightness
profile the sum, in quadrature, of the standard deviation of the mean along the elliptical annulus, and the calibration error
of each band. The profile determination is further described in Sect.~\ref{sec:scalel}.

Besides the bands with $\lambda \ge$~160~$\mu$m the 100 $\mu$m data, when available, was included in the fit
 (or used as an upper limit in pixels 
with S/N < 2). For galaxies with available 70 $\mu$m images, the data were used as an upper limit, in order to avoid
contamination from grains heated by more intense radiation fields that we do not include in our approach. The inclusion
of upper limits in the fit was achieved as in \citet{remy-ruyer13}: first, a fit was performed without 70 $\mu$m data or 
$<2\sigma$ signal at 100 $\mu$m; if the fitted SED exceeded the flux at 70 $\mu$m or $2\sigma$ at 100 $\mu$m, the
fit was repeated including those datapoints. 
Typical average parameters retrieved from the fits are, at $r=0.5 \times r_\mathrm{25}$,
$U=1-4$ and $\Sigma_\mathrm{dust} =0.02-0.1\,{\rm M}_\odot\,{\rm pc}^{-2}$, depending on the galaxy. 
An example of the fits is given in Fig.~\ref{fig:sedfits}.

Given the various reports on the ability of different kinds of fits to retrieve the correct dust masses
(and thus the dust surface mass densities), we also experimented other approaches: 
a single-temperature modified-blackbody (MBB) fit with the averaged absorption cross section of our reference dust model;
a MBB with a power-law absorption cross section with spectral index $\beta=1.5$ and 2, 
scaled on the Milky Way emissivity \citep{bianchi13}; emission spectra from the \citet{draineli07} and 
\citet{compiegne11} dust models. We also tried to estimate the effect of a (maximum) contribution 
from hot dust, by fitting a MBB to all galaxies with available 70~$\mu$m data
after assuming that the entire 70~$\mu$m flux is due to dust at 40 or 50~K (and correcting the
longer wavelength bands accordingly). The dust mass surface density changes from fit to fit, and in 
particular when different dust models are used: values obtained with the \citet{draineli07} emission templates are about 2.7
times larger than those from our reference THEMIS model, and those with the Milky Way-scaled power law
emissivities of \citet{bianchi13} a factor 1.5 times higher\footnote{

The difference is due almost entirely to the difference in the optical properties of the dust models and it does not imply in
itself that a particular model is better or worse than the others. For example, in \citet{bianchi13} the dust masses of galaxies
in the KINGFISH sample were derived assuming that the global {\em Herschel} SED could be described by a single temperature 
modified blackbody and with a power-law absorption cross-section fitting the average properties of the dust model used
in the \citet{draineli07} emission templates, that of \citet{draine03}.
We repeated the same estimate by using the absorption cross-section $k_\lambda = 8.52 \times (250 \mu\mathrm{m}/\lambda)^{1.85}$ cm$^2$/g,
a fit to the average properties of the THEMIS model between 70 and 700 $\mu$m. The ratio of the dust masses derived with the 
two models is 2.6, essentially the same found in this work, despite the analysis is done here for a different sample, for specific positions within each galaxy 
and without assuming a single temperature. The difference is due mostly to the larger average absorption cross-section of the
THEMIS model with respect to \citet{draine03}, a factor 2 at 250~$\mu$m which converts to a factor 2 larger dust masses for  \citet{draine03}; 
and in part to the steeper spectral index of \citet{draine03} ($\beta=2.08$ vs 1.85 in THEMIS) which results, for MBB fits, in smaller 
temperatures and thus in a further increase in the dust mass estimate to reproduce the same SED.
}. 
Nevertheless, we found that the profiles (i.e.\ the gradients) have small differences, with changes in the
fitted scalelength that can be at most large as the error estimate of the fitting procedure (see later in 
Sect.~\ref{sec:scalel} and Table~\ref{tab:scale-lengths-mass}) but in most cases smaller.

A decreasing spectral index for the absorption cross section with radius, and thus possible variations in the 
dust properties along the disk, has been reported for MBB fits to M~31 \citep{smith12b} and M~33 \citep{tabatabaei14}.
Though this change in properties might affect the dust gradients, we did not try fits with varying $\beta$, as the
dust mass determination would also need a knowledge of the unknown variations of the absorption cross section 
normalization with the environment. We only note that an apparent, flatter, absorption cross sections for colder
dust at larger radii can be produced by the mixing due to the different temperatures attained by grains of different 
size and materials \citep{hunt15}. This behaviour, if important in the determination of the dust gradients, 
should be captured by our fitting procedure, that uses the full THEMIS grain distribution. 

Since the aim of this paper is to study the radial profiles and the corresponding scale-lengths, in 
the following figures we only show dust mass profiles obtained with the full THEMIS model.
We instead not comment any further on the absolute values of the dust mass surface density.
A more complete analysis on the dust mass in DustPedia galaxies, including the contribution of grains 
from different environment to the SED composition, will be the subject of future 
works of the collaboration.

\subsection{Stellar mass}
\label{sec:starmass}
Stellar mass is one of the most fundamental parameters describing present-day galaxies.
\citet{gavazzi96a} and \citet{gavazzi96b} pointed out that the structure and star-formation history of disc galaxies are tightly linked with stellar mass.
These findings can be extended to all morphologies as shown by \citet{scodeggio02}.
\citet{bell00} pointed out that the mean stellar mass density of a galaxy might be an even more basic parameter than the total stellar mass in determining
the stellar populations in spiral galaxies.
This conclusion has been confirmed by \citet{kauffmann03} and extended to all morphological types using more than 100~000 galaxies from the SDSS.

To calculate the stellar mass surface density we adopted two different methods. The first method makes use of the MIR images, specifically
the IRAC 3.6 and 4.5~$\mu$m, and the prescription presented in \citet{querejeta15}.
They computed stellar mass maps using the Independent Component Analysis \citep[ICA,][]{meidt12} which is able to disentangle
the contribution of the light from old stars and the dust emission, that can significantly contribute to the observed 3.6~$\mu$m flux.
\citet{querejeta15} also provided a relationship between the effective stellar mass-to-light ratio (M/L) and the observed [3.6]-[4.5] color
(with 0.2 dex of uncertainties), calibrated using their optimal stellar mass estimates and that compares well with other relations
in the literature \citep[e.g.,][]{eskew12,jarrett13,mcgaugh14}.

The second method we adopted is based on the optical stellar images (SDSS and SINGS data) and the prescription of \citet{zibetti09}.
In that paper the authors presented a method to construct spatially resolved maps of stellar mass surface density in galaxies based
on multi-band optical/near IR imaging, and the effective M/L at each pixel is expressed as a function of one or two colors.
We adopted the Table B1 of \citet{zibetti09} and the power-law fit to the M/L ratio as a function of the optical color $(g-i)$ from the \citet{bruzual03} models.
The differences in stellar mass surface density between the profiles obtained with Charlot \& Bruzual~(2007) and \citet{bruzual03} models are consistent within the errors, 
so we report only the results from the Charlot \& Bruzual~(2007) models (Zibetti, private communication).
The calculation of the stellar mass surface density from optical images can be done only for a fraction of galaxies of our sample: for 9 galaxies, we adopted the
\citet{zibetti09}'s procedure using the $(g-i)$ color, being available their SDSS images on the  $g$ and $i$ bands, while for NGC~3621 and NGC~1097,
we used the SINGS $B$ and $R$ images and the corresponding power-law fit.

As shown later (Figs.~\ref{fig:masses-prof} and \ref{fig:masses-prof-app}), stellar mass profiles obtained from the optical and MIR images
are in agreement for the majority of the 9 face-on galaxies for which this comparison can be performed.
For some galaxies (NGC~5194 (M~51), NGC~5457 (M~101), and NGC~2403), the difference of profiles in the absolute scaling
is notable and maybe due to the adopted optical colors.
Similar offsets in stellar profiles are found also in the analysis of the EAGLE simulations \citep[see][]{camps16, trayford16}.

We calculated the optical $(g-i)$ and MIR $([3.6]-[4.5])$ colors using the aperture photometry for the 9 face-on galaxies and we noticed that,
while the MIR colors vary in the range $0.95 - 0.97$, the optical ones go from $\sim$0.74 for NGC~5457 (M~101) up to $\sim$1.26 for NGC~3031 (M~81).
Our conclusion is that for the purpose of comparing the various profiles (dust, gas, stars) it is better to consider the IRAC images,
weakly dependent on the dust with respect to the optical proxies, and the empirical relation of \citet{querejeta15}, that compares well
with the results obtained with other methods in the literature and that is available for our entire sample.

\subsection{Gas mass}
\label{sec:gasmass}
We derived the mass surface density of atomic gas ($\Sigma\rm_{HI}$) from H{\sc i} 21~cm line intensity 
($I{\rm_{21 cm}}$) images, under the assumption of optically thin H{\sc i} emission,  following:
\begin{eqnarray}
\Sigma\rm_{HI} &=& 0.02\,I{\rm_{21 cm}}
\label{mhi}
\end{eqnarray}

\noindent
where $\Sigma\rm_{HI}$ is in units of ${\rm M}_\odot\,{\rm pc}^{-2}$ and $I{\rm_{21 cm}}$ in K~km~s$^{-1}$ \citep[e.g.,][]{schruba11}. 

We derived the mass surface density of molecular gas (\sigmahtwo) from the $^{12}$CO(1--0) or
$^{12}$CO(2--1) line intensity ($I{\rm_{CO(1-0)}}$ or $I{\rm_{CO(2-1)}}$)
images, based on the their availability, by adopting a constant value for the 
CO-to-H$_{2}$ conversion factor expressed in form of $X_{\rm CO}$  
($X_{\rm CO} = N({\rm{H_2}})/I_{\rm {CO}}$ where $N({\rm{H_2}})$ is the molecular gas column density 
in cm$^{-2}$ and $I_{\rm {CO}}$ is the CO line intensity in K~km~s$^{-1}$).
We assumed a value of $X_{\rm CO} = 2.0 \times 10^{20}$ cm$^{-2}$ (K~km~s$^{-1}$)$^{-1}$ \citep[e.g.,][]{bolatto13}.
This value of $X_{\rm CO}$ corresponds to a CO-to-H$_{2}$ conversion factor 
expressed in terms of $\alpha_{CO}$ ($\alpha_{CO} = {\rm M(H_{2})}/{\rm L_{CO}}$ where 
${\rm M(H_{2})}$ is the H$_{2}$ mass in M$_\odot$ and ${\rm L_{CO}}$ the CO line luminosity 
in K~km~s$^{-1}$~pc$^{2}$) of $\alpha_{CO} = 3.2\,{\rm M}_\odot\,{\rm pc}^{-2}$~(K~km~s$^{-1}$)$^{-1}$ 
\citep[][]{narayanan12}.
$X_{\rm CO}$ and $\alpha_{CO}$ are indeed related via 
$X_{\rm CO}~{\rm {[(K~km~s^{-1})^{-1}]}} = 6.3 \times 10^{19} \alpha_{CO}~[{\rm M}_\odot\,{\rm pc}^{-2}\,{\rm {(K\,km\,s^{-1})^{-1}]}}$.  
As is usual in the literature, we  refer to  the  CO-to-H$_{2}$ conversion factor both in terms  of $X_{\rm CO}$ and $\alpha_{CO}$
based on the context.  
In the case of $^{12}$CO(2--1), we assumed a CO line ratio $I_{\rm CO(2-1)}/I_{\rm CO(1-0)}$ = 0.7, 
a typical value in HERACLES and other surveys \citep[e.g.,][]{leroy09,schruba11}.
The derivation of \sigmahtwo\ from $I{\rm_{CO(1-0)}}$ and $I{\rm_{CO(2-1)}}$, under the assumption of optically thick $^{12}$CO 
emission, respectively, is through:
\begin{eqnarray}
\Sigma\rm_{H_{2}} &=& 4.17\,I{\rm_{CO(1-0)}} 
\label{mh2co10}\\
\Sigma\rm_{H_{2}} &=& 5.95\,I{\rm_{CO(2-1)}}
\label{mh2co21}
\end{eqnarray}

\noindent
where \sigmahtwo\ is in units of ${\rm M}_\odot\,{\rm pc}^{-2}$ and $I{\rm_{CO(1-0)}}$ and $I{\rm_{CO(2-1)}}$ 
in K~km~s$^{-1}$.
Equations (\ref{mhi}), (\ref{mh2co10}), and (\ref{mh2co21}) include a factor of 1.36 to account for heavy elements \citep[see][]{schruba11}.
When we refer to total gas (or the surface densities of mass of total gas), we consider the sum of atomic and molecular gas,
$\Sigma_{\rm{tot\,gas}}$ = $\Sigma\rm_{HI}$~+~$\Sigma\rm_{H2}$ (with the helium included).

\subsection{Star-formation rate}  
\label{sec:ssfr}
We derived the surface density of the current SFR ($\Sigma_{\rm SFR}$) combining $\textit{GALEX}$--FUV and WISE 22~$\mu$m  
according to \citet{bigiel08}'s calibration.  
They originally used the \textit{Spitzer} 24~$\mu$m emission to correct the dust attenuation of $\textit{GALEX}$--FUV surface brightness. 
This composite FUV+24~$\mu$m SFR tracer, together with the H$\alpha$+24~$\mu$m one \citep[][]{kennicutt07,calzetti07}, have been extensively used for 
a large number of nearby galaxies \citep[e.g.,][]{rahman11,ford13,momose13}.
    
The 22~$\mu$m flux densities can be also used to correct the FUV luminosity for dust attenuation 
since typical star-forming objects have a 22~$\mu$m emission that should be comparable to, or slightly fainter than, 
the 24~$\mu$m flux density \citep[see for instance,][for published applications]{hao11,cortese12, eufrasio14, huang15}.
We estimated $\Sigma_{\rm SFR}$ from the $\textit{GALEX}$--FUV emission corrected by the WISE 22~$\mu$m one
using the relation presented by \citet{bigiel08} by replacing 24~$\mu$m intensity with the 22~$\mu$m one:
\begin{eqnarray}
\Sigma_{\rm {SFR}} = 3.2 \times 10^{-3} \times I_{22} + 8.1 \times 10^{-2} \times I_{\rm {FUV}}
\label{sfr}
\end{eqnarray}

\noindent
where $\Sigma_{\rm SFR}$ is in units of ${\rm M}_\odot\,{\rm yr}^{-1}\,{\rm kpc}^{-2}$, and $I_{\rm 22}$ and $I_{\rm FUV}$
are the 22~$\mu$m and FUV intensities,  respectively, in units of MJy~sr$^{-1}$.
When the 22~$\mu$m information is not available ($I_{\rm 22}$ = 0), 
Eq.~(\ref{sfr}) reduces to the FUV--SFR calibration by  \citet{salim07}.
This calibration of $\Sigma_{\rm SFR}$ is based on the initial mass function
(IMF) from \citet{calzetti07}, which is the default IMF in STARBURST99 \citep[][]{leitherer99}
consisting of two power laws, with slope $-1.3$ in the range $0.1-0.5$~M$_\odot$
and slope $-2.3$ in the range $0.5-120$~M~$_\odot$.
To convert to the truncated \citet{salpeter55} IMF adopted by, e.g., \citet{kennicutt89}, \citet{kennicutt98a} or \citet{kennicutt07}, 
one should multiply our $\Sigma_{\rm {SFR}}$ by a factor of 1.59.

One caveat in the adopted SFR calibration is that both the 22~$\mu$m and FUV emission might have a contribution 
from stars that are too old to be associated with recent SF \citep[e.g.,][]{kennicutt09}. 
FUV is a tracer sensitive to SF on a timescale of $\sim$100 Myr \citep[e.g.,][]{kennicutt98a,calzetti05,salim07}, while  
22~$\mu$m emission, predominantly due to dust-heating by UV photons from bright young stars, is
sensitive to a SF timescale $<$10~Myr \citep[e.g.,][]{calzetti05,calzetti07,perez-gonzalez06}.
Old stars ($\gtrsim$1~Gyr) are fainter but redder and so emit relatively stronger at 3.6~$\mu$m.
A method to mitigate for the old stars is indeed based on the determination of FUV and 24~$\mu$m emission with respect 
to 3.6~$\mu$m one in regions where SF has ceased, and use this to remove the component of the FUV and 24~$\mu$m 
emission coming from old stars \citep[e.g.,][]{leroy08,ford13}.
However, \citet{leroy08} found that both FUV and 24~$\mu$m bands do appear to be dominated by a young stellar population almost everywhere 
in their sample of 23 galaxies, including spirals and irregulars, 9 of which are in common with our sample.
For this reason we did not apply the correction for the old star component in our sample of spirals.
Vice versa, this correction might be mandatory in the case of elliptical galaxies, dominated by pure populations of very old stars with no evidence of younger stars.
Anyway, since we are not interested in absolute values of the SFR, but on the radial profiles, we do not expect
significant differences on the scale-length values derived from profiles.

\begin{table*}
\caption{\label{tab:radii} 
Inner and outer radii used for exponential fitting for each sample galaxy.
}
\centering
\begin{tabular}{lccccccc}
\hline\hline
Galaxy	&	$r_{\rm in}$	&	$r_{\rm in}$	&	$r_{\rm in}/r_{25}$	&	$r_{\rm out}$(prof.)	&	$r_{\rm out}$	
&	$r_{\rm out}$	&	$r_{\rm out}/r_{25}$	\\
		&	[\arcmin]		&	[kpc]			&		&		&	[\arcmin]	&	[kpc]	&		\\
\hline
NGC~5457 (M~101)	&	2.1	&	4.3	&	0.18	&	70~$\mu$m	&	9.1	&	18.5	&	0.76	\\
NGC~3031 (M~81)	&	5.1	&	5.5	&	0.47	&	70~$\mu$m	&	9.3	&	10.0	&	0.85	\\
NGC~2403		&	4.1	&	4.2	&	0.41	&	12~$\mu$m	&	8.5	&	8.7	&	0.85	\\
IC~342			&	2.1	&	1.9	&	0.21	&	70~$\mu$m	&	9.9	&	8.9	&	0.99	\\
NGC~300			&	3.5	&	2.0	&	0.36	&	100~$\mu$m	&	7.1	&	4.1	&	0.73	\\
NGC~5194 (M~51)	&	2.1	&	4.8	&	0.30	&	*			&	3.9	&	8.9	&	0.56	\\
NGC~5236 (M~83)	&	2.1	&	4.0	&	0.31	&	70~$\mu$m	&	5.7	&	10.8	&	0.85	\\
NGC~1365		&	1.1	&	5.6	&	0.18	&	70~$\mu$m	&	3.5	&	18.0	&	0.58	\\
NGC~5055 (M~63)	&	0.9	&	2.1	&	0.15	&	70~$\mu$m	&	4.7	&	11.3	&	0.80	\\
NGC~6946		&	1.1	&	1.8	&	0.19	&	12~$\mu$m	&	3.1	&	5.1	&	0.54	\\
NGC~925			&	1.1	&	2.8	&	0.21	&	70~$\mu$m	&	4.7	&	11.8	&	0.88	\\
NGC~1097		&	1.0	&	6.3	&	0.21	&	100~$\mu$m	&	4.5	&	25.7	&	0.86	\\
NGC~7793		&	1.1	&	1.2	&	0.21	&	70~$\mu$m	&	4.1	&	4.5	&	0.78	\\
NGC~628 (M~74)	&	1.1	&	2.9	&	0.22	&	70~$\mu$m	&	4.1	&	10.8	&	0.82	\\
NGC~3621		&	1.1	&	2.2	&	0.23	&	70~$\mu$m	&	4.5	&	9.1	&	0.92	\\
NGC~4725		&	2.1	&	8.3	&	0.43	&	70~$\mu$m	&	4.3	&	17.1	&	0.88	\\
NGC~3521		&	1.1	&	3.9	&	0.27	&	70~$\mu$m	&	5.3	&	18.5	&	1.27	\\
NGC~4736 (M~94)	&	0.9	&	1.4	&	0.23	&	70~$\mu$m	&	3.3	&	5.0	&	0.85	\\
\hline
Mean			& 		&		& 	0.27~$\pm$~0.09		&		&		&		&	0.82~$\pm$~0.17\\	
\hline
\hline
\end{tabular}
\tablefoot{
* For NGC~5194 (M~51) the exponential fits stop at a radius of 3\farcm9, lower than $r_{\rm {out}}$(70~$\mu$m),
 because its interacting companion (NGC~5195) affects surface brightness profiles at larger radii.
}
\end{table*}

\section{Surface brightness profiles, and mass and star-formation rate surface density profiles}
\label{sec:prof}
The collected data and the analysis of the surface brightness profiles allowed us to study how 
dust emission relates to the distribution of stellar light. By examining the shape of the surface 
brightness profiles at each wavelength, we can make a comparison that is independent  of 
the absolute calibration.  

Radial surface brightness profiles were extracted from the dust ($\textit{Herschel}$ and WISE) and stellar 
($\textit{GALEX}$, SDSS, SINGS, and IRAC) images. We used elliptical averaging, a 
widely used traditional technique which was recently proven to fare as well as more sophisticate 
approaches  \citep{peters17}.  The adopted centers,  position angles (PA), inclination angles 
($i$), and minor-to-major axis ratios ($b/a$) are listed in Table~\ref{sample}.
The PA and $i$ angles have been determined kinematically and extracted from references listed in Table~\ref{sample}.
The width of each annulus is the same as the pixel size and the radial profile extraction extends up to the largest radius for which 
the surface brightness is larger than $2 \sigma$.
Uncertainties in the surface brightnesses as a function of radius were calculated as the quadrature sum 
of the standard deviation of the mean along the elliptical isophotes and of the uncertainties on the sky subtraction.   

Beside the dust surface density profile, whose derivation we described in Sect.~\ref{sec:dustmass}, we also produced 
mass surface density profiles of the stars and gas (atomic, molecular, and total), and SFR surface density profiles,
from the maps derived as in Sect.~\ref{sec:starmass}, \ref{sec:gasmass} and \ref{sec:ssfr}, respectively.
Uncertainties in the mass profiles as a function 
of radius were calculated as the quadrature sum of the standard deviation along the elliptical isophotes and
the uncertainty on the mass determination.  Uncertainties in the $\Sigma_{\rm SFR}$ profiles as a function 
of radius were calculated as the quadrature sum of the variation along the elliptical isophotes.

\begin{figure*}
\sidecaption
{\includegraphics[width=12.0cm,trim={30 20 60 20},clip]{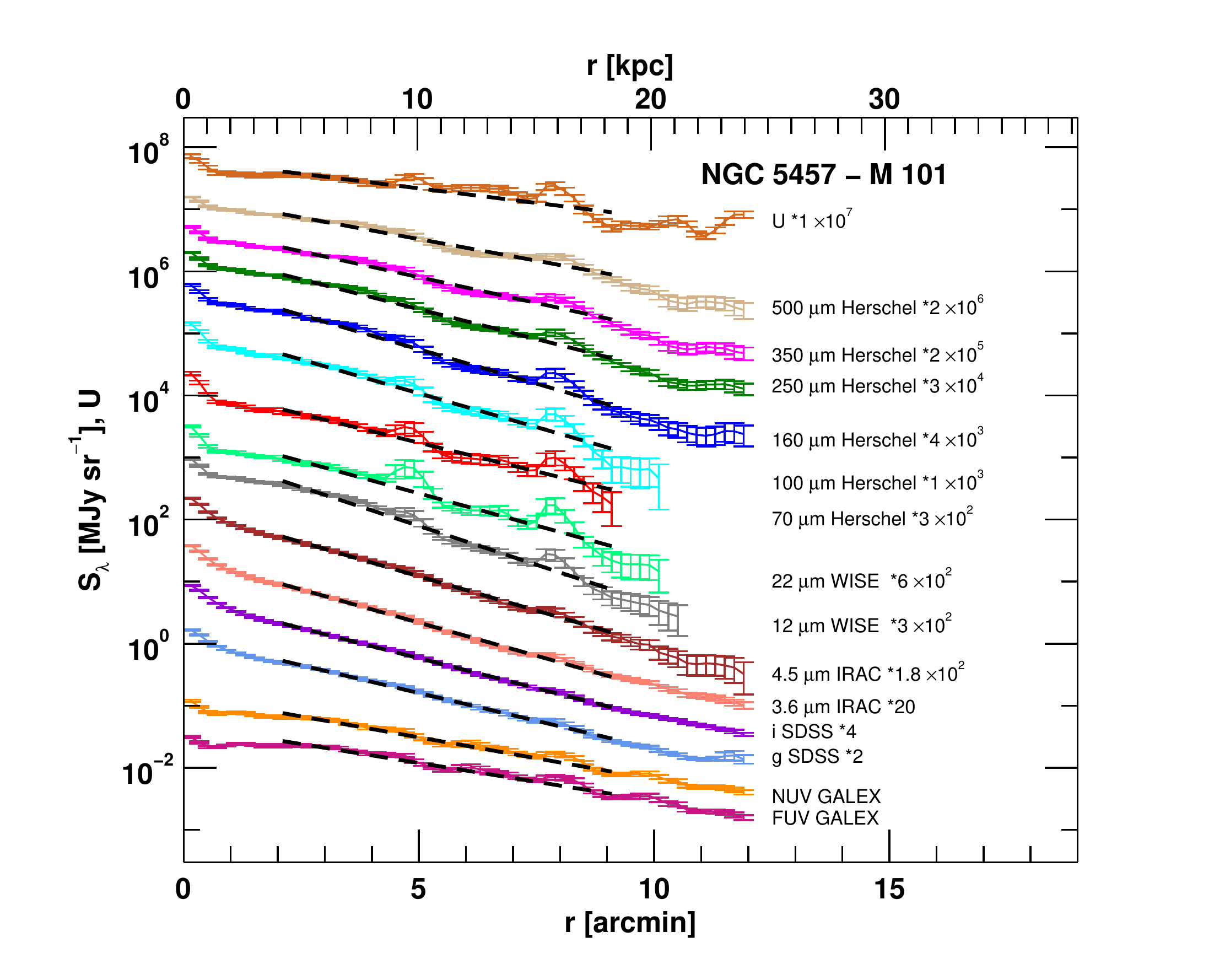}}
\caption{Multi-wavelength surface brightness profiles for the galaxy NGC~5457~(M~101), shown up to $r_{25}$.
The profiles have been shifted for displaying purposes and the corresponding offsets are quoted next to each profile. 
The profiles are arranged in order of decreasing wavelength, from top to bottom.
At the top, the interstellar radiation field intensity  $U$, is shown, in dimensional units.
The black dashed lines are exponential fits performed avoiding the central part of galaxies up to to the 
maximum extension of the 70~$\mu$m profile (shown in red).}
\label{fig:profiles}
\end{figure*}

\begin{figure*}
\includegraphics[width=6.3cm,angle=-90,trim={20 10 20 0},clip]{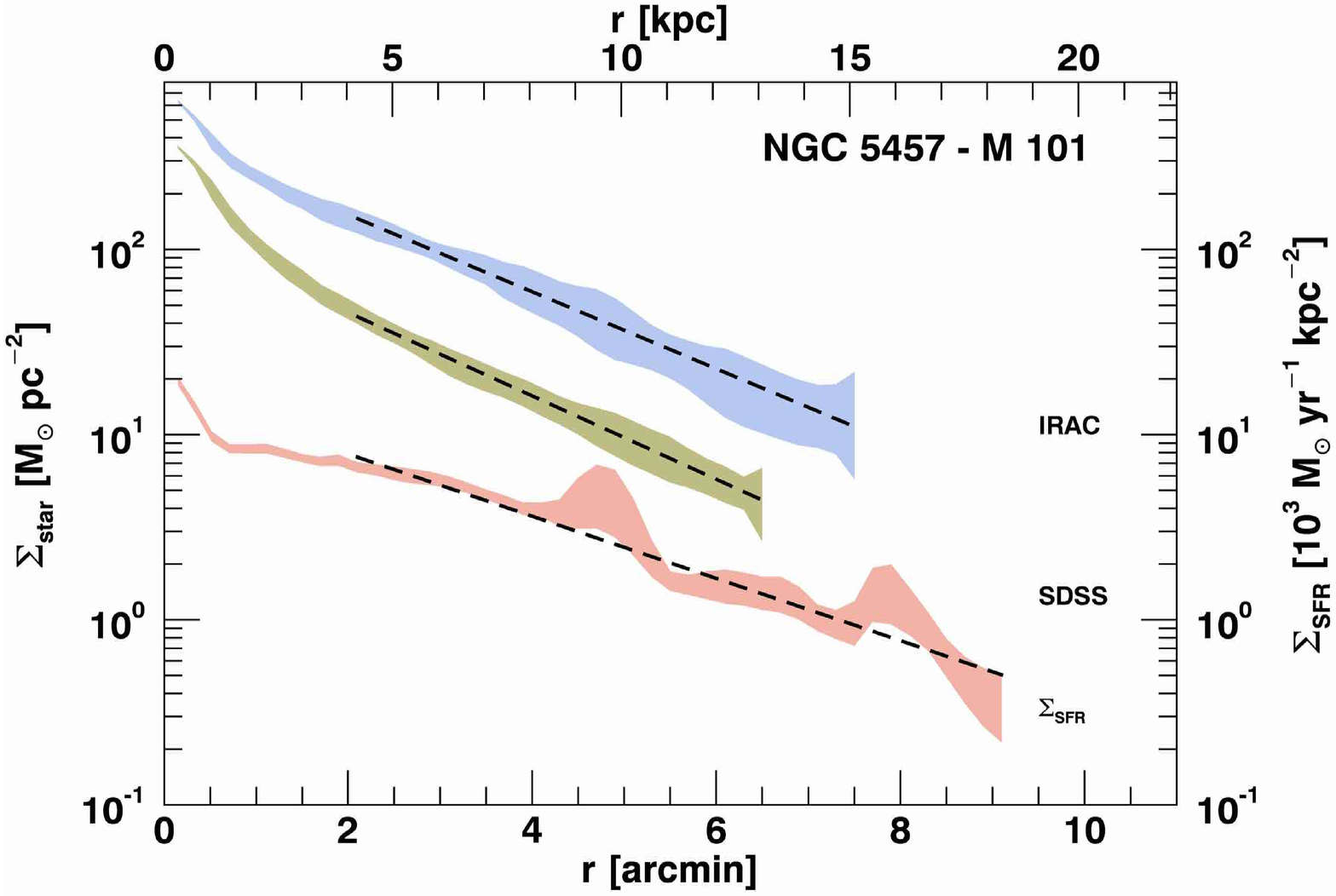}
\hspace{0.5cm}
\includegraphics[width=6.3cm,angle=-90,trim={20 10 20 40},clip]{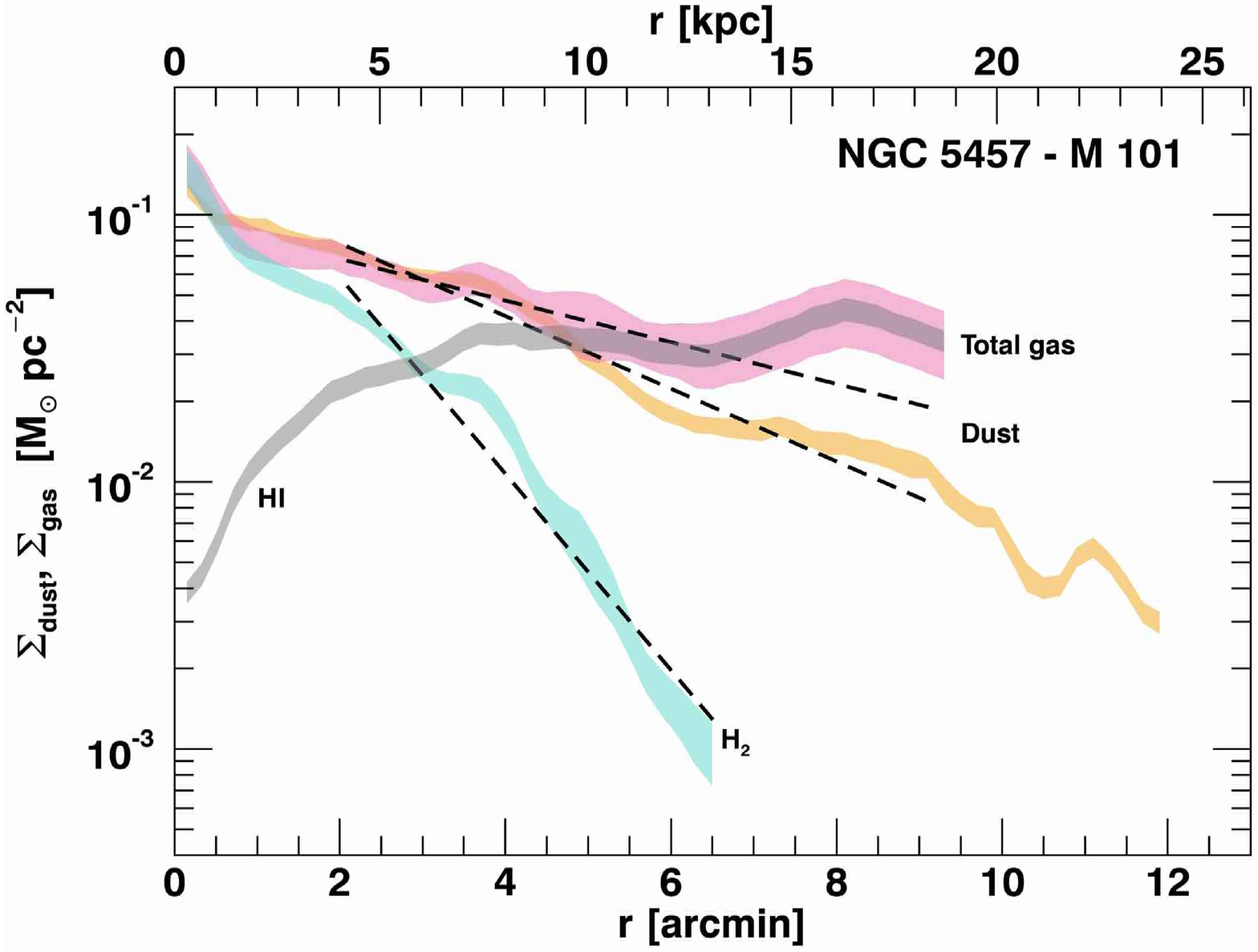}
\caption{\textit{Left panel:} 
Stellar mass surface density profiles from IRAC (blue) and from SDSS or SINGS (green), and SFR surface density profiles (red) 
of the galaxy NGC~5457~(M~101), derived as described in Sect.~\ref{sec:starmass} and \ref{sec:ssfr}, respectively.
The black dashed lines are exponential fits performed avoiding the central part of galaxies in the same radius range 
of Fig.~\ref{fig:profiles}.
\textit{Right panel:} 
Surface density profiles for the mass of dust (orange), molecular gas (turquoise), 
atomic gas (grey), and total gas (pink) of the galaxy NGC~5457~(M~101). 
The gas mass profiles have been corrected for DGR as described in Sect.~\ref{sec:xco}.}
\label{fig:masses-prof}
\end{figure*}

\subsection{Exponential disks and scale-lengths}
\label{sec:scalel}
We fitted the surface brightness profiles of the tracers of dust and stars, the mass surface density 
profiles of dust, stars, molecular gas, and total gas, and the SFR surface density profiles with a simple exponential 
curve, $S = S_0\, {\rm{exp}}(-h/h_0)$ where $h_0$ is the scale-length, and $S_0$ is the surface brightness 
(or mass/SFR surface density) at the radius $h = 0$ (we do not analyze this quantity in the present work).  
The exponential decline with the galactic radius is generally a good approximation for the surface brightness
profiles of stellar disks \citep{vanderkruit11} and has been extended to the description of dust and molecular disks
\citep[e.g.,][]{xilouris99,regan01}.

We excluded from the fit the central part of the surface brightness profiles where the emission might be dominated,
at least for the stellar component, by the presence of a spheroidal bulge.
For each galaxy we determined the starting/inner radius ($r_{\rm in}$) for the fit by eye and once selected this we 
adopted the same $r_{\rm in}$ for all profiles of that galaxy.  Despite the choice tailored on each object, the
resulting $r_{\rm in}$ are not too dissimilar, with an average ratio 
$r_{\rm in}/r_{25} = 0.27$ and a 1-$\sigma$ scatter of 0.09.
For all profiles of a given galaxy the exponential fit extends up to the faintest emission level at 70~$\mu$m surface brightness profile 
(outer radius, $r_{\rm out}(\rm70\,\mu m)$), 
or to those at 12 and 100~$\mu$m when the 70~$\mu$m profile is not available (NGC~300) or it is more extended than the 12~$\mu$m 
(NGC~2403, NGC~6946) and 100~$\mu$m (NGC~1097) profiles.
For NGC~5194~(M~51) the exponential fits stop at a radius lower than $r_{\rm out}(\rm70\,\mu m)$ because its interacting 
companion (NGC~5195) affects surface brightness profiles at larger radii.
Again, the resulting $r_{\rm out}$ are rather uniform, with an average ratio 
$r_{\rm out}/r_{25} = 0.82$ and a 1-$\sigma$ scatter of 0.17.
The small differences in the ranges that are used for the fitting should not be of concern, in particular when the ratios between scale-lengths are analysed.
Table \ref{tab:radii} collects the inner and outer radii used for exponential fitting for each sample galaxy.
Though we do not perform a true bulge-disk decompositions, the derived scale-lengths
still provide a good characterization of the gradients of the various disk components we consider.

An extra source of photons such as an AGN, present in the nuclei of some sample galaxies (see Table~\ref{sample}), 
which basically emits in the whole spectral range we are taking into account, could be in principle influencing the observed profiles. 
There are several reasons why we believe this is not affecting our analysis:
\begin{itemize}
\item the accretion disk mainly emits UV and optical radiation which is very efficiently absorbed by dust (and gas) within the central region of the galaxy; 
\item given the low luminosities of the AGN in the sample, possible radiation leaks to outer regions are negligible with respect to the luminosity of stars at similar wavelengths;
\item the dusty torus, which is instead heated by the accretion disk emission, and would hence emit in the near and mid-IR, is believed to be very weak at these luminosities, 
or even absent \citep[e.g.,][]{gonzalez-martin15};
\item the central regions ($\lesssim$1--2~kpc, see Table~\ref{tab:radii}), which is where a possible influence by AGN emission could be spotted, are not considered in our analysis.
\end{itemize}

Figure~\ref{fig:profiles} shows the example of surface brightness and interstellar radiation field profiles and exponential fits 
for the galaxy NGC~5457~(M~101) and Fig.~\ref{fig:profiles-app} displays those of the entire face-on DustPedia sample.  
The profiles are shown up to $r_{25}$.
Table~\ref{tab:scale-lengths} collects the exponential scale-lengths derived for the multi-wavelength surface brightness profiles.       

Also the stellar mass surface density profiles, SFR surface density profiles and mass surface density profiles of dust, molecular gas, and total gas
have been fitted within the same radius as that of the individual surface brightness profiles. 
Figure~\ref{fig:masses-prof} shows these profiles for the galaxy NGC~5457~(M~101), Fig.~\ref{fig:masses-prof-app} 
displays those of the entire face-on DustPedia sample, and 
Table~\ref{tab:scale-lengths-mass} collects the corresponding exponential scale-lengths.       
The gas mass surface density profiles shown in Figs.~\ref{fig:masses-prof} and \ref{fig:masses-prof-app} 
are corrected for radial variations in DGR and $\alpha_{\rm CO}$ conversion factor, both properties derived 
for each single galaxy according to the prescription described in Sect.~\ref{sec:xco}. 
The adopted values do not affect the derivation of the scale-length for molecular gas.
We do not fit the H{\sc i} gas surface density profiles, as these do not generally follow an exponential distribution.
In general, the difference between H{\sc i} and H$_2$ (or CO) radial distributions in galaxies is striking \citep[e.g.,][]{wong02,bigiel08} and 
H{\sc i} gas alone is extending much beyond the ``optical'' disk, sometimes in average by a factor 2 to 4 (R$_{\rm HI}$ $\sim$ 2--4 R$_{\rm opt}$).
The H{\sc i} gas has very often a small deficiency in the center maybe due to the transformation of the atomic gas in molecular phase in the denser central parts of galaxies.
\citet{bigiel08} find indeed that $\Sigma_{\rm HI}$ saturates at surface density of $\approx$9~${\rm M}_\odot\,{\rm pc}^{-2}$, 
and gas in excess of this value is in the molecular phase both in spirals and in H{\sc i}-dominated galaxies.     
This is the most common trend in galaxies, where the H{\sc i} and CO distributions appear complementary, 
but it is not the general case and all possibilities have been observed, 
including a central gaseous hole, both in CO and H{\sc i}
\citep[like in the Milky Way, M~31, NGC~3031~(M~81), and NGC~3147 in][respectively]{misiriotis06,nieten06,casasola07,casasola08}.

\begin{sidewaystable*}
\caption{\label{tab:scale-lengths} 
Exponential scale-lengths in arcminutes of wavelength expressed in micron measured in radius ranges specified in Table~\ref{tab:radii}.
We listed Pearson's correlation coefficients of each exponential fit, below each derived scale-length and in brackets.
}
\centering
\tiny
\begin{tabular}{lccccccccccccccc}
\hline\hline
Galaxy & $h_{\rm 0.15\,\mu m}$ & $h_{\rm 0.23\,\mu m}$ & $h_{\rm 0.5\,\mu m}$ & $h_{\rm 0.7\,\mu m}$ & $h_{\rm 3.6\,\mu m}$ 
& $h_{\rm 4.5\,\mu m}$ & $h_{\rm 12\,\mu m}$ & $h_{\rm 22\,\mu m}$ & $h_{\rm 70\,\mu m}$  & $h_{\rm 100\,\mu m}$
& $h_{\rm 160\,\mu m}$ & $h_{\rm 250\,\mu m}$ & $h_{\rm 350\,\mu m}$ & $h_{\rm 500\,\mu m}$ & $h_{\rm U}$  \\
&  &  & or & or \\
&  &  & $h_{\rm 0.4\,\mu m}$ & $h_{\rm 0.6\,\mu m}$  \\
&  [$^{\prime}$] & [$^{\prime}$]  & [$^{\prime}$] & [$^{\prime}$] & [$^{\prime}$]   & [$^{\prime}$]  & [$^{\prime}$] &  [$^{\prime}$] & [$^{\prime}$]  & [$^{\prime}$]
& [$^{\prime}$]  & [$^{\prime}$] & [$^{\prime}$]  & [$^{\prime}$] & [$^{\prime}$]   \\
& $(r_{\rm Pearson})$ & $(r_{\rm Pearson})$ & $(r_{\rm Pearson})$ & $(r_{\rm Pearson})$ & $(r_{\rm Pearson})$ & $(r_{\rm Pearson})$ & $(r_{\rm Pearson})$ & $(r_{\rm Pearson})$ & $(r_{\rm Pearson})$ 
& $(r_{\rm Pearson})$ & $(r_{\rm Pearson})$ & $(r_{\rm Pearson})$ & $(r_{\rm Pearson})$ & $(r_{\rm Pearson})$ & $(r_{\rm Pearson})$  \\
\hline
NGC~5457 		& 3.57$\pm$0.05 & 3.22$\pm$0.04 & 2.42$\pm$0.01 & 2.25$\pm$0.01 & 2.04$\pm$0.01 & 1.97$\pm$0.01 & 1.74$\pm$0.02 
			    	& 2.09$\pm$0.04 & 2.35$\pm$0.06 & 1.99$\pm$0.03 & 1.97$\pm$0.02 & 2.25$\pm$0.02 & 2.61$\pm$0.03 & 3.12$\pm$0.04 &  4.62$\pm$0.15  \\ (M~101)
			    	& $(-0.97)$ & $(-0.97)$ & $(-0.95)$ & $(-0.94)$ & $(-0.92)$ & $(-0.93)$ & $(-0.93)$ & $(-0.93)$ & $(-0.94)$ & $(-0.93)$	 & $(-0.93)$
			    	& $(-0.94)$ & $(-0.95)$ & $(-0.96)$ & (-0.91) \\				
NGC~3031 		& 3.76$\pm$0.15 & 3.42$\pm$0.12 & 2.86$\pm$0.03 & 2.78$\pm$0.02 & 2.49$\pm$0.02 & 2.55$\pm$0.02 & 2.41$\pm$0.10 
  			    	& 2.39$\pm$0.14 & 2.21$\pm$0.16  & **	& 2.64$\pm$0.10 & 3.19$\pm$0.18 & 3.74$\pm$0.24 & 4.33$\pm$0.31 & 3.88$\pm$0.26  \\ (M~81)
 			    	& $(-0.99)$ & $(-0.99)$ & $(-0.99)$ & $(-0.98)$ & $(-0.98)$ & $(-0.98)$ & $(-0.98)$ & $(-0.97)$ & $(-0.96)$ & -- & $(-0.97)$ & $(-0.98)$			
			    	& $(-0.98)$ & $(-0.98)$ & (-0.99) \\
NGC~2403 		& 2.73$\pm$0.05 & 2.54$\pm$0.04 & 2.28$\pm$0.03 & 2.35$\pm$0.03 & 2.25$\pm$0.02 & 2.43$\pm$0.02 & 1.46$\pm$0.04 
			    	& 1.81$\pm$0.05 &  1.83$\pm$0.08 & ** 			&  1.53$\pm$0.04 &  1.97$\pm$0.03 &   2.22$\pm$0.04 & 2.44$\pm$0.05 & 1.76$\pm$0.05   \\
			    	& $(-0.92)$ & $(-0.92)$	& $(-0.91)$ & $(-0.91)$ & $(-0.90)$ & $(-0.89)$ & $(-0.86)$ & $(-0.82)$ & $(-0.82)$ & -- & $(-0.88)$ & $(-0.91)$
			    	& $(-0.93)$ & $(-0.94)$ & (-0.95) \\
IC~342 	    		& * & 8.76$\pm$0.22 & ** & ** & 3.59$\pm$0.02 & 3.67$\pm$0.02 & 3.78$\pm$0.06 & 3.04$\pm$0.05 & 3.81$\pm$0.11 & 3.52$\pm$0.06 & 3.90$\pm$0.09 & 4.52$\pm$0.19 & 5.11$\pm$0.28 & 5.96$\pm$0.43 & 7.61$\pm$0.30 \\
			    	& -- & $(-0.94)$ & -- & -- & $(-0.98)$ & $(-0.98)$ & $(-0.98)$ & $(-0.96)$ & $(-0.97)$ & $(-0.98)$ & $(-0.98)$ & $(-0.98)$ & $(-0.98)$ & $(-0.98)$ & $(-0.92)$  \\	
NGC~300   		& 3.73$\pm$0.14 & 3.42$\pm$0.11 & **	& ** & 2.85$\pm$0.03 & 2.75$\pm$0.04 & 1.79$\pm$0.06  & 1.54$\pm$0.09 	
			    	& ** & 1.41$\pm$0.10 & 1.68$\pm$0.08  & 2.74$\pm$0.08 & 3.05$\pm$0.11 	& 3.58$\pm$0.13 & 1.71$\pm$0.05     \\
             		    	& $(-0.91)$ & $(-0.93)$ & -- & -- & $(-0.96)$ & $(-0.95)$ & $(-0.89)$ & $(-0.88)$ & -- & $(-0.90)$ & $(-0.90)$ & $(-0.90)$ & $(-0.90)$ & $(-0.89)$ & $(-0.96)$ \\	
NGC~5194 		& 1.34$\pm$0.06 	& 1.18$\pm$0.05 	& 1.46$\pm$0.05  	& 1.46$\pm$0.06  	& 1.30$\pm$0.06   	& 1.31$\pm$0.06  	& 1.12$\pm$0.04  	&  0.99$\pm$0.04  
			    	& 0.98$\pm$0.05 	& ** 				& 1.07$\pm$0.04 	& 1.18$\pm$0.04 	& 1.23$\pm$0.05 	& 1.29$\pm$0.05 	& 3.52$\pm$0.44  	\\ (M~51)		    
			    	& $(-0.91)$ 		& $(-0.94)$		& $(-0.99)$		& $(-0.98)$		& $(-0.97)$ 		& $(-0.97)$		& $(-0.98)$		& $(-0.96)$ 
			    	& $(-0.96)$		& --				& $(-0.98)$		& $(-0.98)$		& $(-0.98)$ 		&  $(-0.94)$ 		& $(-0.87)$  \\
NGC~5236	 	& 1.06$\pm$0.01  & 0.99$\pm$0.01 & **	& ** & 1.60$\pm$0.01 & 1.56$\pm$0.01 & 1.09$\pm$0.01 & 1.01$\pm$0.01 	
				& 0.82$\pm$0.01 & ** &  1.05$\pm$0.01 	& 1.15$\pm$0.01 & 1.23$\pm$0.01 & 1.32$\pm$0.02 & 2.00$\pm$0.05 \\ (M~83)
				& $(-0.96)$ & $(-0.96)$ & -- & -- & $(-0.95)$ & $(-0.95)$ & $(-0.94)$ & $(-0.90)$ & $(-0.90)$ & -- & $(-0.95)$ & $(-0.97)$	
				& $(-0.97)$ & $(-0.98)$ & $(-0.97)$\\
NGC~1365	 	& 2.11$\pm$0.14 & 1.74$\pm$0.09 & ** & ** & 0.97$\pm$0.02 & 1.01$\pm$0.02 & 0.91$\pm$0.02 & 0.82$\pm$0.01 
				& 0.78$\pm$0.03 &  0.82$\pm$0.02 & 0.85$\pm$0.02 & 0.96$\pm$0.03 & 1.08$\pm$0.04 & 1.21$\pm$0.05 & 1.46$\pm$0.06  \\
				& $(-0.84)$ & $(-0.91)$ & -- & -- & $(-0.97)$ & $(-0.96)$ & $(-0.98)$ & $(-0.93)$ & $(-0.95)$ & $(-0.96)$ & $(-0.98)$	& $(-0.95)$ & $(-0.96)$ & $(-0.95)$ & $(-0.94)$ \\
NGC~5055	 	& 1.65$\pm$0.02 	& 1.46$\pm$0.01 & 1.53$\pm$0.01 & 1.49$\pm$0.01 & 1.28$\pm$0.01 & 1.26$\pm$0.01 & 1.22$\pm$0.01 
			    	& 1.15$\pm$0.01 	& 0.94$\pm$0.02 & 0.99$\pm$0.01 & 1.10$\pm$0.01 & 1.28$\pm$0.01  & 1.40$\pm$0.01 & 1.52$\pm$0.02 & 2.42$\pm$0.09 \\ (M~63)
			    	& $(-0.98)$ & $(-0.94)$ & $(-0.91)$ & $(-0.89)$ & $(-0.89)$ & $(-0.89)$ & $(-0.94)$ & $(-0.94)$ & $(-0.88)$ & $(-0.90)$	 & $(-0.92)$ & $(-0.94)$			                     & $(-0.95)$			& $(-0.96)$ & $(-0.97)$\\
NGC~6946	 	& 10.01$\pm$3.25 & 4.53$\pm$0.53 & **	& ** & 1.71$\pm$0.03 & 1.75$\pm$0.04 & 1.89$\pm$0.40  & 1.81$\pm$0.14 
				& 1.44$\pm$0.03 & 1.82$\pm$0.07 & 2.00$\pm$0.07 & 2.27$\pm$0.09 & 2.43$\pm$0.11 & 2.60$\pm$0.14 & 3.26$\pm$0.12  \\
				& $(-0.71)$ & $(-0.95)$ & -- & -- & $(-0.98)$ & $(-0.99)$ & $(-1.00)$ & $(-0.98)$ & $(-0.98)$ & $(-0.99)$ & $(-1.00)$ 
				& $(-0.99)$			& $(-1.00)$			& $(-1.00)$ & $(-0.98)$ \\ 
NGC~925		 	& 2.88$\pm$0.13 & 2.57$\pm$0.10 & ** & ** & 1.59$\pm$0.03 & 1.65$\pm$0.03 & 1.56$\pm$0.05  
				& 2.05$\pm$0.10 & 1.67$\pm$0.11 & 1.71$\pm$0.08 & 1.82$\pm$0.06 & 2.03$\pm$0.07 & 2.29$\pm$0.07 
				& 2.59$\pm$0.10 & 3.30$\pm$0.29    \\
				& $(-0.97)$ & $(-0.96)$ & -- & -- & $(-0.93)$ & $(-0.93)$ & $(-0.93)$ & $(-0.91)$ & $(-0.93)$ & $(-0.92)$ & $(-0.94)$ 
				& $(-0.95)$ & $(-0.97)$ & $(-0.98)$ & $(-0.93)$  \\	
NGC~1097	 	& 2.14$\pm$0.09 & 1.81$\pm$0.06 & 1.47$\pm$0.03 & 1.28$\pm$0.02 & 1.24$\pm$0.02 & 1.24$\pm$0.02  & 1.09$\pm$0.02 
				& 0.89$\pm$0.01 & 0.73$\pm$0.02 & 0.76$\pm$0.02 & 1.07$\pm$0.03 & 1.34$\pm$0.03 & 1.53$\pm$0.04 & 1.73$\pm$0.06
				& 1.47$\pm$0.04  \\
				& $(-0.91)$ & $(-0.91)$ & $(-0.89)$ & $(-0.87)$	& $(-0.86)$ & $(-0.86)$ & $(-0.90)$	& $(-0.87)$ & $(-0.85)$ & $(-0.87)$ & $(-0.89)$								& $(-0.90)$			& $(-0.89)$			& $(-0.89)$ & $(-0.95)$\\		
NGC~7793	  	& 1.45$\pm$0.01 & 1.38$\pm$0.01 & ** & ** & 1.18$\pm$0.02 & 1.14$\pm$0.02 & 1.31$\pm$0.06
				& 1.21$\pm$0.04 & 1.30$\pm$0.05 & 1.29$\pm$0.03 & 1.41$\pm$0.02 & 1.63$\pm$0.03 & 1.88$\pm$0.03 & 2.17$\pm$0.05
				& 1.97$\pm$0.10 \\
				& $(-0.99)$ & $(-0.99)$ & -- & -- & $(-0.98)$ & $(-0.98)$ & $(-0.98)$	& $(-0.97)$ & $(-0.99)$ & $(-0.98)$ & $(-0.99)$	& $(-1.00)$								& $(-1.00)$			& $(-1.00)$  & $(-1.00)$\\
NGC~628		 	&  2.02$\pm$0.04 & 1.72$\pm$0.03  	& 1.32$\pm$0.01 & 1.18$\pm$0.01 & 1.08$\pm$0.01 & 1.17$\pm$0.01 & 1.13$\pm$0.02 
				& 1.04$\pm$0.02 & 1.25$\pm$0.06 	& 1.12$\pm$0.03  & 1.25$\pm$0.02 	& 1.38$\pm$0.02  & 1.61$\pm$0.03  	& 1.89$\pm$0.04
				& 1.96$\pm$0.09   \\ (M~74)
				& $(-0.99)$			& $(-0.99)$			& $(-0.97)$			& $(-0.94)$			& $(-0.95)$			
				& $(-0.95)$			& $(-0.97)$
				& $(-0.98)$			& $(-0.97)$			& $(-0.97)$			& $(-0.98)$			& $(-0.98)$			
				& $(-0.99)$			& $(-0.99)$ & $(-0.98)$\\
NGC~3621 		& 1.41$\pm$0.03 & 1.29$\pm$0.02 & 1.48$\pm$0.10 &  1.37$\pm$0.09 & 0.97$\pm$0.01 & 0.99$\pm$0.01 & 0.95$\pm$0.02 
			    	& 0.92$\pm$0.02 & 0.76$\pm$0.02 & 0.79$\pm$0.02 & 0.90$\pm$0.01 & 1.03$\pm$0.01 & 1.15$\pm$0.02 & 1.27$\pm$0.02 & 1.81$\pm$0.07    \\
			    	& $(-0.98)$ & $(-0.97)$ & $(-0.90)$ & $(-0.90)$ & $(-0.90)$ & $(-0.90)$ & $(-0.88)$ & $(-0.90)$ & $(-0.89)$ & $(-0.89)$ & $(-0.91)$ & $(-0.93)$					    & $(-0.95)$ & $(-0.96)$ & $(-0.96)$ \\
NGC~4725	 	& 0.93$\pm$0.02 & 0.96$\pm$0.02 & 1.21$\pm$0.02 & 1.22$\pm$0.01 & 1.11$\pm$0.01 & 1.10$\pm$0.01 & 0.82$\pm$0.02  
				& 0.79$\pm$0.04  & 0.81$\pm$0.06  &  0.71$\pm$0.03 & 0.84$\pm$0.02 & 0.95$\pm$0.02 & 1.10$\pm$0.04 & 1.27$\pm$0.05 
				& 1.39$\pm$0.07  \\
				& $(-0.97)$ & $(-0.97)$ & $(-0.97)$ & $(-0.96)$ & $(-0.95)$ & $(-0.95)$ & $(-0.96)$ & $(-0.95)$ & $(-0.94)$ & $(-0.96)$ & $(-0.97)$								& $(-0.98)$			& $(-0.98)$			& $(-0.99)$ & $(-0.98)$\\
NGC~3521	 	& 2.16$\pm$0.08 & 1.59$\pm$0.05 & 1.36$\pm$0.06 & 1.29$\pm$0.06 & 1.09$\pm$0.04 & 1.07$\pm$0.04 & 0.98$\pm$0.02 
				& 0.93$\pm$0.02 & 0.84$\pm$0.02 & 0.87$\pm$0.02 & 0.99$\pm$0.02 & 1.13$\pm$0.02 & 1.25$\pm$0.03 & 1.38$\pm$0.03
				& 2.29$\pm$0.08 \\
				& $(-0.98)$ & $(-0.97)$ & $(-0.94)$ & $(-0.94)$ & $(-0.92)$ & $(-0.92)$ & $(-0.94)$ & $(-0.93)$ & $(-0.92)$ & $(-0.93)$ & $(-0.94)$								& $(-0.95)$ & $(-0.95)$ & $(-0.96)$ & $(-0.97)$\\		
NGC~4736	 	& 0.68$\pm$0.01 & 0.67$\pm$0.01 & 0.91$\pm$0.01 & 0.92$\pm$0.01 & 0.83$\pm$0.01 & 0.85$\pm$0.01 & 0.72$\pm$0.01 
				& 0.61$\pm$0.01 & 0.36$\pm$0.01 & 0.46$\pm$0.01 & 0.60$\pm$0.01 & 0.73$\pm$0.01 & 0.76$\pm$0.01 & 0.75$\pm$0.01 
				& 1.13$\pm$0.05 \\ (M~94)
				& $(-0.75)$ & $(-0.76)$ & $(-0.86)$ & $(-0.87)$ & $(-0.85)$ & $(-0.85)$ & $(-0.80)$ & $(-0.77)$ & $(-0.75)$ & $(-0.78)$ & $(-0.81)$								& $(-0.84)$ & $(-0.85)$ & $(-0.85)$ & $(-0.87)$\\
\hline
\hline
\end{tabular}
\tablefoot{
* Curve not fitted by an exponential fit within the selected range in radius;
** Profile not available.
}
\end{sidewaystable*}

\begin{sidewaystable*}
\caption{\label{tab:scale-lengths-mass} 
Exponential scale-lengths in arcminutes for dust, gas and star mass surface density profiles and SFR surface density profiles in the same range in radius
used for fitting of surface brightness profiles. 
Pearson's correlation coefficients of each exponential fit are listed.
}
\centering
\begin{tabular}{lcccccc}
\hline\hline
Galaxy &  $h_{\Sigma{\rm {dust}}}$	& $h_{\Sigma{\rm {H2}}}$ 	& $h_{\Sigma{\rm {tot\,gas}}}$ & $h_{\Sigma{\rm {star\,IRAC}}}$	& $h_{\Sigma{\rm {star\,SDSS}}}$ &  $h_{\Sigma{\rm {SFR}}}$ \\
&  [$^{\prime}$] & [$^{\prime}$]  & [$^{\prime}$]  &  [$^{\prime}$] & or   & [$^{\prime}$]  \\
&			&			&			&			& $h_{\Sigma{\rm {star\,SINGS}}}$ & \\
&			&			&			&			&  [$^{\prime}$] \\
& $(r_{\rm Pearson})$ & $(r_{\rm Pearson})$ & $(r_{\rm Pearson})$ & $(r_{\rm Pearson})$ & $(r_{\rm Pearson})$ & $(r_{\rm Pearson})$ \\
\hline
NGC~5457 (M~101)$^{1}$ 	& 3.19 $\pm$ 0.06 	& $1.18 \pm 0.03$ 						& 5.56 $\pm$ 0.85 	& $2.09 \pm 0.13$	& $1.93 \pm 0.11$	& $ 2.59 \pm  0.08$   \\
						& $ (-0.94)$		& $ (-0.96)$							& $(-0.74)$ 		& $(-0.95)$ 		& $(-0.95)$ 		& $(-0.95)$  \\
NGC~3031 (M~81)  			& 6.20 $\pm$ 0.47  	& **									& *****			& $2.64 \pm 0.21$ 	& $2.51 \pm 0.25$	& $2.86 \pm  0.24$ 	\\
						& (-0.96)			& --  									& --   			& $(-0.98)$		& $(-0.97)$ 		& $(-0.98)$ 	\\
NGC~2403 				& 6.04  $\pm$ 0.51   	& ***  								& 7.51 $\pm$ 0.34  	& $2.06 \pm 0.18$	& $2.46 \pm 0.34$ 	& $2.22 \pm  0.09$   \\
						& $(-0.90)$ 		& --									& $(-0.94)$		& $(-0.92)$		& $(-0.91)$ 		& $(-0.86)$ \\
IC~342          				& 6.36 $\pm$ 0.13  & *	   								& * 			& $3.77 \pm  0.33$ 	& **** 			& $3.0\ \pm 0.08$	\\
						& (-0.98) 			& -- 									& -- 				& $(-0.99)$ 		& -- 				& $(-0.98)$ \\
NGC~300 				& *	 			& **** 								& ****   			& $6.16 \pm 0.92$	& ****			& $3.50 \pm 0.21$ \\
						& --				& --									& --    			& $(-0.95)$ 		& --				& $(-0.92)$\\
NGC~5194 (M~51) 			& 1.41  $\pm$ 0.06	&  0.75 $\pm$ 0.02   						& 1.55 $\pm$ 0.08	& $1.27 \pm 0.05$	& $1.24 \pm 0.06$ 	& $1.04 \pm 0.06$ \\
						& $(-0.98)$		& $(-0.97)$							& $(-0.98)$		& $(-0.97)$  		& $(-0.77)$		& $(-0.96)$\\	
NGC~5236 (M~83) 			& 1.64 $\pm$ 0.03 	& **** 								& ****			& $1.66 \pm 0.13$	& ****			& $ 1.04 \pm 0.02$    \\
						& $(-0.99)$ 		& --									& -- 				&$(-0.96)$ 		& --				& $(-0.92)$\\
NGC~1365          			& 1.55 $\pm$ 0.06	& **** 								& **** 			& $0.97 \pm 0.16$ 	& **** 			& $0.92 \pm 0.03$   \\
						& $(-0.93)$ 		& -- 									& -- 				& $(-0.97)$ 		& -- 				& $(-0.95)$\\
NGC~5055 (M~63) 			& 1.78 $\pm$ 0.04	& $1.19 \pm 0.02$						& 2.60 $\pm$ 0.07 	& $ 1.24 \pm 0.06$	& $1.27 \pm 0.08$	& $1.20 \pm 0.02$   \\
						& $(-0.97)$		& $(-0.98)$ 							& $(-0.97)$		& $(-0.89)$ 		& $(-0.86)$ 		& $(0.94)$ \\
NGC~6946          			& 2.83 $\pm$ 0.22	& 1.11 $\pm$ 2.15 $\times 10^{-3}$ 			& 2.81 $\pm$ 0.04  	& $1.47 \pm 0.09$ 	& ****			& $1.73 \pm 0.11$\\
						& $(-0.99)$ 		& $(-0.95)$ 							& $(-1.00)$		& $(-0.95)$ 		& --				& $(-0.97)$\\
NGC~925         				& 3.30 $\pm$ 0.19	& 1.33 $\pm$ 0.18 						& *				& $1.46 \pm  0.12$	& ****			& $ 2.34 \pm 0.18$ \\
						& $(-0.98)$ 		& $(-0.87)$ 							& -- 				& $(-0.93)$		& --				& $(-0.93)$\\
NGC~1097        			& 2.94 $\pm$ 0.15	& *	 								& **** 			& $1.13 \pm 0.12$	& $0.93 \pm 0.10$	& $1.00 \pm 0.02$ \\
						& $(-0.87)$ 		& -- 									& -- 				& $(-0.85)$ 		& $(-0.81)$ 		& $(-0.87)$    \\
NGC~7793        			& 2.75 $\pm$ 0.14 	& **** 								& **** 			& $1.40 \pm 0.08$ 	& **** 			& $1.43 \pm 0.06$  \\
						& $(-1.00)$ 		& -- 									& -- 				& $(-0.97)$ 		& -- 				& $(-0.98)$    \\
NGC~628 (M~74)  			& 2.20 $\pm$ 0.09 	& 0.75 $\pm$0.02  						& 1.29 $\pm$ 0.08  	& $1.04 \pm 0.05$ 	& $0.91 \pm 0.05$ 	& $1.24 \pm 0.05$\\
						& $(-0.99)$ 		& $(-0.92)$ 							& $(-0.95)$ 		& $(-0.94)$ 		& $(-0.88)$ 		& $(-0.99)$\\
NGC~3621       				& 1.57 $\pm$ 0.04	& ****  								& **** 			& $0.95 \pm 0.06$	& ****			& $0.98 \pm 0.03$   \\	
						& $(-0.97)$ 		& --  									& -- 				& $(-0.90)$		& --				& $(-0.91)$ \\
NGC~4725     				& 1.78 $\pm$ 0.09	& 0.51 $\pm$ 0.06						& 1.56 $\pm$ 0.33   	& $1.11 \pm 0.09$	& $ 1.23 \pm 0.12$	& $ 0.84 \pm  0.05$\\
						& $(-0.99)$ 		& $(-0.93)$ 							& $(-0.98)$ 		& $(-0.95)$ 		& $(-0.94)$ 		& $(-0.97)$\\
NGC~3521  				& 1.70 $\pm$ 0.03 	& 0.87 $\pm$ 0.03						& 1.69 $\pm$ 0.03 	& $1.22 \pm 0.34$ 	& $1.33 \pm 0.46$	& $1.01 \pm 0.03$    \\
						& $(-0.93)$		& $(-0.99)$ 							& $(-0.95)$ 		& $(-0.86)$ 		& $(-0.84)$ 		& $(-0.88)$\\
NGC~4736 (M~94) 			& 1.11 $\pm$ 0.03	& 0.47 $\pm$ 0.02						& 1.10 $\pm$ 0.04  	& $0.76 \pm 0.57$	& $0.92 \pm 0.08$ 	& $0.61 \pm 0.17$  \\
						& $(-0.90)$ 		& $(-0.93)$ 							& $(-0.95)$ 		& $(-0.86)$		& $(-0.92)$ 		& $(-0.77)$ \\
\hline
\hline
\end{tabular}
\tablefoot{
* Curve not fitted by an exponential function;
** $^{12}$CO(1--0) emission line, used to derive H$_2$ content, has been observed but not detected; 
*** Too small coverage in radius;
**** Profile not available;  
***** Since $^{12}$CO(1--0) emission line has not been detected the total has profile has not been derived. 
}
\end{sidewaystable*}

\subsection{Results}
\label{sec:sl}
\label{sec:sl-whole}
\begin{figure*}
\center
\includegraphics[trim={30 60 30 60},angle=270,clip,width=12cm]{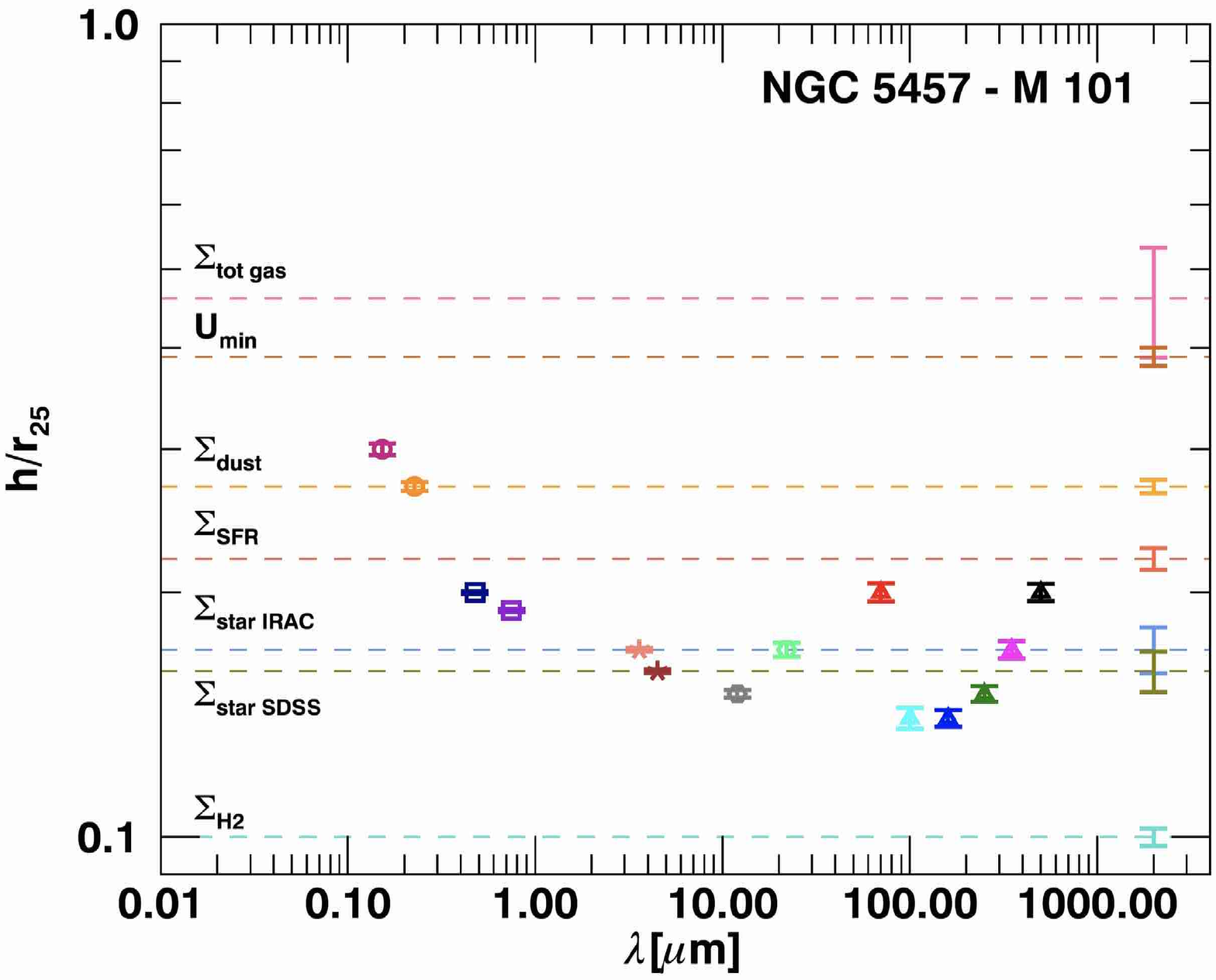}
\caption{
Scale-length fitted to the surface brightness profiles from UV to sub-mm and normalized with respect to $r_{\rm 25}$, $h/r_{\rm 25}$, 
as a function of wavelength for the galaxy NGC~5457~(M~101). 
The scale-lengths of mass (of dust, gas, and stars) and SFR surface density profiles and the scale-length of $U$ profile, normalized with respect to $r_{\rm 25}$,
are plotted as with horizontal lines because they are not associated with a wavelength. 
They and the corresponding error bars are drawn at $\lambda = 2000$~$\mu$m.
}
\label{fig:sl-wl}
\end{figure*}

Figures~\ref{fig:sl-wl} and \ref{fig:sl-wl-app} show the scale-lengths fitted to the surface brightness profiles from UV to sub-mm, 
normalized with respect to $r_{\rm 25}$ ($h/r_{\rm 25}$), as a function of wavelength, for NGC~5457~(M~101) and     
the entire face-on DustPedia sample, respectively. 
Together with the scale-lengths of surface brightness profiles, we also plotted those of $U$ profiles, 
and of mass and SFR surface density profiles. 
Since $U$ profiles and mass and SFR surface density profiles are not associated with a given wavelength, 
we plotted their scale-lengths marked with horizontal lines 
and the corresponding error bars are drawn at $\lambda = 2000$~$\mu$m.

\begin{figure*}
\center
\includegraphics[trim={30 60 30 60},angle=270,clip,width=12cm]{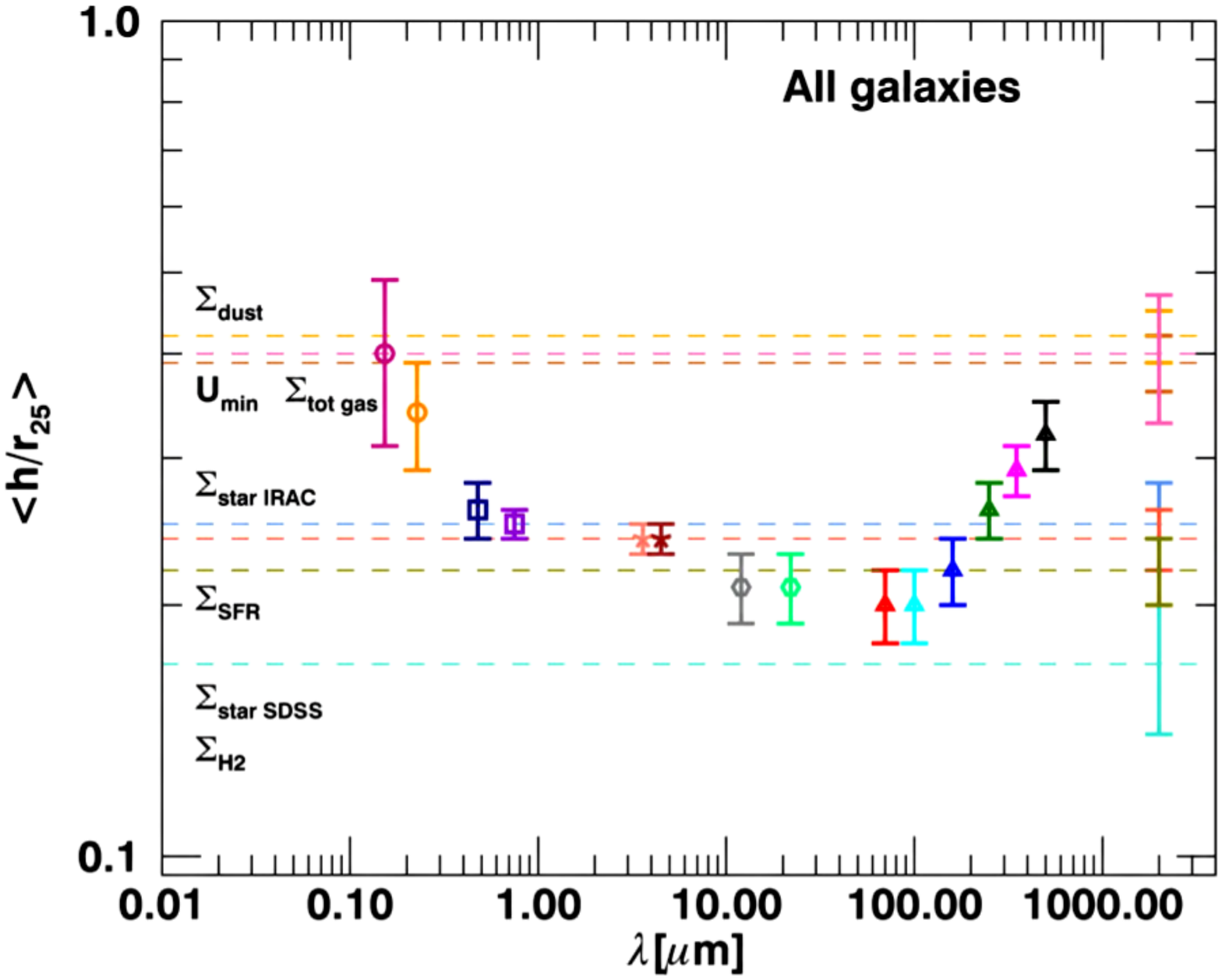}
\caption{
Mean scale-length normalized with respect to $r_{\rm 25}$ of all sample galaxies as a function of wavelength.
Scale-lengths of mass (of dust, gas, and stars) and SFR surface density profiles, the scale-length of $U$ profile, and the corresponding error bars are drawn 
as in Fig.~\ref{fig:sl-wl}.}
\label{fig:sl-wl-all}
\end{figure*}

The analysis performed for the whole sample shows a variety of behaviours in terms of exponential scale-lengths 
despite the fact that the investigated galaxies 
belong to the same subclass of objects (i.e., nearby, face-on, large, spiral galaxies) and that all images have been homogeneously treated. 
Therefore, each  galaxy  has  its  own trend of the scale-lengths.
This is likely due to the fact that the derived scale-lengths are affected by the peculiarities of a given galaxy, like, for
example, the local maxima in the spiral arms. 
Nevertheless, we identified some common behaviours in our sample: they are highlighted in Fig.~\ref{fig:sl-wl-all}, showing the mean 
$r_{25}$-normalized scale-lengths, <$h/r_{\rm 25}$>, over the whole sample, where differences due to the peculiarities 
of individual galaxies are all smeared out.
The mean values of $r_{25}$-normalized scale-lengths and scale-length ratios for selected bands are collected in Table~\ref{tab:sl}.
 
From Fig.~\ref{fig:sl-wl-all} it can be seen that the mean $r_{25}$-normalized stellar surface brightness scale-lengths decrease 
from the UV to the NIR by a factor $\sim$1.5, with values of <$h_{\rm NUV}$/$r_{25}$>~$\sim0.34$ and 
<$h_{\rm 4.5\,\mu m}$/$r_{25}$>~$\sim0.24$, respectively.
This decrease continues in the MIR dust-dominated bands and the mean $r_{25}$-normalized scale-length reach a minimum at 70~$\mu$m of $\sim$0.20.
From 70~$\mu$m to the sub-mm, instead, there is a steady increase up to the 500~$\mu$m scale-length with a value of
<$h_{\rm 500\,\mu m}$/$r_{25}$>~$\sim0.32$.  
There are individual variations to these common trends:
the UV scale-lengths in some galaxies are smaller than those in the optical and NIR (e.g., NGC~5194~(M~51)); in a few cases they are
much larger than in the common trends (e.g., NGC~6946, where unfortunately the intermediated optical SDSS/SINGS data are absent); the
steady increase in the FIR can start at wavelengths longer than 70~$\mu$m (e.g., NGC~5457~(M~101)).     

The scale-lengths of the dust mass surface density profiles are larger than those of the surface brightness profiles at 500~$\mu$m
(<$h_{\Sigma{\rm {dust}}}$/$h_{\rm 500\,\mu m}$>~$\sim$~1.3). 
They are generally larger than most of the other surface-brightness scale-lengths, and than those of the stellar mass surface density profiles. 
On average, the dust mass surface density scale-length is about 
1.8 times the scale-length of the stellar mass surface density derived from IRAC data and of  
the 3.6~$\mu$m surface brightness.
Only for NGC~300 the dust mass surface density gradient is so flat that the exponential profile was not fitted.
Only the average FUV scale-length can be as large as the dust mass surface density gradient, though the scatter in the former is large. Also,
the interstellar radiation field intensity has a gradient comparable to that of the dust mass.
SDSS and IRAC stellar mass profiles have comparable scale-lengths, and similar to those of the corresponding surface brightness profiles 
from which they have been derived ($g$ and $i$ bands from SDSS, $B$ and $R$ bands from SINGS, and 3.6 and 4.5~$\mu$m bands from IRAC). 
The scale-lengths of the SFR surface density are comparable to those of the stellar 3.6~$\mu$m emission and of the stellar mass distribution.

As for the gas, the H$_2$ surface density profiles have typically smaller scale-lengths than the NIR surface brightness profiles
(e.g., $h_{\Sigma{\rm {H2}}}$/$h_{\rm 12\,\mu m}$>~$\sim0.8$) and to IRAC stellar mass surface density profiles, and thus, on average, 
a factor $\sim$2.3 smaller than the dust mass ones. 
We stress that these results are based on H$_2$ gas profiles typically less extended than the others. 
The total gas mass surface density profiles, instead, have scale-lengths slightly larger than the dust (<$h_{\Sigma{\rm {tot\,gas}}}$/$h_{\Sigma{\rm {dust}}}$>~$\sim1.1$), 
though the result is influenced by the assumed value for $\alpha_{\rm CO}$, and thus by the assumed relative contribution of atomic and molecular
components to the total.
We will discuss this issue further in the next Section.

\begin{figure}
\center
\includegraphics[width=9.5cm]{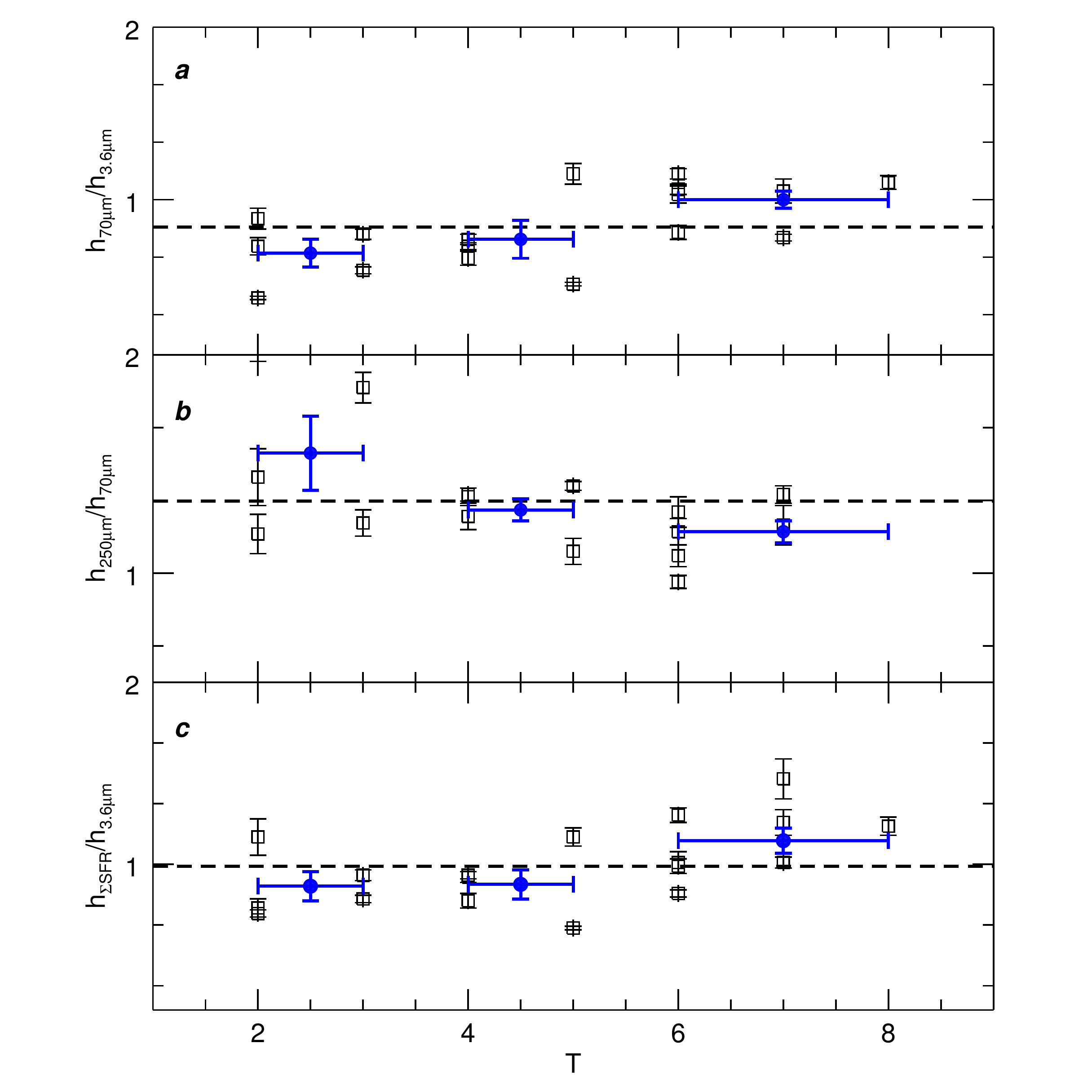}
\caption{Ratios of the exponential scale-lengths as a function of Hubble stage T:
\textit{a)}~70~$\mu$m/3.6~$\mu$m, \textit{b)}~250~$\mu$m/3.6~$\mu$m, \textit{c)}~$\Sigma_{\rm {SFR}}$/3.6~$\mu$m. 
In each panel, the (blue) large circles with error bars are the means over small ranges of Hubble stages (2--3, 4--5, 6--8) as indicated by the horizontal 
error bars, while the vertical error bars are the standard deviations.
The horizontal dashed lines show the mean scale-lengths by averaging all galaxies.
}
\label{fig:sl-dust-star-gas}
\end{figure}

We also studied the trends of the scale-lengths as a function of Hubble stage T. 
In general, the scatter is large and apparent trends are always feeble, within the scatter.  
Fig.~\ref{fig:sl-dust-star-gas} shows the trends for exponential scale-length ratios that are less weak. 
The ratio between the scale-lengths of the $\textit{Herschel}$ bands and the 3.6~$\mu$m scale-length tends 
to increase from earlier to later types, with the stronger trend being that for 
$h_{\rm 70\,\mu m}$/$h_{\rm 3.6\,\mu m}$ (panel a in  Fig.~\ref{fig:sl-dust-star-gas}). 
The value for $h_{\rm 70\,\mu m}$ instead, tends to be smaller than that at longer sub-mm wavelength 
and shown by the decreasing $h_{\rm 250\,\mu m}$/$h_{\rm 70\,\mu m}$ with increasing T (panel b). 
Among the surface density ratios, only  $h_{\Sigma{\rm {SFR}}}$/$h_{\rm 3.6\,\mu m}$ shows a
weak trend of increasing SFR scale-length, with respect to stars, when moving to later types (panel c).

\subsection{Discussion and comparison with the literature}
The mean stellar scale-lengths we find in this work  (e.g., $h_{\rm 3.6\,\mu m}/r_{25} \sim 0.24$) are consistent within the scatter
to those found from detailed two-dimensional bulge-disk decompositions of spiral galaxy images \citep[0.25--0.30, e.g.,][]{giovanelli94,moriondo98,hunt04}. 
Also, the decrease in the surface brightness scale-lengths from the UV to the NIR is in line with several previous works in the literature 
\citep{peletier94,dejong96,pompei97,moriondo98,macarthur03,moellenhoff04,taylor05,fathi10}. The color gradient has been explained with an intrinsic gradient in
the stellar populations, resulting in a blueing of the disks at larger distance from the center \citep{dejong96,cunow04,macarthur04} compatible with an inside-out 
formation scenario \citep{munoz07}. 
Other authors, instead, claim that the effect is mostly due to preferential reddening in the inner regions of a galaxy
\citep{beckman96,pompei97,moellenhoff06} or that dust has, at least, a significant role together with stars in shaping the gradient \citep{munoz11}.

According to \citet{alton98}, we interpret the steady increase in the surface brightness scale-length from 70 to 
500~$\mu$m as the result of a cold-dust temperature gradient with galactocentric distance combined with a scale-length 
of the dust disk larger than the stellar (see also \citealt{davies99}). 
A such cold-dust temperature gradient is consistent with a dust heating by a diffuse interstellar radiation field gradually 
decreasing with galactocentric distance, while a more {\em clumpy} stellar emission (and dust absorption/emission) would 
have resulted in flatter FIR/sub-mm profiles \citep[as shown, though for edge-on systems, by][]{bianchi08}.
The diffuse field could arise for a widespread component of old stars, and thus the derived trend might 
enforce the idea that dust heating and emission are not related to recent SF.
However, it could also arise from younger stars, if parental clouds have a sufficient leakage of radiation. 
There might not be a simple answer to the question:
indeed, the analysis of sub-mm colour maps by \citet{bendo15} has shown that old and young stars might
contribute in different ways to dust emission at different wavelengths, depending on each galaxy peculiar
situation.  
The flat trend in the derived radiation field, with $h_U \approx h_\mathrm{FUV}$, seems to suggest a
dominant contribution of SF to dust heating. The result is puzzling, though, as one would expect a non
negligible contribution to heating from older stars, and thus a scalelength in between $h_\mathrm{FUV}$ 
and $h_{\rm 3.6\,\mu m}$; also the FUV gradient might be significantly flattened by dust extinction. We
cannot exclude that the result is due to systematics effect in the modelling, due to the non-linear
dependence of the modelled SED on U, combined with the temperature mixing along the line of sight.
We only note that we found little differences in $h_U$ estimates if we use the pixel-by-pixel procedure
for SED fitting, which should be less affected by temperature mixing (See Sect.~\ref{sec:dustmass}).

For the dust emission surface brightness profiles we find trends similar to those of \citet{alton98}  (i.e. a steady increase of 
the scale-length for larger wavelengths), though with different and much less extreme scale-length ratios, as a result of the 
dramatic improvement in resolution from the IRAS/ISO data to the {\em Herschel} ones. 
While \citet{alton98} found that the 200~$\mu$m scale-length was about 30\% larger than that in the $B$-band 
(with a mean scale-length ratio of $\sim$1.3), we derive here that the 160 and 250~$\mu$m scale-lengths
are of the same order than the 3.6~$\mu$m scale-length (<$h_{\rm 160\,\mu m}$/$h_{\rm 3.6\,\mu m}$>~$\sim$~0.9, 
<$h_{\rm 250\,\mu m}$/$h_{\rm 3.6\,\mu m}$>~$\sim$~1.1) 
and thus are smaller than those in the $B$-band, for the wavelength trends we discussed. 
In addition, while in \citet{alton98} the 200~$\mu$m scale-length is 80\% longer than at 60~$\mu$m (with a mean ratio of $\sim$1.8), 
we find that the 160 and 250~$\mu$m scale-lengths are 10 and 30\% larger, respectively, than that at 70~$\mu$m (<$h_{\rm 160\,\mu m}$/$h_{\rm 70\,\mu m}$>~$\sim$~1.1, 
<$h_{\rm 250\,\mu m}$/$h_{\rm 70\,\mu m}$>~$\sim$~1.3).   

As highlighted in the previous section, the dust mass surface density we obtained through SED fitting to the {\em Herschel} data 
has an average scale-length that is 
$\sim$1.8 times the stellar one.
This dust-mass/stars scale-length ratio is slightly higher than those found by \citet{xilouris99} (1.4$\pm$0.2) and \citet{bianchi07} (1.5$\pm$0.5) 
with radiative transfer fits of the dust extinction lanes in edge-on spirals, and compatible with that derived by \citet{degeyter14} 
(1.8$\pm$0.8, also in this case by using radiative transfer fits).

\citet{munoz09a} derived the dust surface density profiles of SINGS galaxies by fitting {\em Spitzer} SED profiles with the \citet{draineli07} method. 
After fitting the profile with an exponential curve, they found that the dust scale-lengths in the sample are 1.1 the stellar scale-lengths derived from 3.6~$\mu$m 
data\footnote{The common resolution of the analysis of \citet{munoz09a} at 3.6 and 160 $\mu$m and of \citet{hunt15} at 3.6 and 250 $\mu$m is that of the 
\textit{Spitzer}/MIPS instrument at 160 $\mu$m maps (FWHM=38"), very close to that of Herschel/SPIRE at 500~$\mu$m maps, adopted here.}. 
From the median value of the $r_{25}$-normalized dust scale-length (0.29) one can obtain the $r_{25}$-normalized 3.6~$\mu$m scale-length (0.26), which is in excellent 
agreement with what we derived here with our smaller sample (<$h_{\rm 3.6\,\mu m}$/$r_{25}$>~$\sim0.24$). 
Instead, the dust surface surface density scale-length is smaller than what is inferred from fits to
edge-on galaxies and almost half of what we derive here from fitting the surface density maps. 
However, this does not necessarily indicate a disagreement between our work and that of \citet{munoz09a}.
As recognized by the same authors, this is most likely the result of the longest wavelength available in their SED,  
the 160~$\mu$m data from {\em Spitzer}: without a full coverage of the thermal emission peak and 
beyond, they could not probe the full temperature gradient with the emission from colder dust, and thus their
dust mass scale-lengths might be simply a reflection of the surface brightness scale-lengths at 160~$\mu$m. 

Exponential scale-lengths of the  surface brightness profiles at 3.6 and 250~$\mu$m for the KINGFISH sample
have been determined by \citet{hunt15}. At 250~$\mu$m, they found $h_{250\,\mu m}/r_{25} = 0.35$, a value larger
than our determination and that of \citet{munoz09a}. The main reason for the difference seems to be the
radial range used by \citet{hunt15}: to avoid any contribution from the bulge, even in early type spirals,
they fitted data for $r/r_{\rm opt}\geq0.6$ and up to $r/r_{\rm opt}=1.5$. This range has a small overlap 
with that used in our analysis, which refers to the inner disks and whose choice was dictated by the necessity
of a uniform coverage for all the wavelengths we considered. In fact, for the 12 galaxies we have in common with
the KINGFISH sample (which are essentially from the same \textit{Herschel} and \textit{Spitzer} datasets), we find that
the profiles of \citet{hunt15} are similar to our own for the inner disks, and tend to flatten in the range used in their
work. 

The methodology in the analysis of \citet{hunt15} might also be responsible for the larger estimate of the scale-length at
3.6~$\mu$m: they find $h_{3.6\,\mu m}/r_{25} = 0.37$, a value which, similarly to that at 250~$\mu$m, is larger than those obtained
for stars by other authors \citep[e.g.,][]{giovanelli94,moriondo98,hunt04} and on SINGS galaxies by \citet{munoz09a}.
Part of the differences might be due to the method of sky subtraction in \citet{hunt15}. Nevertheless, 
the authors claim that the homogeneity in the image and profile extraction processing at different wavelengths 
should obviate possible biases that could affect their conclusions (as well as ours and those of \citealt{munoz09a}),
in particular on scale-length ratios. Indeed,  \citet{hunt15} find a mean ratio between the 250~$\mu$m scale-length and 
the 3.6~$\mu$m one of about 1, the same value we derive here. We cannot extend our comparison to other
tracers of cold dust emission, since exponential fits at longer sub-mm wavelengths 
are not available in \citet{hunt15}.

More complex is the comparison with the recent analysis of \citet{smith16}. They stacked azimuthally averaged profiles for 
a large sample of low-inclination HRS galaxies and detect dust emission up to a distance of $2 \times r_{25}$. From the 
stacked {\em Herschel} profile of the largest 45 objects in their sample, they derived the radial variation of the dust SED, 
and a dust mass surface density after fitting a MBB to the SED in each radial bin. Though results are provided
for fits up to $2 \times r_{25}$, the exponential behaviour of their profiles within the inner $r_{25}$ is not much 
different from the full one and thus could be compared with ours. For the common resolution 
of 500~$\mu$m data, they provide gradients for the dust mass surface density, the stellar mass surface density and the 
SFR, normalized to $r_{25}$. From their gradients we derived a $r_{25}$-normalized scale-length for the dust mass surface 
density of 0.25. Not only we find a larger dust mass $r_{25}$-normalized scale-length (0.42) but also a larger ratio between
the dust and  500~$\mu$m scale-lengths (1.33), while the 500~$\mu$m stacked profile presented in their Fig.~2 has almost 
the same scale-length as the dust mass. 

In principle, the result of \citet{smith16} could be explained with a much shallower temperature gradient. Indeed, the sub-mm profiles shown in their  
Fig.~2 gives a slightly smaller $h_{500\,\mu m}/h_{250\,\mu m}$ ratio (1.15 vs. 1.25,
determined from their de-convolved exponential fits for the same radial range used in the current work). 
However, the scale-length of the average temperature gradient for $r/r_{25} < 1$, derived from their Fig.~4,  is $h_{\rm T}/r_{25} \approx 1.8$,
which, assuming ${U \approx T^{4+\beta}}$ \citep[][they used $\beta=2$]{hunt15}, 
converts to
$h_U/r_{25} \approx 0.3$, again smaller than our determination. Thus, their dust heating gradient is steeper than ours,
and should have produced a larger ratio for $h_{500\,\mu m}/h_{250\,\mu m}$ and a larger $h_{\Sigma_\mathrm{dust}}/h_{500\,\mu m}$ than in this work,
which is the contrary of what they derive. Clearly, the complex procedure adopted for the stacked analysis makes it complicate 
to understand the reason of these discrepancy.  We only note that we get discrepant results also for the SFR surface 
density scale-length which in our case has a value in 
between those of the UV and NIR surface brightness distributions used to derive it (0.24); while in theirs it is smaller (0.16)
than in the NIR (our determination), despite the similarity of the methods used to derive it. Instead, their stellar mass surface 
density has a scale-length similar to our own (they have 0.24). 
Thus, they obtain a dust to stellar mass scale-length ratio of $\sim$1, in contrast with our previous discussion.
 
The trend of increasing 70~$\mu$m scale-length with respect to 3.6~$\mu$m and 250~$\mu$m emission 
for increasing Hubble stage T can instead be the result of a residual bulge contribution which have escaped our simple
1-D analysis and masking of the inner regions: a more concentrated heating source could result in a
steeper gradient to dust heating for earlier type galaxies, as seen in the models of \citet{bianchi08}.
The weak increasing trend of the $h_{\Sigma {\rm SFR}}$/$h_{\rm 3.6\,\mu m}$ ratio moving to the later types is expected in the framework inside-out scenario 
of galaxy disks. 
As show in the grid of chemical evolution models of \citet{molla05} (see for instance their Figure 7, panel b), 
at a given galaxy mass, galaxies with different SF efficiencies (that can be translated into different morphological types) 
result in different SF histories. 
Spiral galaxies of later type, thus having lower SF efficiency, tend to have very low SFR in the past and to have the peak of their SFR at present time, 
while early type spiral (with higher SF efficiencies) had it few Gyr after their formation. 
At the later times, the SFR of early type spirals is maintained constant. 
Thus, we expect differences in the scale-lengths of galaxies of different morphological type due to their different SF histories. 
In particular, we expect that  in the late type galaxies the SFR might be more spatially extended at present time than in the past because 
they have just reached or they are still reaching the peak of their SFR. 
On the other hand, earlier type galaxies are in stationary phase after the peak of the SFR. 
Consequently the comparison of the scale-lengths are essentially unchanged with time.  

The dependence on the wavelength flattens for the MIR dust emission bands, because of a less strong dependence (a linear one) 
on the interstellar radiation field of dust emission from very small grains subject to stochastic heating \citep{draineli07}. 
At 12~$\mu$m, we find an average scale-length with respect to 3.6~$\mu$m emission, <$h_{\rm 12\,\mu m}$/$h_{\rm 3.6\,\mu m}$>,
of 0.9 which is close to what has been derived at 8~$\mu$m, another band dominated by emission features, 
by \citeauthor{regan06} (\citeyear{regan06}; 0.97, as derived from their tabulated data).

The large scale-lengths found in this work for the dust mass surface brightness imply, if there is no disk
truncation, that $\approx$30$\%$ of the dust mass resides in the outskirts of the disk at $r>r_{25}$, as
opposed to only $\approx$10$\%$ for the stellar mass. 
Together with the previous results from radiative transfer fits of edge-on galaxies, this
poses the problem of how dust can migrate from its origin in evolved stellar atmospheres and SNe or, as
we found here when comparing dust and molecular gas, from the denser parts of the ISM where it is
supposed to accrete most of its mass. 
Several qualitative explanations for the large dust scale-lengths have been proposed: the dust could trace an ``unseen'' molecular gas component at large radii, or
grains could be transported from the optical disk, through the action of macro turbulence, or thanks to interactions with nearby galaxies \citep[for a review, see][]{sauvage05}. 
For the well studied case of the edge-on galaxy NGC~891 the dust emission detected at  $r\approx r_{25}$
\citep{popescu03} is likely associated to non-pristine gas from the inner disk perturbed by the
nearby companion UGC~1807 \citep{bianchi11}; however, the detected feature constitutes only
a minor fraction of the total dust mass of the galaxy.

Despite the galaxies of our sample are classified as interacting (see Sect.~\ref{sec:sample}), we cannot
invoke a similar mechanism to explain our results: our dust mass scale-lengths, as well as all the other
measurements, refer only to the inner disk for  $r < r_{25}$ and there is no indication (but also no
proof of the contrary) that the observed behaviour extends beyond the optical disk. 
Despite the increased sensitivity of the {\em Herschel} telescope and other instruments, the dust mass fraction beyond $r_{25}$
cannot be easily derived for individual galaxies; instead, stacking analysis seems to suggest that
stars and dust follow the same distribution up to  $2 \times r_{25}$ \citep[][ though this result, as we
discussed earlier, is not consistent with that of this work for $r < r_{25}$]{smith16}. 
A possible mechanism that could explain our findings is a change in the typical lifetimes of grains against destruction by shocks,
with a longer lifetime at larger radii \citep{sauvage05}. 
Because of the similarity between UV and dust scale-lengths, it is tempting to associate this mechanism to the inside-out galaxy formation scenario that
is used to explain, together with dust extinction, the larger scale-lengths of radiation from the younger stellar
component \citep{munoz09a}.
The planned radiative transfer analysis of DustPedia galaxies will shed a light on the true contribution of dust extinction to the shaping of the UV and optical
appearance of spiral galaxies, and provide a detailed description of the disk surface brightness across the SED of dust emission.
This will allow us to analyze in detail why some galaxies present variations to common trends of scale-length vs.\ wavelength
(e.g., NGC~5457 (M~101), NGC~5194 (M~51), and NGC~6946).

As for the molecular gas, its scale-length is on average smaller than that of the optical-NIR bands (<$h_{\Sigma{\rm {H2}}}$/$h_{\rm 3.6\,\mu m}$>$\sim$0.7, 
<$h_{\Sigma{\rm {H2}}}$/$h_{\rm 12\,\mu m}$>$\sim$0.8). 
\citet{leroy09} found that the average molecular gas scale-length is only 5\% smaller than the 3.6~$\mu$m stellar scale-length and the SFR scale-length, while 
\citet{regan01,regan06} and \citet{bendo10} that molecular gas and optical-NIR bands have profiles with comparable scale-lengths. 
The comparison between the scale-lengths of molecular gas and MIR emission features could be conditioned 
by using different $^{12}$CO transitions and MIR bands.  
For the molecular gas we used $^{12}$CO(1--0) and $^{12}$CO(2--1) emission lines, \citet{regan06} the $^{12}$CO(1--0) emission line, 
and \citet{bendo10} the $^{12}$CO(3--2) emission line, 
while for the MIR emission features we analysed the images at 12~$\mu$m, while \citet{regan06} and \citet{bendo10} those at 8~$\mu$m.  
It is well known that each $^{12}$CO transition traces different physical gas properties. 
The kinematic temperature of molecular gas is typically $\sim$10 K \citep{scoville87}, which is above the level
energy temperature of 5.5~K for the $J = 1$ level of the $^{12}$CO, but below the temperatures of 16.5~K and 33~K for the $J = 2$
and $J = 3$ levels, respectively. 
This implies that a slight change in gas kinematic temperature is sufficient to affect the excitation
for the $^{12}$CO(2--1) and $^{12}$CO(3--2) emission lines. 
In addition, the different critical densities of the three $^{12}$CO transitions ($\sim$10$^3$~cm$^{-3}$, $\sim$2~$\times$~10$^4$~cm$^{-3}$,
and $\sim$7~$\times$~10$^4$~cm$^{-3}$ for $^{12}$CO(1--0), $^{12}$CO(2--1), and $^{12}$CO(3--2), respectively) 
make their line ratios sensitive to local gas density. 
The emissions at 8 and 12~$\mu$m are instead attributable to the same kind of transitions in C-H bonds in the proposed carriers of the MIR features,
like, e.g., PAH molecules \citep{draine11} or nanoparticles of amorphous hydrocarbons \citep{jones13}.    
Being the main result of the current analysis that the dust mass surface density scale-length is about 1.8 times the stellar one, a significative fraction of the
refractory materials must reside in the atomic gas. 
Based on these results, in the next Section we will attempt at deriving the
relative contribution of H$_2$ and H{\sc i} to the ISM, and the $\alpha_{\rm {CO}}$ conversion factor.

\begin{table}[!ht]
\caption{\label{tab:sl} 
Mean values over the entire galaxy sample of the $r_{25}$-normalized scale-lengths ($h/r_{25}$) of $U$, of the surface brightness profiles from FUV to 500~$\mu$m, 
and of the surface density profiles of mass (of dust, gas, and star) and SFR.
Other mean ratios between scale-lengths are collected.
}
\centering
\begin{tabular}{lcc}
\hline\hline
ratio & value & n. galaxies \\
\hline
<$h_U$/$r_{25}$>  									& $ 0.39 \pm 0.03 $   & 18\\
<$h_{\rm FUV}$/$r_{25}$> 							& $ 0.40  \pm 0.09 $ & 17 	\\
<$h_{\rm NUV}$/$r_{25}$>  							& $ 0.34  \pm 0.05 $ & 18 	\\
<$h_{\rm 0.5\,\mu m}$/$r_{25}$>  						& $ 0.26  \pm 0.02 $ & 11 	\\
<$h_{\rm 0.7\,\mu m}$/$r_{25}$>  						& $ 0.25  \pm 0.01 $ & 11 	\\
<$h_{\rm 3.6\,\mu m}$/$r_{25}$>  						& $ 0.24 \pm 0.01 $  & 18 	\\
<$h_{\rm 4.5\,\mu m}$/$r_{25}$> 						& $ 0.24 \pm 0.01 $  & 18 	\\
<$h_{\rm 12\,\mu m}$/$r_{25}$>						& $ 0.21 \pm 0.02 $  & 18 	\\
<$h_{\rm 22\,\mu m}$/$r_{25}$> 						& $ 0.21 \pm 0.02 $  & 18 	\\
<$h_{\rm 70\,\mu m}$/$r_{25}$> 						& $ 0.20 \pm 0.02 $ 	& 17	\\
<$h_{\rm 100\,\mu m}$/$r_{25}$> 						& $ 0.20  \pm 0.02 $ & 14 	\\
<$h_{\rm 160\,\mu m}$/$r_{25}$> 						& $ 0.22  \pm 0.02 $ & 18	\\
<$h_{\rm 250\,\mu m}$/$r_{25}$> 						& $ 0.26 \pm 0.02 $  & 18 	\\
<$h_{\rm 350\,\mu m}$/$r_{25}$> 						& $ 0.29 \pm 0.02 $  & 18	\\
<$h_{\rm 500\,\mu m}$/$r_{25}$> 						& $ 0.32 \pm 0.03 $  & 18 	\\
<$h_{\Sigma{\rm {dust}}}$/$r_{25}$> 					& $ 0.42 \pm 0.03 $  & 17 	\\
<$h_{\Sigma{\rm {H2}}}$/$r_{25}$>						& $ 0.17 \pm 0.03 $  & 9 	\\
<$h_{\Sigma{\rm {tot\,gas}}}$/$r_{25}$>					& $ 0.40 \pm 0.07 $  & 9 	\\
<$h_{\Sigma{\rm {star\,IRAC}}}$/$r_{25}$>				& $ 0.25 \pm 0.03 $  & 18	\\
<$h_{\Sigma{\rm {star\,SDSS}}}$/$r_{25}$	>				& $ 0.22 \pm 0.02 $  & 10 	\\
<$h_{\Sigma{\rm {SFR}}}$/$r_{25}$>						& $ 0.24 \pm 0.02 $  & 18	\\
<$h_{\rm 12\,\mu m}$/$h_{\rm 3.6\,\mu m}$> 				& $ 0.90 \pm 0.03 $ & 18 \\ 
<$h_{\rm 70\,\mu m}$/$h_{\rm 3.6\,\mu m}$> 				& $ 0.84 \pm 0.05 $ & 17 \\ 
<$h_{\rm 160\,\mu m}$/$h_{\rm 3.6\,\mu m}$>    			& $ 0.91 \pm 0.05 $ & 18 \\ 
<$h_{\rm 250\,\mu m}$/$h_{\rm 3.6\,\mu m}$>    			& $ 1.07 \pm 0.05 $ & 18 \\ 
<$h_{\rm 500\,\mu m}$/$h_{\rm 3.6\,\mu m}$>    			& $ 1.34 \pm 0.07 $  & 18 \\ 
<$h_{\rm 160\,\mu m}$/$h_{\rm 70\,\mu m}$>    			& $ 1.14 \pm 0.05 $ &  17 \\ 
<$h_{\rm 250\,\mu m}$/$h_{\rm 70\,\mu m}$>    			& $ 1.32 \pm 0.06 $ &  17 \\ 
<$h_{\rm 500\,\mu m}$/$h_{\rm 160\,\mu m}$>   			& $ 1.48 \pm 0.05 $ & 18\\
<$h_{\Sigma{\rm {dust}}}$/$h_{\rm 3.6\,\mu m}$>			& $ 1.77 \pm 0.12 $ & 17  \\
<$h_{\Sigma{\rm {dust}}}$/$h_{\rm 500\,\mu m}$>   			& $ 1.33 \pm 0.09 $ & 17\\
<$h_{\Sigma{\rm {dust}}}$/$h_{\rm {H2}}$>   				& $ 2.27 \pm 0.30 $ & 9\\
<$h_{\Sigma{\rm {dust}}}$/$h_{\Sigma{\rm {star\,IRAC}}}$>	& $ 1.80 \pm 0.13 $ & 17\\
<$h_{\Sigma{\rm {H2}}}$/$h_{\rm 3.6\,\mu m}$>			& $ 0.73 \pm 0.11 $ & 9\\
<$h_{\Sigma{\rm {H2}}}$/$h_{\rm 12\,\mu m}$>				& $ 0.78 \pm 0.14$ & 9 \\
<$h_{\Sigma{\rm {tot\,gas}}}$/$h_{\Sigma{\rm {dust}}}$>		& $ 1.11 \pm 0.16 $ & 9 \\
<$h_{\Sigma{\rm {star\,IRAC}}}$/$h_{\rm 3.6\,\mu m}$>   		& $ 1.06 \pm 0.07  $ & 18\\
<$h_{\Sigma{\rm {SFR}}}$/$h_{\rm 3.6\,\mu m}$>   			& $ 0.99  \pm  0.05 $ & 18\\
\hline
\hline
\end{tabular}
\end{table}

\section{Dust and gas mass profiles to solve for single CO-to-H$_2$ conversion factors}
\label{sec:xco}
As already mentioned in Sect.~\ref{sec:gasmass}, a major uncertainty in the derivation of the molecular gas content in a galaxy is the poorly-known
$\alpha_{\rm {CO}}$ conversion factor. 
We developed a method to solve for the $\alpha_{\rm CO}$ conversion factor per galaxy by using dust and gas mass profiles.

Recently, for resolved galaxies with available \textit{Herschel} data it has been shown that it is possible to derive $\alpha_{\rm {CO}}$ independently 
of the dust mass estimate and DGR \citep[][]{sandstrom13}.
The idea of \citet{sandstrom13} is that spatially resolved measurements of $\Sigma_{\rm dust}$, $\Sigma_{\rm HI}$, and $\Sigma_{\rm H2}$ (from CO observations) 
allow one to solve for $\alpha_{\rm {CO}}$ and the DGR over regions smaller than the typical scale over which the DGR varies (i.e., one region is well represented by a single DGR value) 
and it covers a range of CO/H{\sc i} ratios.   
The peculiarity of their approach is the absence of assumptions about the values of $\alpha_{\rm {CO}}$ and the DGR and their dependence on metallicity 
or other parameters.
Another approach has been developed by \citet{magrini11}. 
Using typical properties of Galactic dust and gas, they converted the DGR profile
into a profile of the oxygen abundances for the metals in solid state; then they compared it to the gas metallicity gradients and studied the dependency of the $\alpha_{\rm {CO}}$ conversion factor on metallicity.

The method we developed to solve for the $\alpha_{\rm {CO}}$ conversion factor per galaxy is a simplified version of that of \citet{magrini11} 
and it is based on the following assumptions:
$i)$ the DGR and the dust optical properties are the same in both the atomic and molecular gas phases;
$ii)$ dust and gas are well mixed;
$iii)$ the DGR is invariant with respect to the metallicity gradient or it follows the same trend of the metallicity gradient;
$iv)$ the fraction of mass in ionized gas is negligible.

As early as the mid-1980s, \citet{hildebrand83} suggested that a good way to estimate the mass and the distribution of the ISM in a galaxy might be
from the optical depth of the sub-mm continuum emission from the dust. 
In later years, experimental studies based on laboratory samples of dust grain species have instead shown that there is less scatter between species around 
100~$\mu$m than in the sub-mm \citep[e.g.,][]{coupeaud11}. 
Dust grains are robust and found in all phases of the ISM.
Several works therefore used the dust emission to infer the gas distribution \citep[e.g.,][]{guelin93,guelin95,boselli02,corbelli12,eales12,sandstrom13,groves15}.
The  amount  of  gas  is usually measured using H{\sc i} and  CO  surveys, 
however \citet{smith12a} have found that for early-type galaxies (E/S0) the ISM was detected for 50$\%$ of objects through its dust
emission but only 22$\%$ through its CO emission, in addition to the uncertainty associated with $\alpha_{\rm {CO}}$.

From the practical point of view, in order to solve for the $\alpha_{\rm {CO}}$ conversion factor per galaxy,  
we first determined the dust, atomic, molecular and total gas mass profiles following prescriptions of Sect.~\ref{sec:masses},
in particular by assuming the constant $\alpha_{\rm CO}$ value of $3.2\,{\rm M}_\odot\,{\rm pc}^{-2}$~(K~km~s$^{-1}$)$^{-1}$.
Secondly, based on our assumptions and in the hypothesis of a constant, global value for the DGR, we can write:
\begin{eqnarray}
\Sigma_{\rm dust}& = &{\rm DGR} (\Sigma_{\rm HI} + \Sigma_{\rm H2}) \\
                             & = &{\rm DGR}  (\Sigma_{\rm HI} +  \alpha_{\rm CO}\, I_{\rm CO})
\label{eq:DGR}
\end{eqnarray}
Finally, we fitted Eq.~(\ref{eq:DGR}) to the data and derived global values for ${\rm DGR}$ and $\alpha_{CO}$ for each galaxy. 

We applied the same approach described above also taking into account that the DGR is an indicator of the amount of metals 
that get locked up in the dust through the stellar yields. 
Thus, a dependence on the metallicity gradient is expected. 
A correlation has been widely shown between the DGR and the gas-phase oxygen abundance \citep[e.g.,][]{issa90,lisenfeld98,edmunds01,james02,hirashita02,draine07b}: 
the DGR decreases with radius, following the trend with metallicity.
We therefore corrected atomic and molecular gas mass profiles for metallicity thanks to the radial metallicity gradient available 
for the entire galaxy sample (see Sect.~\ref{sec:metallicity}).   
To do this, we applied Eq.~(\ref{eq:DGR}) with the following additional assumption:
\begin{eqnarray}
{\rm log(DGR)} = {\rm log( DGR_0)} + C_{\rm O/H} \times (r/r_{25})
\label{eq:DGR-z}
\end{eqnarray}
\noindent
(i.e.\ the same gradient as for metallicity given by Eq.~(\ref{eq:oh})) with  ${\rm DGR}_0$ the
value at $r_0=0$.  
Thus, we also derived global values for DGR and $\alpha_{\rm CO}$ ``corrected'' for metallicity gradient ($\alpha_{\rm {CO}}$$_{\rm (O/H)}$). 
Again, we do not comment on the  ${\rm DGR}_0$ values.

The right panels of Fig.~\ref{fig:masses-prof} and \ref{fig:masses-prof-app} show the mass surface density profiles of the dust, 
atomic, molecular, and total gas corrected by using the derived ${\rm DGR}$ and $\alpha_{CO}$ according to Eq.~(\ref{eq:DGR}) 
without the correction for metallicity gradient of Eq.~(\ref{eq:DGR-z}).
In the case of one available gas component (H{\sc i} or H$_2$), we corrected it by using the derived ${\rm DGR}$.   
Given the uncertainties in the dust mass derivation and since this work is focused on the study of the gradients, 
we do not attach any particular importance to the values of   ${\rm DGR}$.
We use these values to scale the profiles in Figs.~\ref{fig:masses-prof} and \ref{fig:masses-prof-app}.
The estimated $\alpha_{\rm CO}$ values are instead more directly related to the relative gradients of the various ISM components.

The $\alpha_{\rm CO}$ values derived with or without the correction for metallicity gradient are collected in Table~\ref{tab:xco}. 
In this table, column (1) gives the galaxy names, column (2) the metallicity gradients with uncertainty from \citet{pilyugin14}, 
column (3) the oxygen abundances at the radius of $0.7 \times r_{25}$ that is equivalent  to the solar radius in our Galaxy,
columns (4) and (5) $\alpha_{\rm CO}$ values derived without and with the correction for metallicity gradient, respectively, 
and columns (6) and (7) mean $\alpha_{\rm CO}$ values derived from 
\citet{sandstrom13} for the galaxies in common with our sample for the entire galaxies and the inner galactic kpc, respectively.

\begin{sidewaystable*}
\caption{Metallicity information and derived $\alpha_{\rm {CO}}$ values for face-on DustPedia galaxies with data both for atomic and molecular gas, and
molecular gas profiles fitted by an exponential function$^{1}$. 
\label{tab:xco} }
\begin{tabular}{lcccccccc}
\hline\hline
Galaxy 	& (O/H) Gradient$^{2}$	& 12 + log (O/H)$_{0.7r25}$$^{3}$ & $\alpha_{\rm {CO}}$$^{5}$ &  $\alpha_{\rm {CO}}$$_{\rm (O/H)}$$^{6}$ & $\alpha_{\rm {CO, \ S13}}$$^{7}$ 
& $\alpha_{\rm {CO, \ cen, \ S13}}$$^{8}$ \\
& [dex~$r_{\rm 25}$$^{-1}$]	&								& [M$_\odot$~pc$^{-2}$~(K km s$^{-1}$)$^{-1}$] 	& [M$_\odot$~pc$^{-2}$~(K km s$^{-1}$)$^{-1}$] & [M$_\odot$~pc$^{-2}$~(K km s$^{-1}$)$^{-1}$] 
& [M$_\odot$~pc$^{-2}$~(K km s$^{-1}$)$^{-1}$] \\
(1)						& (2)									& (3)										& (4) & (5) & (6) & (7) \\
\hline
NGC 5457 (M 101)	&	$-0.840 \pm 0.029$ 	&	$     	8.12	\pm 0.03	$	&	3.63  	& 3.18  	& $2.3 - 2.9$ & $0.35 \pm 0.20 $ 	\\
NGC 2403		&	$-0.524 \pm 0.043$	&	$	8.11	\pm 0.05	$	& 	2.97		& 2.42  							\\
IC 342			&	$-0.513 \pm 0.087$	&	$	8.47	\pm 0.10	$	& 	0.48		& 0.27 							\\
NGC 5194 (M 51)	&	$-0.324 \pm 0.053$	&	$	8.65	\pm 0.06	$	&	3.55		& 2.17  							\\
NGC 5055 (M 63)	&	$-0.490 \pm 0.031$	&	$	8.53	\pm 0.04	$	& 	1.83   	& 1.45   	& $3.7 - 4.0$ &  $ 1.00 \pm 0.25 $ 	\\
NGC 6946		&	$-0.337 \pm 0.099$	&	$	8.48	\pm 0.12 	$	& 	0.28		& 0.26 	& 	$1.8 - 2.0$ & $ 0.40 \pm 0.31 $	\\
NGC 925			&	$-0.473 \pm 0.037$	&	$	8.15	\pm 0.04	$	&	9.06		& 7.71 	& 10.0	\\
NGC 628 (M 74)	&	$-0.572 \pm 0.027$	&	$      8.38	\pm 0.03 	$	&	6.38		& 4.80 	& 	$3.9 - 5.1$ & $ 2.24 \pm 0.24 $	 \\
NGC 4725		&	$-0.472 \pm 0.853$	&	$	8.50	\pm 0.93 	$	& 	0.26		& 0.25 	& 	$1.2 - 1.8$	& $ 0.72 \pm 0.71 $ \\
NGC 3521		&	$-0.666 \pm 0.096$	&	$	8.36	\pm 0.10	$	& 	7.78  	& 4.05 	& $7.3 - 7.6$	\\
NGC 4736 (M 94)	&	$-0.066 \pm 0.170$	&	$	8.52	\pm 0.17	$	&	0.37 		& 0.37  	& $1.0 - 1.1$	& $ 0.28 \pm 0.17 $ \\
\hline\hline
\end{tabular}
\tablefoot{
1) NGC~3031 (M~81) is not listed in this table, although it has data both in atomic and molecular gas because its molecular gas profile is not
fitted by an exponential function;
2) metallicity gradients from \citet{pilyugin14};
3) derived oxygen abundances at the radius of $0.7 \times r_{\rm 25}$;
4) and 5) $\alpha_{\rm {CO}}$ values derived neglecting metallicity gradients and taking into account metallicity gradients following prescriptions given in Sect.~\ref{sec:xco};
6) and 7) $\alpha_{\rm {CO}}$ values derived from \citet{sandstrom13} for the entire galaxy and the inner kpc of the galaxy, respectively.}
\end{sidewaystable*}

The $\alpha_{\rm {CO}}$ values are in the range $(0.3 - 9)$~M$_\odot$~pc$^{-2}$~(K km s$^{-1}$)$^{-1}$ both 
without and with the correction of H{\sc i} and H$_2$ profiles for metallicity.   
The $\alpha_{\rm {CO}}$ determinations obtained taking into account the metallicity gradient are always  
lower than (or equal to) those derived neglecting the metallicity gradient. 
The higher discrepancy between the two $\alpha_{\rm {CO}}$ derivations ($\alpha_{\rm {CO}}$/$\alpha_{\rm {CO}}$$_{\rm (O/H)}$$\sim$1.9) 
is found for NGC~3521, the galaxy with the second most steep metallicity gradient in our sample,  
$\sim$0.7~dex~$r_{\rm 25}$$^{-1}$.
This could be due to the very small number of oxygen abundances (9) used to derive the metallicity gradient for NGC~3521
\citep[see Fig.~1 of ][]{pilyugin14}, giving statistically weak metallicity gradients with respect to those of other galaxies 
(e.g., NGC~628 and NGC~5457~(M~101) where the metallicity gradients are derived from several tens of oxygen abundances).
The salient point is that the correction of gas profiles for metallicity gradient has no significant influence on the $\alpha_{\rm {CO}}$ 
value at least for the present galaxy sample.  
This happens although O/H gradients of our sample range from $-0.066$~dex~$r_{25}$ (almost flat) 
to $-0.840$~dex~$r_{25}$ \citep[very steep, see for instance][for examples of flat and steep 
metallicity gradients]{magrini16}.
Because of this, we did not explore the possibility of a further dependence of the $\alpha_{\rm CO}$ value on the metallicity gradient. 

The almost invariance of the $\alpha_{\rm {CO}}$ value by adopting a fixed DGR or a DGR depending on the metallicity gradient
can mean that $\alpha_{\rm {CO}}$ is mainly constrained by the central DGR value in the galaxy 
or that the DGR values derived from both approaches are different to give the same resulting $\alpha_{\rm {CO}}$.   
For our sample the explanation could be a compromise of both causes: in the central regions dominated by molecular gas the DGR values  
are almost independent on metallicity gradient, while in outer parts rich of atomic  gas the DGR values seem to be metallicity gradient dependent. 
The final result is that $\alpha_{\rm {CO}}$ is independent on the metallicity gradient for our sample.

\subsection{CO-to-H$_2$ values:  comparison with the literature}
\label{sec:literature}
Many authors derived $\alpha_{\rm {CO}}$ using a variety of techniques:
measuring total gas masses from $\gamma$-ray emission plus a model for the cosmic ray distribution
\citep[e.g.,][]{strong96,abdo10}, using the observed velocity dispersion and size of the molecular cloud to obtain virial masses
\citep[e.g.,][]{solomon87,wilson95,bolatto08,gratier12}, and modeling multiple molecular gas lines with varying optical depths and
critical densities \citep[e.g.,][]{weiss01,israel09a,israel09b}.
We do not summarize results of the literature, but we focus on the comparison between our derivations of $\alpha_{\rm {CO}}$ and those of \citet{sandstrom13}
who solved for $\alpha_{\rm {CO}}$ with a method similar to ours and for eight galaxies present in our sample.

In four (NGC~5457 (M101), NGC~3521, NGC~925, NGC~628 (M74)) of the eight galaxies in common with \citet{sandstrom13}, 
our $\alpha_{\rm {CO}}$ derivations referred to the galactic region within $r_{\rm 25}$ (or minor if gas mass profiles are not available until to $r_{25}$)
are similar to (for 3 on 4 galaxies consistent with) their values within $r_{25}$ (see column (6) in Table~\ref{tab:xco}).  
For the other four galaxies (NGC~5055 (M63), NGC~4736 (M94), NGC~6946, NGC~4725), there is instead an evident discrepancy between 
the two $\alpha_{\rm {CO}}$ determinations, with our values lower than those of \citet{sandstrom13} within $r_{25}$ but more similar to 
their determinations for galaxy centers (see column (7) in Table~\ref{tab:xco}). 
\citet{sandstrom13} indeed found relatively flat $\alpha_{\rm {CO}}$ profiles as a function of galactocentric radius aside from in the galaxy centers 
(within the inner $\sim$$0.1 \times r_{\rm 25}$) that have $\alpha_{\rm {CO}}$ values lower than the galaxy mean ones by factors of 2 -- 3 (or more).
 
The reason of the discrepancy for four galaxies in common with the sample of \citet{sandstrom13} is difficult to establish, likely
due to differences between the two methods for deriving $\alpha_{\rm {CO}}$ 
(e.g., the work of Sandstrom et al. 2013 is based on circular regions while we used radially averaged profiles),
and to how different methods correlate with the ISM's properties of individual galaxies.
Local environmental conditions (metallicity, galactic dynamics, ISM pressure, interstellar radiation field strength, 
optical depth of the ISM, gas temperature, dust properties) can produce variations in $\alpha_{\rm {CO}}$ 
\citep{sandstrom13}. 
Additionally, our sample contains some galaxies defined as SF or H{\sc ii} region dominated (e.g., NGC~5457 (M~101), NGC~6946),     
and others with signatures of AGN or LINER activity (e.g., NGC~4736 (M~94)).

\subsection{Caveats about the derivation of CO-to-H$_2$ conversion factor}
\label{sec:limits}
The method we developed to derive $\alpha_{\rm {CO}}$ could not fit the dust and total gas profiles for the entire galaxy sample
(see Figs.~\ref{fig:masses-prof} and \ref{fig:masses-prof-app} for the shape of dust and gas profiles of the single galaxies).
The assumption that the DGR is not a function of atomic/molecular phase could, for instance, be the 
cause of the low $\alpha_{\rm {CO}}$ values derived for NGC~4736 (M~94) and NGC~6946. 
In NGC~4736 (M~94), the H{\sc i} gas profile has indeed a behaviour similar to the H$_2$ one, with a central peak and 
a strong decrease towards the outer parts, while the H{\sc i} gas distribution is typically low in the inner parts and 
it increases at large radius.
In NGC~6946, the gas profiles show expected trends but the H$_2$ one is strongly peaked in the center with respect to the increasing 
of the H{\sc i} distribution in the outer parts.
Implicit with our assumption on the DGR is the assumption that the dust properties do not change significantly across the different environments
that are best sampled by the {\em Herschel} resolution. In the current method, and in the analogous works from the literature we referred to,
the dust mass is derived using the same optical properties, those of the diffuse medium, regardless of the (unknown) contribution of dust
associated to molecular or atomic gas in a specific position within a galaxy. The assumption could be valid if the evolution of dust in denser environments
(and in particular the changes in optical properties of grains resulting from coagulation and accretion; see, e.g., \citealt{kohler15}) are restricted
to regions with a small filling factor with respect to the global distribution of molecular gas. On this assumption rely not only the entire analysis of
this paper, as stated in Sect. 5.1, but also the many dust mass determinations in the literature: only detailed modelling of the various phases of
the ISM in a galaxy, of the optical properties of dust in each phase, and of the local heating conditions will help in the future to understand if the
use of a single, diffuse-medium, dust model still provide a reasonable estimate, or instead introduce a strong bias, in the determination of the dust mass
surface density or global values.

The galaxy NGC~628 (M~74) gives the best results of our method.
First, its scale-length ratio $h_{\Sigma{\rm {tot\,gas}}}$/$h_{\Sigma{\rm {dust}}}$ is around the unity and this means that the assumption that 
the total gas profile follows the dust one is reasonable for this galaxy.
Secondly, dust and gas profiles are available for the same radius range.
Under these conditions the method we developed gives a $\alpha_{\rm {CO}}$ consistent with determinations from other authors \citep[e.g.,][]{sandstrom13}. 

The scale-length ratio <$h_{\Sigma{\rm {tot\,gas}}}$/$h_{\Sigma{\rm {dust}}}$> averaged over the galaxy sample is $\sim$1.1, and the excess,
although small, with respect to the unity could suggest that dust and gas are not so well mixed for all galaxies, differently from what we assumed in our method to derive $\alpha_{\rm {CO}}$.
This should simply mean that the DGR varies non-linearly with metallicity, as in the outskirts H$_2$ is negligible.
For a large (126) sample of local Universe galaxies covering a 2~dex metallicity range, \citet{remy-ruyer14} found that the chemical evolution of galaxies 
(especially for low-metallicity galaxies), traced by their DGR, strongly depends on their local internal conditions and individual histories. 
Finally, the scale-length ratio <$h_{\Sigma{\rm {tot\,gas}}}$/$h_{\Sigma{\rm {dust}}}$> larger than 1 could also be due to the fact that for some galaxies 
the H$_2$ gas profile is available for a smaller radius with respect to the dust and H{\sc i} profiles (e.g., NGC~5055 (M~63), NGC~2403).

\section{Conclusions}
\label{sec:conclusions}  
We studied the radial distribution of dust, stars, gas, and SFR in a sub-sample of 18 spiral face-on galaxies extracted from the DustPedia sample.
The selection was based on galaxies imaged over their whole extent with both PACS and SPIRE in $\textit{Herschel}$, with $1 \le$ T $\le 8$, 
(d/D)$_\mathrm{sub-mm} \ge 0.4$, and D$_\mathrm{sub-mm}\ge 9\arcmin$. 
This study exploited the multi-wavelength (UV/sub-mm) DustPedia database, together with 
millimeter and centimeter maps extracted from other sources and information on metallicity abundances available in the literature.
We fitted the surface brightness profiles of the tracers of dust and stars, the mass surface density profiles of dust, stars, molecular gas, and total gas, and the SFR surface density 
with a simple exponential curve and derived their exponential scale-lengths. 

We identified average trends of the scale-lengths with wavelengths:
the scale-length decreases from UV to 70~$\mu$m, and  
this effect can be due to an intrinsic gradient in the stellar populations with a blueing galaxy disk at larger distance from the center,
to a preferential reddening in the inner regions of a galaxy, or to the combined role of dust and stars in shaping the gradient. 
The scale-length increases again from 70 to 500~$\mu$m, most likely because of the cold-dust temperature gradient with galactocentric 
distance. 
The scale-lengths of the dust mass surface density are larger than those of the surface brightness profiles at 500~$\mu$m, 
generally larger than any of the other surface-brightness scale-lengths and than the stellar mass surface density distributions. 
On average, the dust surface density scale-length is about 1.8 times the stellar density one derived from IRAC data 
or the 3.6~$\mu$m surface brightness scale-length, a result in agreement with radiative transfer analysis of dust extinction
in edge-on galaxies.
This result could be explained by a change in the typical lifetimes of grains against destruction by shocks,
with a longer lifetime at larger radii.
We also found that UV and dust scale-lengths are similar and this could be associated with the inside-out galaxy formation scenario that
is used to explain, together with dust extinction, the larger scale-lengths of radiation from the younger stellar
component.
As for the gas, the H$_2$ surface density profiles have typically scale-lengths $\sim$20\% smaller than 
the NIR surface brightness, and -on average- a factor 2.3 smaller than the dust ones. 
The total gas surface density has $\sim$10\% flatter profiles than dust, though the result is influenced 
by the $\alpha_{\rm CO}$ value.

By studying the trends of the scale-lengths as a function of Hubble stage T, 
we found that the ratio between the scale-lengths of the $\textit{Herschel}$ bands 
and the 3.6~$\mu$m scale-length tends to increase from earlier to later types, 
and this trend is particularly evident for the ratio $h_{\rm 70\,\mu m}$/$h_{\rm 3.6\,\mu m}$.
The value for $h_{\rm 70\,\mu m}$ instead tends to be smaller than that at longer sub-mm wavelengths 
with ratios between longer sub-mm wavelengths and 70~$\mu$m that decrease with increasing T 
(see $h_{\rm 250\,\mu m}$/$h_{\rm 70\,\mu m}$).  
These trends could be the result of a bulge contribution, escaped our simple
1-D analysis, to dust heating for earlier type galaxies.     
Among the surface density ratios, only  $h_{\Sigma{\rm {SFR}}}$/$h_{\rm 3.6\,\mu m}$ shows a
weak trend of increasing SFR scale-length, with respect to stars, towards later types, compatible with 
longer infall time scales for late type galaxies.

We developed a method to solve for single $\alpha_{\rm CO}$ conversion factor for each sample galaxy with both atomic and molecular gas data, 
by assuming that the total gas profile mimics the dust one.
For the derivation of $\alpha_{\rm CO}$, we also took taken into account the dependence of DGR on metallicity. 
We found values of $\alpha_{\rm CO}$ in the range $(0.3 - 9)$~M$_\odot$~pc$^{-2}$~(K km s$^{-1}$)$^{-1}$ both 
with and without the correction of H{\sc i} and H$_2$ profiles for metallicity.   
The determinations of $\alpha_{\rm {CO}}$  obtained taking into account the metallicity gradient are always  
lower than (or equal to) those derived neglecting the metallicity gradient. 
Anyway, the correction of gas profiles for metallicity gradient has no significant influence on the $\alpha_{\rm {CO}}$ value 
at least for our galaxy sample with mean metallicity ranging from 12~+~log(O/H)~$ \sim 8.3$ to 12~+~log(O/H)~$ \sim 8.7$.

As an extension of this work, in our next papers in the frame of the DustPedia project, 
we are about to compare single S{\'e}rsic decomposition parameters retrieved 
for the whole DustPedia sample galaxies for the WISE W1 and \textit{Herschel} observations. 
Thus, we will be able to directly compare  effective radii of the stellar population and dust.

\begin{acknowledgements}
We  thank  an  anonymous  referee  for  useful  comments
and  suggestions  that  improved  the  quality  of  the  manuscript. 
DustPedia is a collaborative focused research project supported by the European Union under the Seventh Framework Programme (2007-2013) call (proposal no. 606824). 
The participating institutions are: Cardiff University, UK; National Observatory of Athens, Greece; Ghent University, Belgium; Universit\'{e} Paris Sud, France; 
National Institute for Astrophysics, Italy and CEA (Paris), France. 
J.~F. acknowledges the financial support from UNAM-DGAPA-PAPIIT IA104015 grant, Mexico.
This research made use of the NASA/IPAC Extragalactic Database (NED), which is operated by the Jet Propulsion Laboratory, 
California Institute of Technology, under contract with the National Aeronautics and Space Administration. 
We acknowledge the usage of the HyperLeda database (http://leda.univ-lyon1.fr).
This research has made use of the NASA/ IPAC Infrared Science Archive, which is operated by the Jet Propulsion Laboratory, 
California Institute of Technology, under contract with the National Aeronautics and Space Administration.
This publication makes use of data products from the Two Micron All Sky Survey, which is a joint project of the University of Massachusetts and 
the Infrared Processing and Analysis Center/California Institute of Technology, funded by the National Aeronautics and Space Administration and the National Science Foundation.
This research made use of the VizieR catalogue access tool, CDS, Strasbourg, France.
\end{acknowledgements}

%----------------

\begin{appendix}

\section{Figures}

\begin{figure*}[!ht]
\centering
{\includegraphics[width=8.0cm]{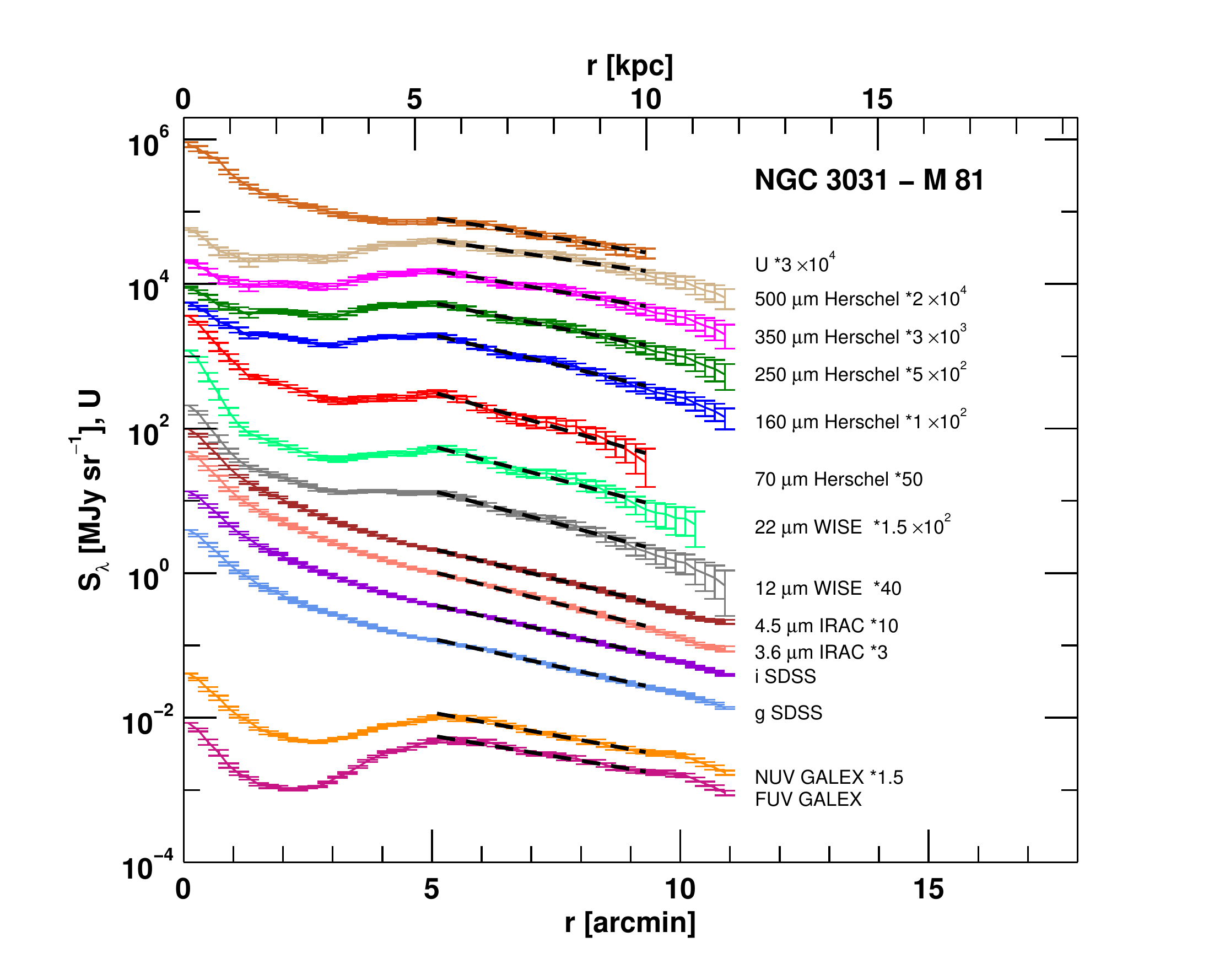}}
{\includegraphics[width=8.0cm]{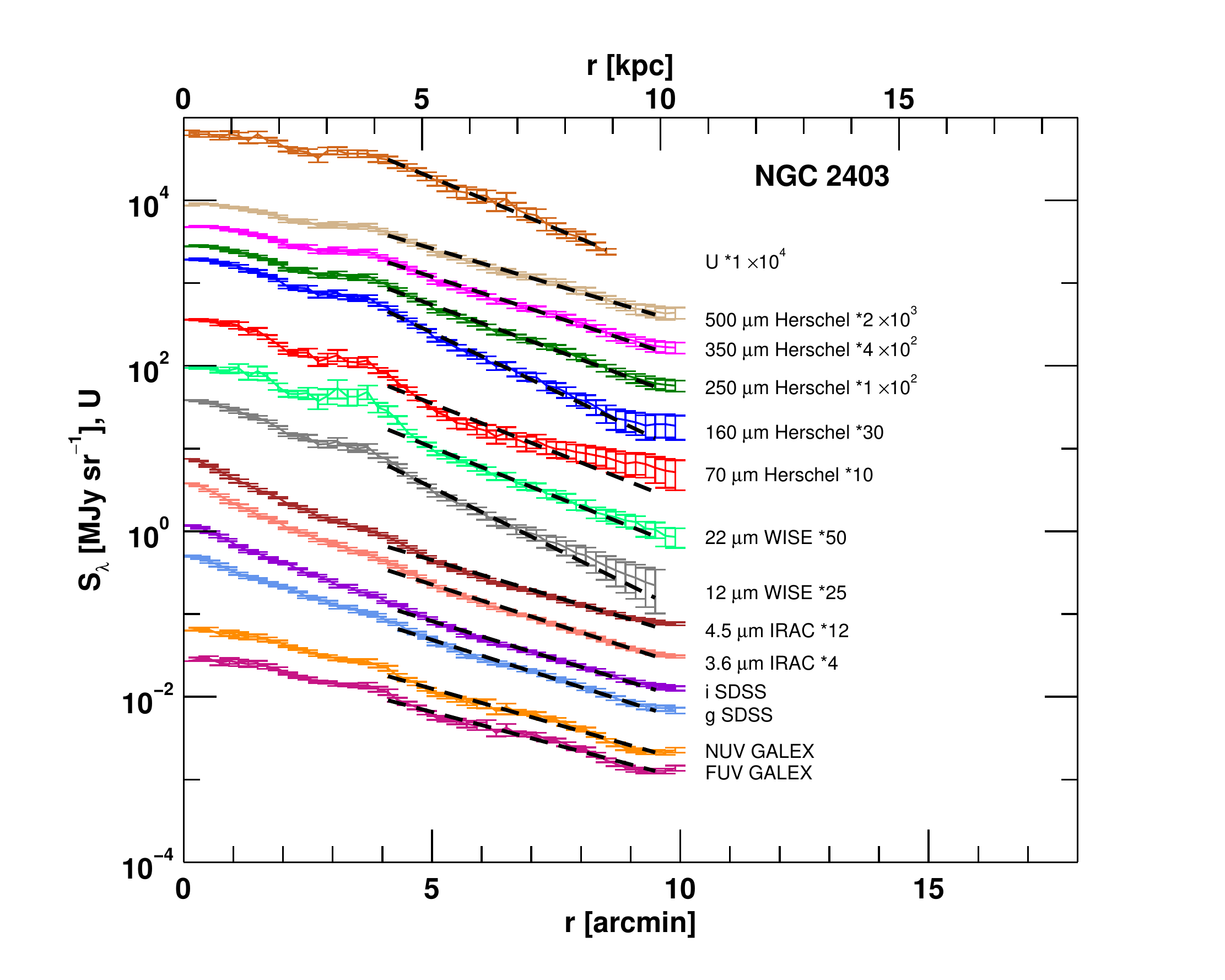}}\\
{\includegraphics[width=8.0cm]{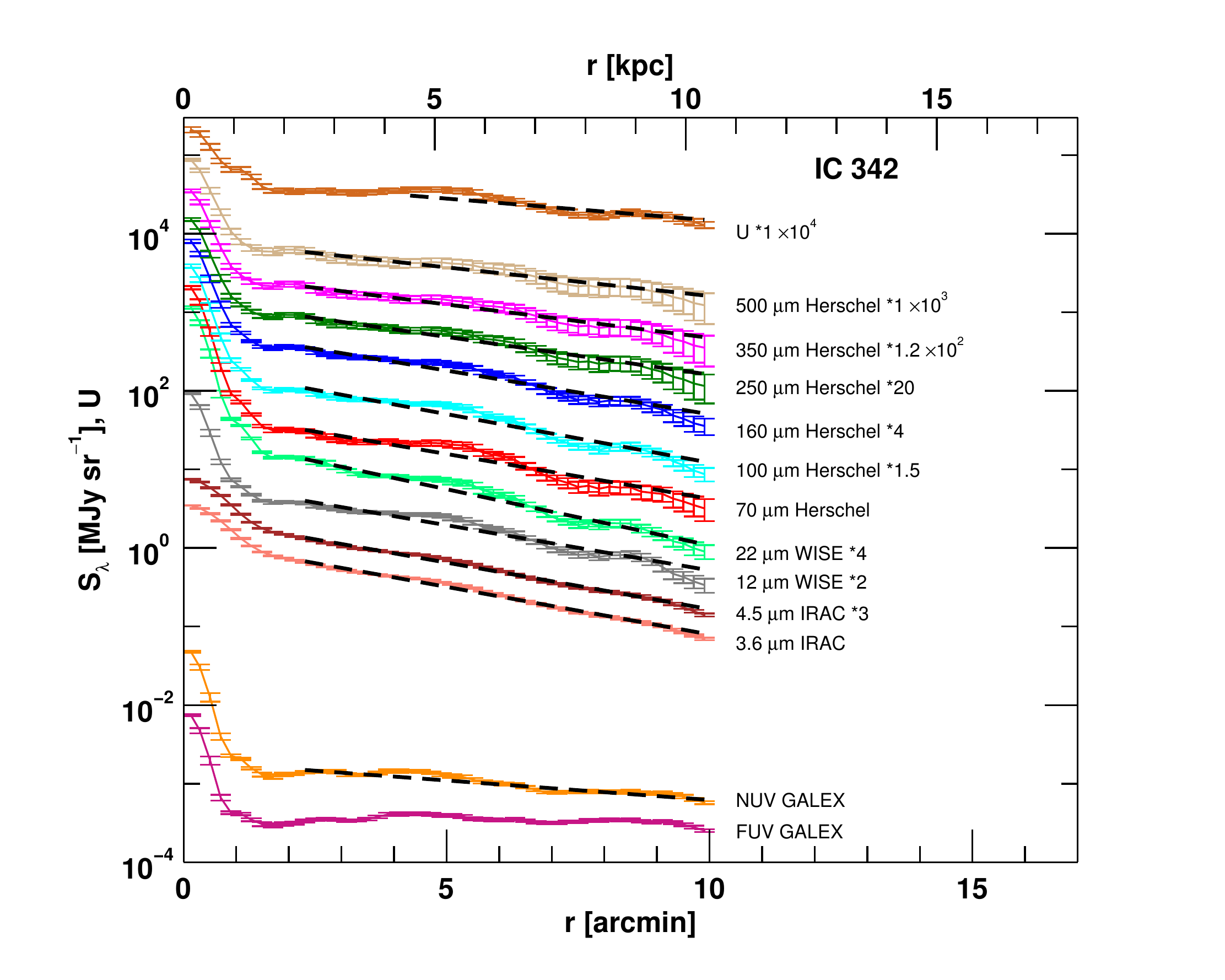}}
{\includegraphics[width=8.0cm]{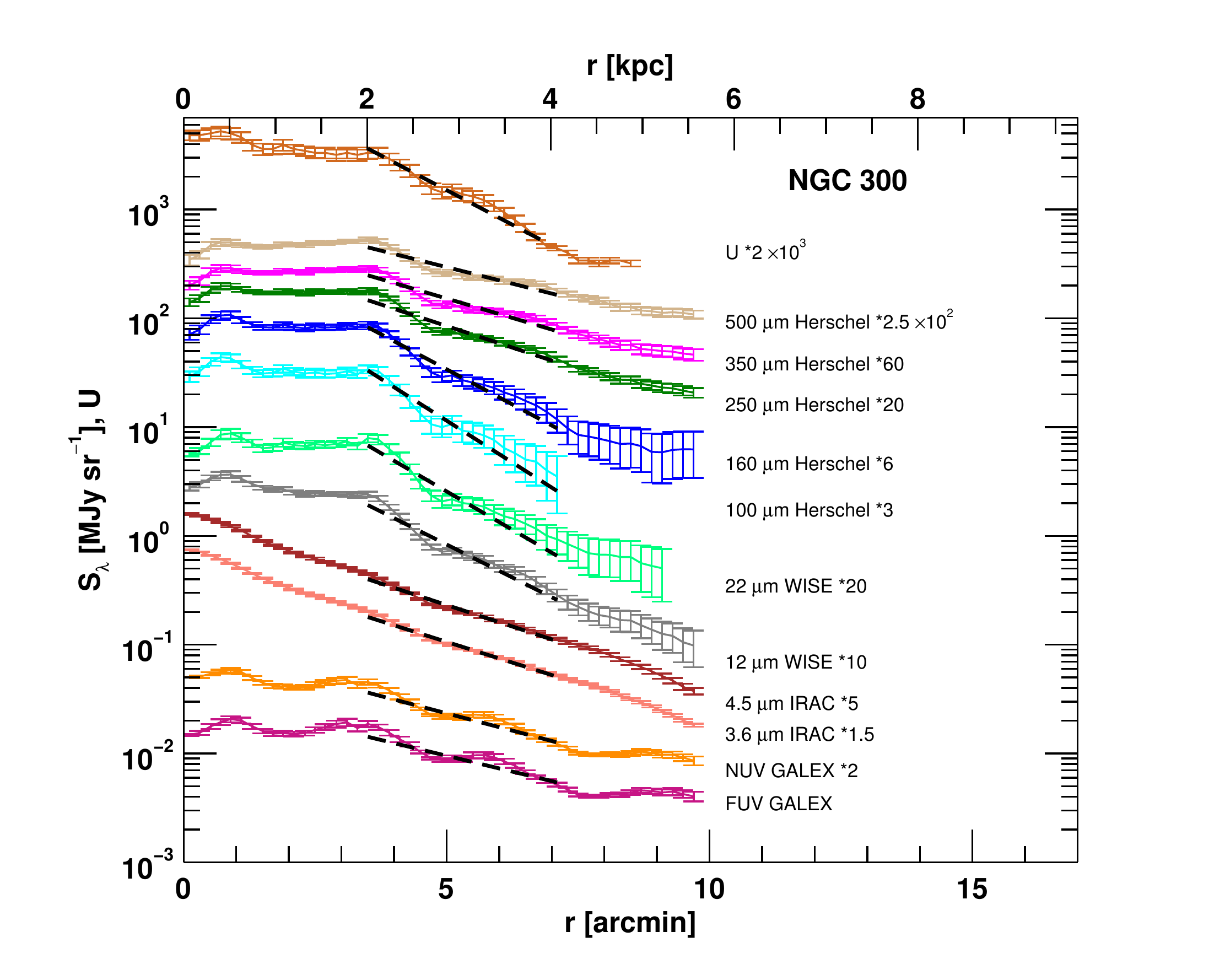}}\\
{\includegraphics[width=8.0cm]{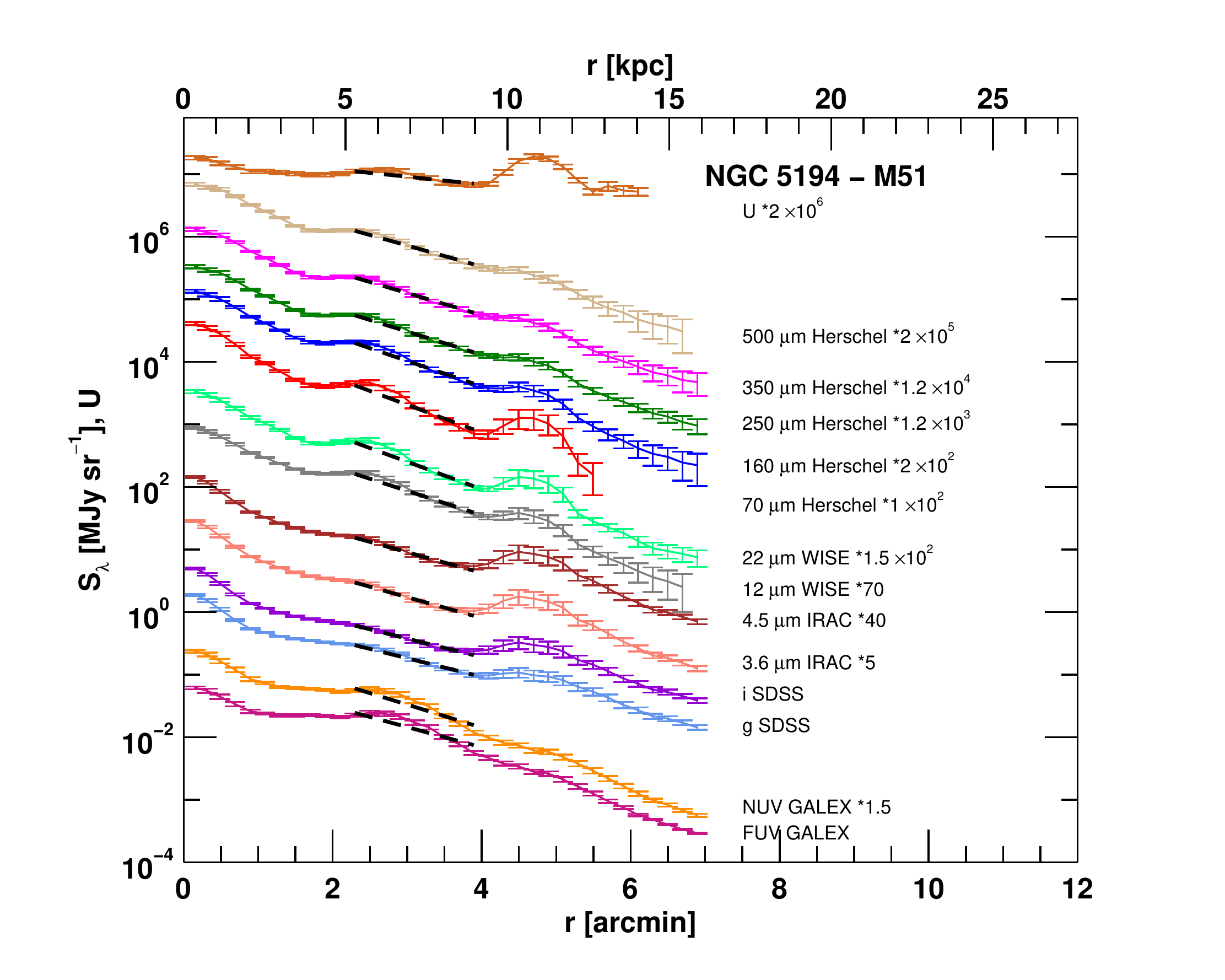}}
{\includegraphics[width=8.0cm]{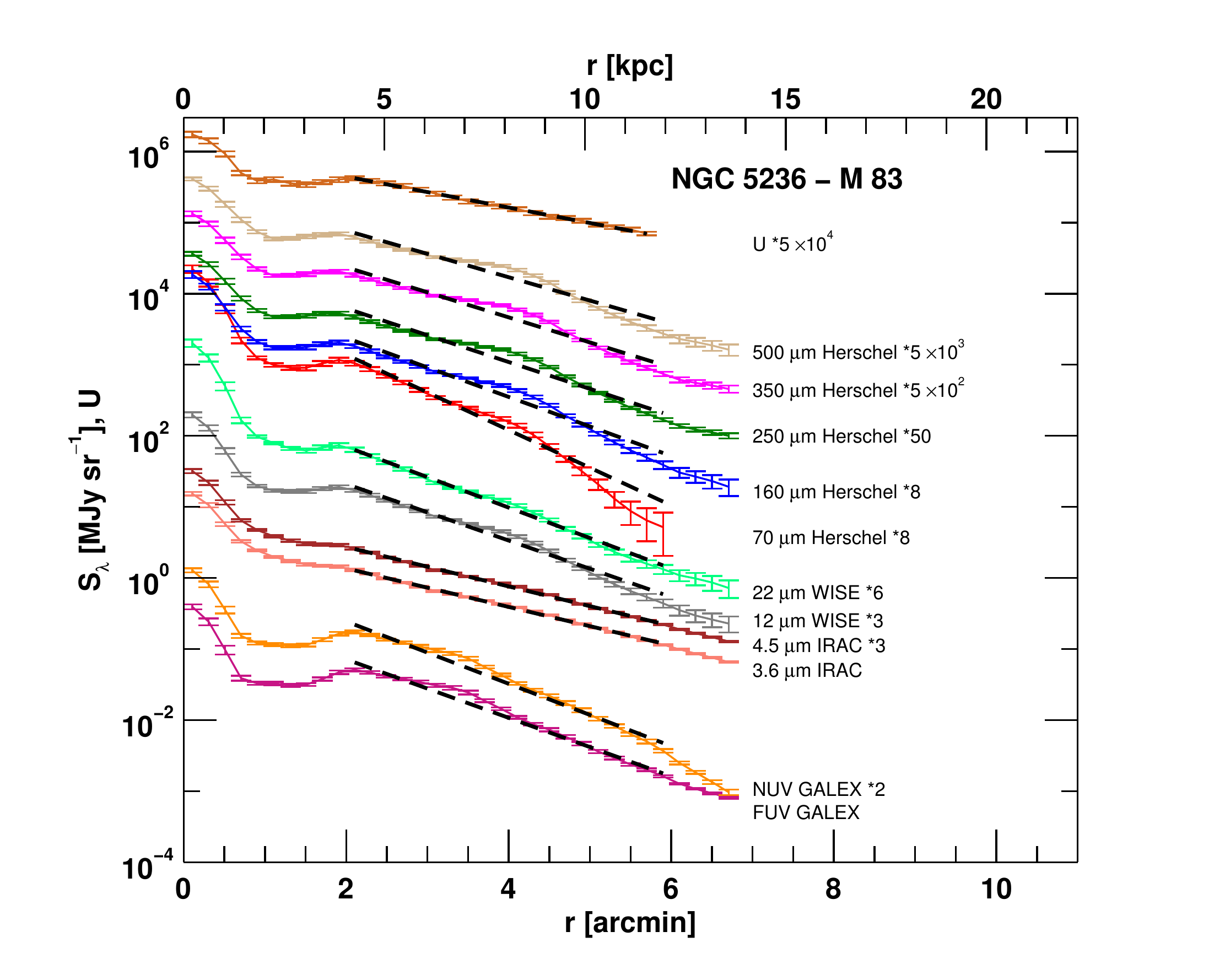}}
\caption{Multi-wavelength surface brightness and $U$ profiles for the face-on DustPedia sample, except for the galaxy NGC~5457~(M~101) already 
displayed in Fig.~\ref{fig:profiles}, shown up to $r_{25}$.
The profiles are shown as in Fig.~\ref{fig:profiles}.
The black dashed lines are exponential fits performed avoiding the central part of galaxies up to to the 
maximum extension of the 70~$\mu$m profile (or the 12~$\mu$m profile for NGC~6946 and the 100~$\mu$m 
profiles for NGC~1097 and NGC~300).}
\label{fig:profiles-app}
\end{figure*}

\begin{figure*}[!ht]
\centering
{\includegraphics[width=8.0cm]{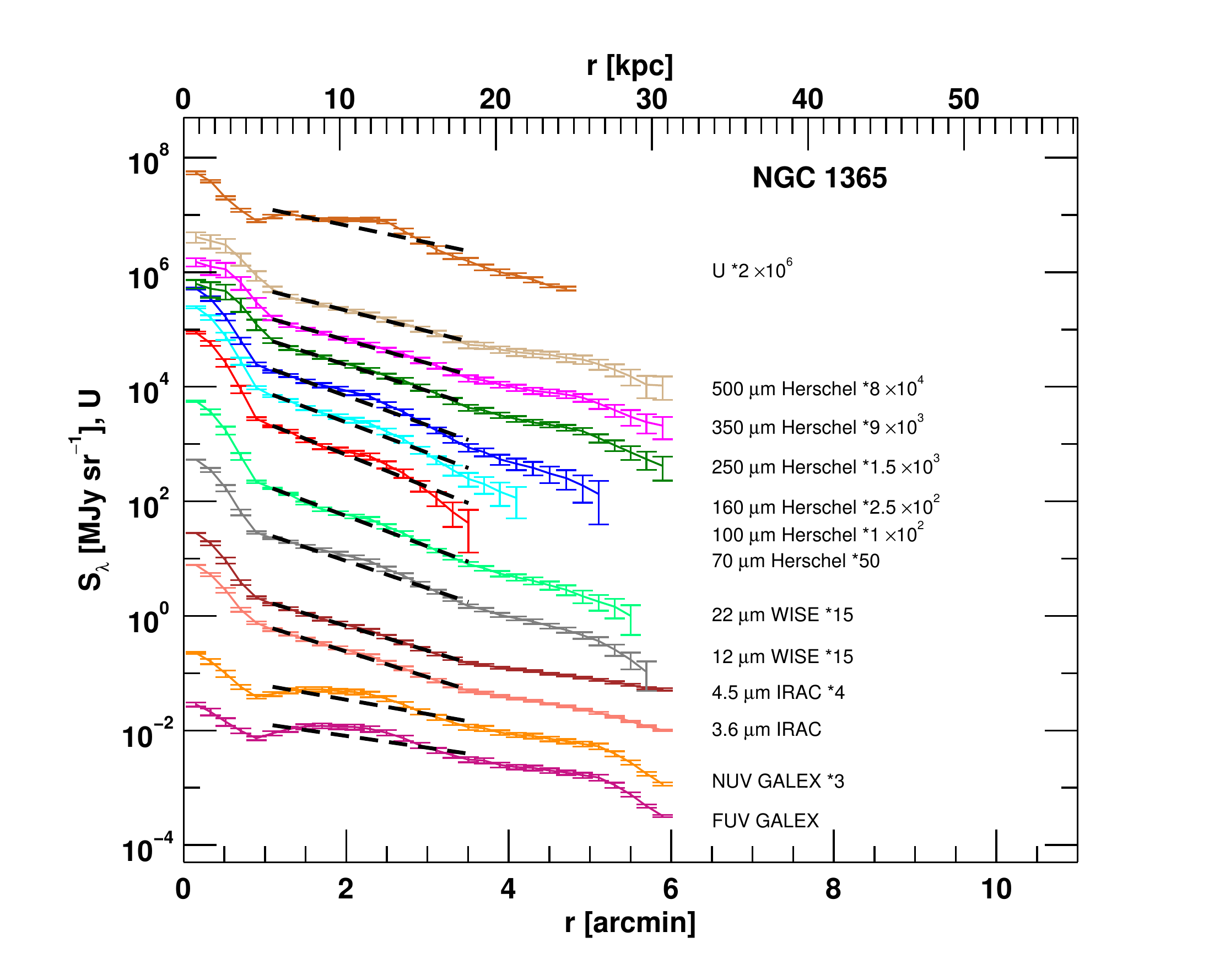}}
{\includegraphics[width=8.0cm]{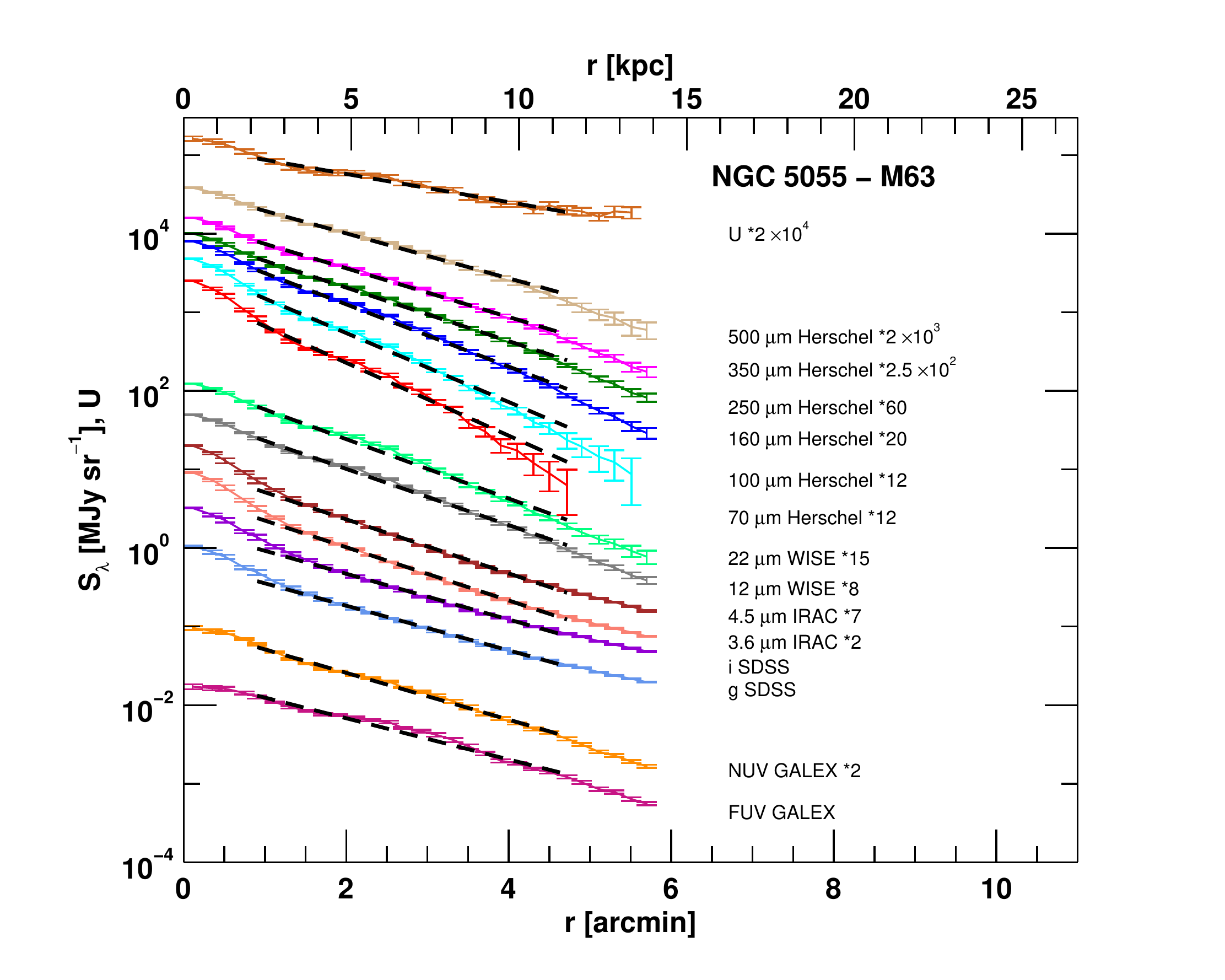}}\\
{\includegraphics[width=8.0cm]{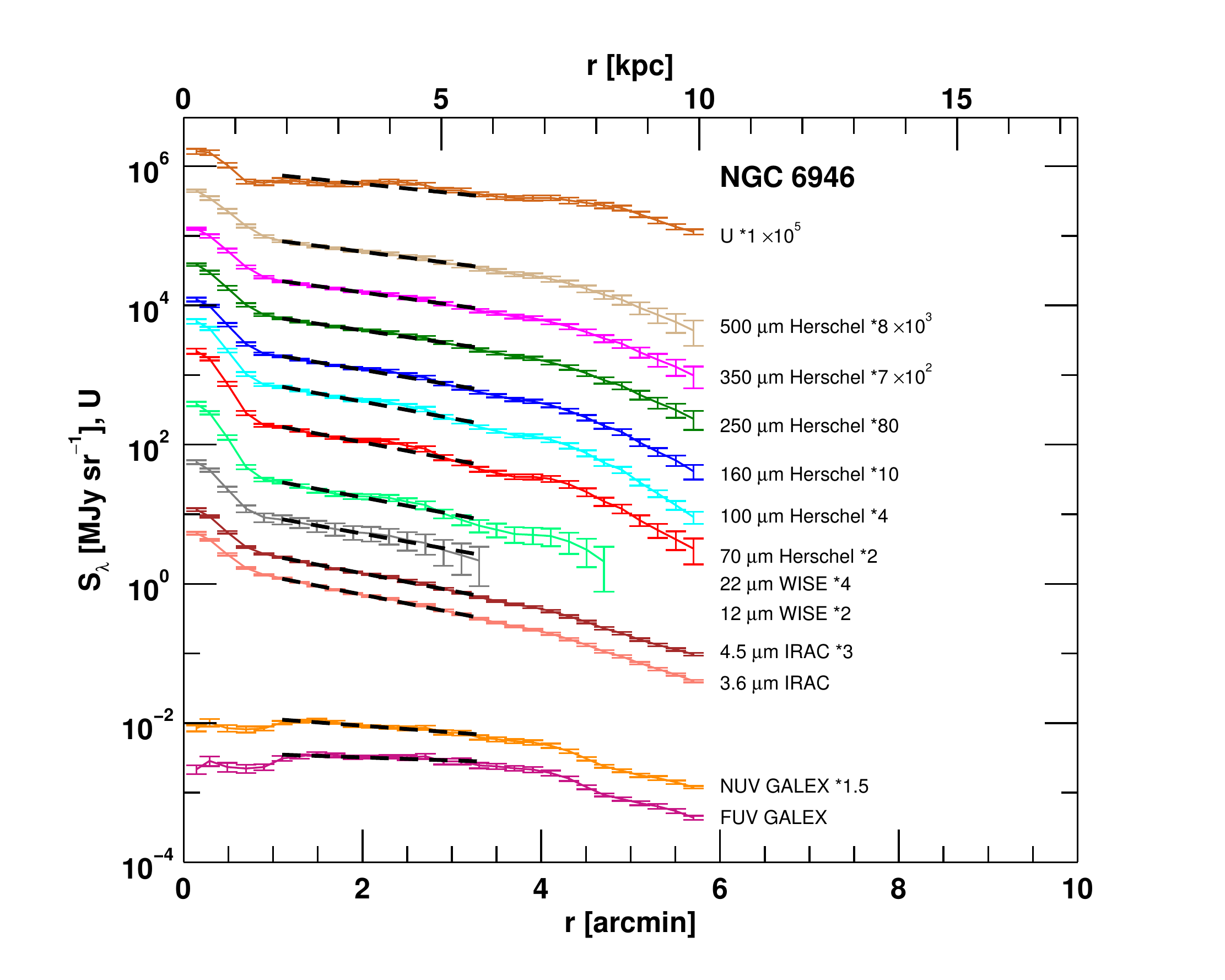}}
{\includegraphics[width=8.0cm]{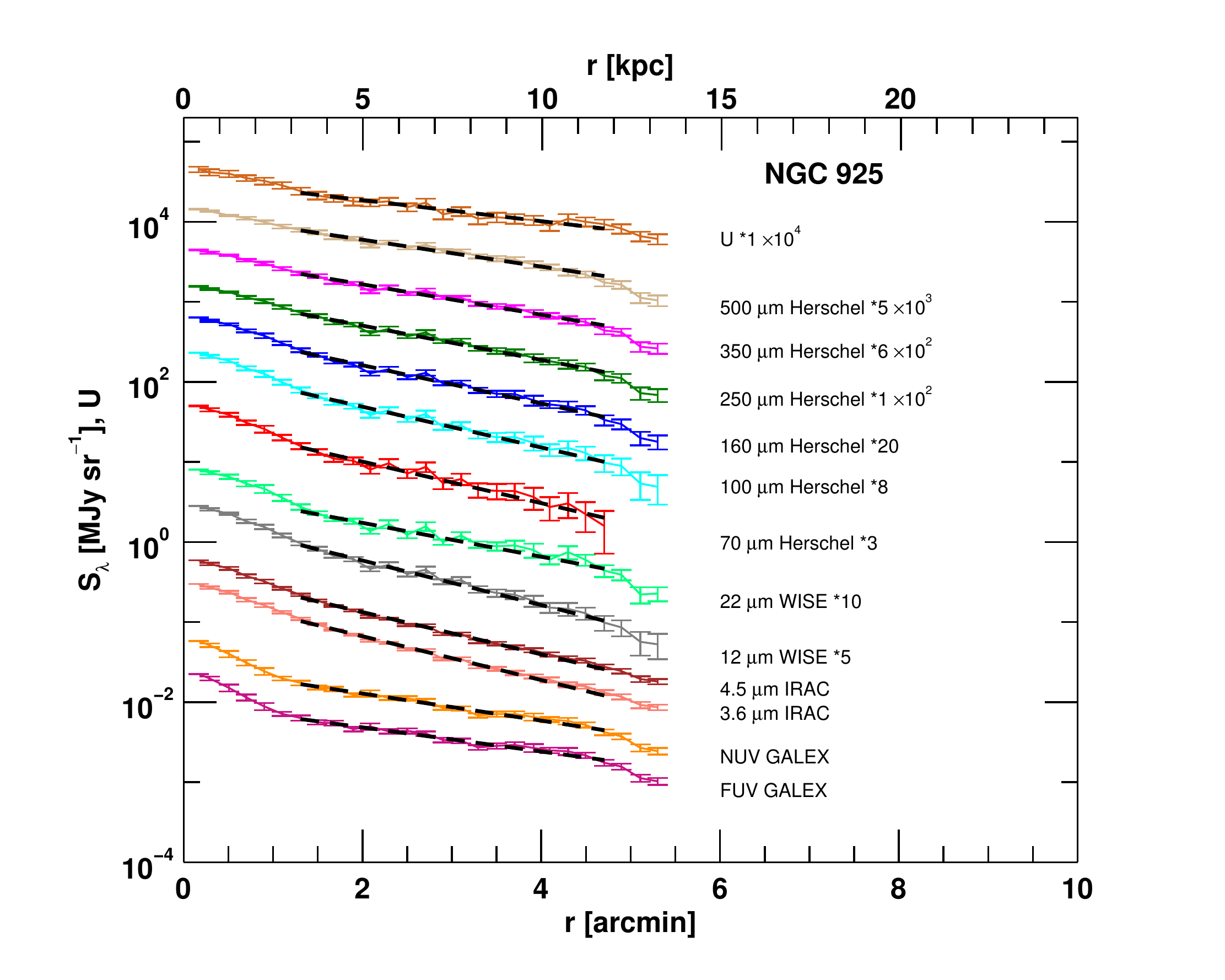}}\\
{\includegraphics[width=8.0cm]{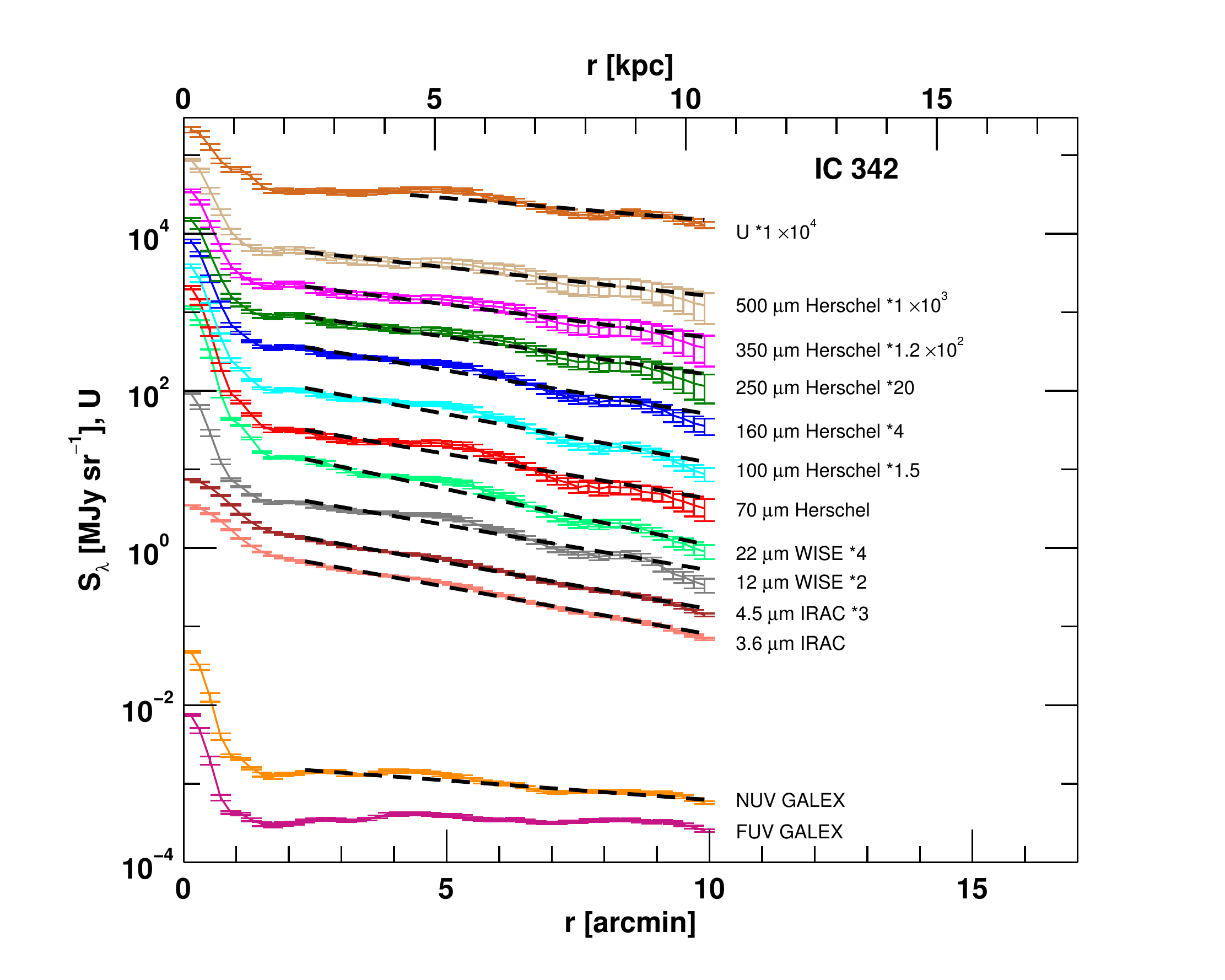}}
{\includegraphics[width=8.0cm]{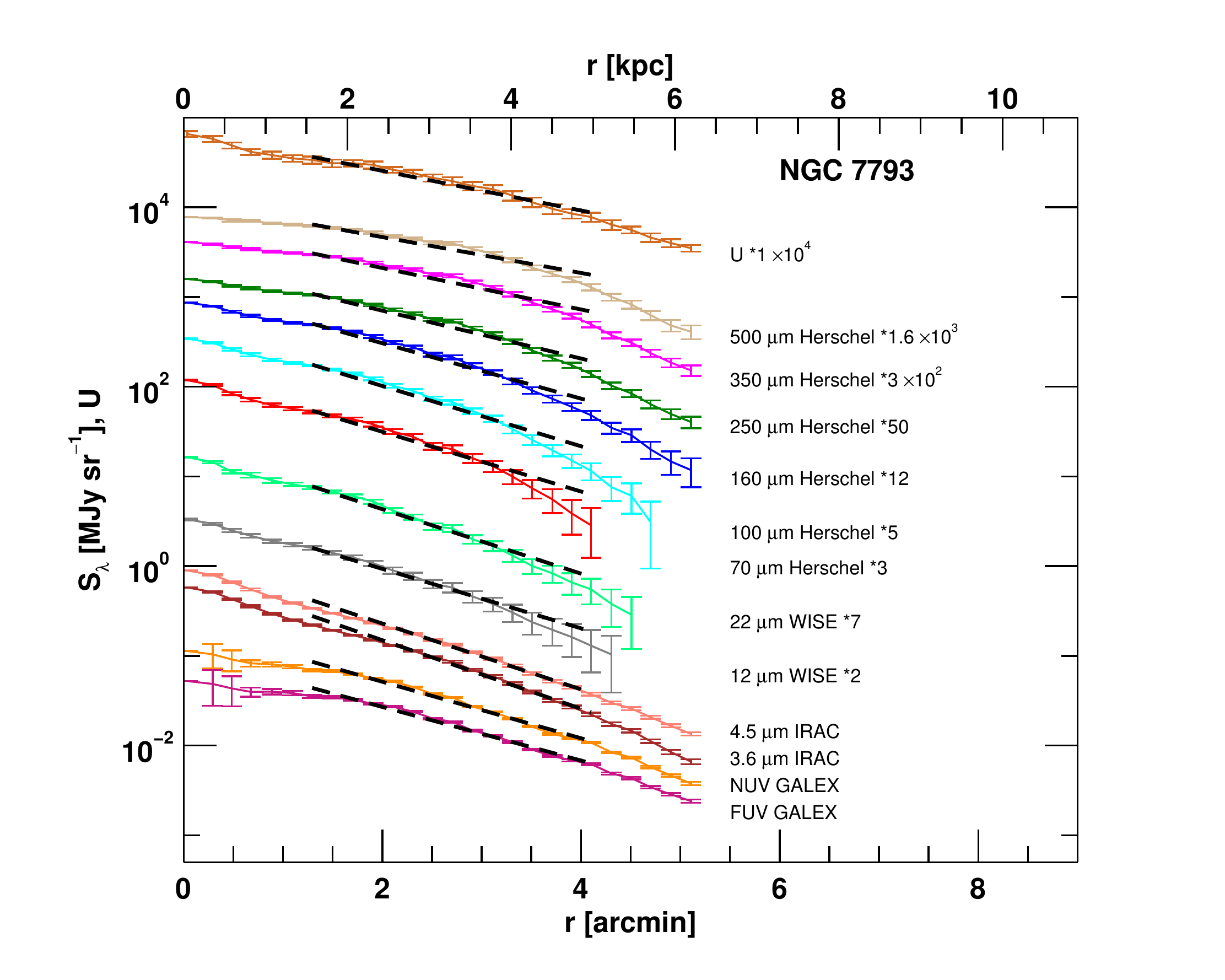}}
\caption*{Figure~\ref{fig:profiles-app} (continued)}
\end{figure*}

\begin{figure*}[!ht]
\centering
{\includegraphics[width=8.0cm]{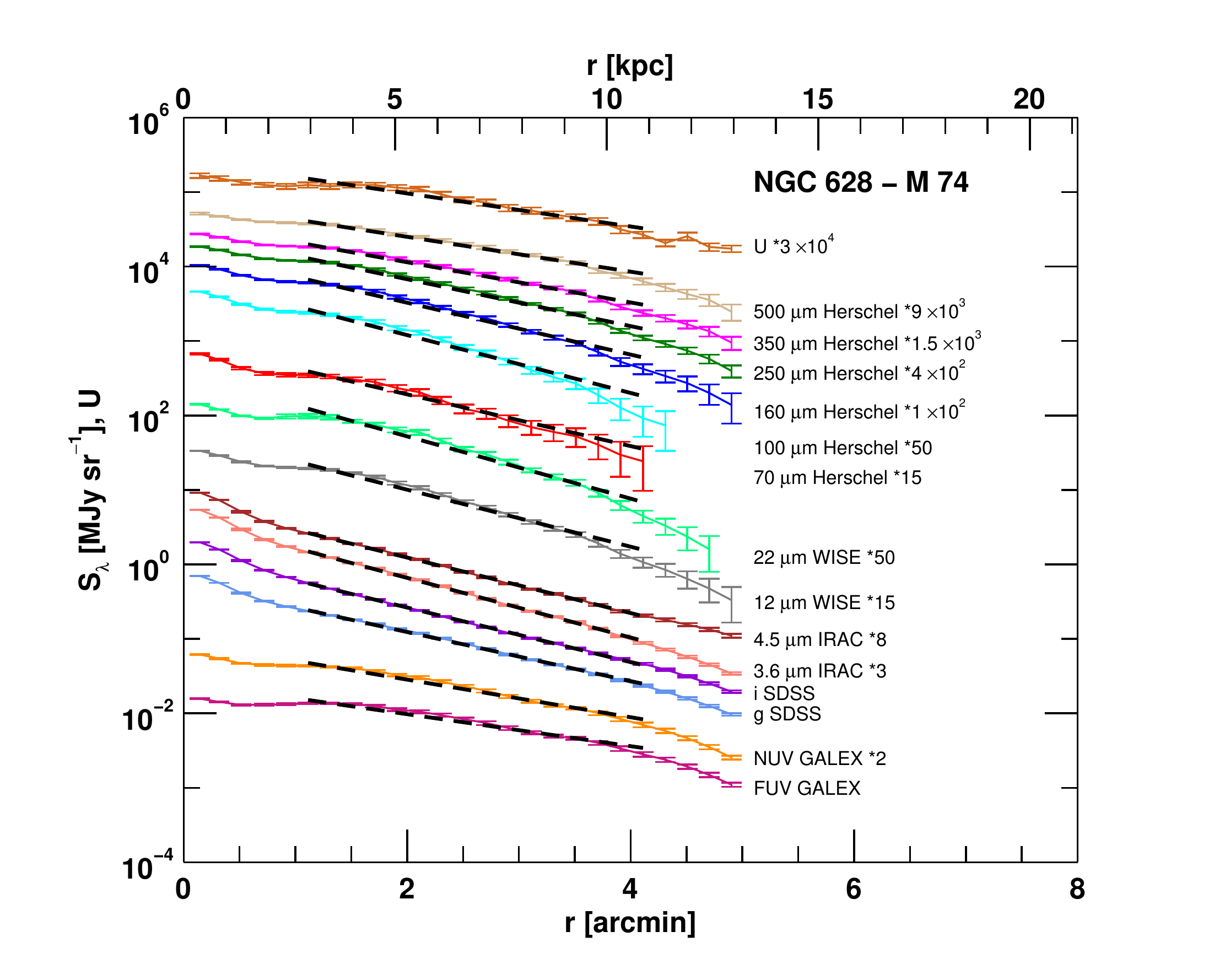}}
{\includegraphics[width=8.0cm]{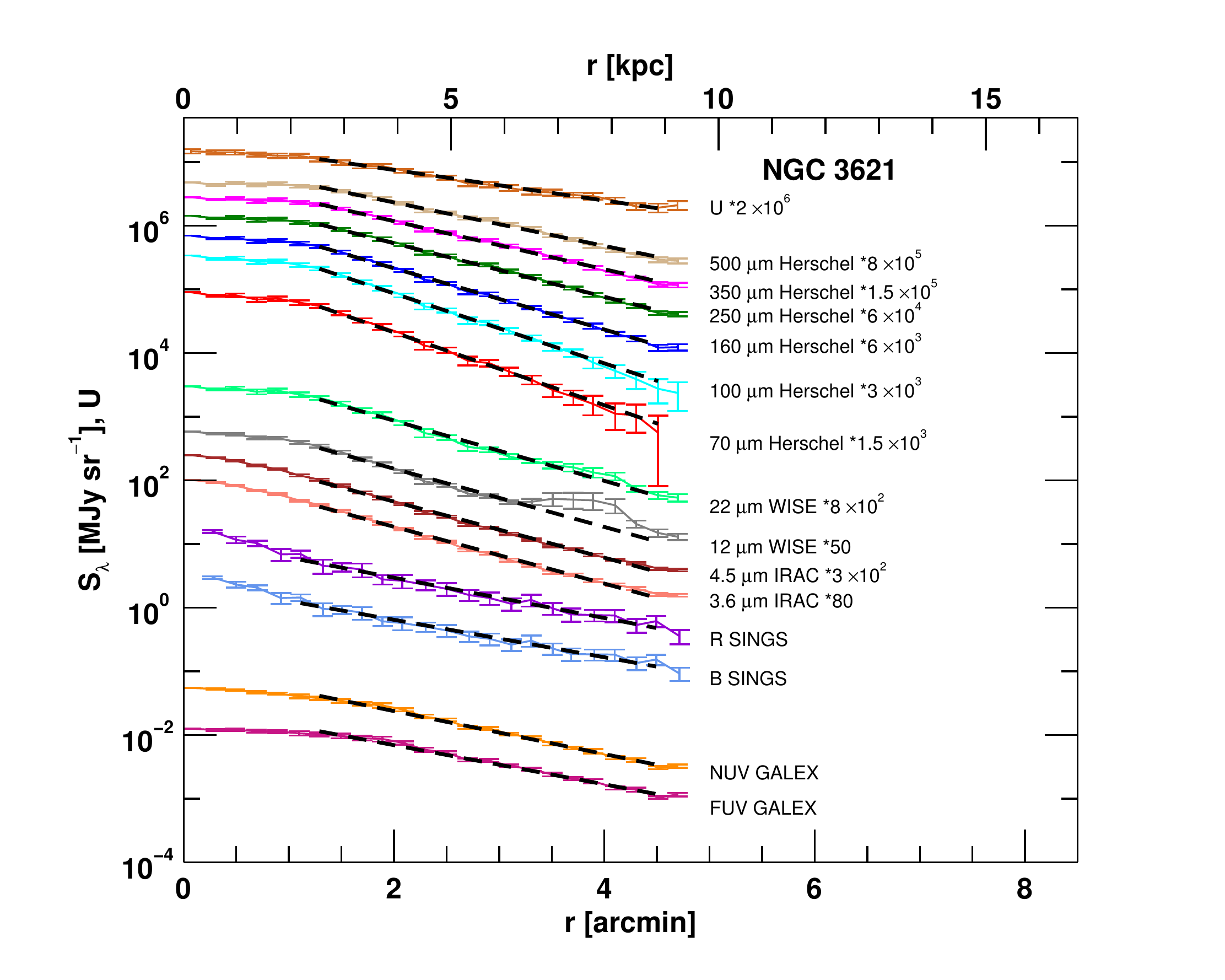}}\\
{\includegraphics[width=8.0cm]{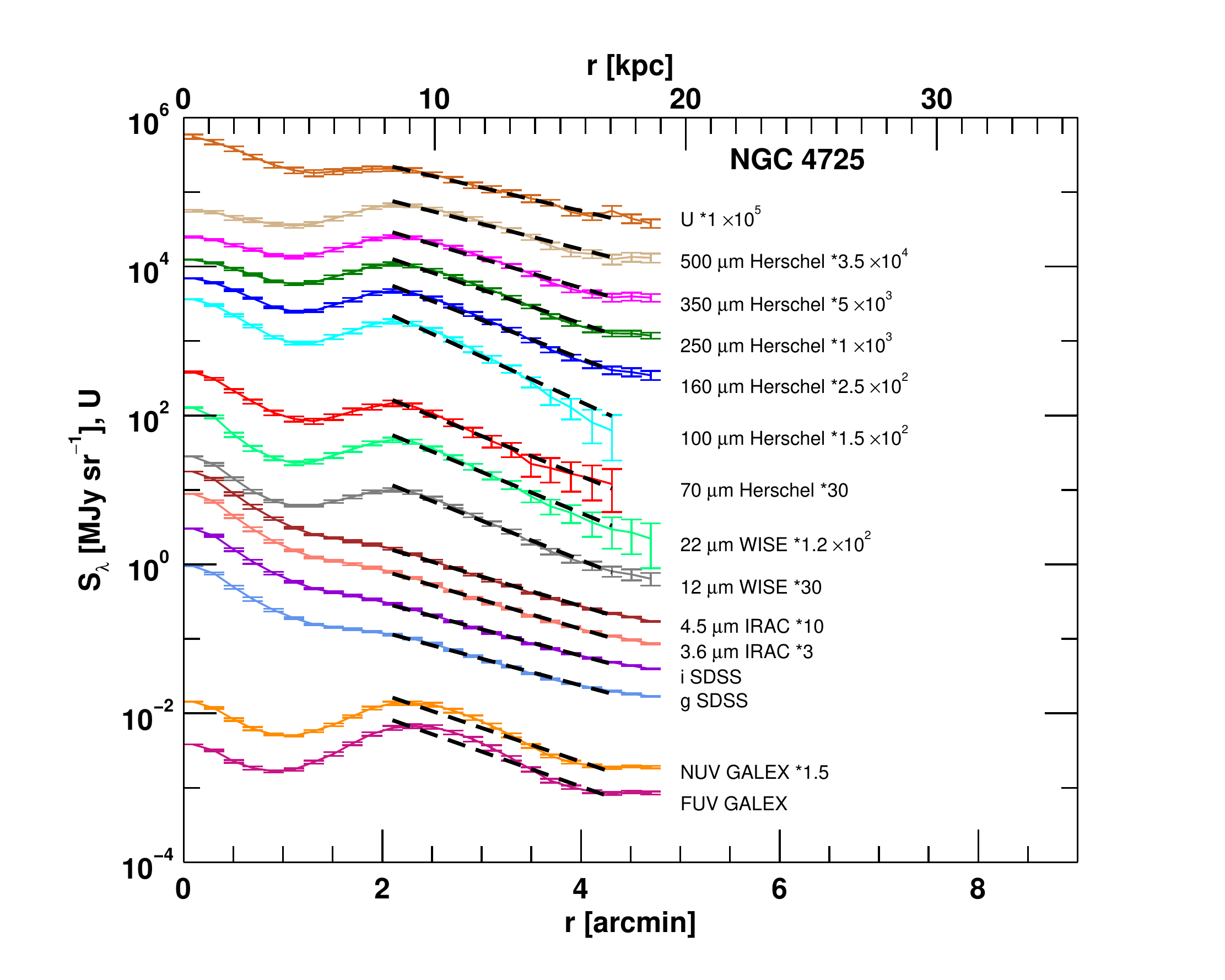}}
{\includegraphics[width=8.0cm]{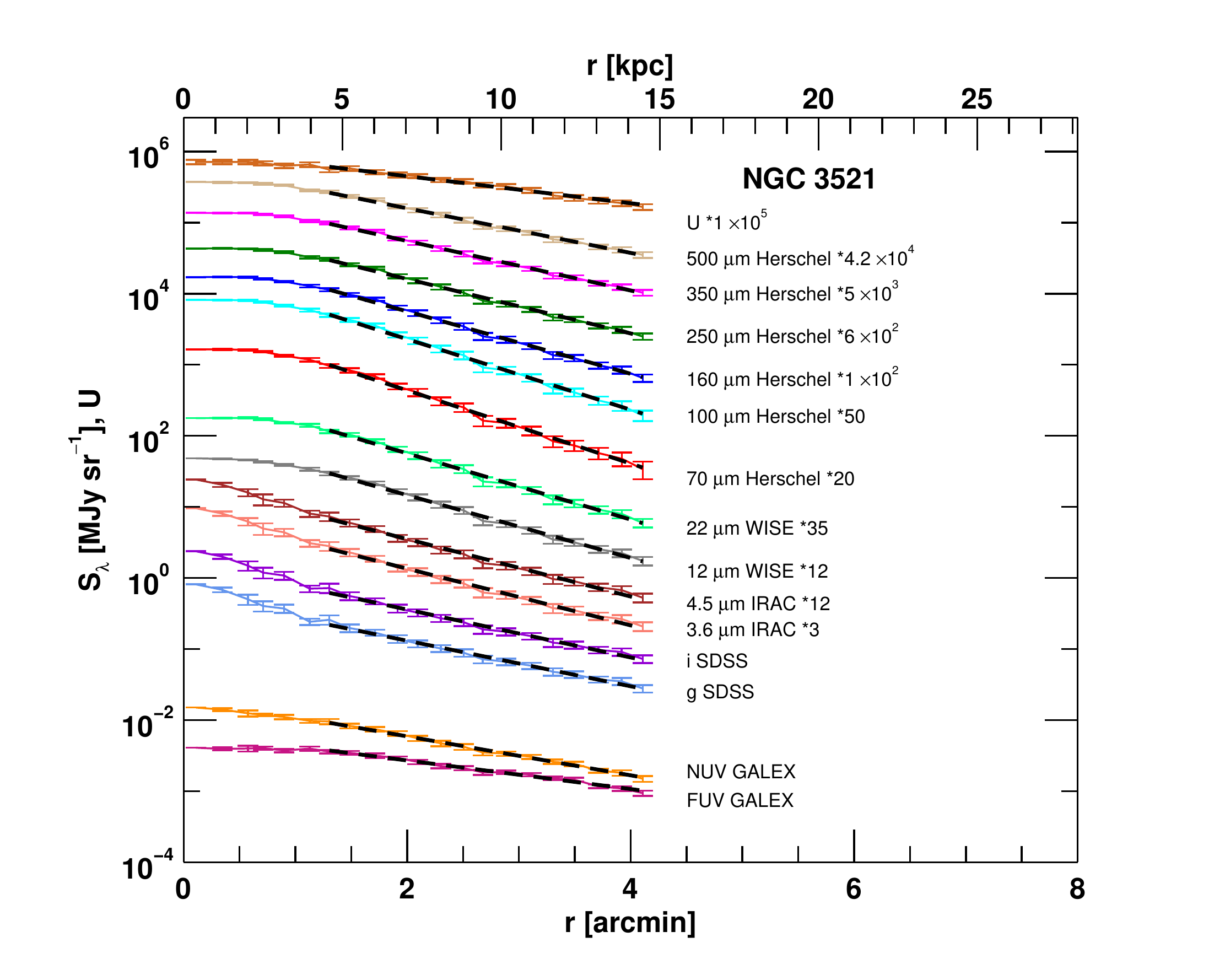}}\\
{\includegraphics[width=8.0cm]{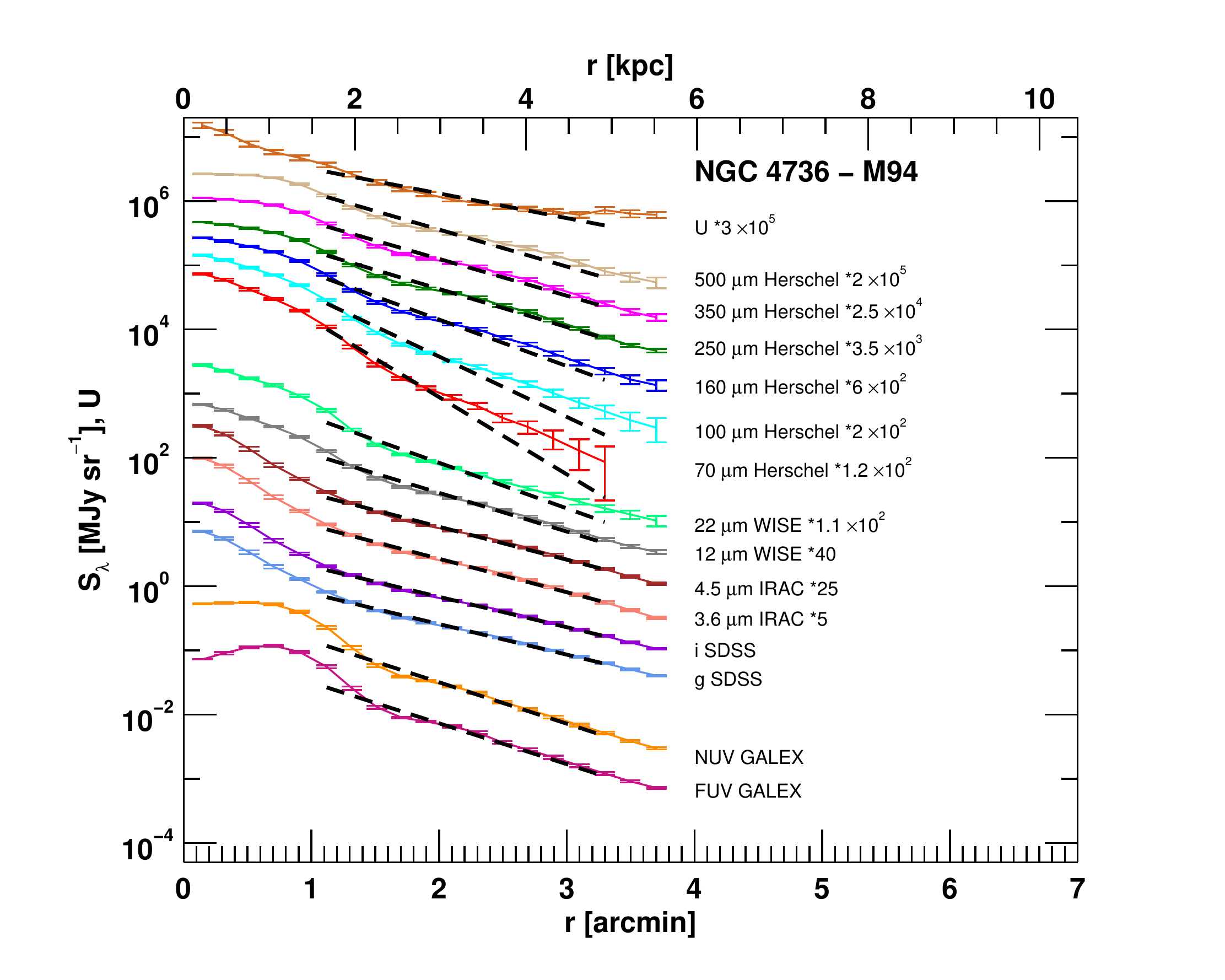}}
\caption*{Figure~\ref{fig:profiles-app} (continued)}
\end{figure*}

\begin{figure*}[!ht]
\centering
%{\includegraphics[width=6.0cm,angle=-90]{n3031-m81-sstar-ssfr_2.pdf}}
{\includegraphics[width=6.0cm,angle=-90]{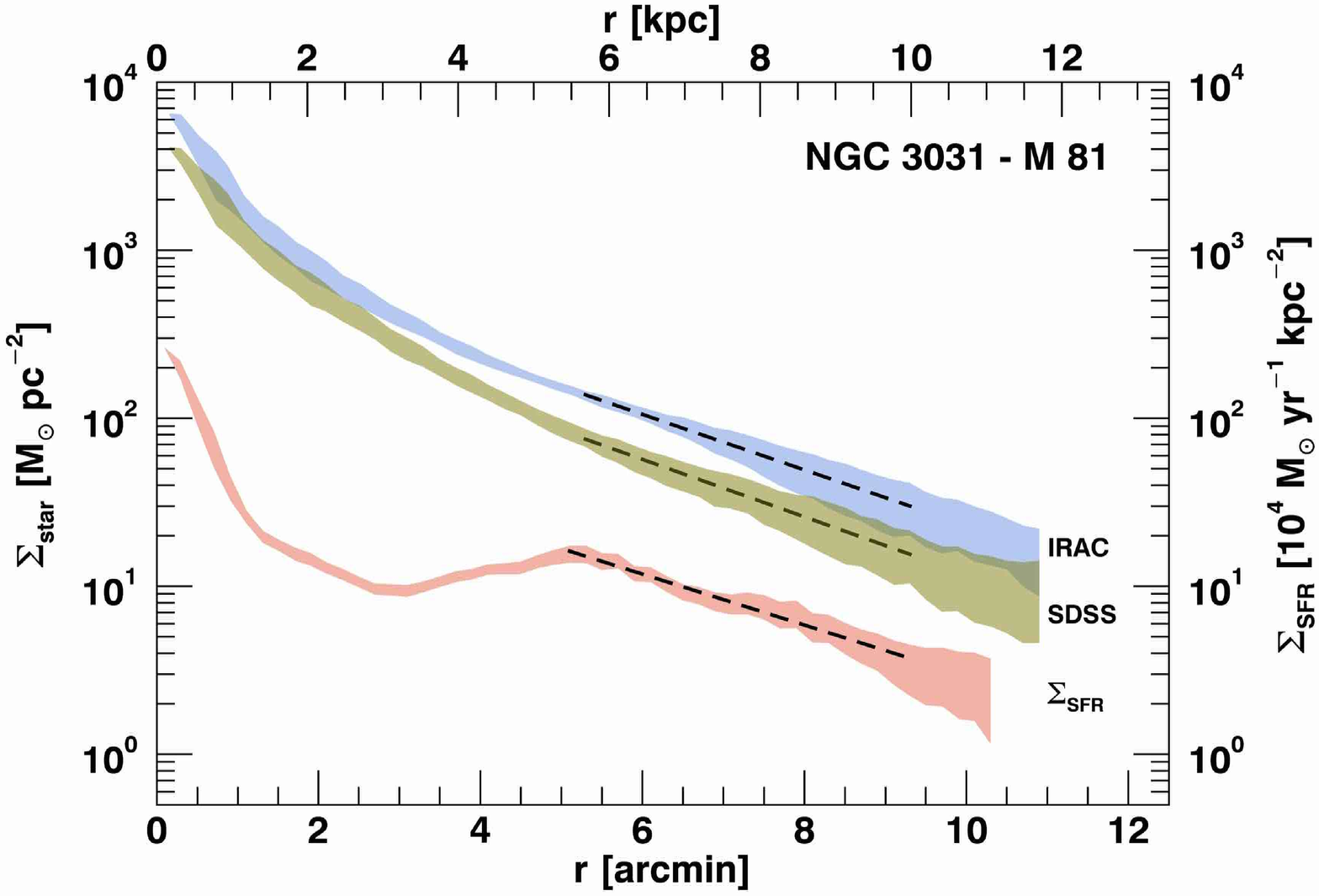}}
\hspace{0.5cm}
%{\includegraphics[width=6.0cm,angle=-90]{n3031-m81-smasses_2.pdf}}\\
{\includegraphics[width=6.0cm,angle=-90]{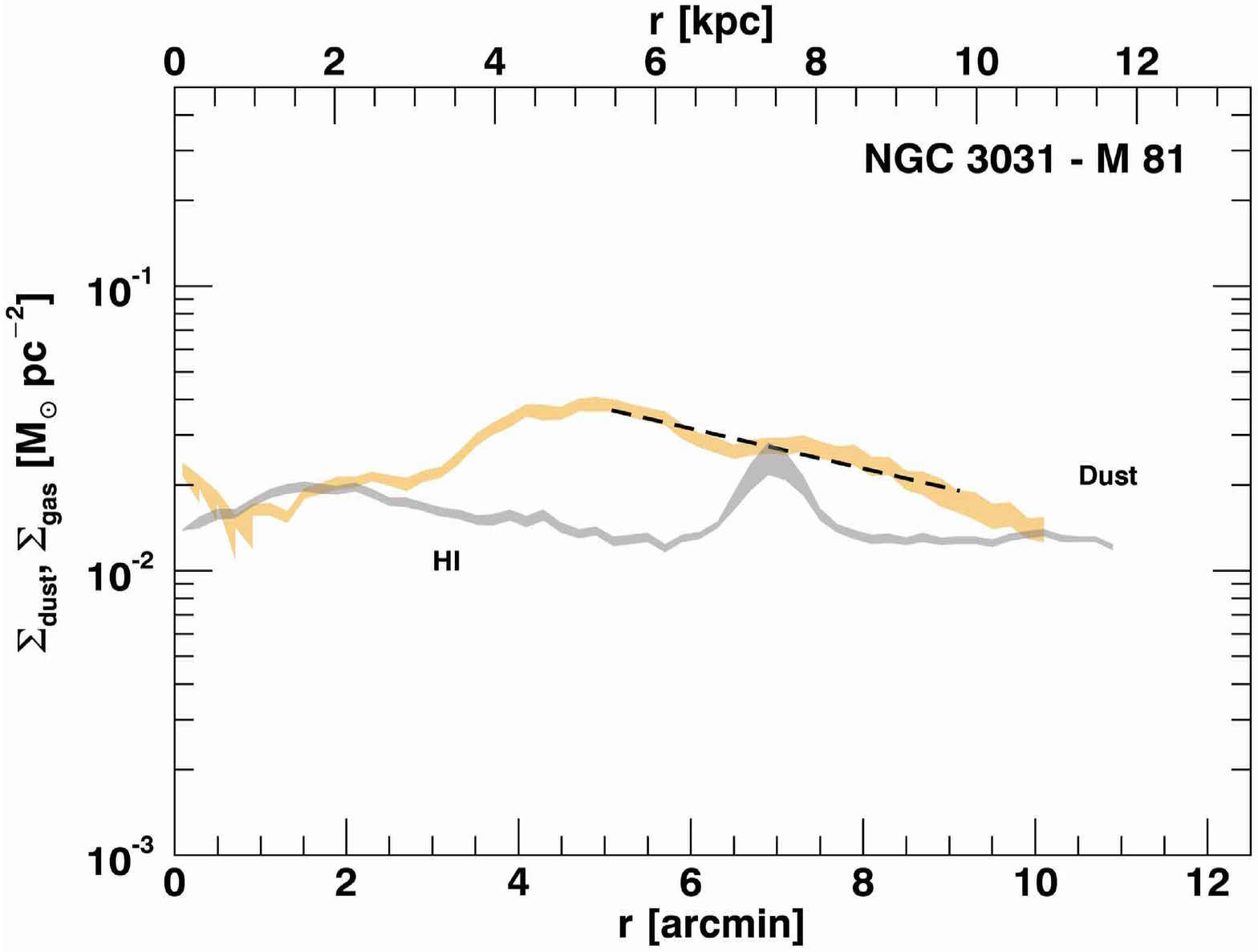}}\\
%{\includegraphics[width=6.0cm,angle=-90]{n2403-sstar-ssfr_2.pdf}}
{\includegraphics[width=6.0cm,angle=-90]{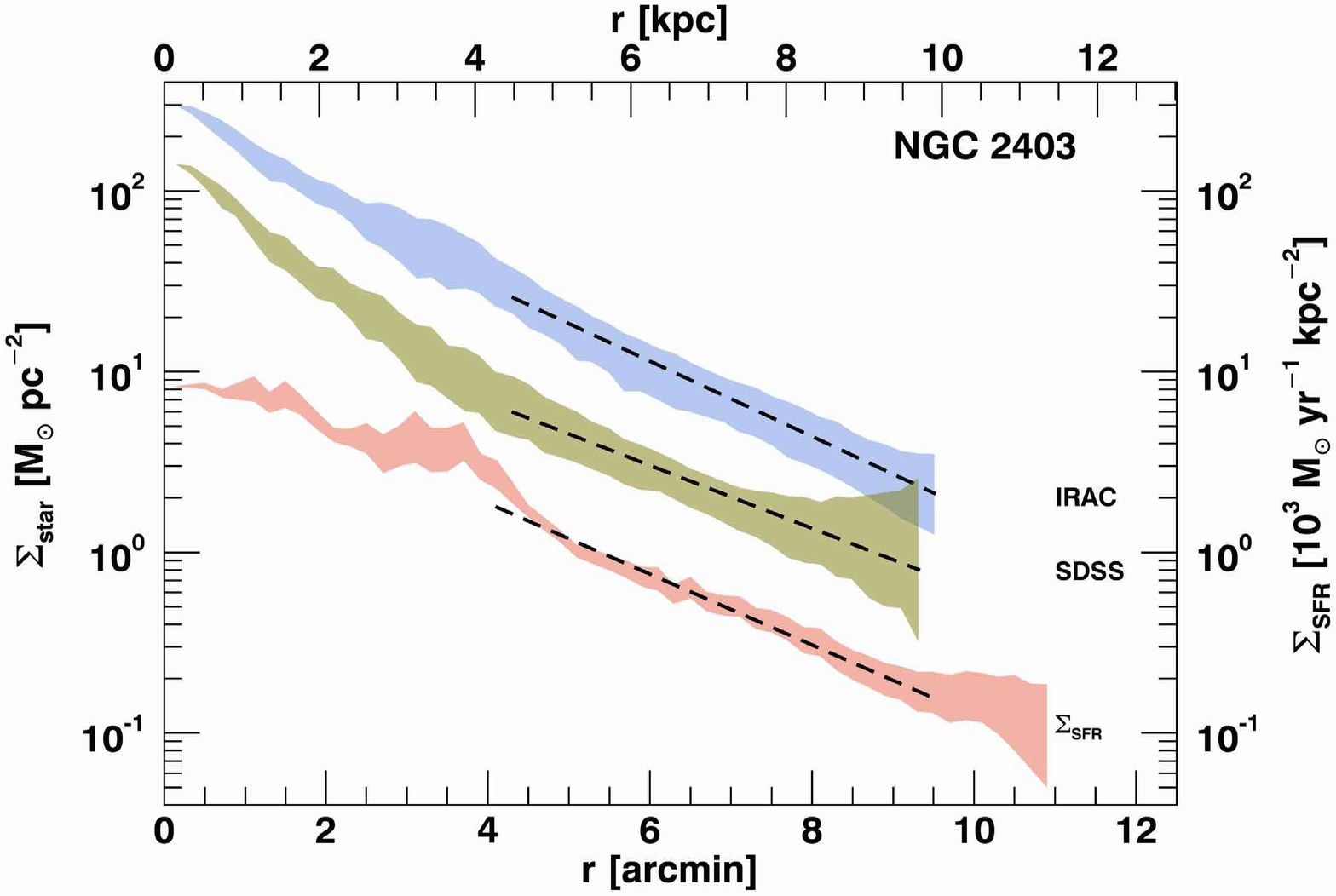}}
\hspace{0.5cm}
%{\includegraphics[width=6.0cm,angle=-90]{n2403-smasses_2.pdf}}\\
{\includegraphics[width=6.0cm,angle=-90]{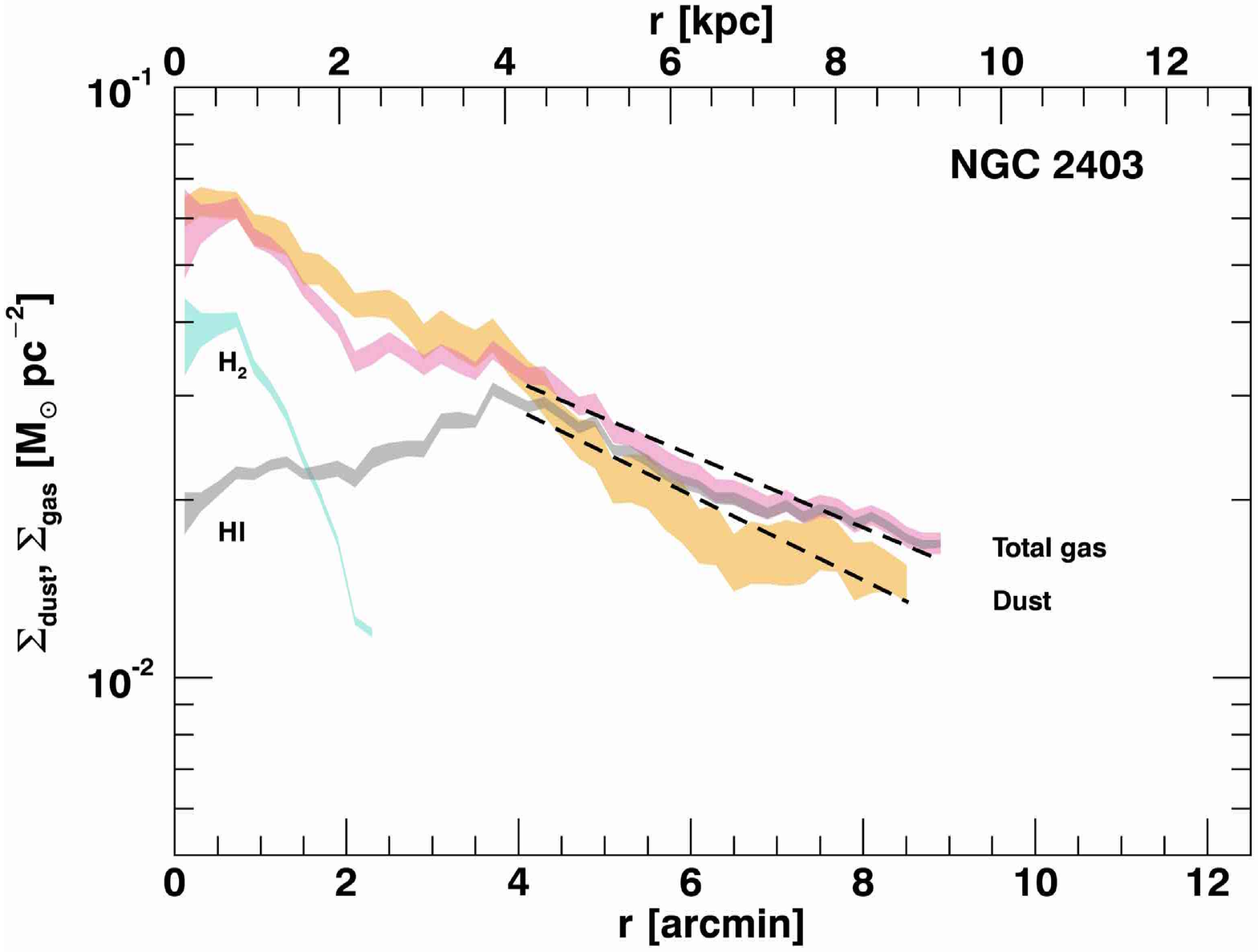}}\\
%{\includegraphics[width=6.0cm,angle=-90]{ic342-sstar-ssfr_2.pdf}}
{\includegraphics[width=6.0cm,angle=-90]{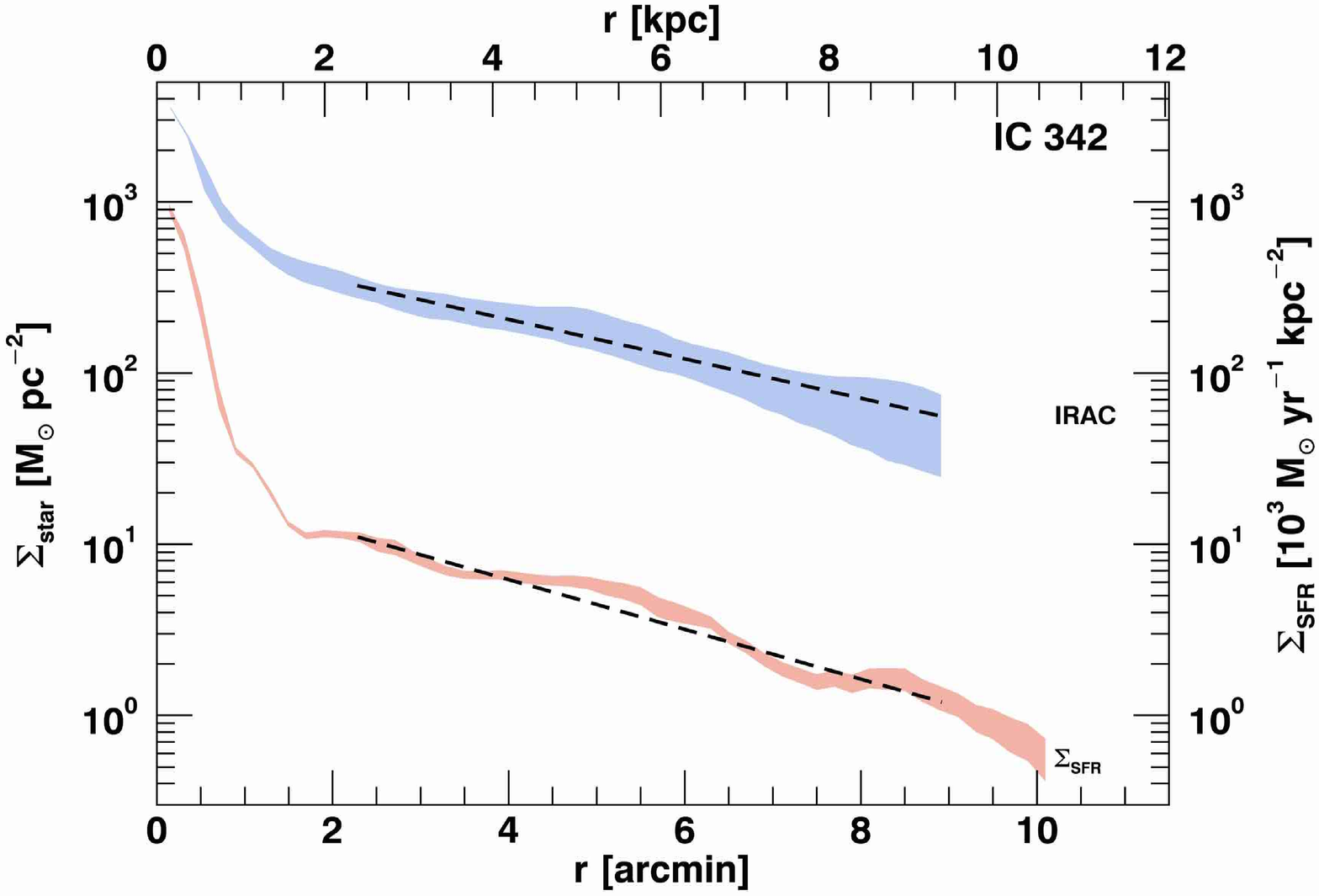}}
\hspace{0.5cm}
%{\includegraphics[width=6.0cm,angle=-90]{ic342-smasses_2.pdf}}\\
{\includegraphics[width=6.0cm,angle=-90]{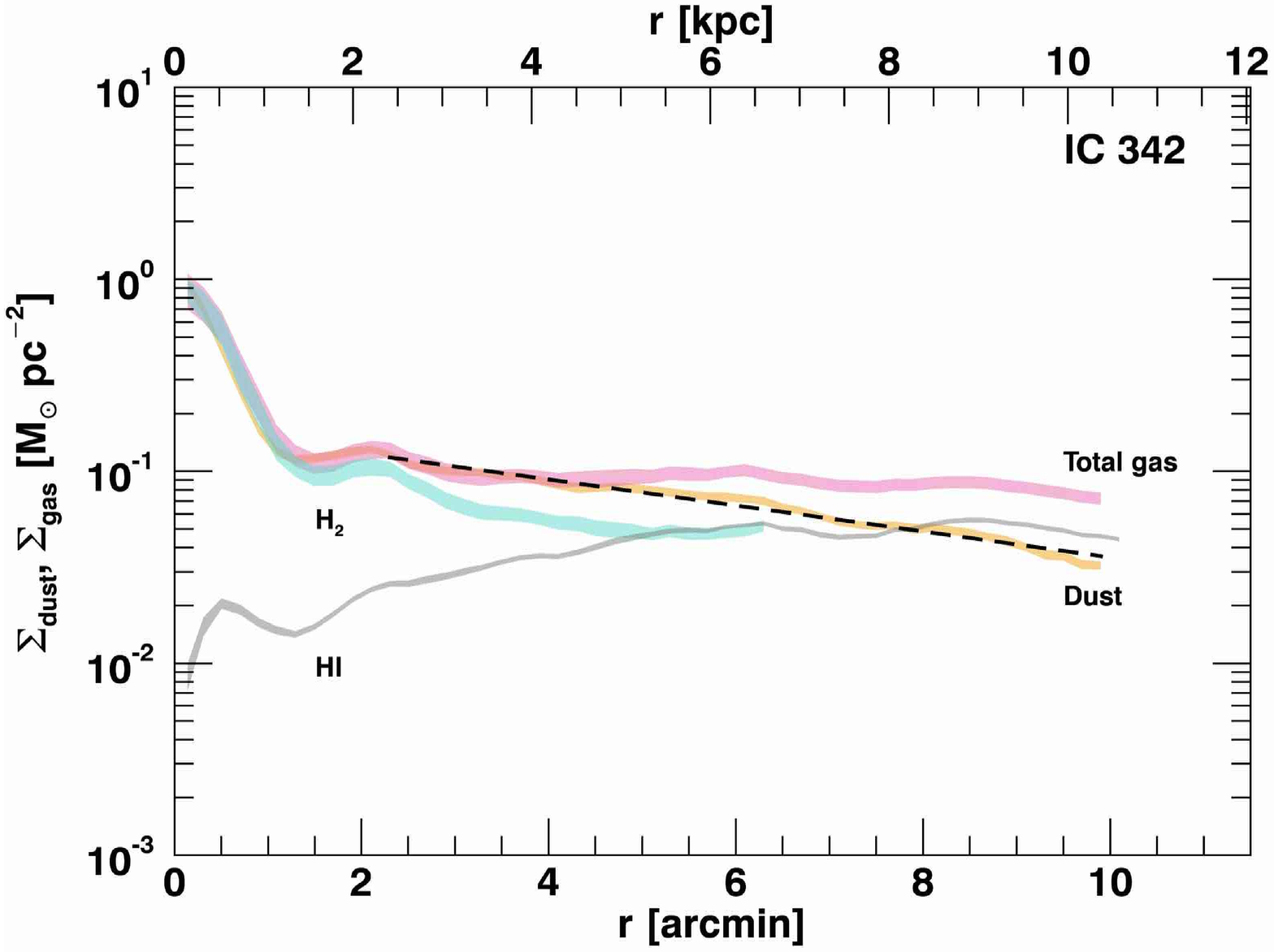}}\\
\caption{\textit{Left panel:} 
Stellar mass surface density profiles for the face-on DustPedia sample, except for the galaxy NGC~5457~(M~101) already displayed in the left panel 
of Fig.~2.
The profiles are shown as in the left panel of Fig.~2. 
The black dashed lines are exponential fits performed avoiding the central part of galaxies in the same radius ranges 
of Fig.~\ref{fig:profiles-app} or up to the profile is available.
For NGC~3621 and NGC~1097 stellar mass surface density profiles derived from SINGS are plotted without error bars since these profiles have too large uncertainties to be 
displayed in a logarithmic scale.
\textit{Right panel:}
surface density profiles for the mass of dust, molecular gas, atomic gas, and total gas for the entire DustPedia face-on sample, except for NGC~5457~(M~101) already shown
in the right panel of Fig.~2.
The profiles are shown as in the right panel of Fig.~2. 
}
\label{fig:masses-prof-app}
\end{figure*}

\begin{figure*}[!ht]
\centering
%{\includegraphics[width=6.0cm,angle=-90]{n300-sstar-ssfr.pdf}}
{\includegraphics[width=6.0cm,angle=-90]{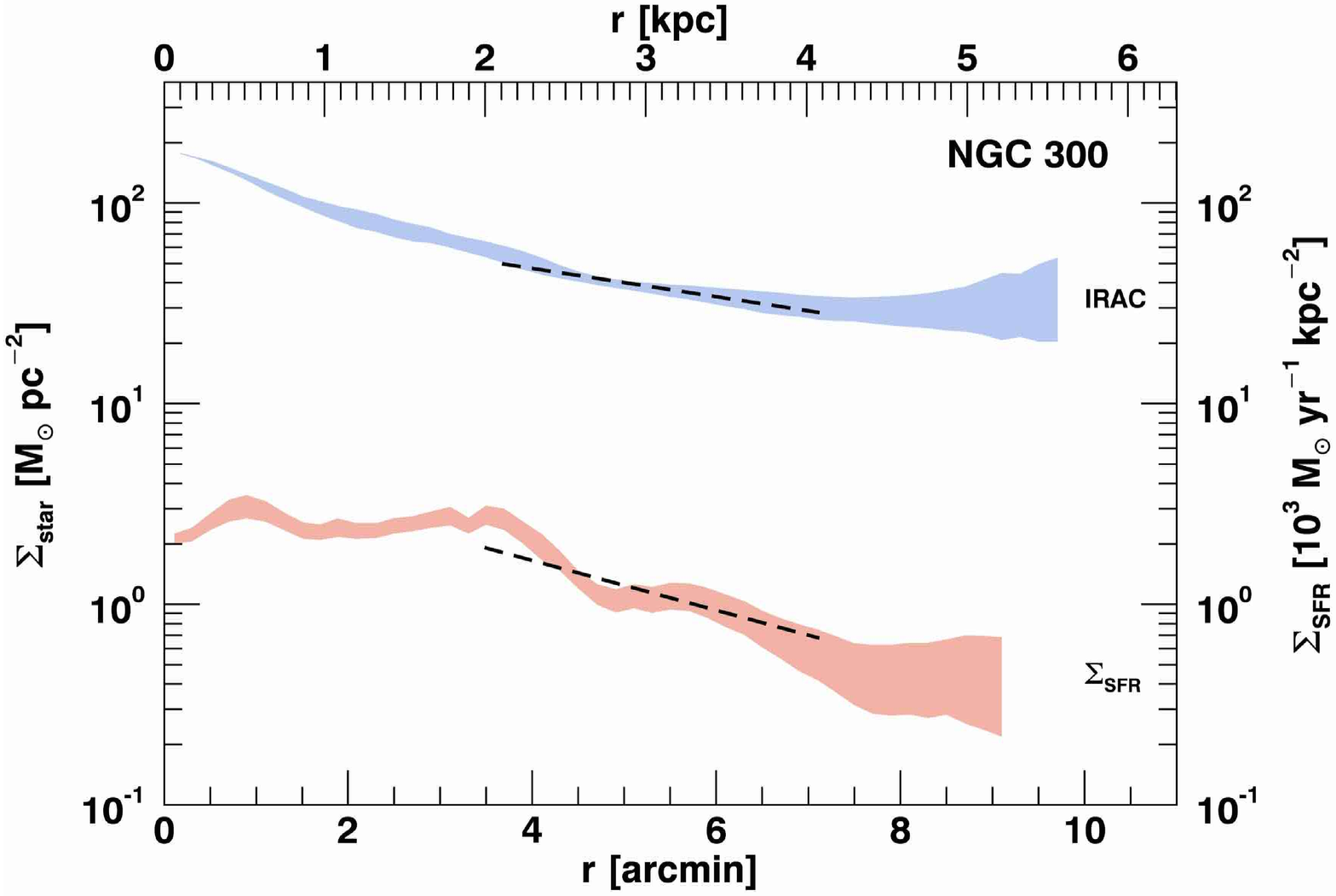}}
\hspace{0.5cm}
%{\includegraphics[width=6.0cm,angle=-90]{n300-smasses_2.pdf}}\\
{\includegraphics[width=6.0cm,angle=-90]{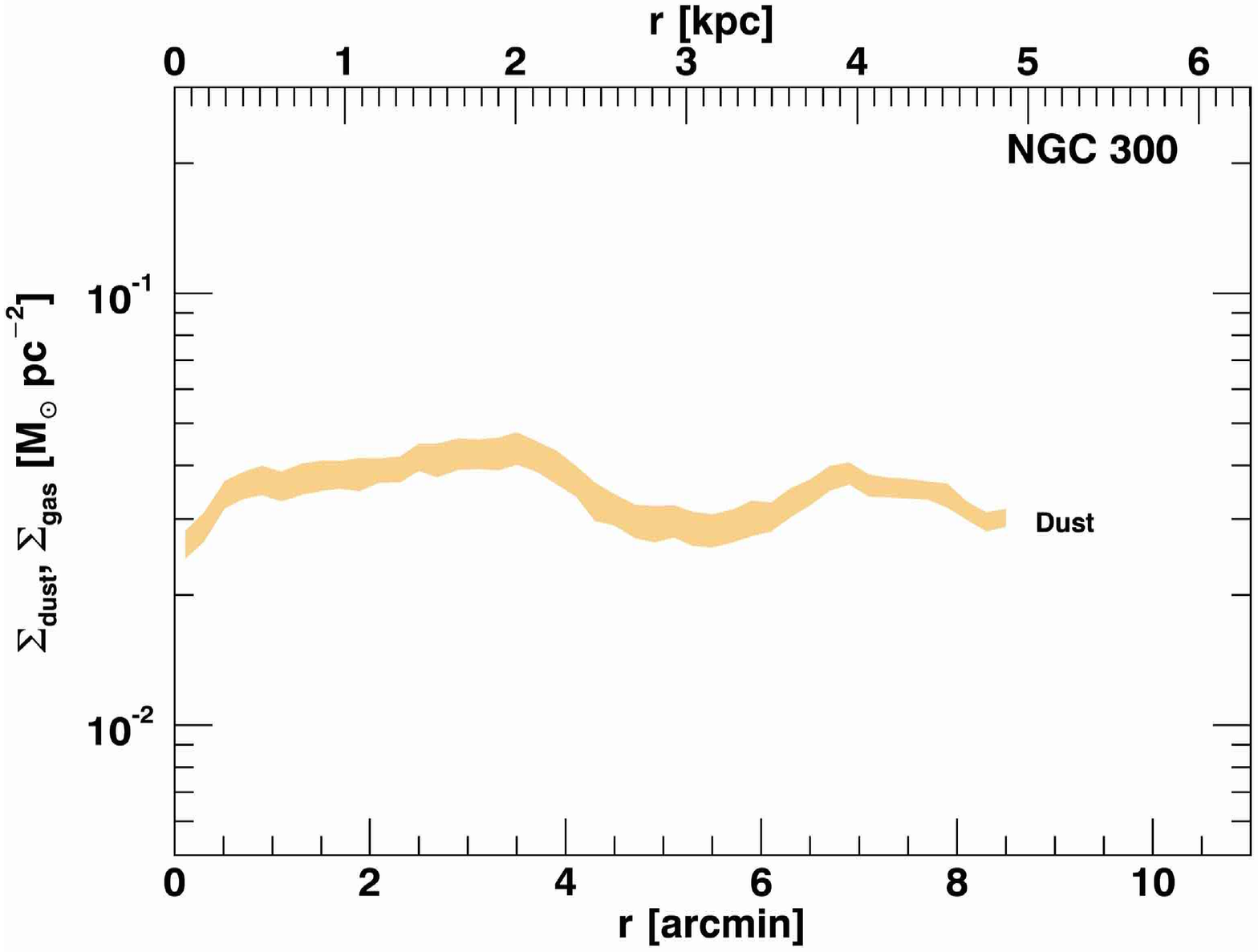}}\\
%{\includegraphics[width=6.0cm,angle=-90]{n5194-m51-sstar-ssfr.pdf}}
{\includegraphics[width=6.0cm,angle=-90]{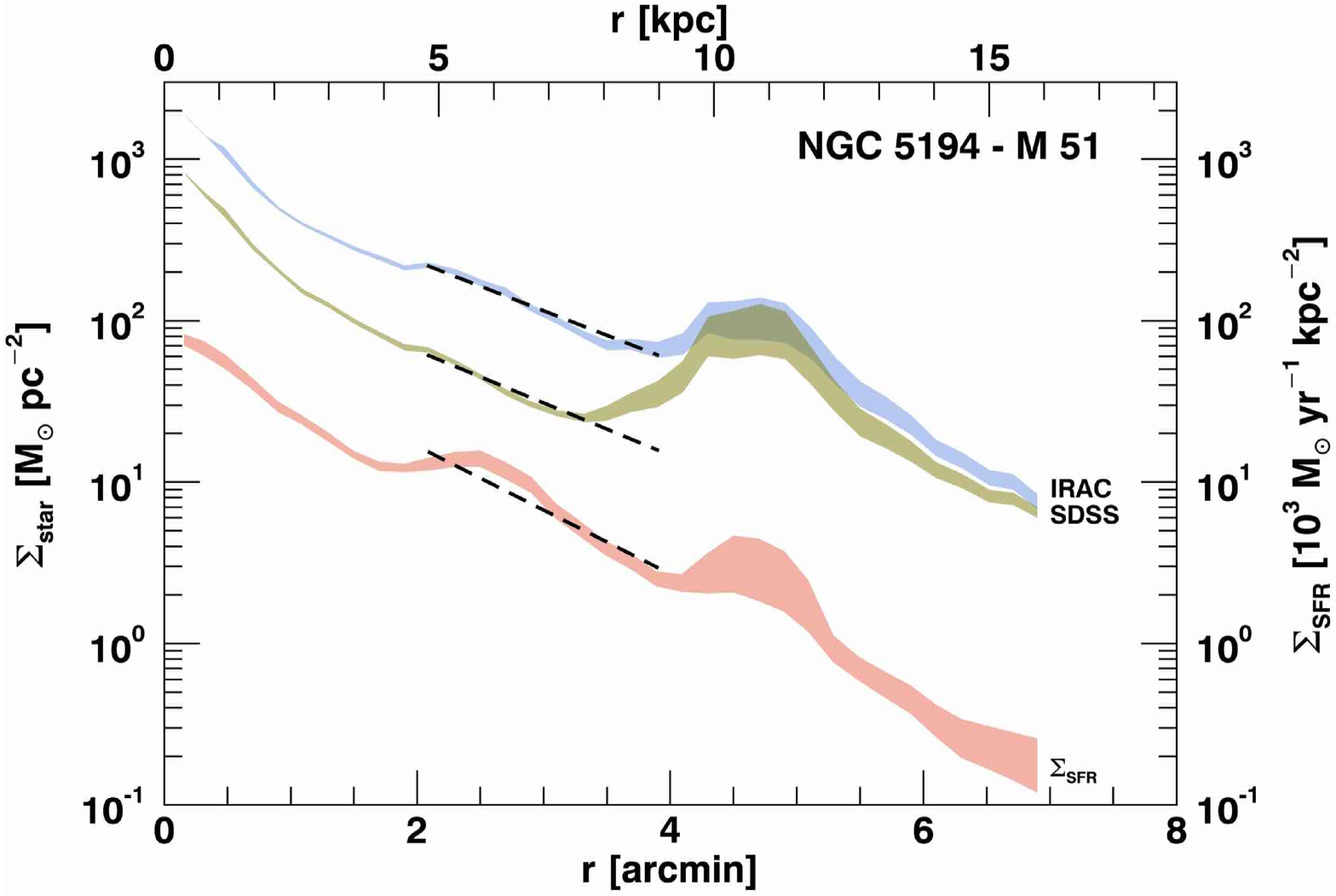}}
\hspace{0.5cm}
%{\includegraphics[width=6.0cm,angle=-90]{n5194-m51-smasses_2.pdf}}\\
{\includegraphics[width=6.0cm,angle=-90]{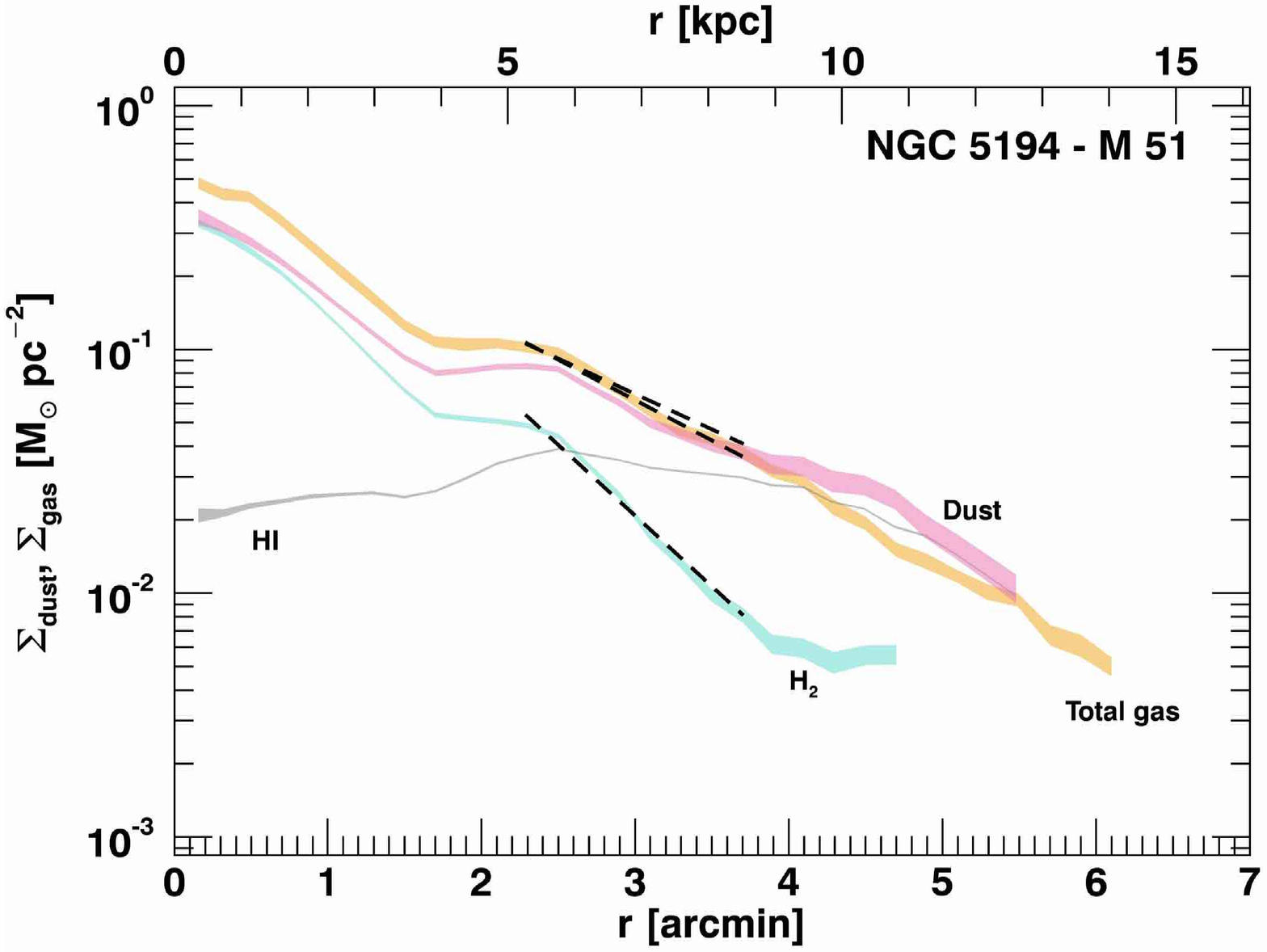}}\\
%{\includegraphics[width=6.0cm,angle=-90]{n5236-m83-sstar-ssfr.pdf}}
{\includegraphics[width=6.0cm,angle=-90]{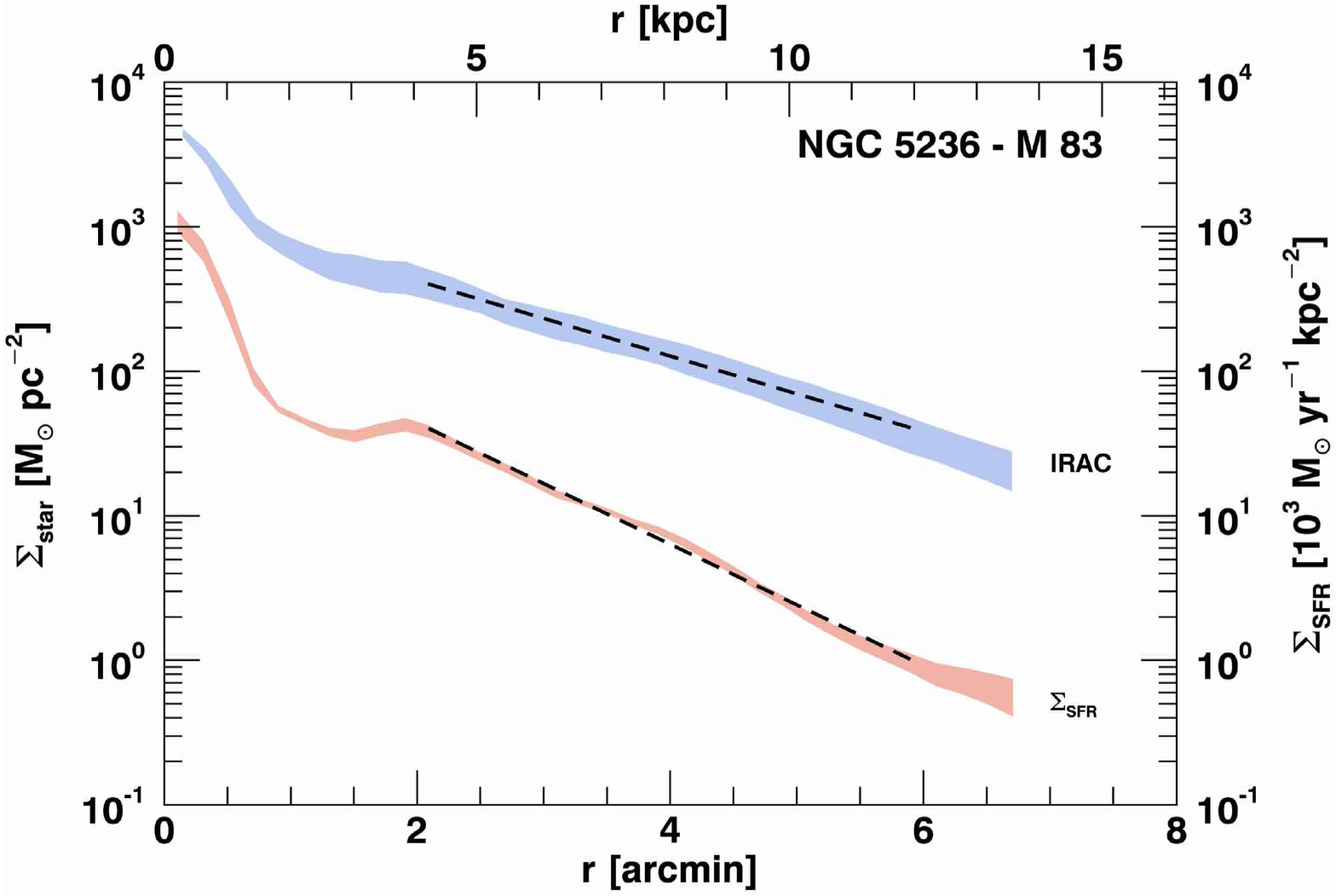}}
\hspace{0.5cm}
%{\includegraphics[width=6.0cm,angle=-90]{n5236-m83-smasses_2.pdf}}\\
{\includegraphics[width=6.0cm,angle=-90]{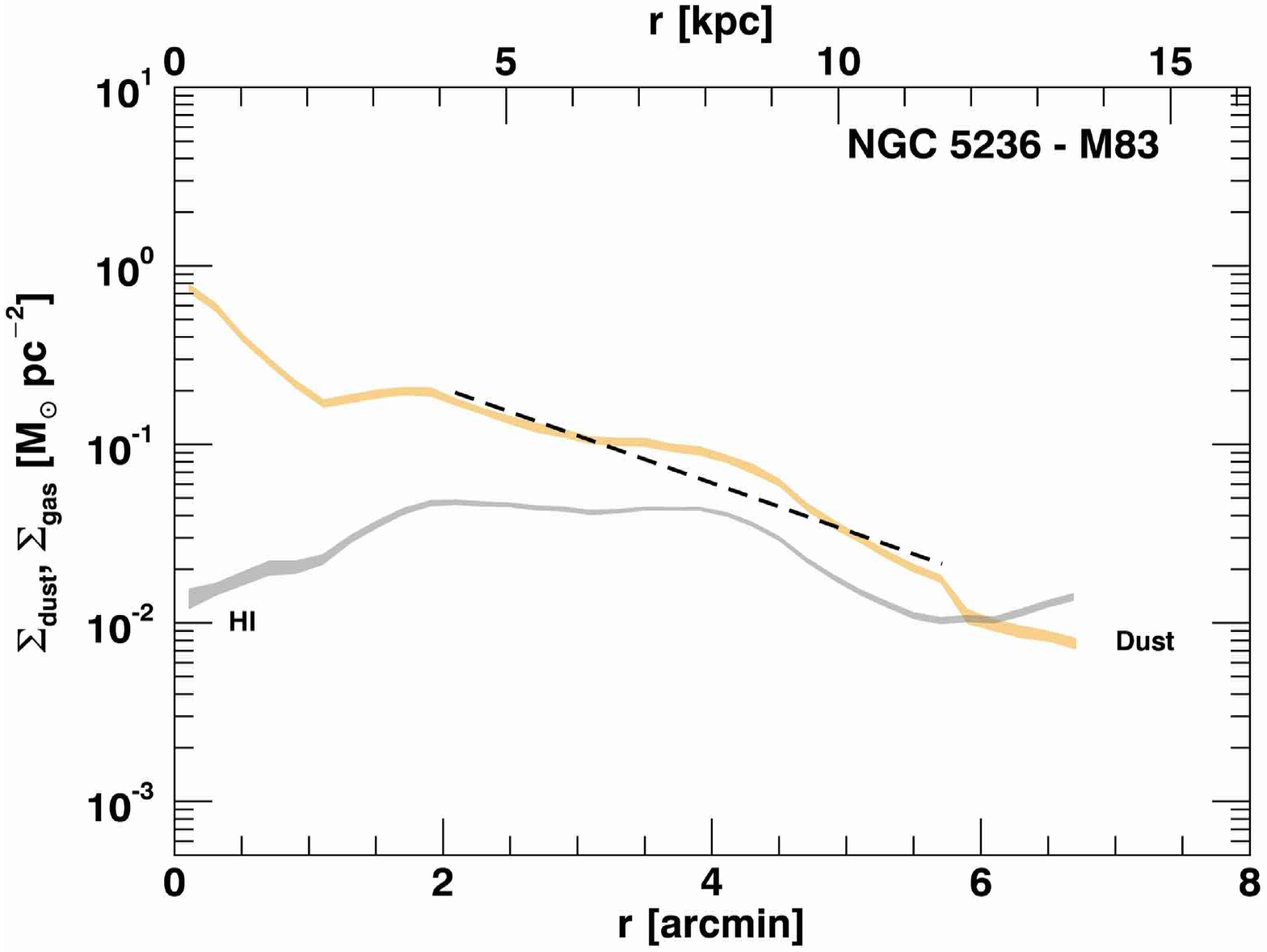}}\\
%{\includegraphics[width=6.0cm,angle=-90]{n1365-sstar-ssfr.pdf}}
{\includegraphics[width=6.0cm,angle=-90]{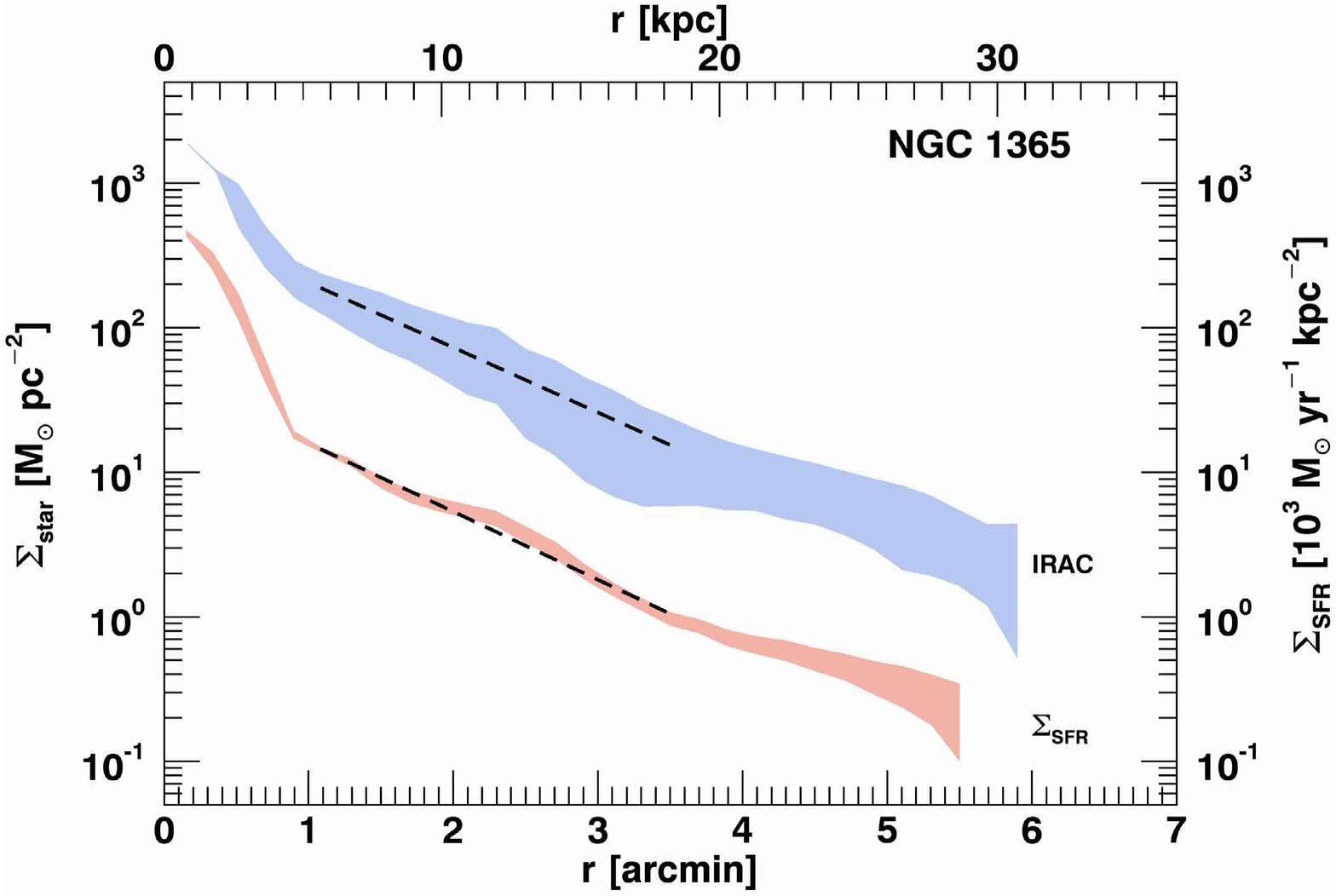}}
\hspace{0.5cm}
%{\includegraphics[width=6.0cm,angle=-90]{n1365-smasses_2.pdf}}\\
{\includegraphics[width=6.0cm,angle=-90]{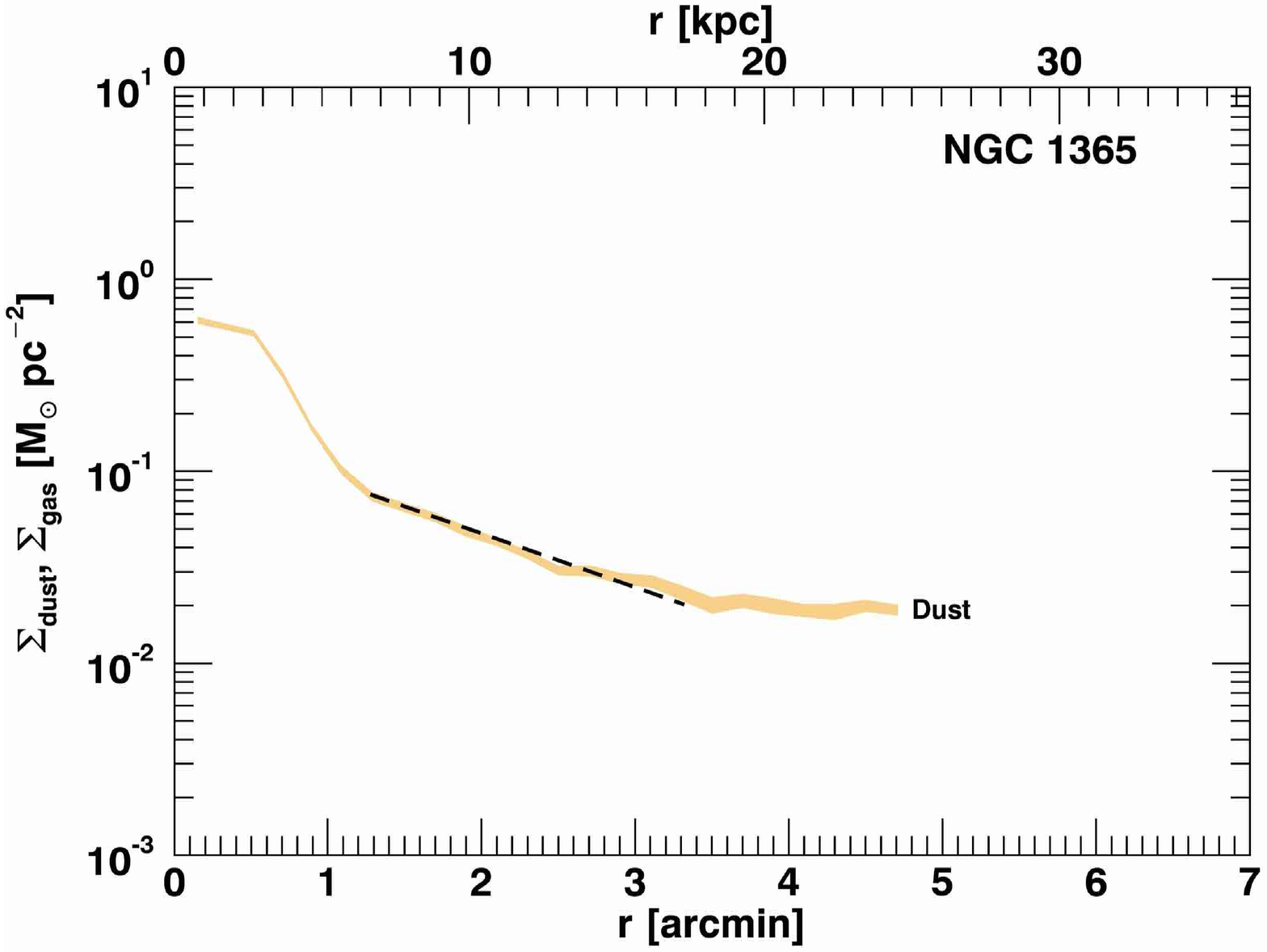}}\\
\caption*{Figure~\ref{fig:masses-prof-app} (continued)}
\end{figure*}

\begin{figure*}[!ht]
\centering
%{\includegraphics[width=6.0cm,angle=-90]{n5055-m63-sstar-ssfr.pdf}}
{\includegraphics[width=6.0cm,angle=-90]{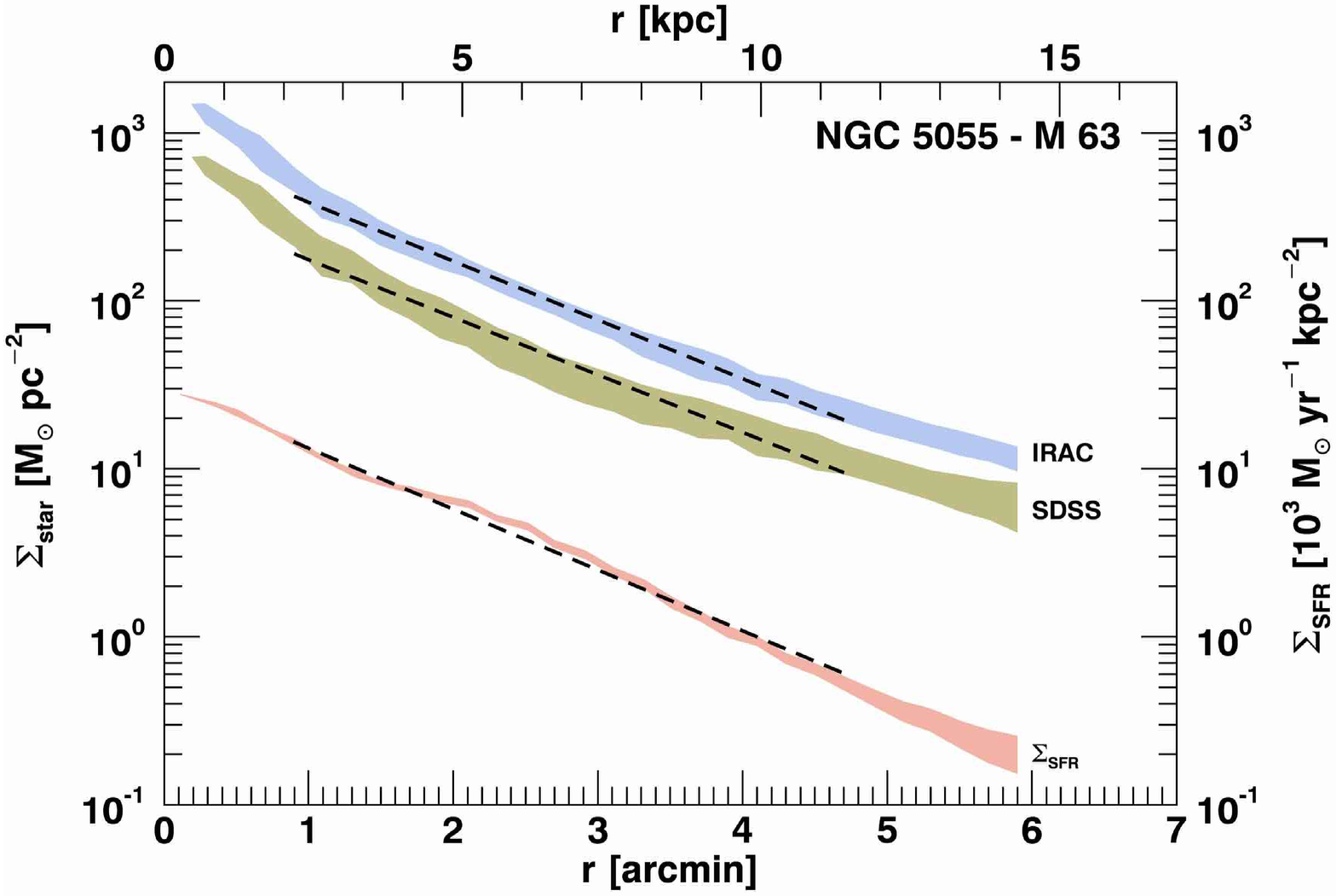}}
\hspace{0.5cm}
%{\includegraphics[width=6.0cm,angle=-90]{n5055-m63-smasses_2.pdf}}\\
{\includegraphics[width=6.0cm,angle=-90]{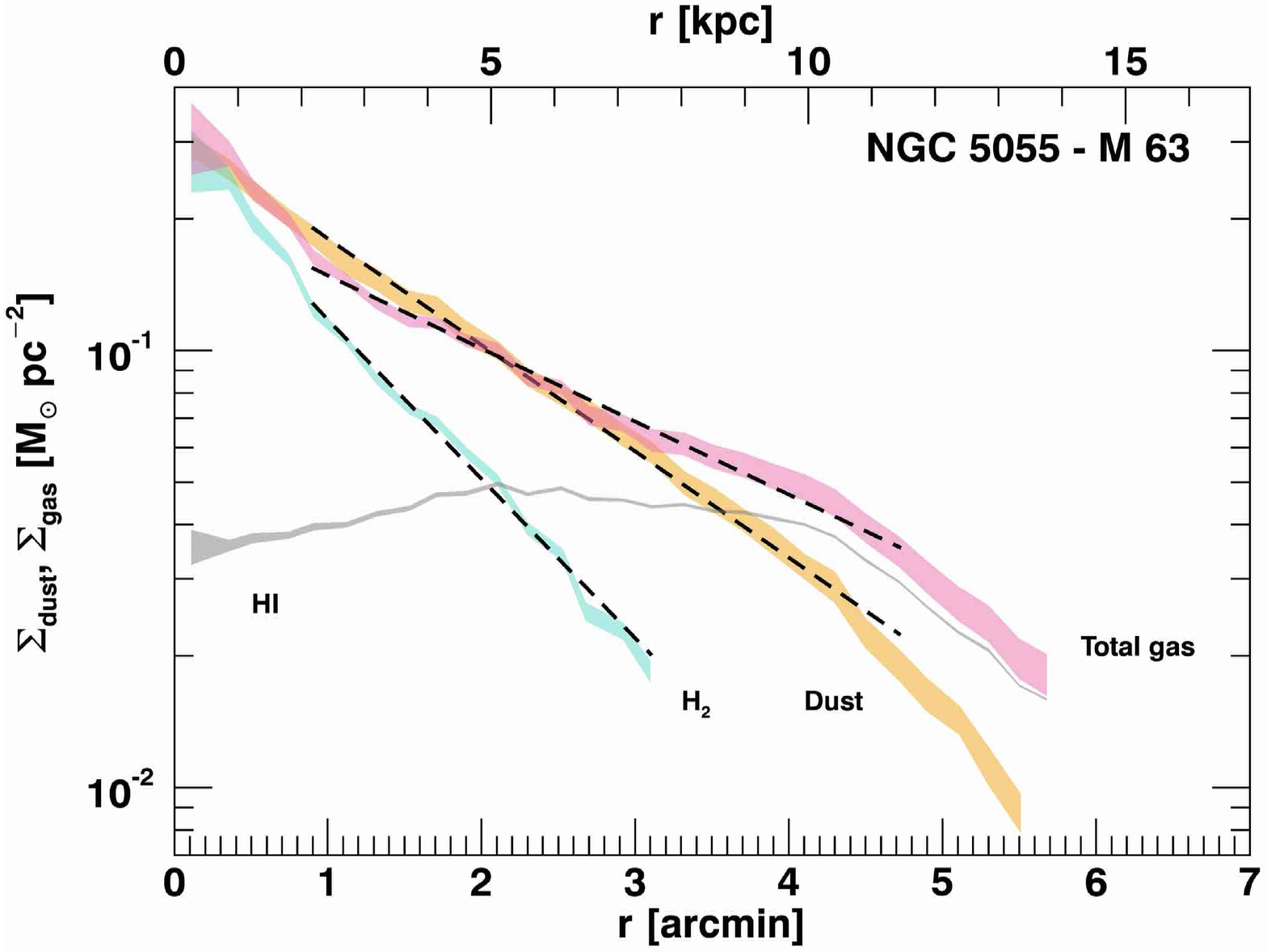}}\\
%{\includegraphics[width=6.0cm,angle=-90]{n6946-sstar-ssfr.pdf}}
{\includegraphics[width=6.0cm,angle=-90]{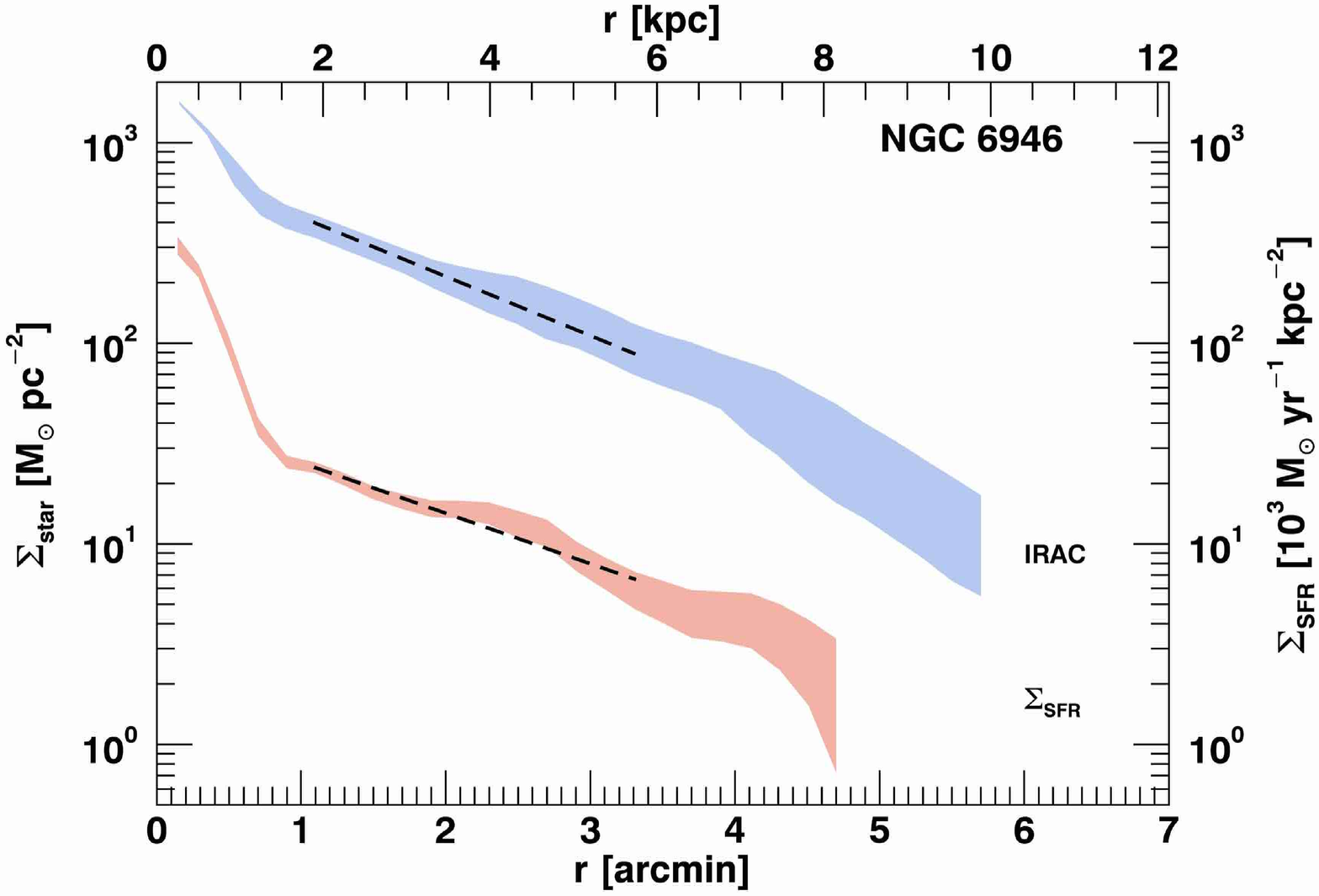}}
\hspace{0.5cm}
%{\includegraphics[width=6.0cm,angle=-90]{n6946-smasses_2.pdf}}\\
{\includegraphics[width=6.0cm,angle=-90]{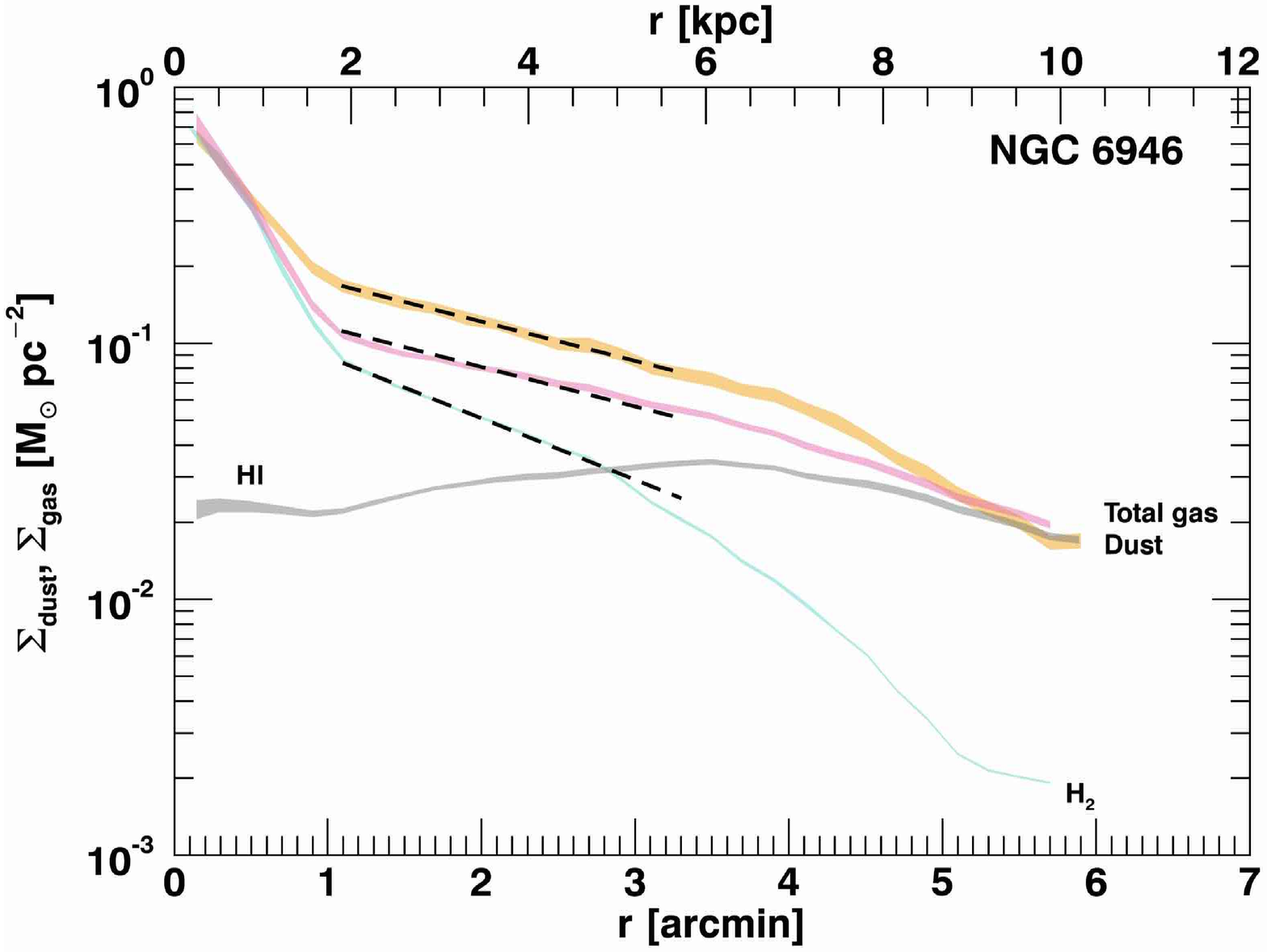}}\\
%{\includegraphics[width=6.0cm,angle=-90]{n925-sstar-ssfr.pdf}}
{\includegraphics[width=6.0cm,angle=-90]{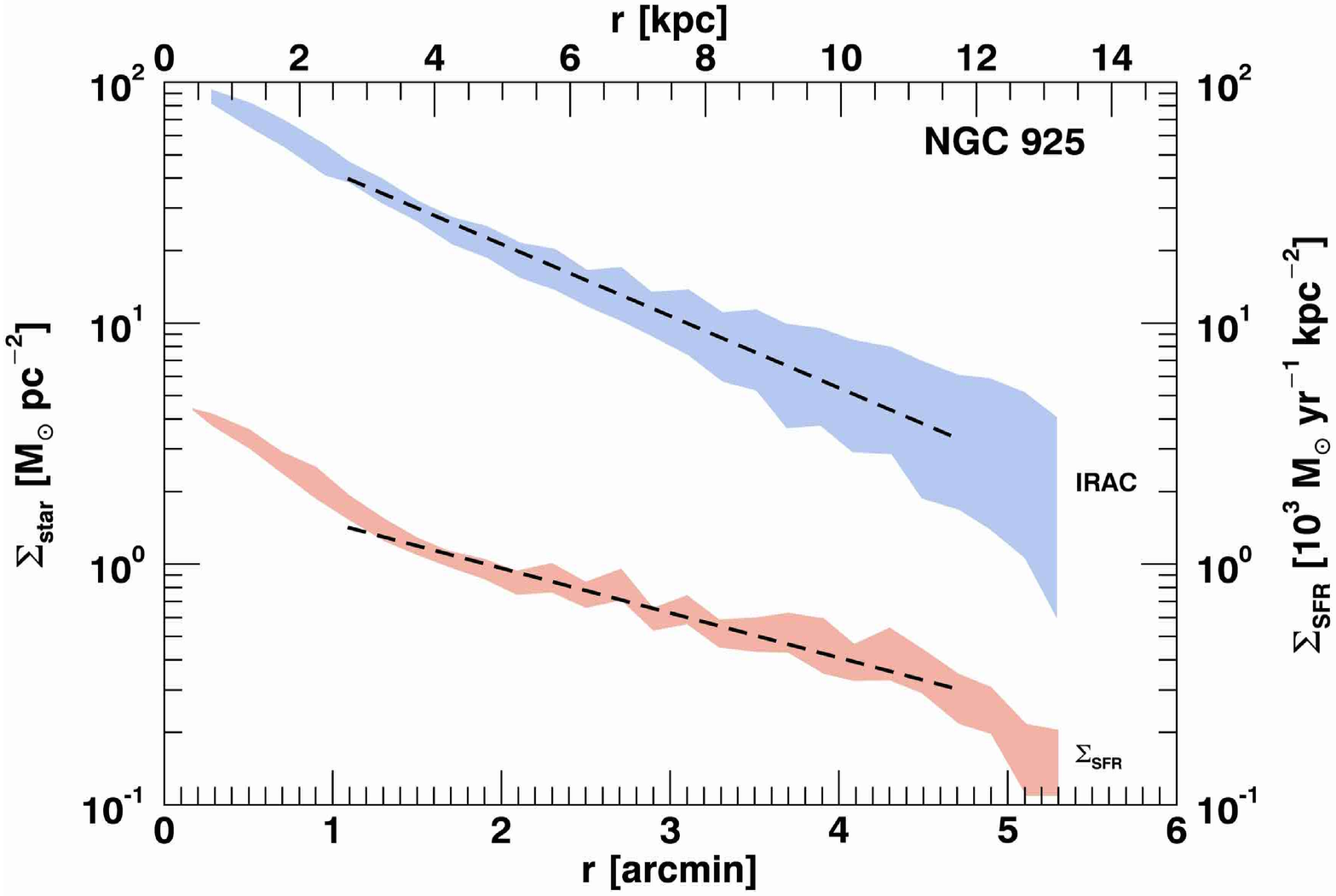}}
\hspace{0.5cm}
%{\includegraphics[width=6.0cm,angle=-90]{n925-smasses_2.pdf}}\\
{\includegraphics[width=6.0cm,angle=-90]{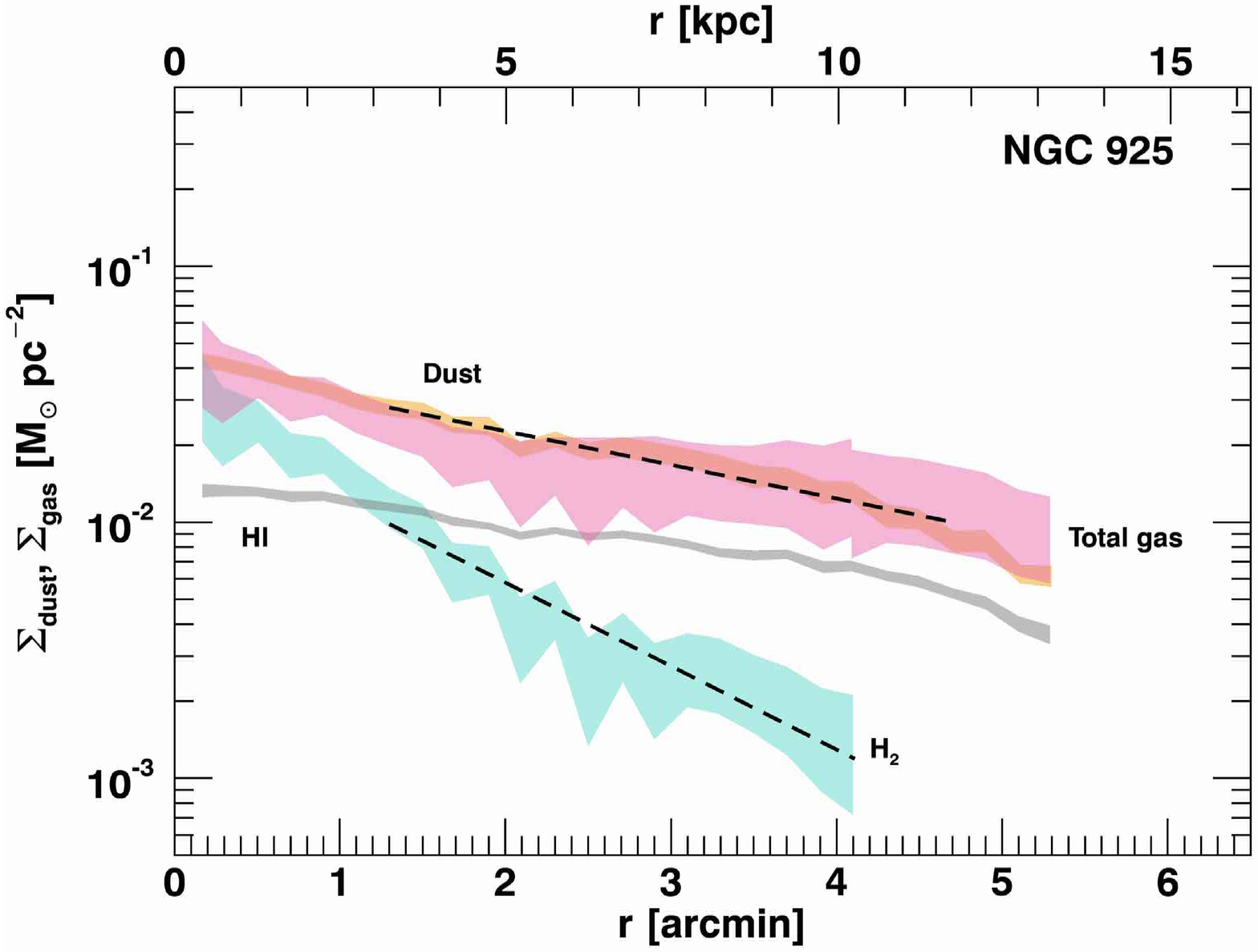}}\\
%{\includegraphics[width=6.0cm,angle=-90]{n1097-sstar-ssfr.pdf}}
{\includegraphics[width=6.0cm,angle=-90]{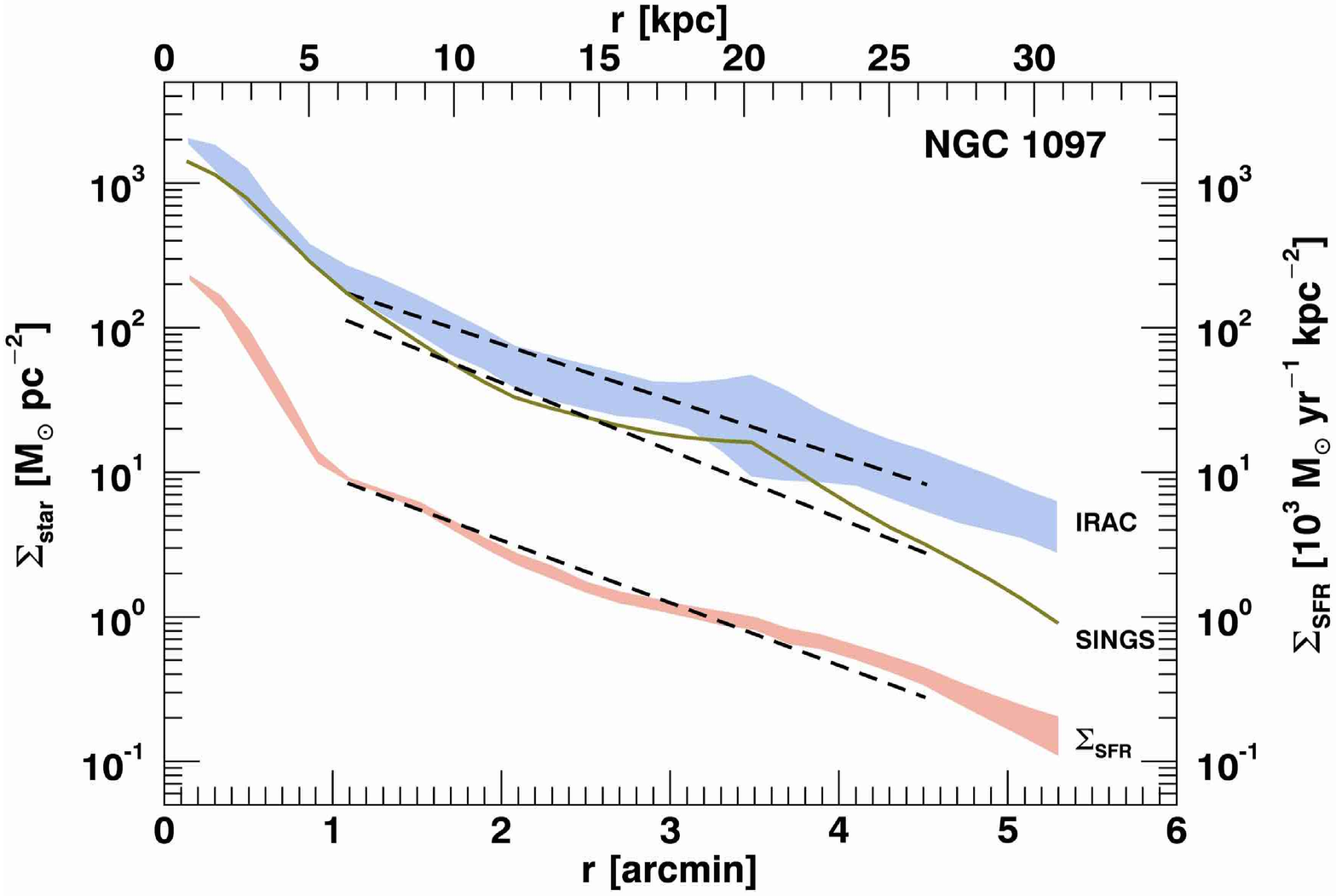}}
\hspace{0.5cm}
%{\includegraphics[width=6.0cm,angle=-90]{n1097-smasses_2.pdf}}\\
{\includegraphics[width=6.0cm,angle=-90]{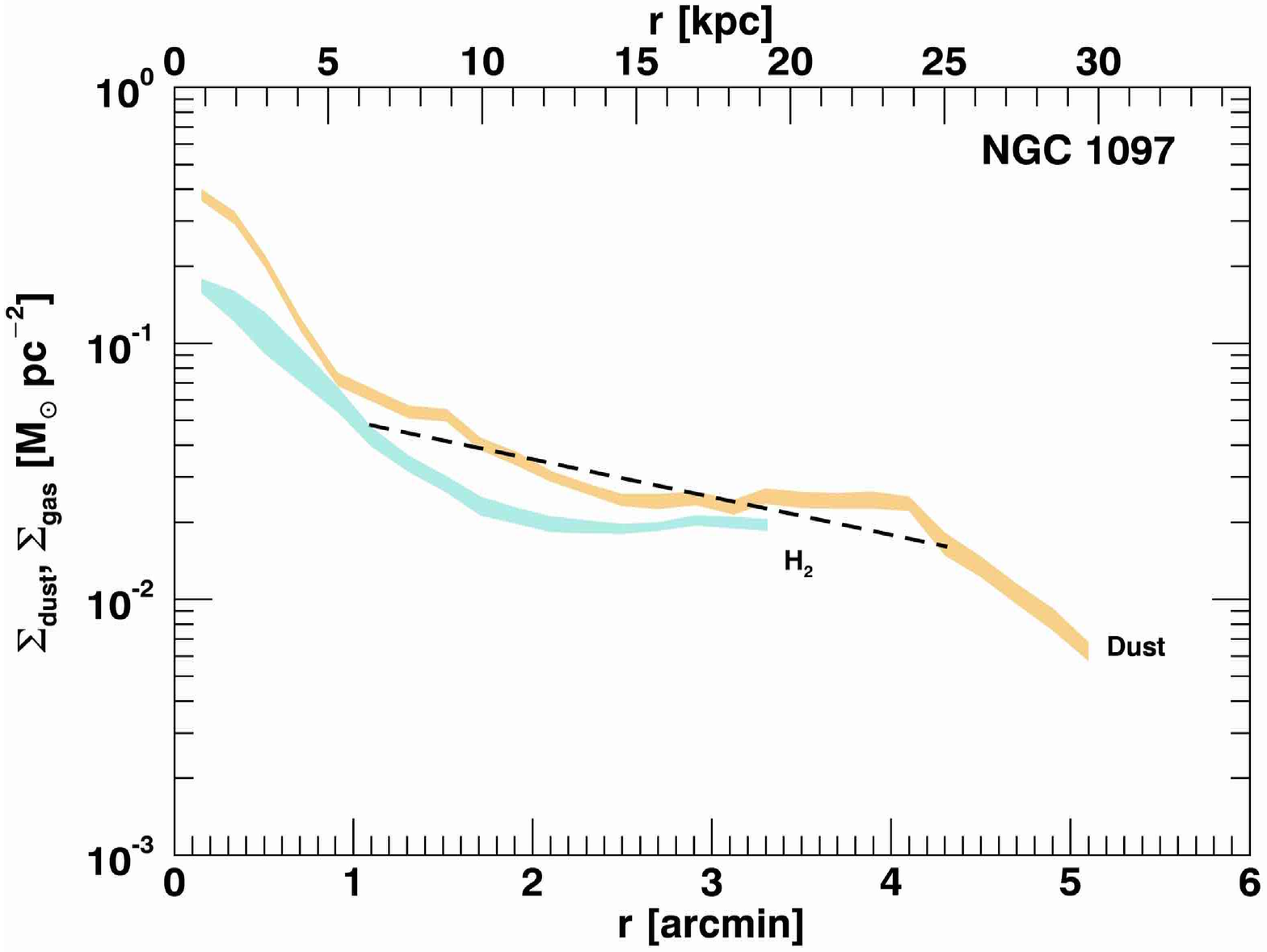}}\\
\caption*{Figure~\ref{fig:masses-prof-app} (continued)}
\end{figure*}

\begin{figure*}[!ht]
\centering
%{\includegraphics[width=6.0cm,angle=-90]{n7793-sstar-ssfr.pdf}}
{\includegraphics[width=6.0cm,angle=-90]{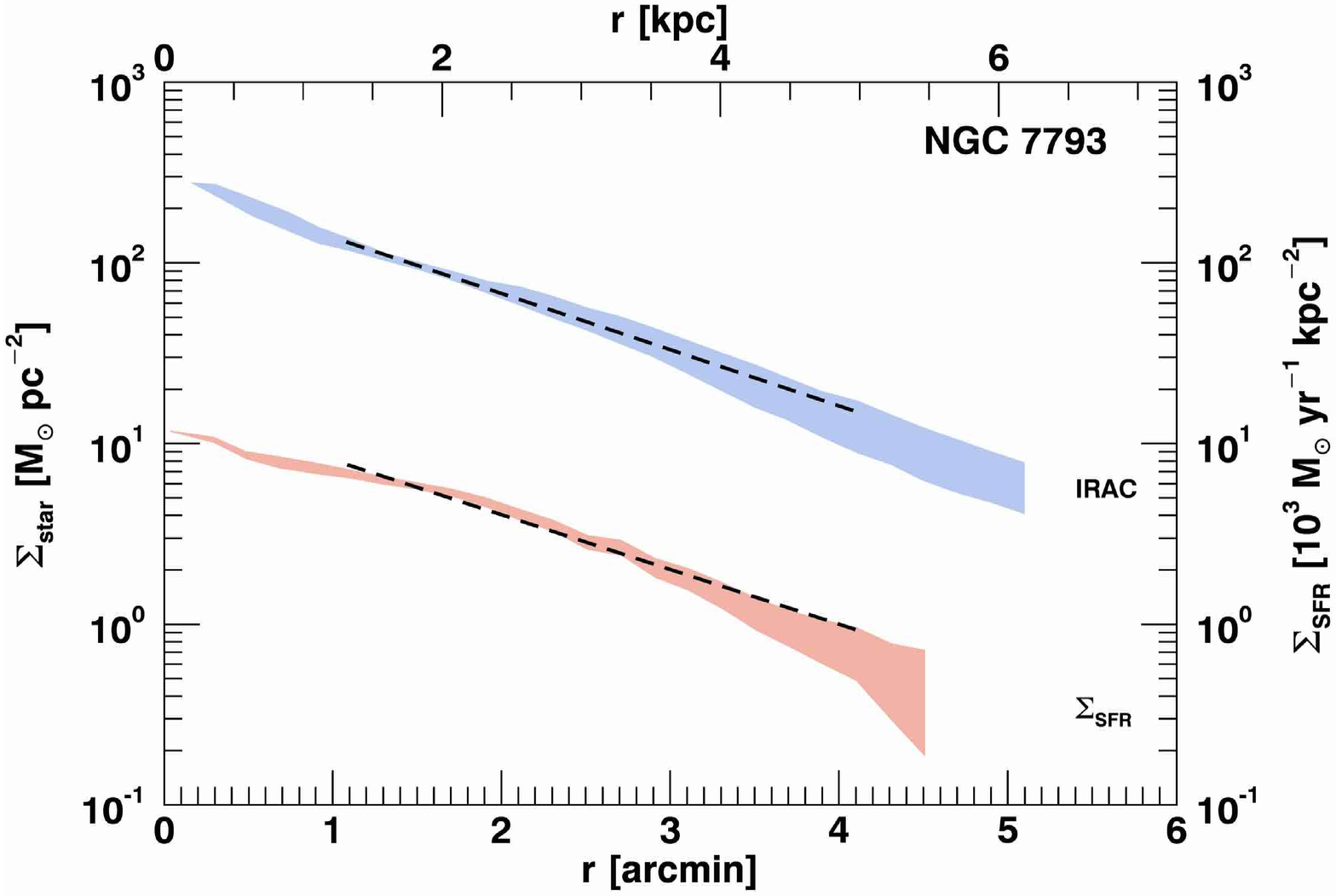}}
\hspace{0.5cm}
%{\includegraphics[width=6.0cm,angle=-90]{n7793-smasses_2.pdf}}\\
{\includegraphics[width=6.0cm,angle=-90]{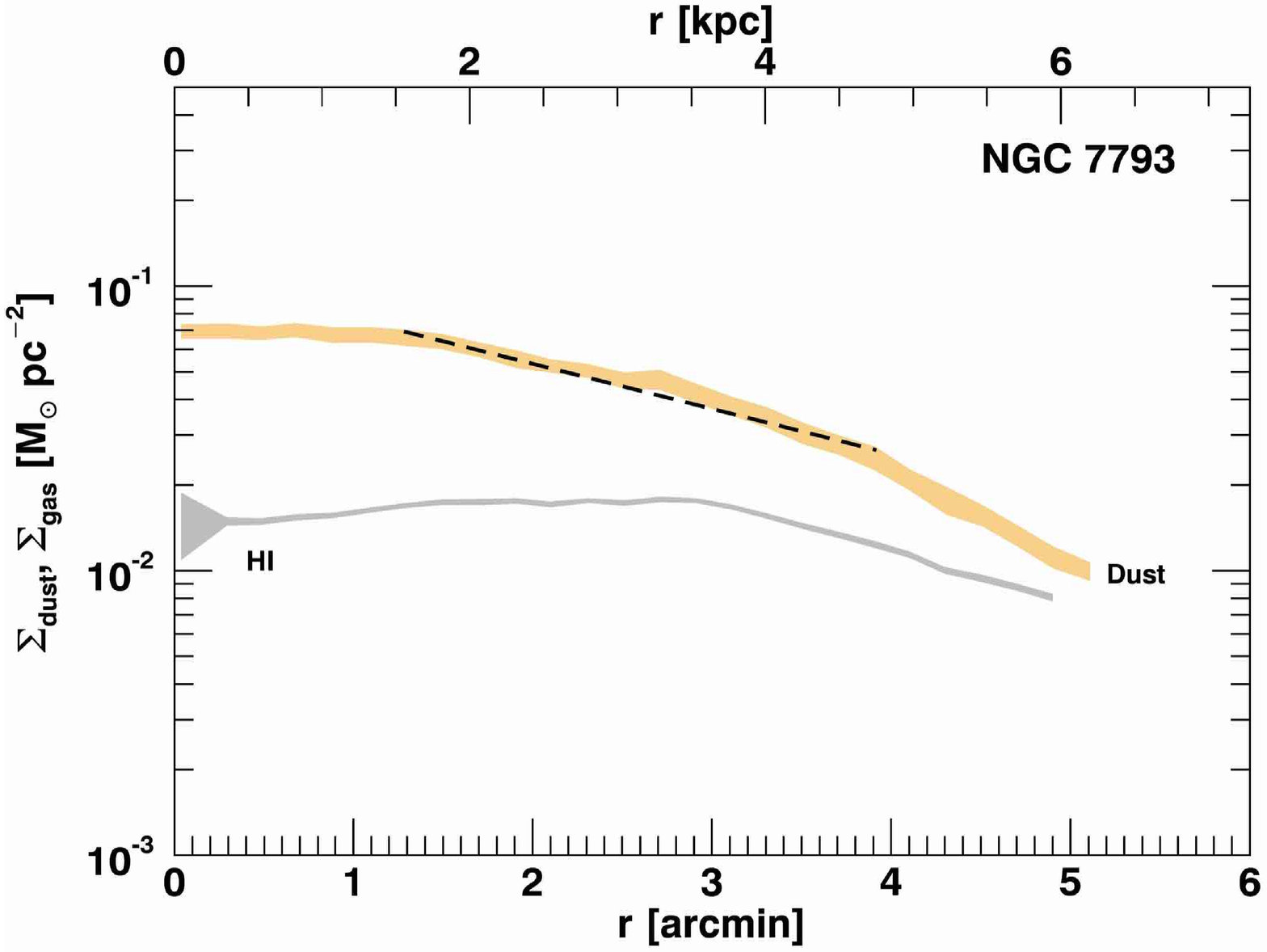}}\\
%{\includegraphics[width=6.0cm,angle=-90]{n628-m74-sstar-ssfr.pdf}}
{\includegraphics[width=6.0cm,angle=-90]{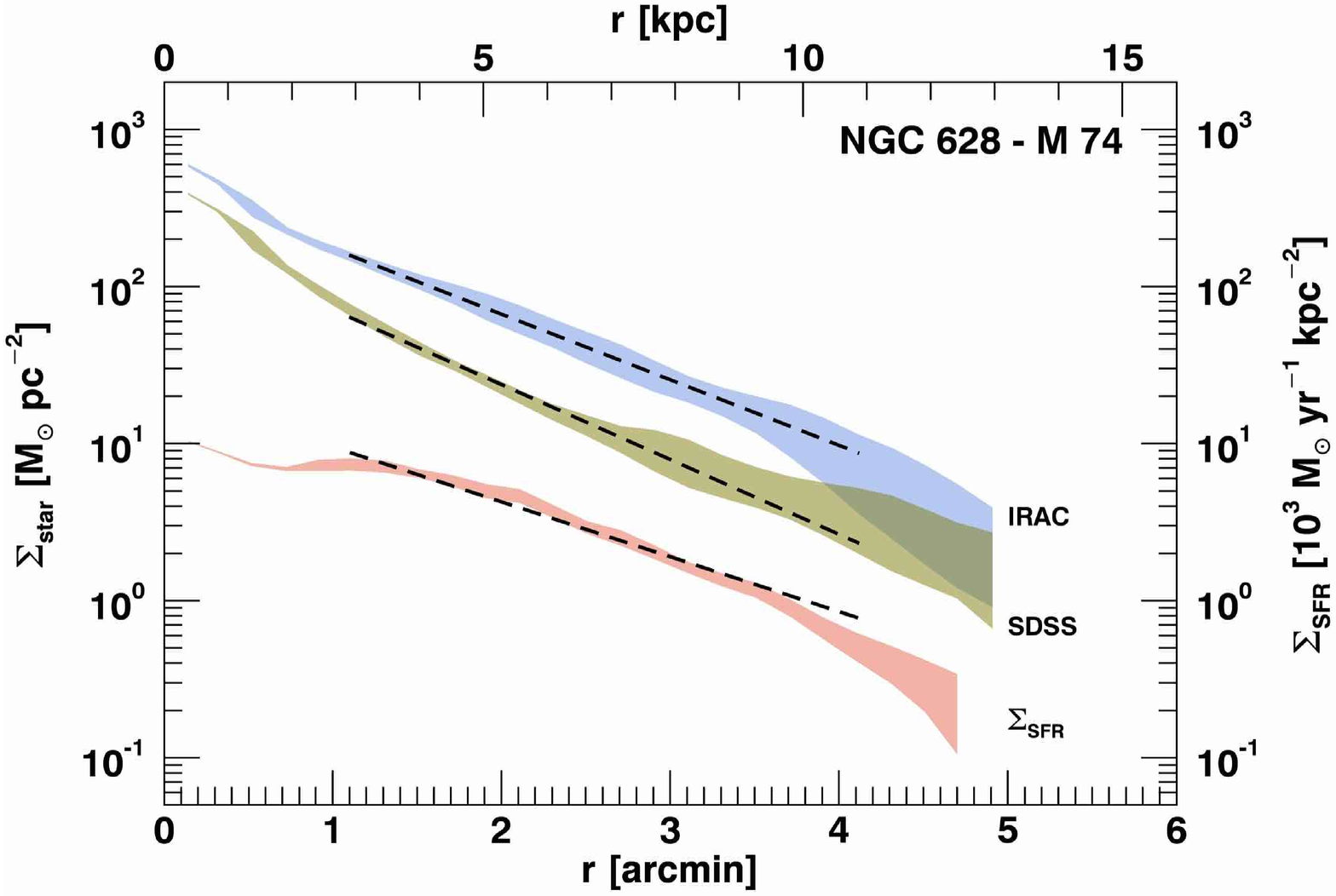}}
\hspace{0.5cm}
%{\includegraphics[width=6.0cm,angle=-90]{n628-m74-smasses_2.pdf}}\\
{\includegraphics[width=6.0cm,angle=-90]{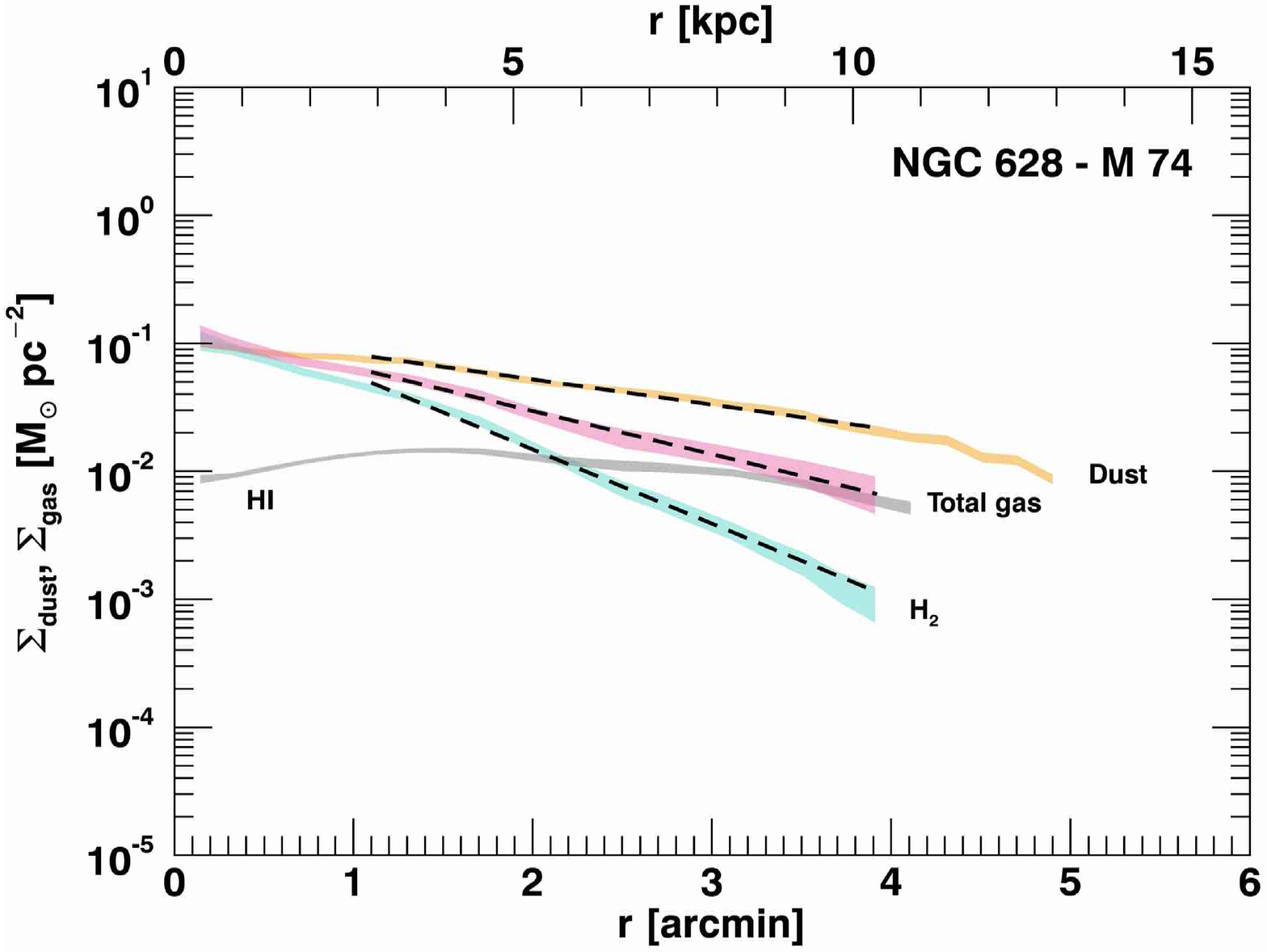}}\\
%{\includegraphics[width=6.0cm,angle=-90]{n3621-sstar-ssfr.pdf}}
{\includegraphics[width=6.0cm,angle=-90]{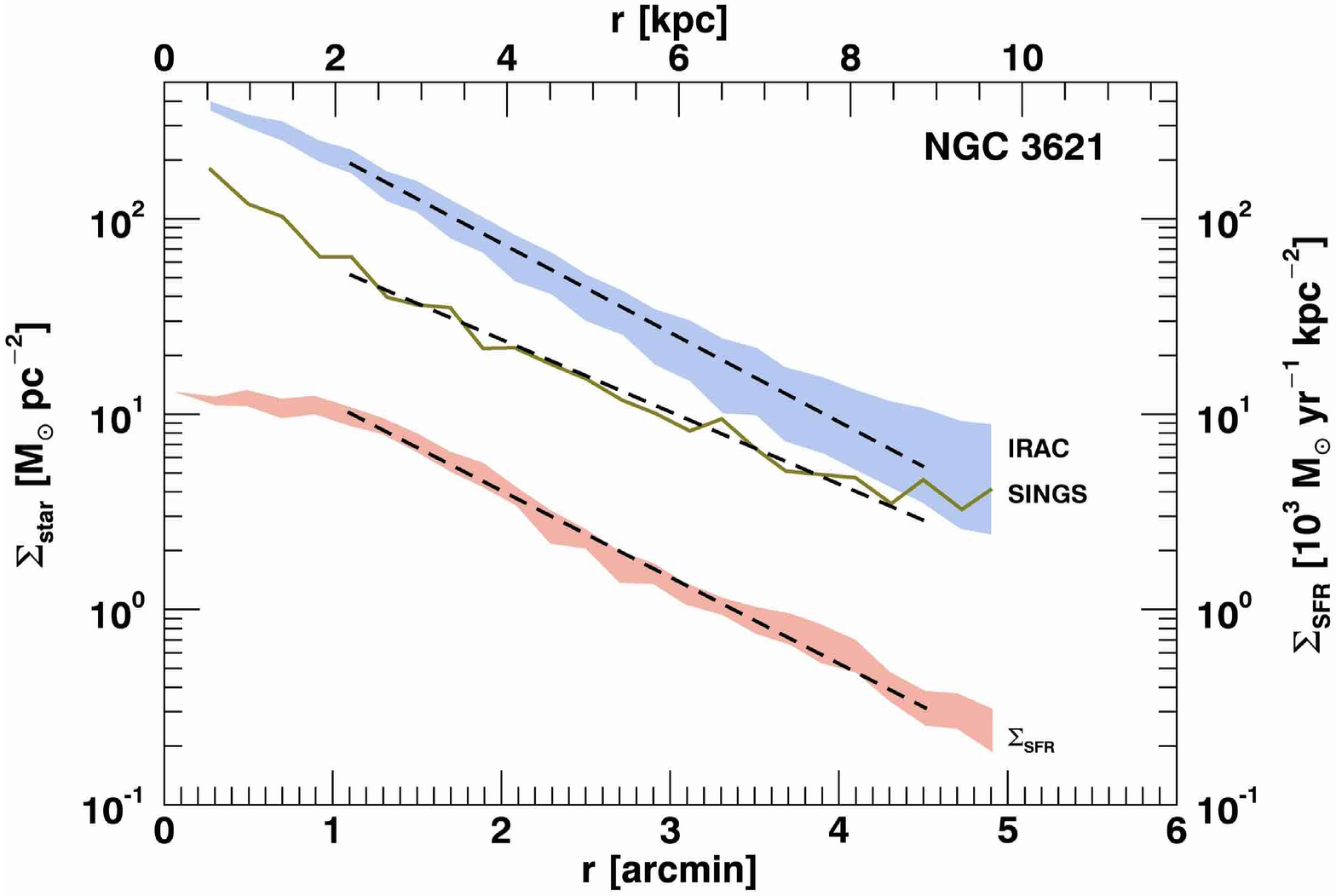}}
\hspace{0.5cm}
%{\includegraphics[width=6.0cm,angle=-90]{n3621-smasses_2.pdf}}\\
{\includegraphics[width=6.0cm,angle=-90]{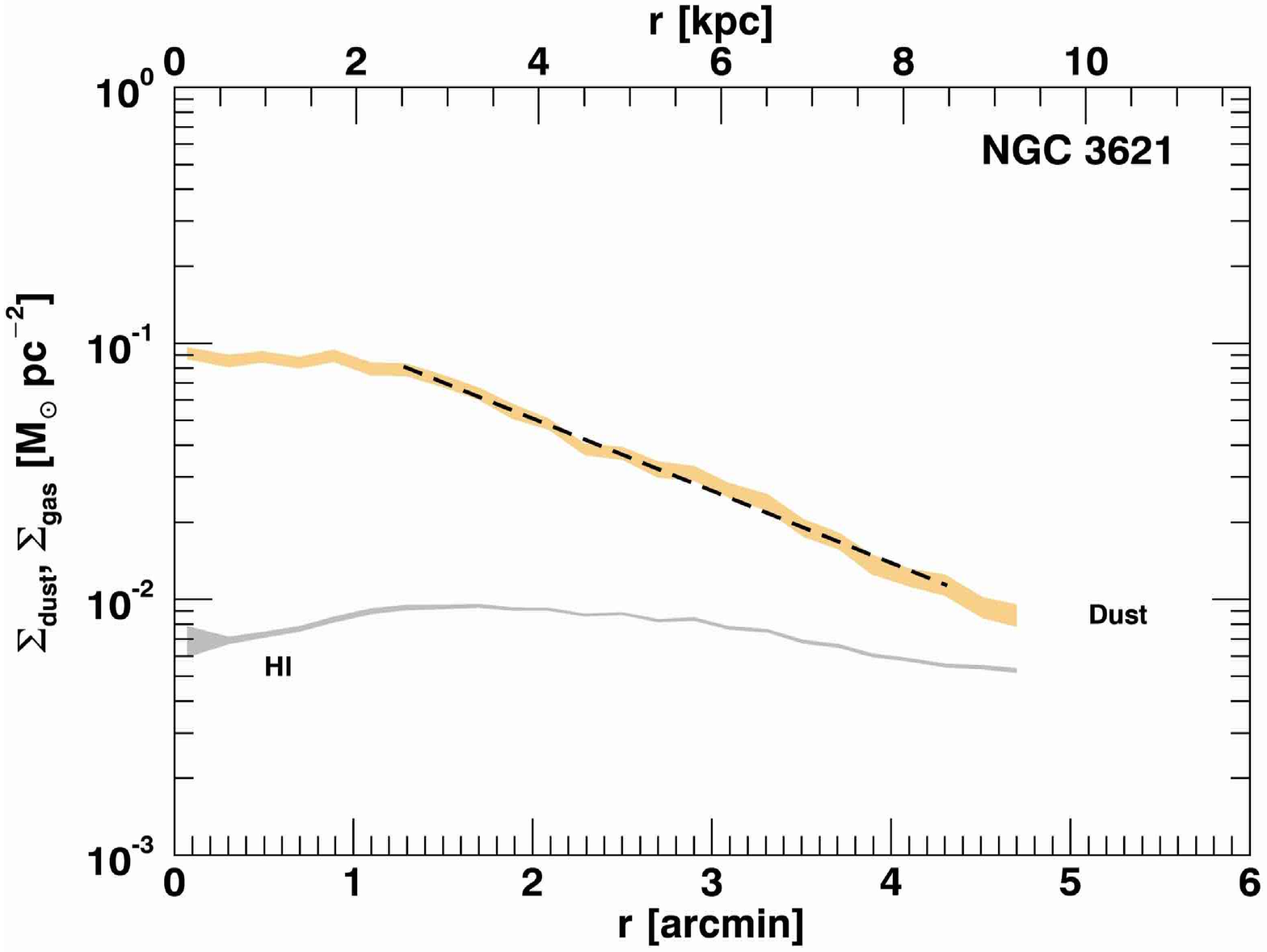}}\\
%{\includegraphics[width=6.0cm,angle=-90]{n4725-sstar-ssfr.pdf}}
{\includegraphics[width=6.0cm,angle=-90]{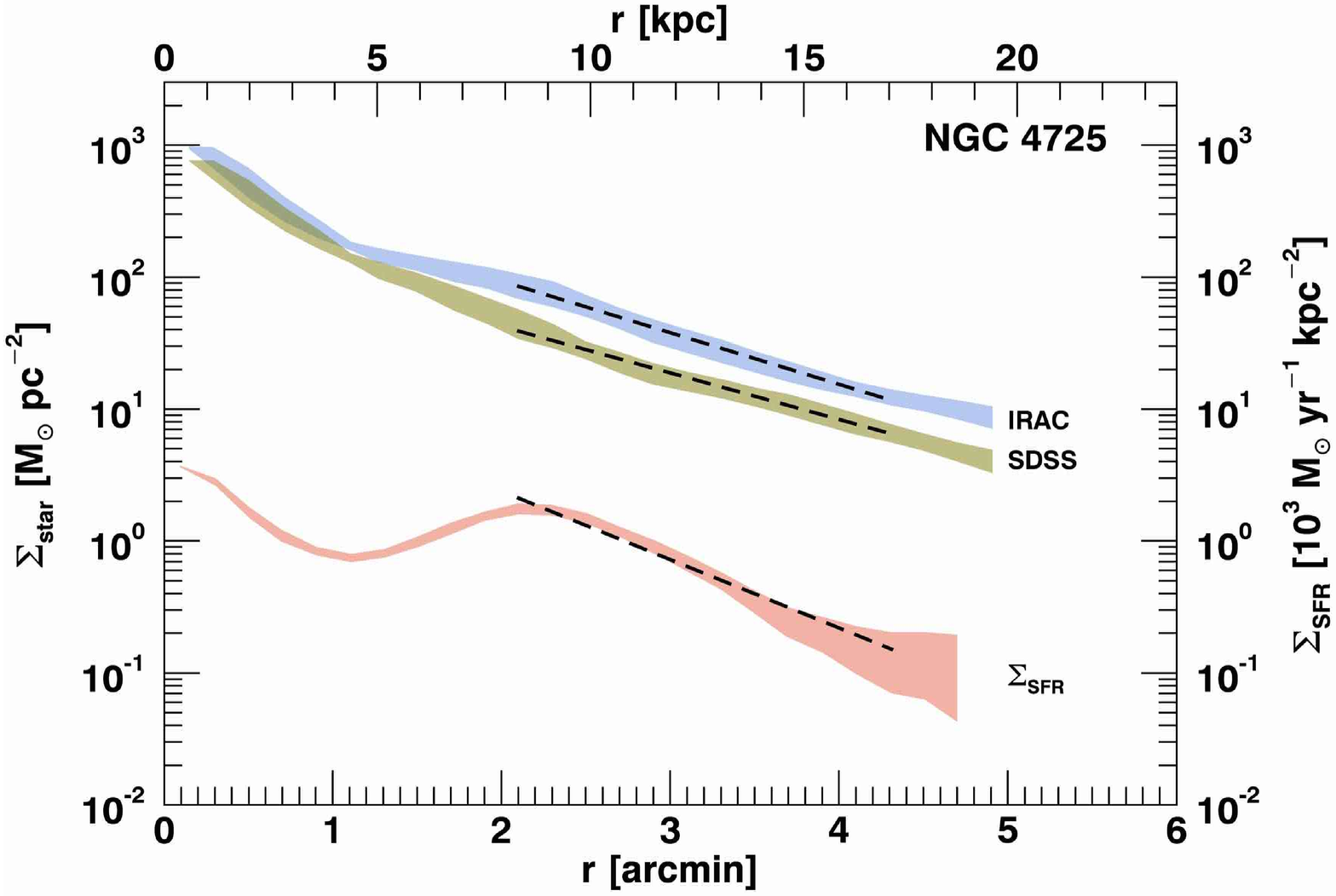}}
\hspace{0.5cm}
%{\includegraphics[width=6.0cm,angle=-90]{n4725-smasses_2.pdf}}\\
{\includegraphics[width=6.0cm,angle=-90]{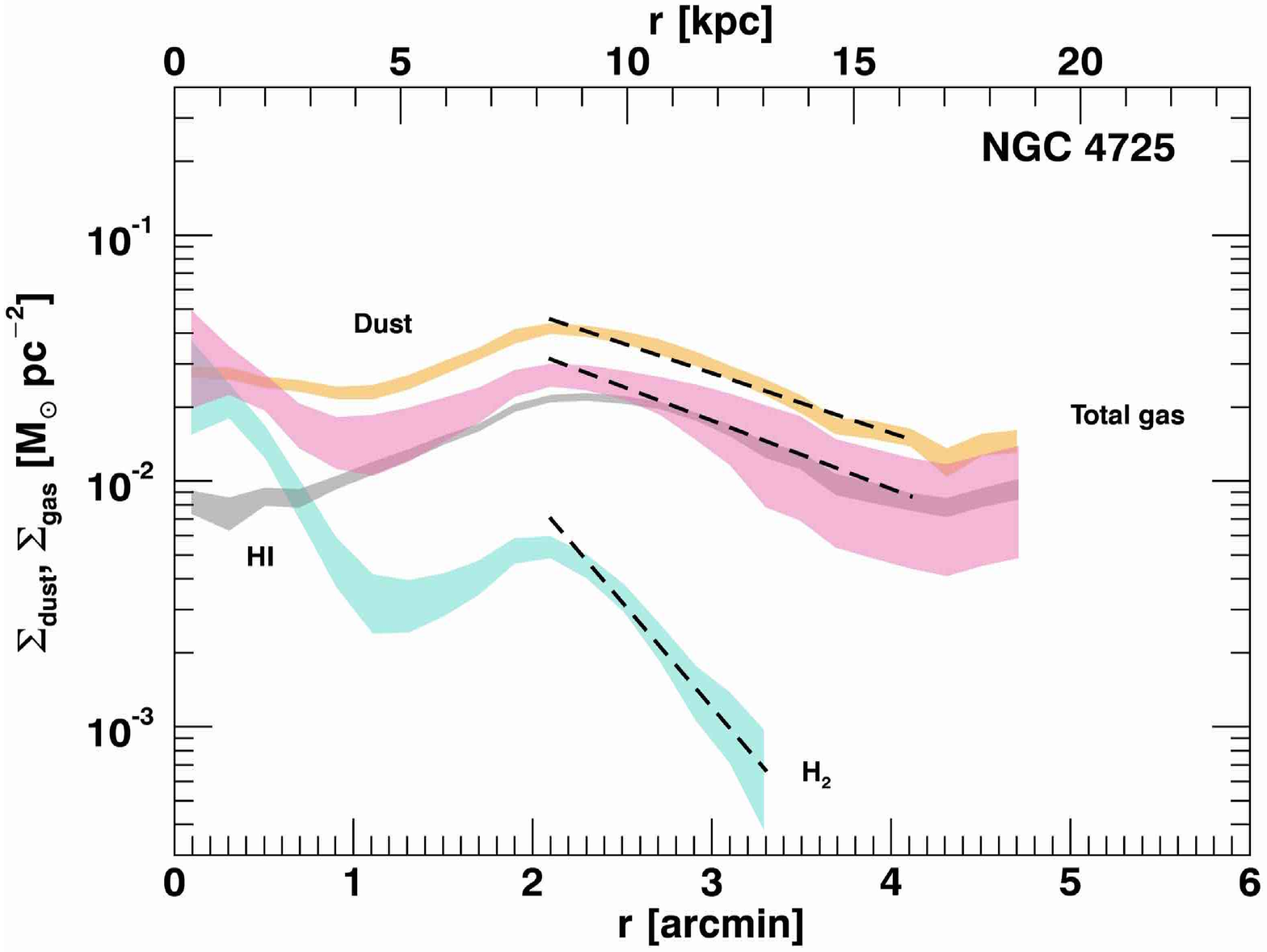}}\\
\caption*{Figure~\ref{fig:masses-prof-app} (continued)}
\end{figure*}

\begin{figure*}[!ht]
\centering
%{\includegraphics[width=6.0cm,angle=-90]{n3521-sstar-ssfr.pdf}}
{\includegraphics[width=6.0cm,angle=-90]{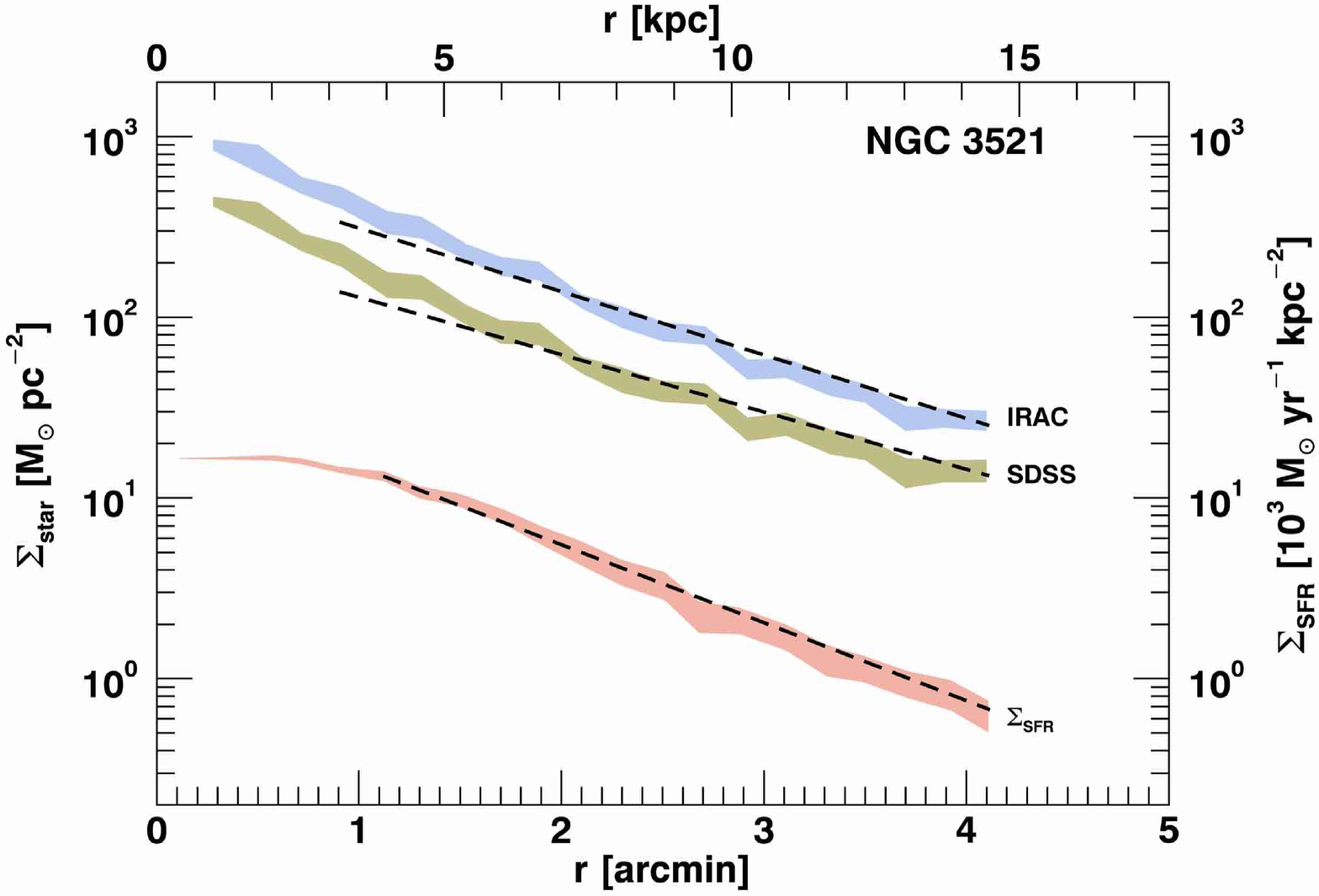}}
\hspace{0.5cm}
%{\includegraphics[width=6.0cm,angle=-90]{n3521-smasses_2.pdf}}\\
{\includegraphics[width=6.0cm,angle=-90]{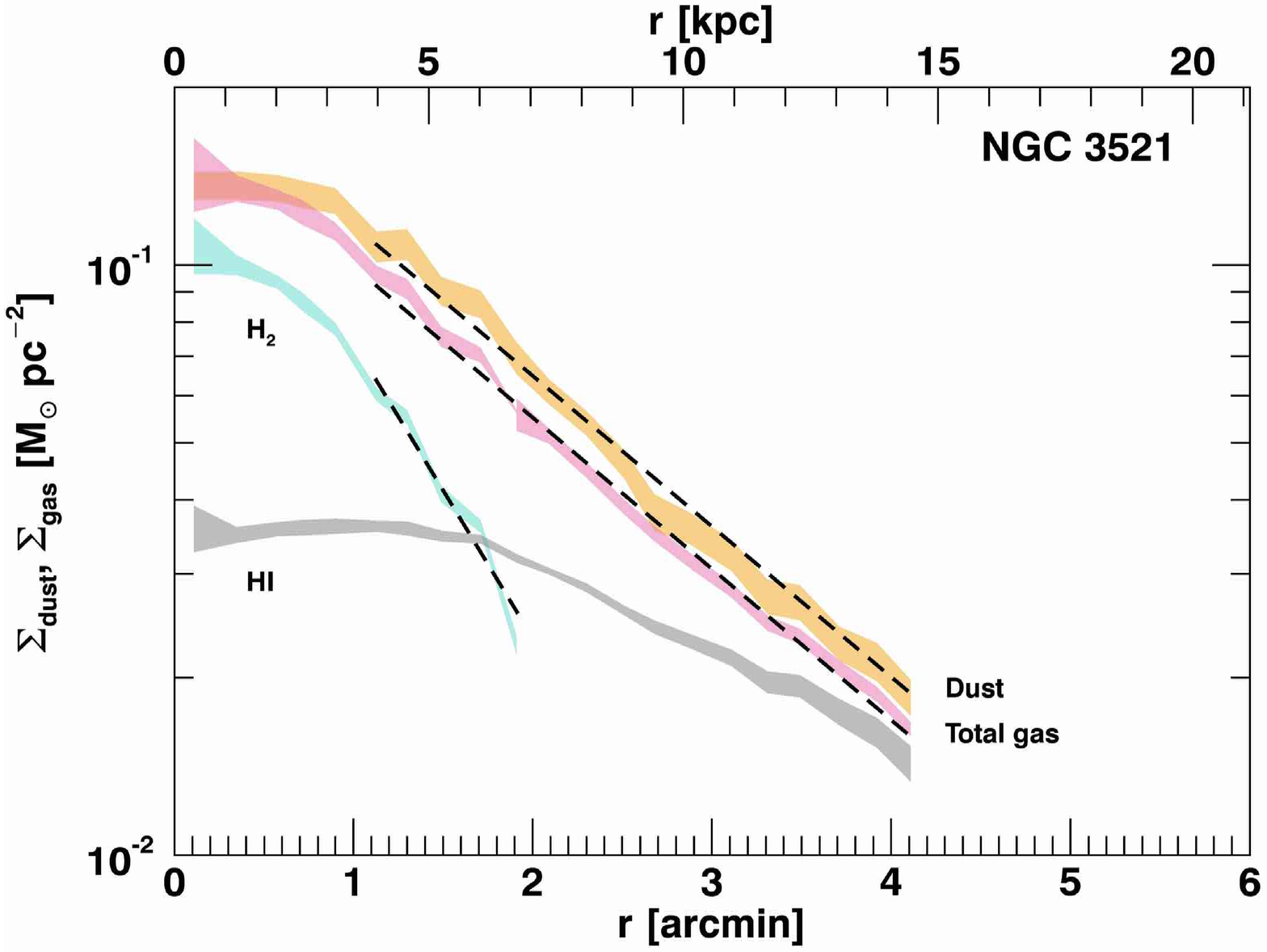}}\\
%{\includegraphics[width=6.0cm,angle=-90]{n4736-m94-sstar-ssfr.pdf}}
{\includegraphics[width=6.0cm,angle=-90]{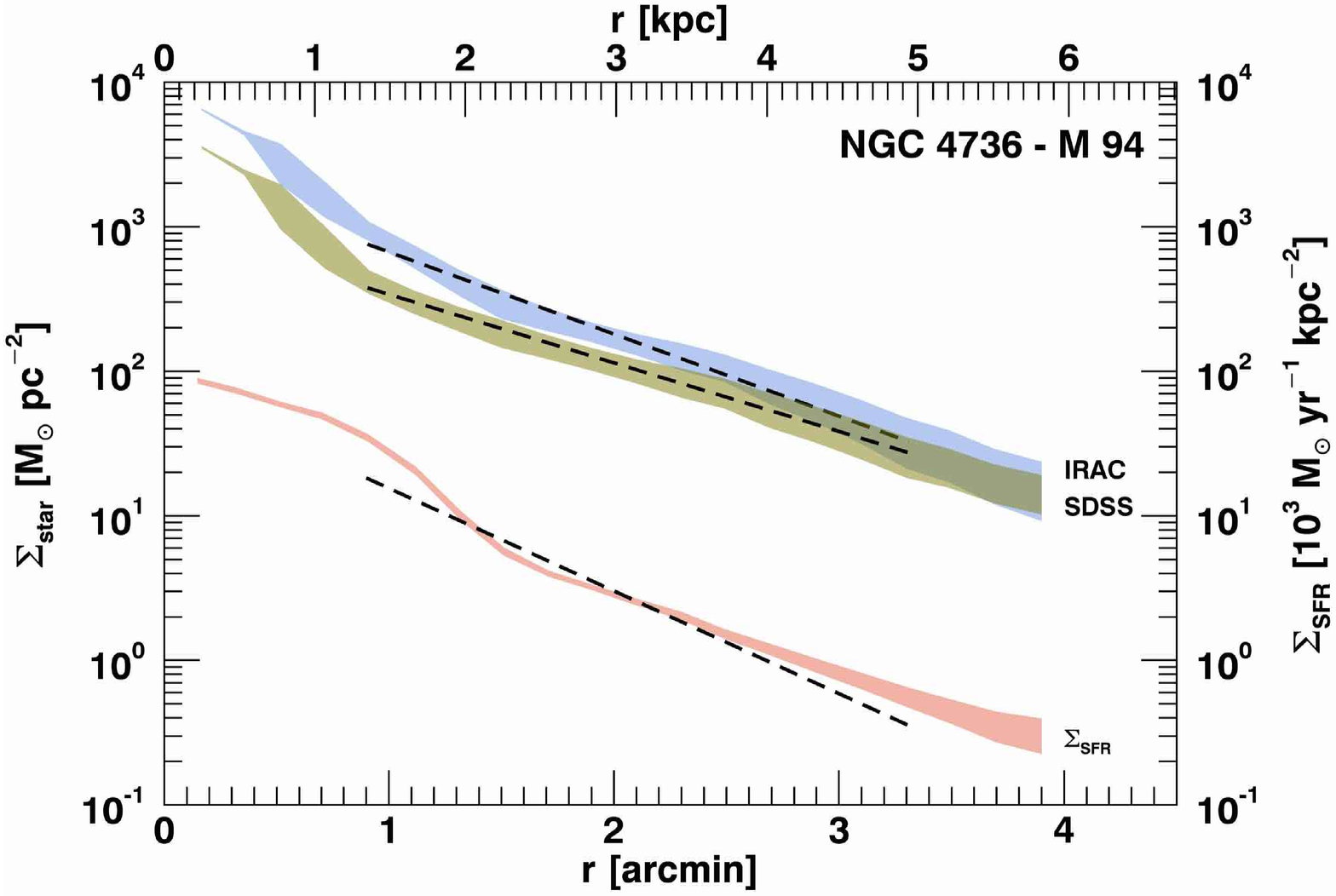}}
\hspace{0.5cm}
%{\includegraphics[width=6.0cm,angle=-90]{n4736-m94-smasses_2.pdf}}\\
{\includegraphics[width=6.0cm,angle=-90]{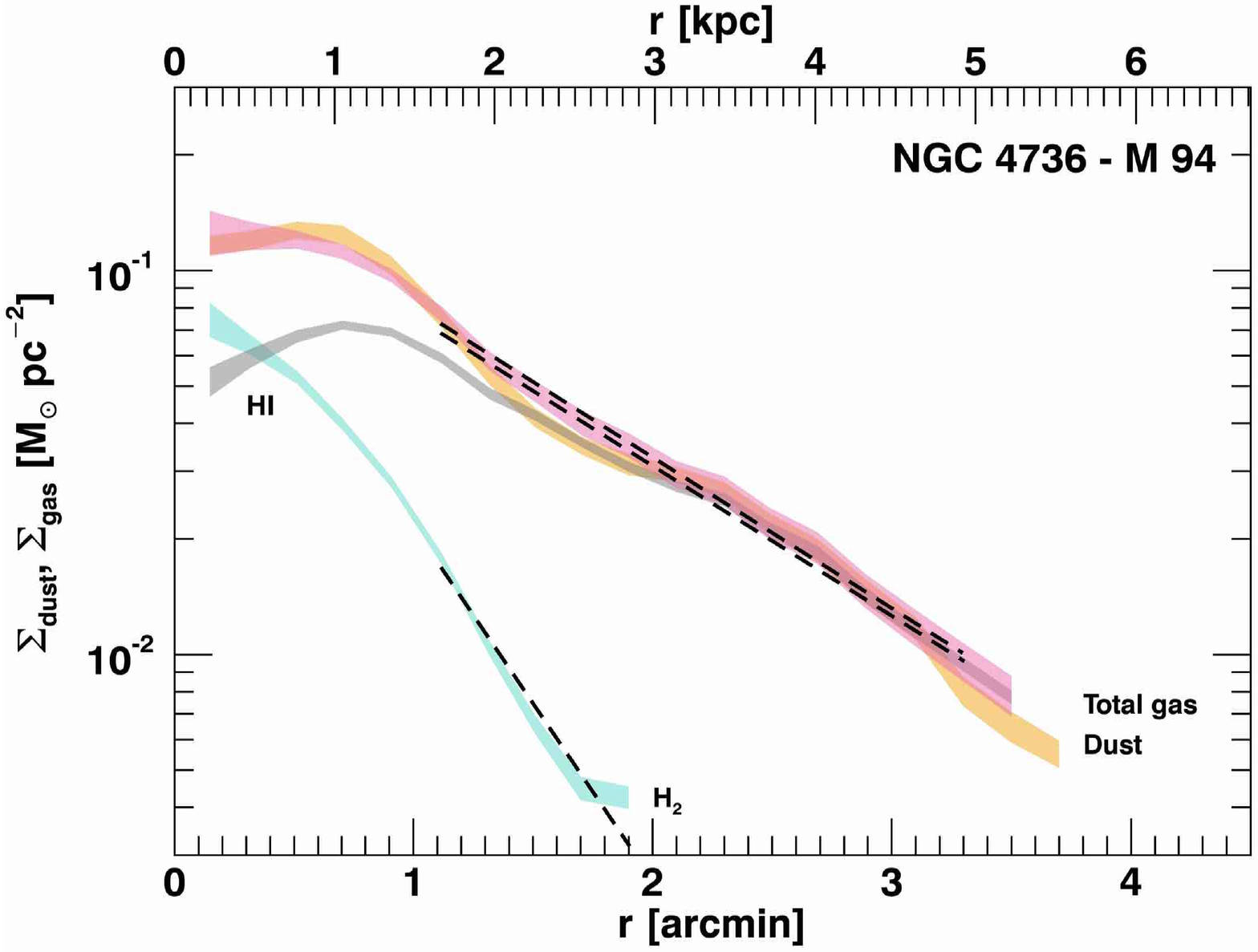}}\\
\caption*{Figure~\ref{fig:masses-prof-app} (continued)}
\end{figure*}

\begin{figure*}
%{\includegraphics[width=4.5cm,angle=-90]{n3031-m81-sl_r25_w02-Umin3.pdf}}
{\includegraphics[width=4.5cm,angle=-90]{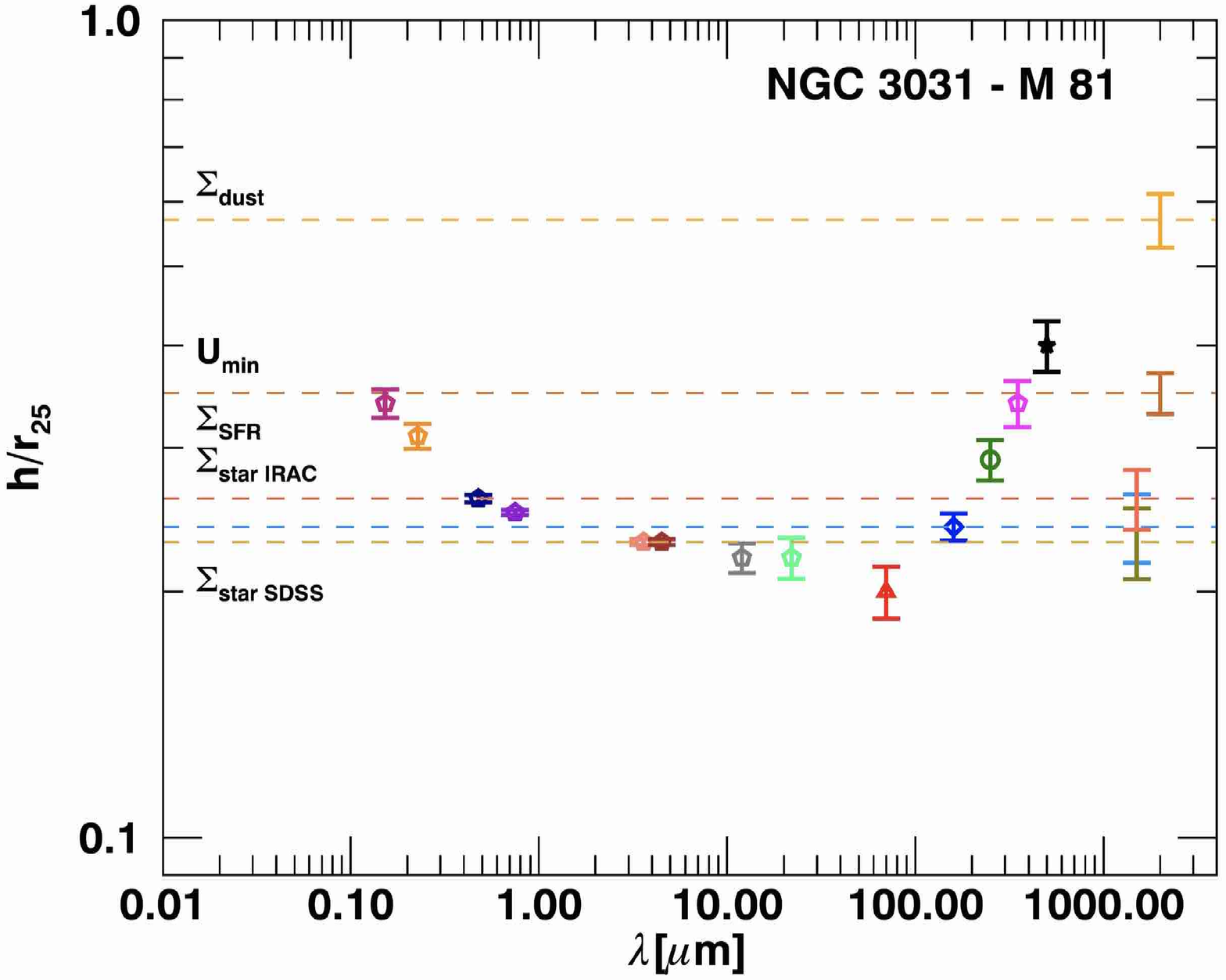}}
\hspace{-0.6cm}
%{\includegraphics[width=4.5cm,angle=-90]{n2403-sl_r25-w02-Umin3.pdf}}
{\includegraphics[width=4.5cm,angle=-90]{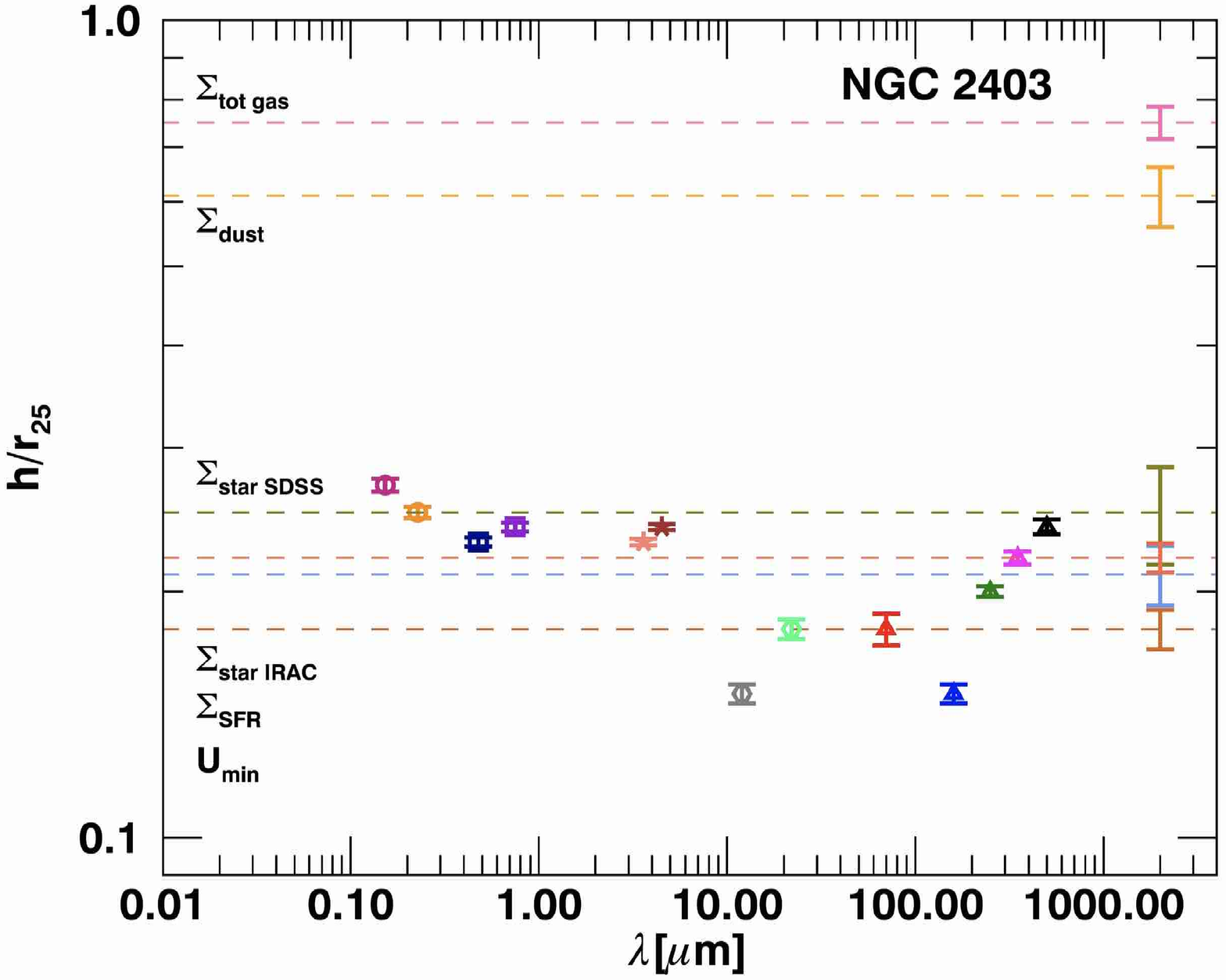}}
\hspace{-0.6cm}
%{\includegraphics[width=4.5cm,angle=-90]{ic342-sl_r25_w02-Umin3.pdf}}\\
{\includegraphics[width=4.5cm,angle=-90]{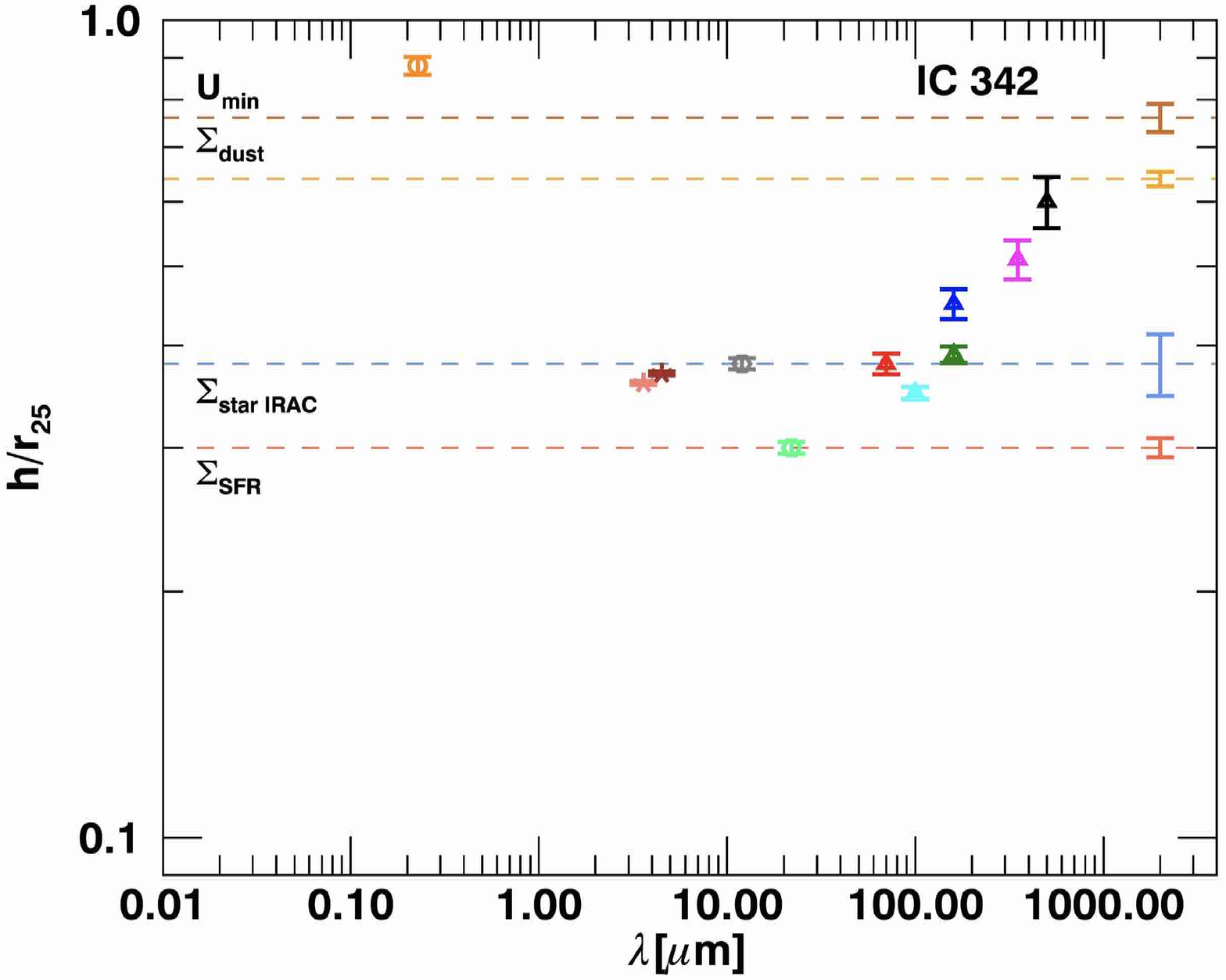}}\\
\hspace{-0.6cm}
%{\includegraphics[width=4.5cm,angle=-90]{n300-sl_r25_w02-Umin3.pdf}}
{\includegraphics[width=4.5cm,angle=-90]{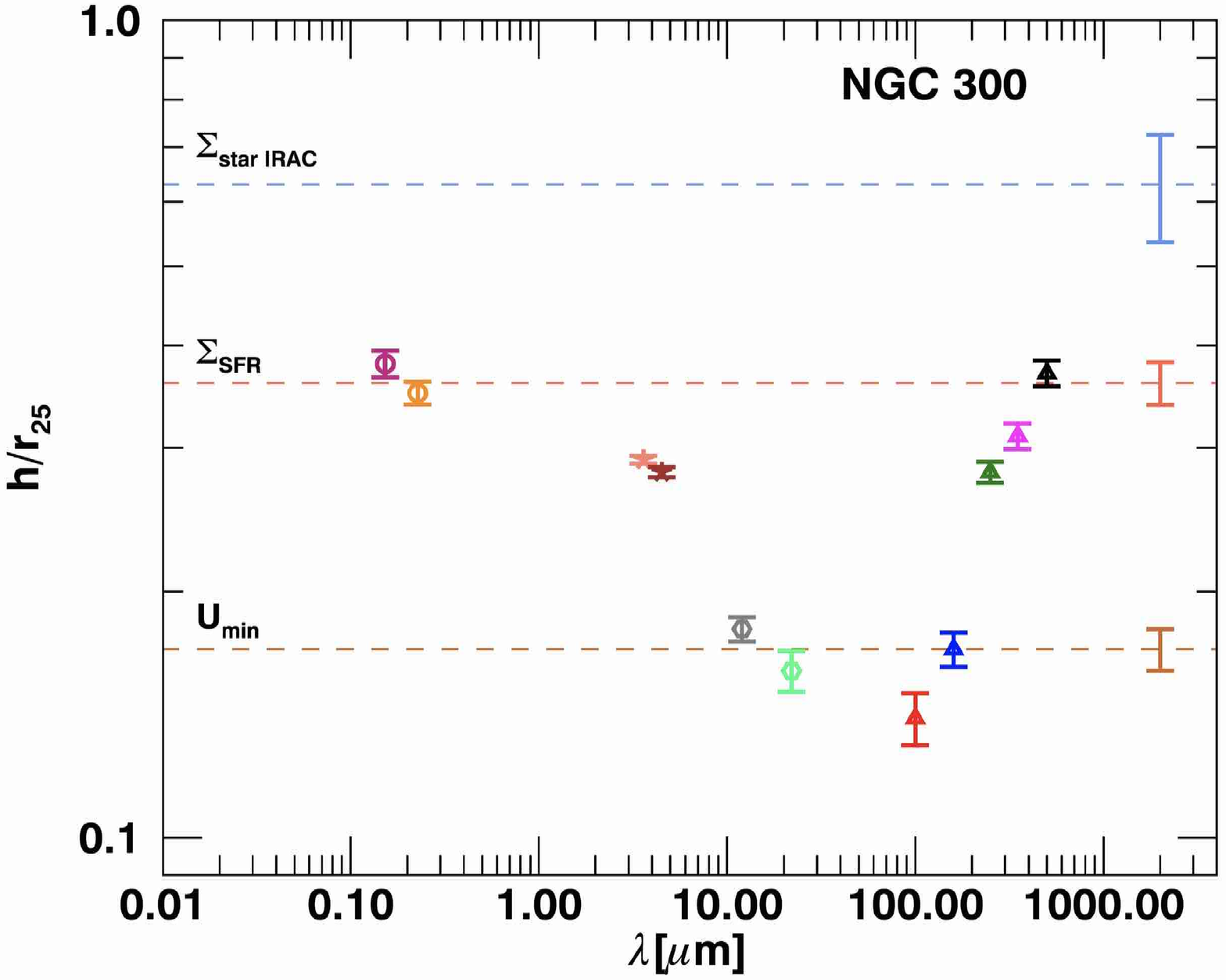}}
\hspace{-0.6cm}
%{\includegraphics[width=4.5cm,angle=-90]{n5194-m51-sl_r25_w02-Umin3.pdf}}
{\includegraphics[width=4.5cm,angle=-90]{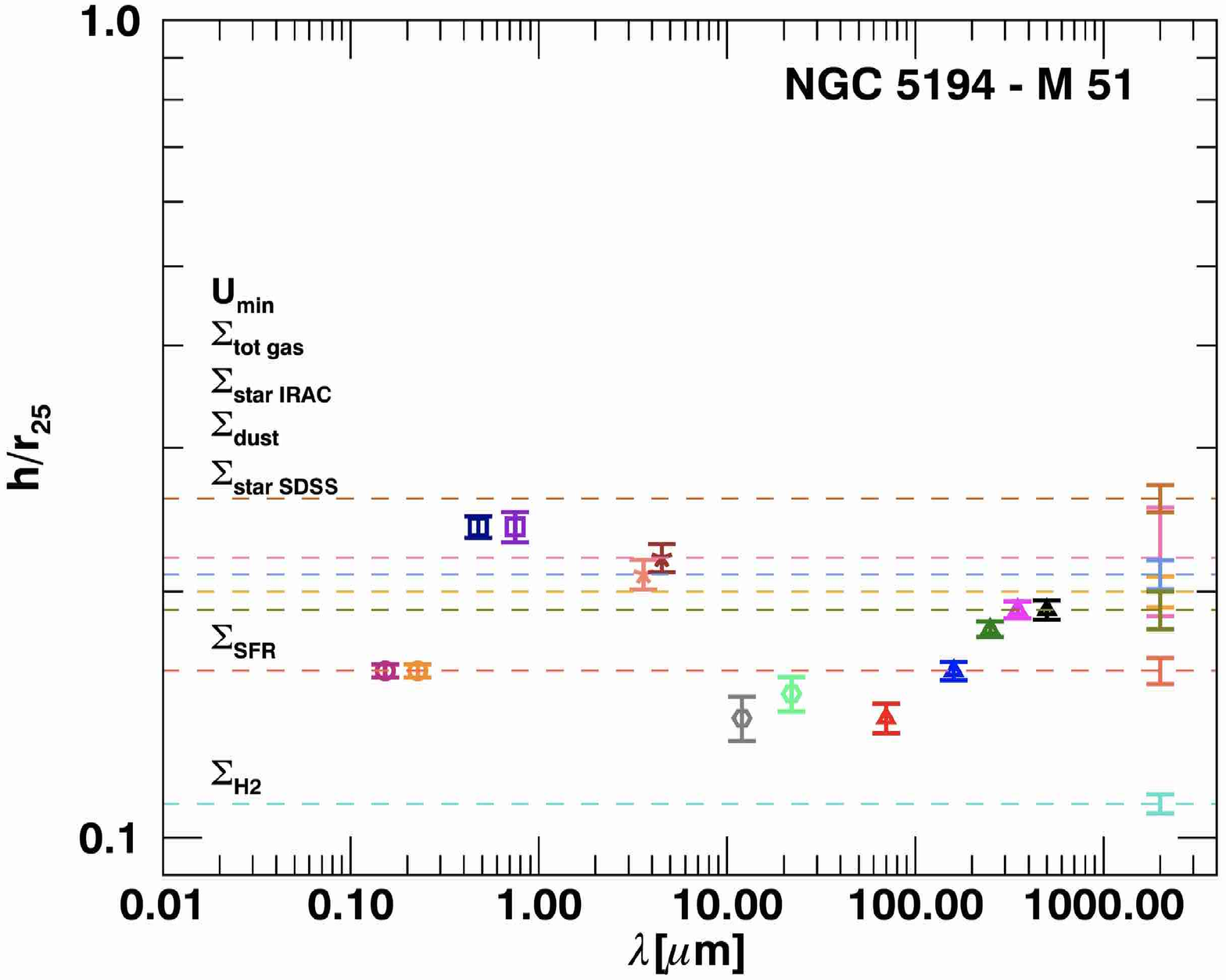}}
\hspace{-0.6cm}
%{\includegraphics[width=4.5cm,angle=-90]{n5236-m83-sl_r25-w02-Umin3.pdf}}\\
{\includegraphics[width=4.5cm,angle=-90]{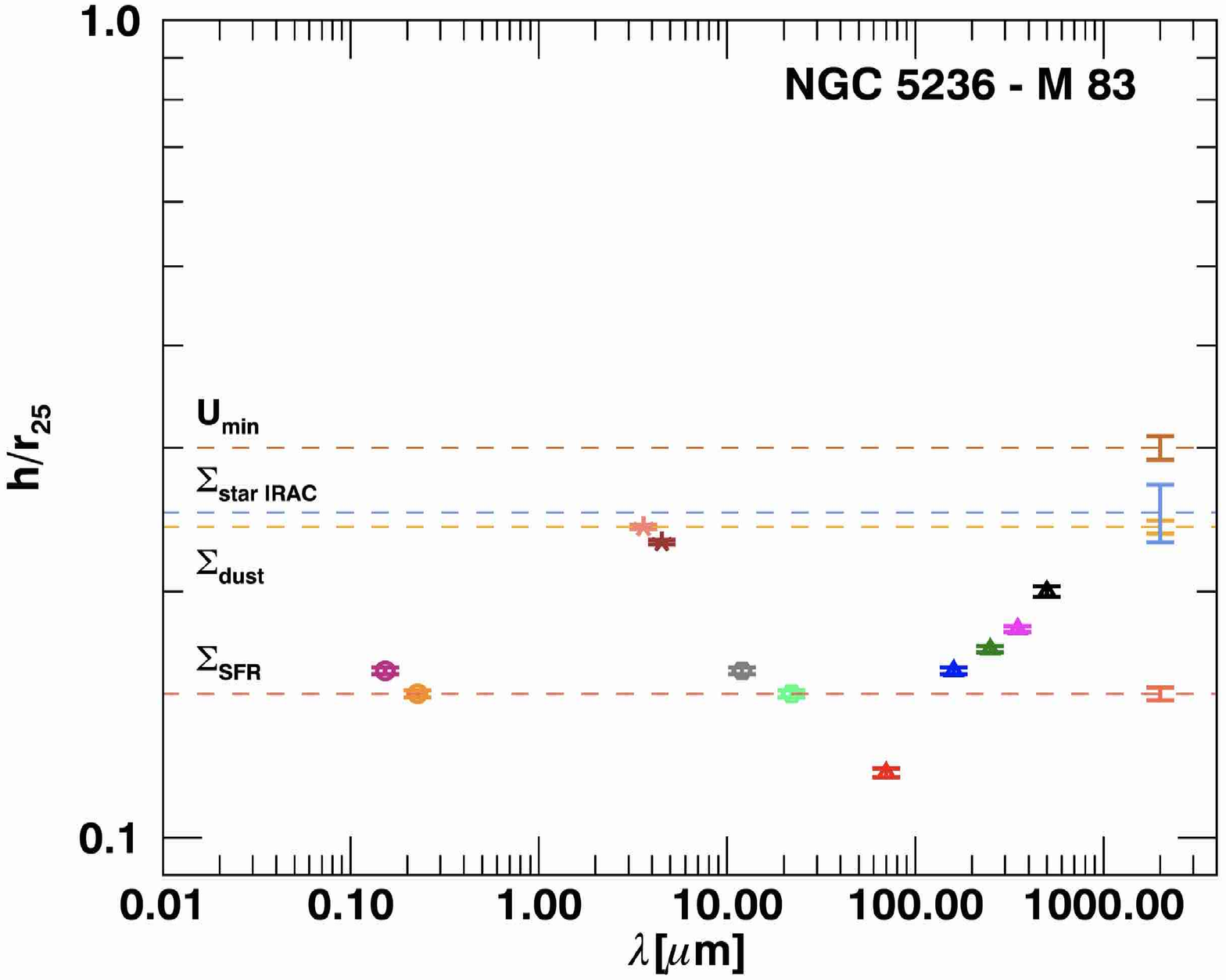}}\\
\hspace{-0.6cm}
%{\includegraphics[width=4.5cm,angle=-90]{n1365-sl_r25_w02-Umin3.pdf}}
{\includegraphics[width=4.5cm,angle=-90]{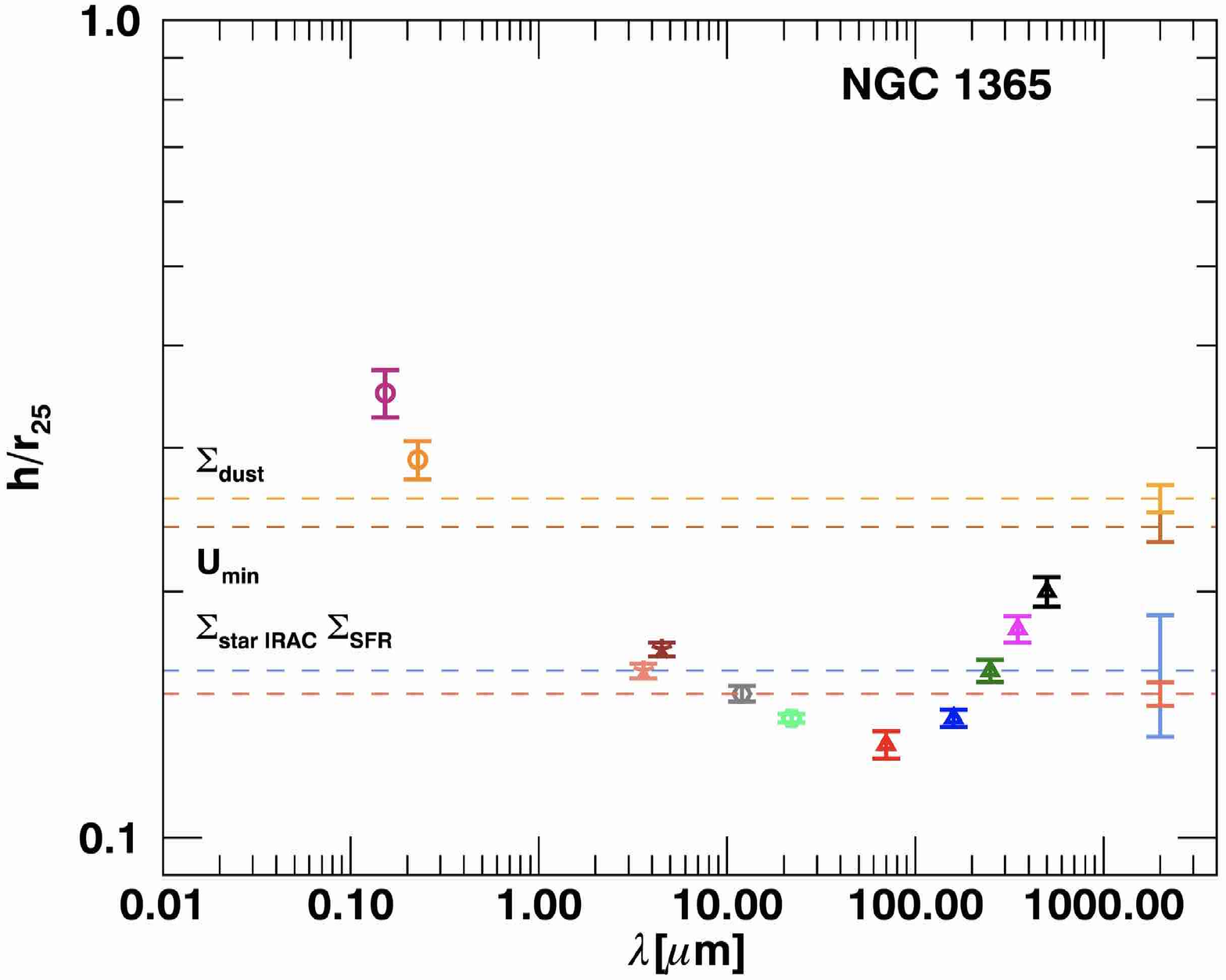}}
\hspace{-0.6cm}
%{\includegraphics[width=4.5cm,angle=-90]{n5055-m63-sl_r25_w02-Umin3.pdf}}
{\includegraphics[width=4.5cm,angle=-90]{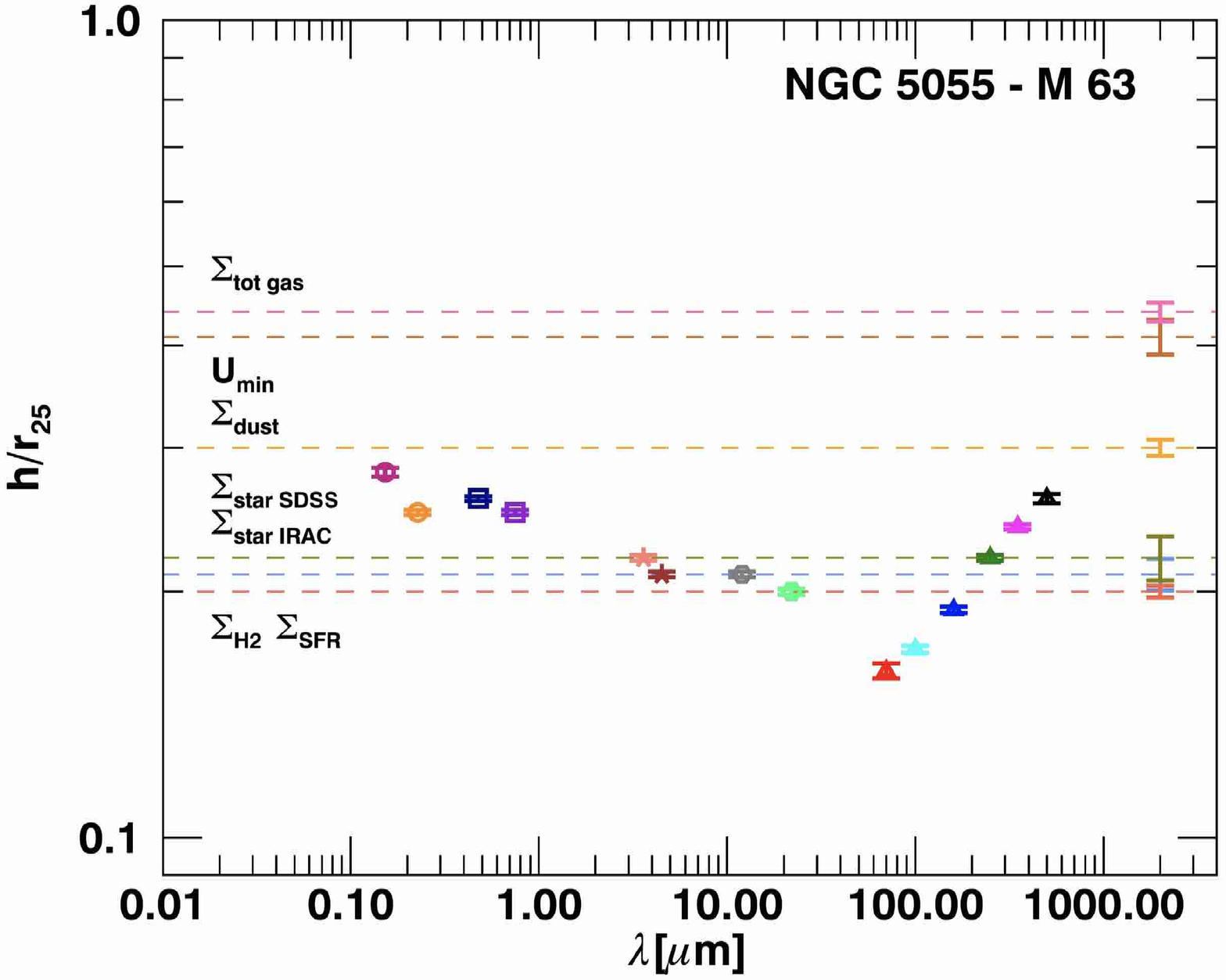}}
\hspace{-0.6cm}
%{\includegraphics[width=4.5cm,angle=-90]{n6946-sl_r25_w02-Umin3.pdf}}\\
{\includegraphics[width=4.5cm,angle=-90]{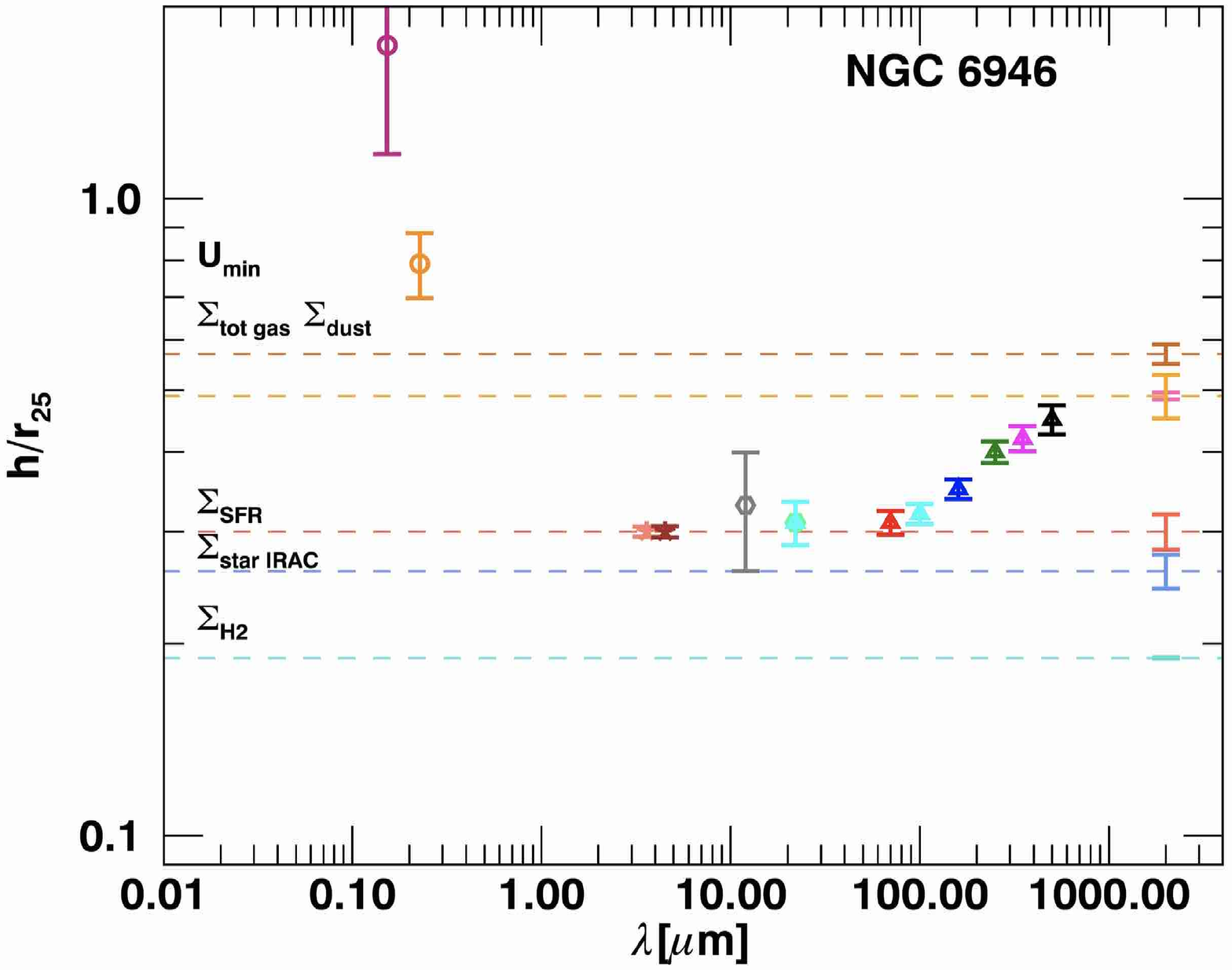}}\\
\hspace{-0.6cm}
%{\includegraphics[width=4.5cm,angle=-90]{n925-sl_r25_w02-Umin3.pdf}}
{\includegraphics[width=4.5cm,angle=-90]{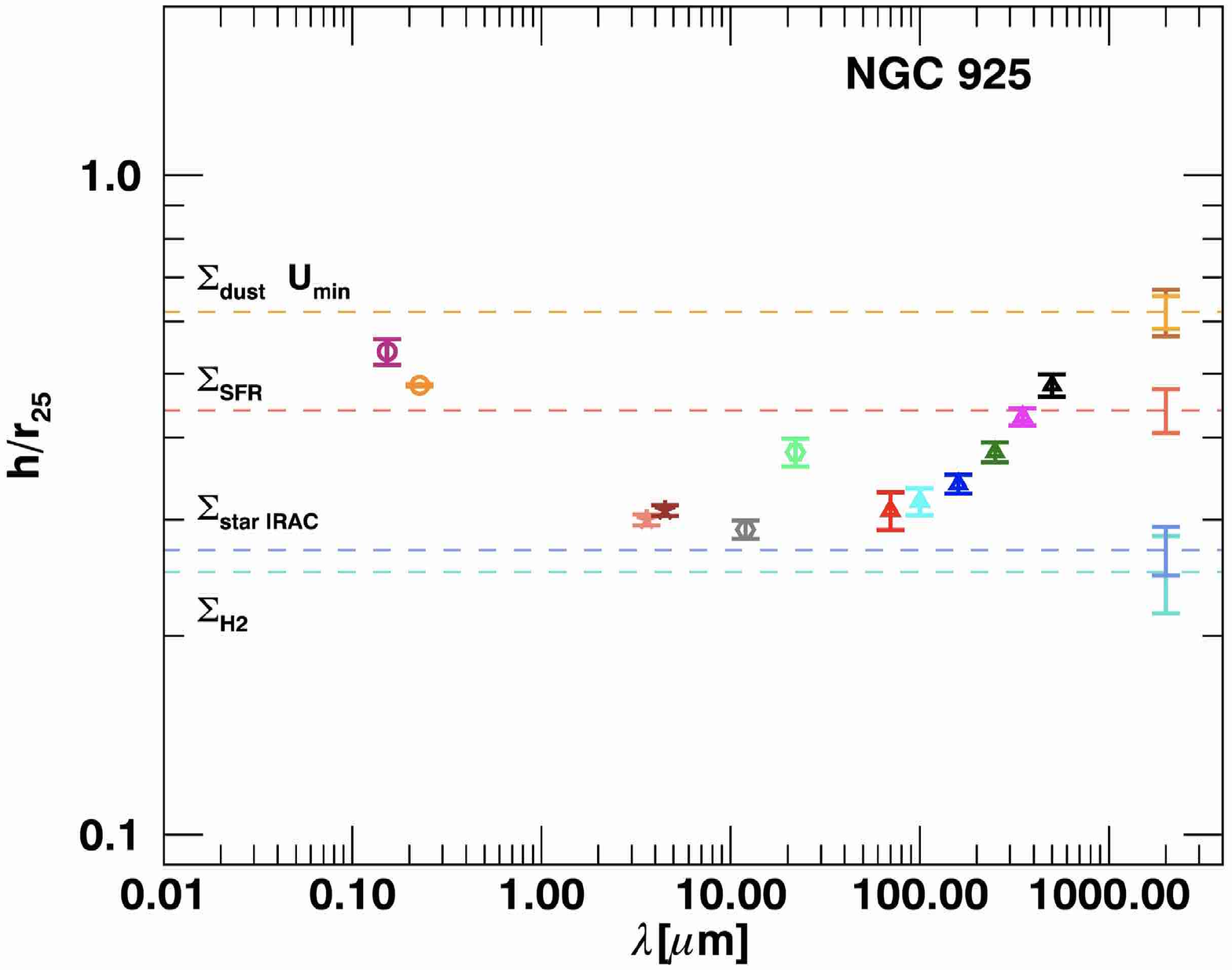}}
\hspace{-0.6cm}
%{\includegraphics[width=4.5cm,angle=-90]{n1097-sl_r25-w02-Umin3.pdf}}
{\includegraphics[width=4.5cm,angle=-90]{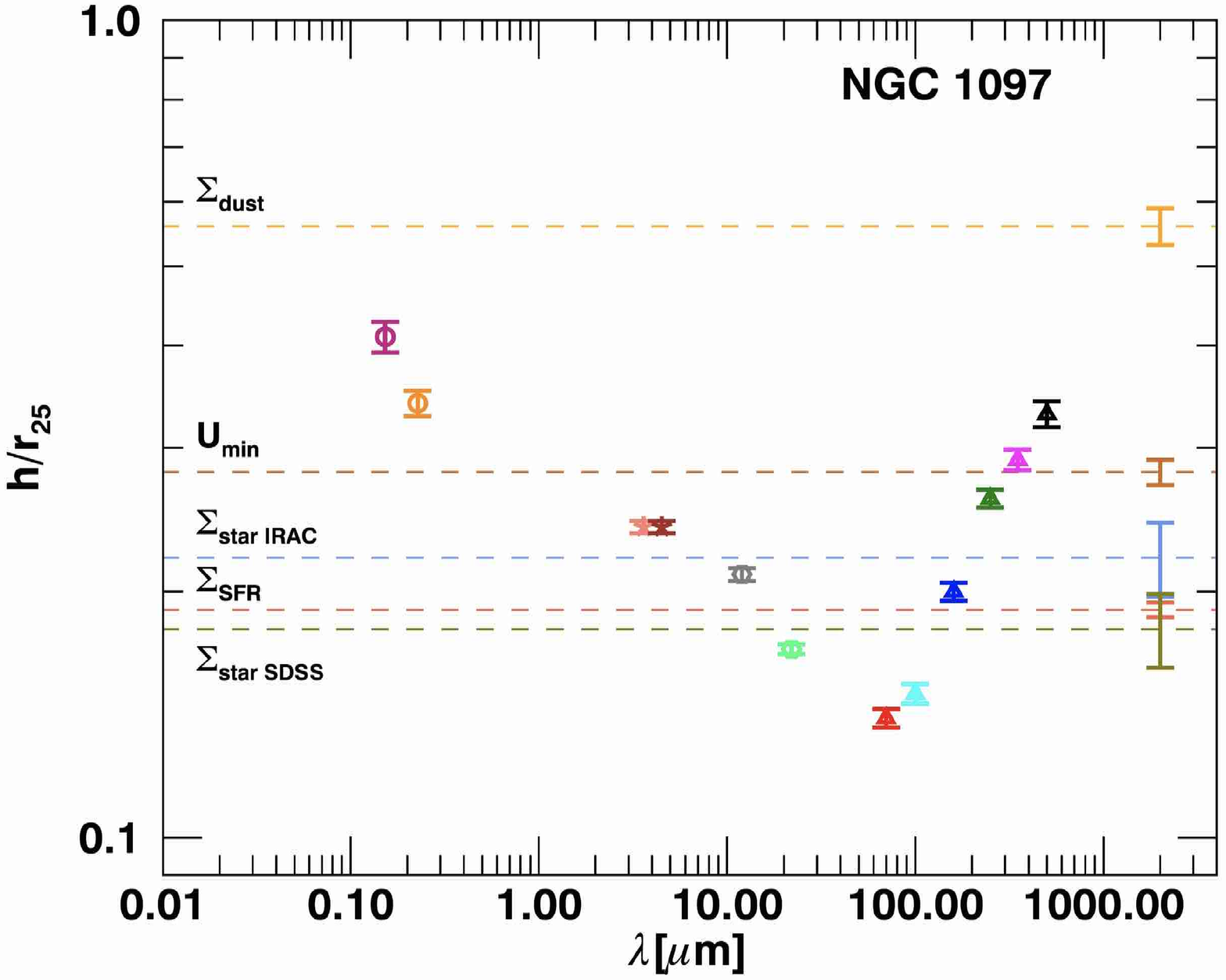}}
\hspace{-0.6cm}
%{\includegraphics[width=4.5cm,angle=-90]{n7793-sl_r25_w02-Umin3.pdf}}\\
{\includegraphics[width=4.5cm,angle=-90]{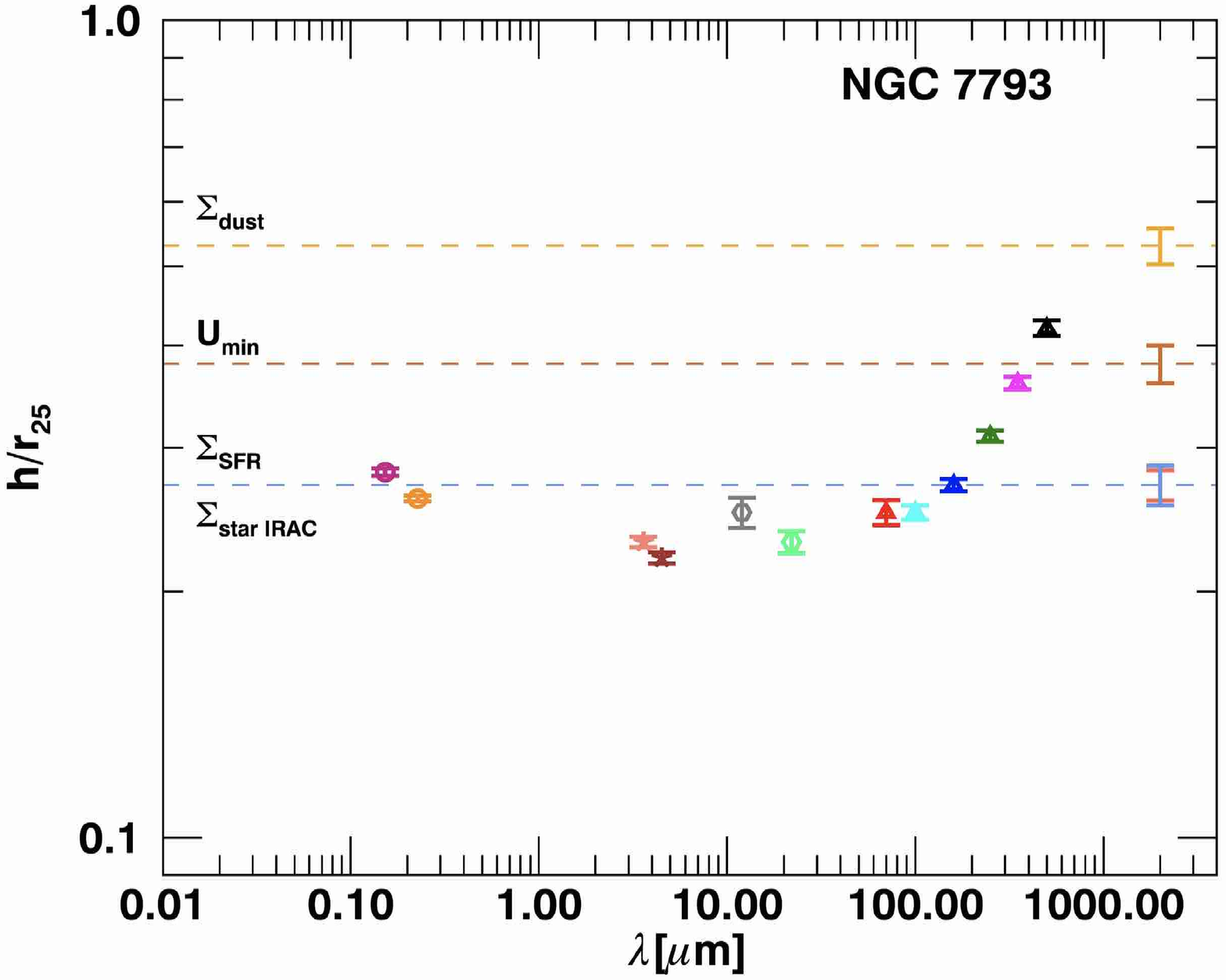}}\\
\hspace{-0.6cm}
%{\includegraphics[width=4.5cm,angle=-90]{n628-m74-sl_r25-w02-Umin3.pdf}}
{\includegraphics[width=4.5cm,angle=-90]{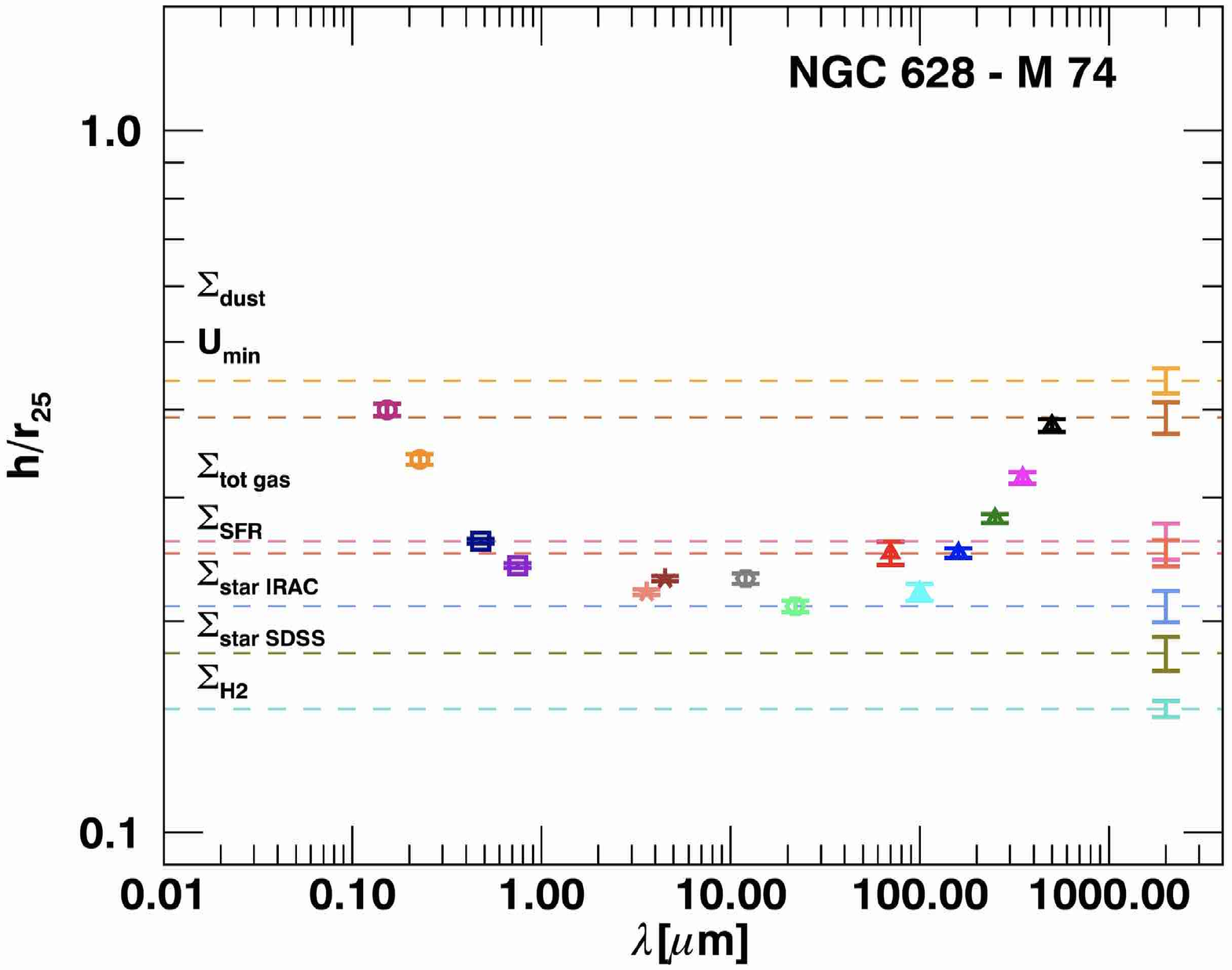}}
\hspace{-0.6cm}
%{\includegraphics[width=4.5cm,angle=-90]{n3621-sl_r25_w02-Umin3.pdf}}
{\includegraphics[width=4.5cm,angle=-90]{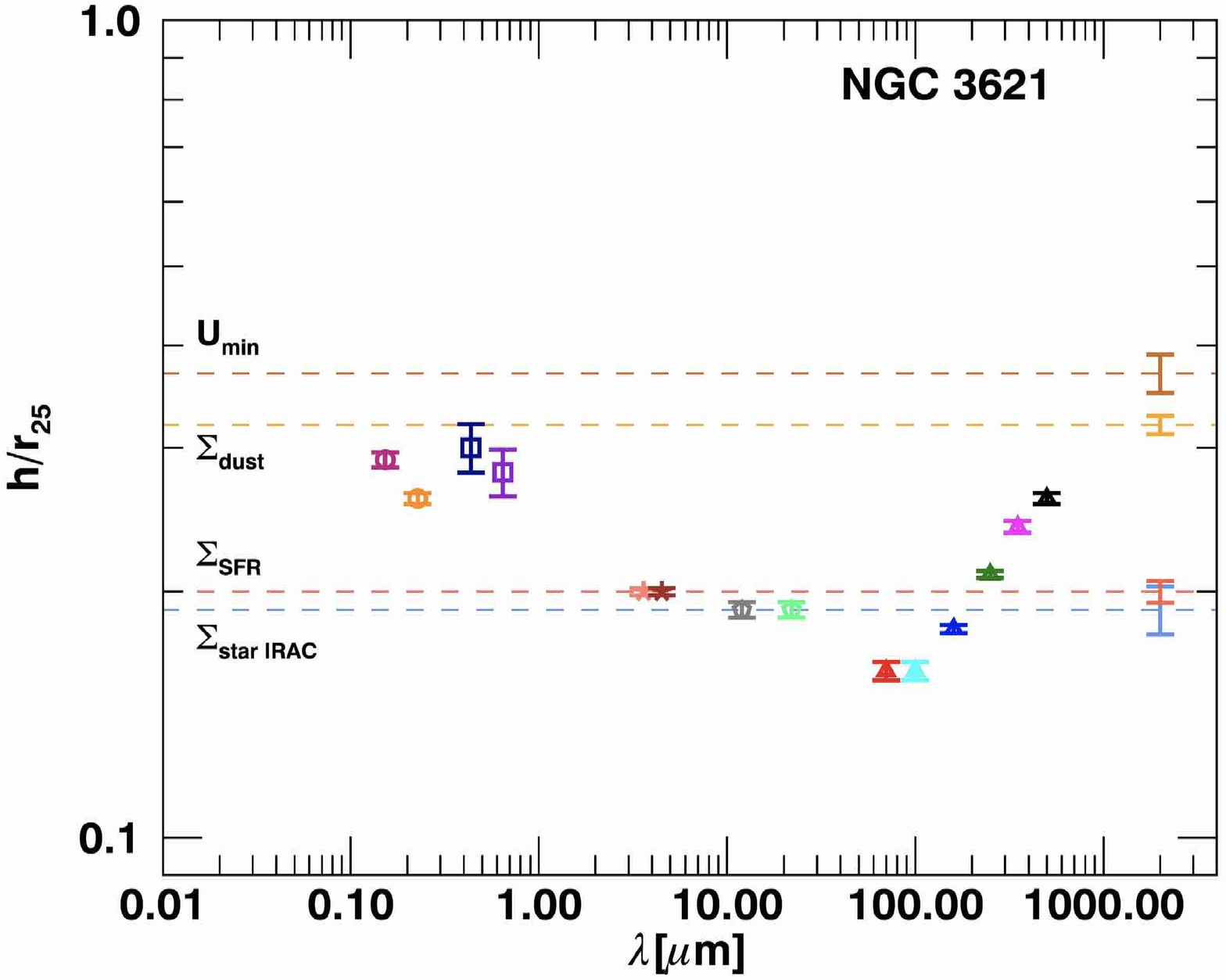}}
\hspace{-0.6cm}
%{\includegraphics[width=4.5cm,angle=-90]{n4725-sl_r25-w02-Umin3.pdf}}
{\includegraphics[width=4.5cm,angle=-90]{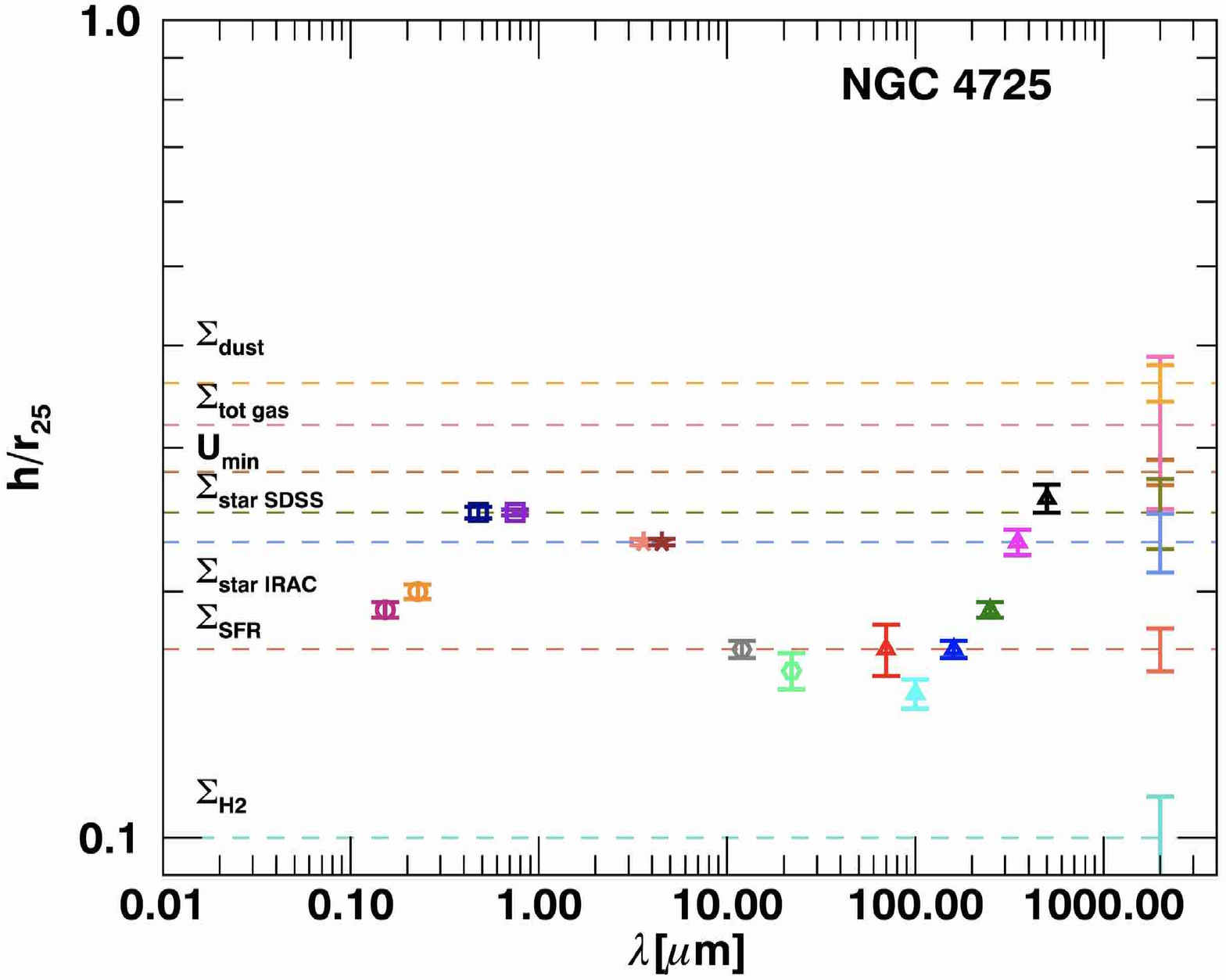}}
\caption{Scale-lengths of surface brightness profiles from UV to sub-mm, normalized with respect $r_{25}$, as a function of wavelength 
for the face-on DustPedia sample, except for the galaxy NGC~5457~(M~101) already shown in Fig.~\ref{fig:sl-wl}. 
Scale-lengths of mass (of dust, gas, and stars) and SFR surface density profiles, scale-length of $U$, and 
the corresponding error bars are drawn as in Fig.~\ref{fig:sl-wl}.}
\label{fig:sl-wl-app}
\end{figure*}

\begin{figure*}
\centering
%{\includegraphics[width=4.5cm,angle=-90]{n3521-sl_r25_w02-Umin3.pdf}}
{\includegraphics[width=4.5cm,angle=-90]{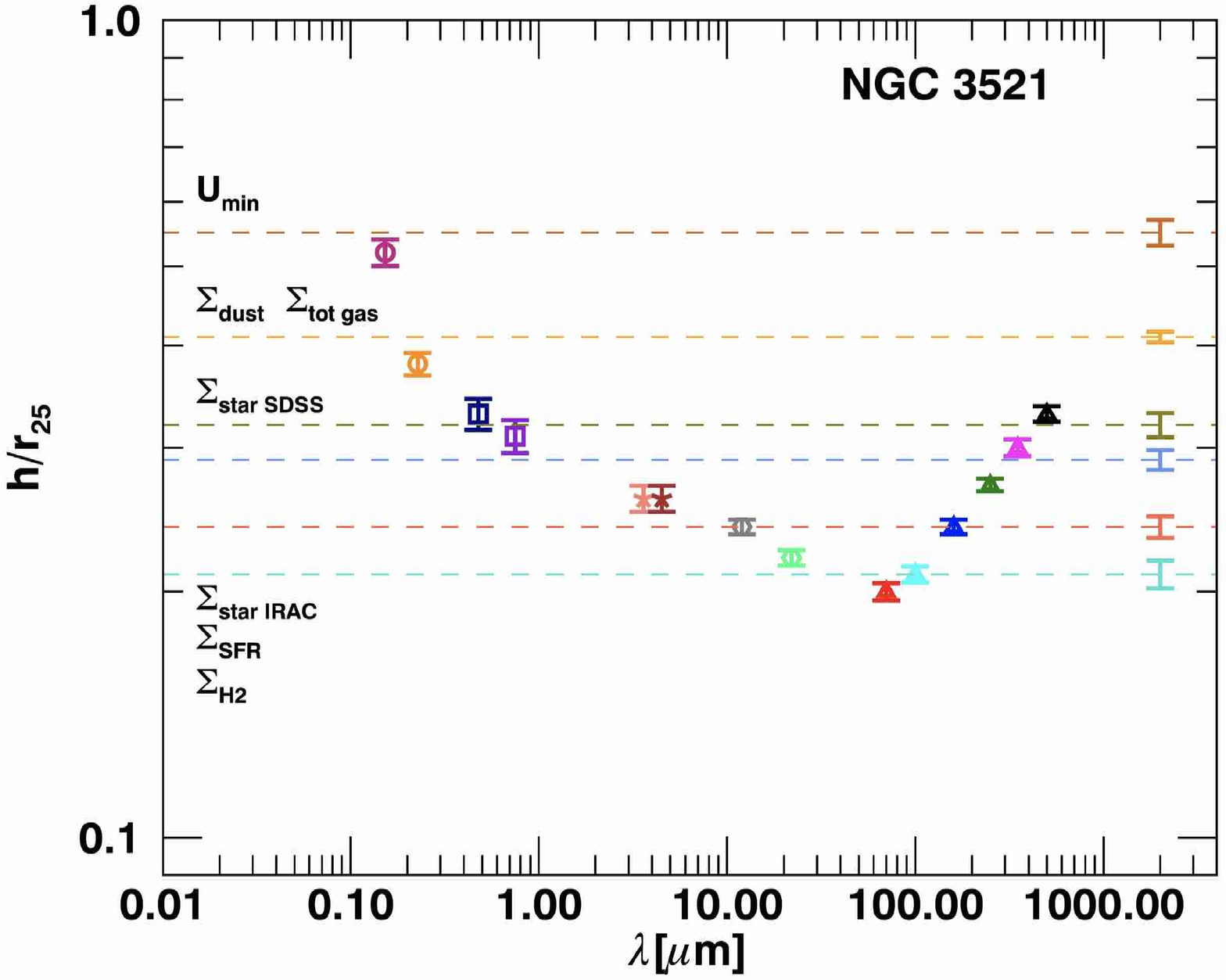}}
%{\includegraphics[width=4.5cm,angle=-90]{n4736-m94-sl_r25-w02-Umin3.pdf}}
{\includegraphics[width=4.5cm,angle=-90]{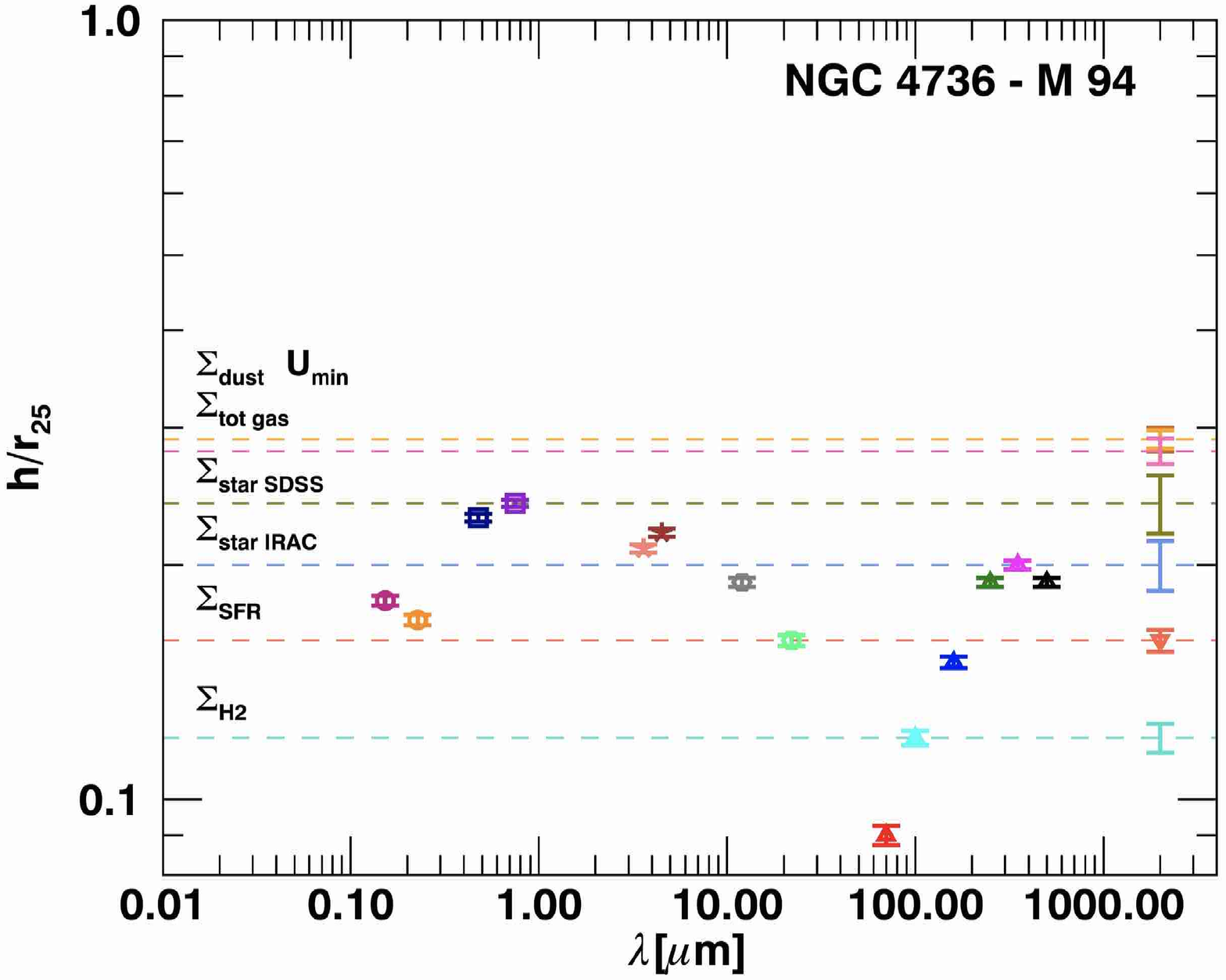}}
\caption*{Figure~\ref{fig:sl-wl-app} (continued)}
\end{figure*}

\end{appendix}

\end{document}